\documentclass[iop]{emulateapj}

\usepackage{amsmath}
\usepackage{color}

\newcommand{\msun}{M$_{\sun}$}
\newcommand{\myr}{M$_\sun$~yr$^{-1}$} 

\newcommand{\ha}{H$\alpha$}
\newcommand{\hb}{H$\beta$}
\newcommand{\nii}{[N{\sc II}]}
\newcommand{\oii}{[O{\sc II}]}
\newcommand{\oiii}{[O{\sc III}]}

\newcommand{\kms}{km\,s$^{-1}$} 
\newcommand{\rchi}{$\tilde{\chi}$$^{2}$}

\newcommand{\sigmaave}{$\sigma_{\text{ave}}$}
\newcommand{\sigmamean}{$\sigma_{\text{mean}}$}
\newcommand{\sigmaoned}{$\sigma_{\text{1D}}$}
\newcommand{\vshear}{$v_{\text{shear}}$}
\newcommand{\s}{$S'_{0.5}$}

\newcommand\pasa{\ref@jnl{PASA}}

\newenvironment{Contfigure}{%
\addtocounter{figure}{-1}%
\begin{figure*}}{%
\end{figure*}}

\shorttitle{Resolved kinematics of z$\sim$1 galaxies}
\shortauthors{Mieda et al. 2016}

\begin{document}
\title{IROCKS: Spatially resolved kinematics of \lowercase{z}$\sim$1 star forming galaxies}
\author{Etsuko Mieda\altaffilmark{1,2,*}, 
        Shelley A. Wright\altaffilmark{3,4}, 
        James E. Larkin\altaffilmark{5}, 
        Lee Armus\altaffilmark{6}, 
        St\'ephanie Juneau\altaffilmark{7}, 
        Samir Salim\altaffilmark{8}, 
        Norman Murray\altaffilmark{9}}

\altaffiltext{1}{Dunlap Institute for Astronomy \& Astrophysics, University of Toronto, 50 St. George street, Toronto, ON, M5S 3H4 Canada}
\altaffiltext{2}{Department of Astronomy \& Astrophysics, University of Toronto, 50 St. George street, Toronto, ON, M5S 3H4 Canada}
\altaffiltext{3}{Department of Physics, University of California, San Diego, 9500 Gilman Drive, La Jolla, CA 92093 USA}
\altaffiltext{4}{Center for Astrophysics and Space Sciences, University of California, San Diego, 9500 Gilman Drive, La Jolla, CA 92093 USA}
\altaffiltext{5}{Department of Physics and Astronomy, University of California, Los Angeles, CA 90095 USA}
\altaffiltext{6}{Spitzer Science Center, California Institute of Technology, 1200 E. California Blvd., Pasadena, CA 91125 USA}
\altaffiltext{7}{CEA-Saclay, DSM/IRFU/SAp, F-91191 Gif-sur-Yvette, France}
\altaffiltext{8}{Department of Astronomy, Indiana University, Bloomington, IN 47404, USA}
\altaffiltext{9}{Canadian Institute for Theoretical Astrophysics, University of Toronto, 60 St. George Street, Toronto, ON M5S 3H8, Canada}
\altaffiltext{*}{Send correspondence to Etsuko Mieda: mieda@dunlap.utoronto.ca}

\begin{abstract}
We present results from IROCKS (Intermediate Redshift OSIRIS Chemo-Kinematic Survey) for sixteen $z \sim 1$ and one $z \sim 1.4$ star-forming galaxies. All galaxies were observed with OSIRIS with the laser guide star adaptive optics system at Keck Observatory. We use rest-frame nebular \ha{} emission lines to trace morphologies and kinematics of ionized gas in star-forming galaxies on sub-kiloparsec physical scales. We observe elevated velocity dispersions ($\sigma$ $\gtrsim$ 50 km s$^{-1}$) seen in $z>1.5$ galaxies persist at $z \sim 1$ in the integrated galaxies. Using an inclined disk model and the ratio of $v/\sigma$, we find that 1/3 of the $z \sim 1$ sample are disk candidates while the other 2/3 of the sample are dominated by merger-like and irregular sources. We find that including extra attenuation towards HII regions derived from stellar population synthesis modeling brings star formation rates (SFR) using \ha{} and stellar population fit into a better agreement. We explore properties of compact \ha{} sub-component, or ''clump,'' at $z \sim 1$ and find that they follow a similar size-luminosity relation as local HII regions but are scaled-up by an order of magnitude with higher luminosities and sizes. Comparing the $z \sim 1$ clumps to other high-redshift clump studies, we determine that the clump SFR surface density evolves as a function of redshift. This may imply clump formation is directly related to the gas fraction in these systems and support disk fragmentation as their formation mechanism since gas fraction scales with redshift.
\end{abstract}


\section{Introduction}\label{intro}
Star formation plays a crucial, but poorly understood, role in regulating the growth and formation of distant galaxies over a wide range of mass scales \citep[e.g.,][]{governato:07, hopkins:12, wurster:13, agertz:13, muratov:15}. Characterizing star formation at high redshift is challenging since both high signal-to-noise ratio (SNR) and high-angular resolution observations are required to resolve the kinematics, chemical abundances, and outflow and/or shocks in individual star forming regions. The latest observations of distant galaxies (z $\gtrsim$ 1) have shown high velocity dispersions in their star forming regions, suggesting that there are strong energetics and large turbulences present, which may be driven by radiation pressure \citep[e.g.,][]{murray:10}, cold flow \citep[e.g.,][]{genel:10}, and/or supernovae \citep[e.g.,][]{joung:06}. While we are rapidly compiling values for the global parameters (e.g., luminosity, color, star formation rate (SFR), gas/dust content, and stellar mass) of high redshift galaxies, there is still a gap in our knowledge of processes that regulate galaxy growth and evolution even at modest redshifts of $z \sim 1$ to 2. 

Ground-breaking observations with integral field spectrographs (IFSs) coupled to adaptive optics (AO) (e.g., OSIRIS at Keck and SINFONI at VLT) have probed the dynamical processes of individual high-redshift (z \textgreater 1.5) star-forming galaxies on kiloparsec scales \citep[see review by][]{glazebrook:13}. IFS data of high-redshift systems provide valuable insight into the assembly and star formation properties of early systems. IFS studies at z $\sim$ 2 have shown mounting evidence that a large fraction (between 1/3 and 1/2) of high-z star-forming galaxies (\textgreater 10 \myr) are in rotating disk systems \citep{forster:11a, forster:11}, while the rest are irregular or interacting. In general, velocity dispersions seen in the gas of early disk candidates are much higher than expected and may imply strong feedback in the form of energy being injected into the interstellar gas \citep[e.g.,][]{newman:14}. However, IFS+AO observations of $z \sim 1.5$ galaxies have found systems with lower SFR which are consistent with rotationally stable disks with lower intrinsic velocity dispersions \citep[e.g.,][]{wright:09, wisnioski:11}, which may indicate an evolution in the settling of disks. Seeing-limited slit-based spectroscopic observations have also shown that most galaxies have large V/$\sigma$ values at $z \sim 1$ while only a small fraction of galaxies have high V/$\sigma$ values at $z \sim 2$, implying a rapid evolution of disks in this 5 Gyr period \citep{kassin:12}.

IFS observations of $z \gtrsim 1.5$ galaxies have shown that the most luminous star-forming galaxies have turbulent velocity dispersions, and the sites of star formation occur in large ($\gtrsim$ 1kpc) ''clumps'' or concentrated''complexes'' of star formation. These star forming ''clumps'' are embedded in the rotation curves of these turbulent disks at high-redshift and share similar velocity dispersions, and the observations suggest that clumps form at sites of disk instability \citep{genzel:11, newmans:13}. WFC3 slitless grism observations have measured the properties of $z\sim1$ star-forming regions, which have large \ha{} sizes and fluxes, indicating a large variation in the \ha{} sizes within the population \citep[half light radii of 1-15 kpc;][]{nelson:12, nelson:13, wuyts:12, lang:14}. Scaling relations of high-redshift star forming clumps relating \ha{} size, velocity dispersion, luminosity, and mass have been explored \citep{wisnioski:11,wisnioski:12,genzel:11, livermore:12, livermore:15}. Each of these studies are able to relate the size-luminosity and size-velocity dispersion in these systems, and some find that there are luminosity offsets of high-redshift clumps compared to local HII regions \citep{jones:10, livermore:12}. Other IFS studies find that high-redshift clumps follow a similar trend and power law to that of local HII regions \citep{wisnioski:12}, and investigations of whether there is redshift evolution between these samples have been explored and have contrary implications \citep{livermore:15}. There are various observational selection differences between all IFS samples, from lensed systems to non-lensed to varying redshift and mass bins, but the number of high-redshift IFS observations is limited.

Thus far, IFS plus AO observations of high-redshift galaxies have been limited to z$\sim$1.5 and z$\sim$2 where the prominent \ha{} emission line is redshifted into the H and K bands where AO performance is better and instruments are more sensitive. In 2010, a new, powerful, center-launching laser guide star (LGS) AO was installed on Keck-I \citep{chin:10, chin:12}. In 2012, our team installed a new grating on OSIRIS and increased its sensitivity by a factor of 1.5 to 2 \citep{mieda:14}. With these factors combined, we are now capable of observing large samples of ''normal'' $z \sim 1$ galaxies with an IFS + AO. Selection of targets is still limited by available tip-tilt (TT) stars, but this criterion does not bias our sample selection. In this paper, we present the first result of the Intermediate Redshift OSIRIS Chemo-Kinematic Survey (IROCKS), an AO enhanced IFS study of $z \sim 1$ star-forming galaxies using OSIRIS at the Keck-I telescope. We focus on the kinematics and morphological properties of $z\sim1$ galaxies traced by \ha{} emission.

This paper is organized as follows. In Section \ref{sec_obs}, we detail our sample selection, OSIRIS observations, and data reduction. We present morphology, kinematics, and disk fitting results in Sections \ref{morph}, \ref{sec_kine_map}, and \ref{kine_model}, respectively. Gas and dynamical mass estimates are described in Section \ref{mass}. In Section \ref{sec_clump}, we introduce our definition of clumps and describe their properties. Finally, we summarize our survey in Section \ref{dis}. Throughout this paper, we adopt the concordance cosmology with $\Omega_m$ = 0.306, $\Omega_\Lambda$ = 0.692, and $H_0$ = 67.8 km s$^{-1}$ Mpc$^{-1}$ \citep{planck:14}, where 1 arcsecond is 8.2 kpc at z = 1. For comparisons with other cosmologies used in other IFS high redshift galaxy studies, comoving distances are different from the Planck cosmology by \textless 3\%.

\section{Observation and data reduction}\label{sec_obs}
\subsection{Sample Selection}\label{selection}
We select $z \sim 1$ galaxies in several well-studied fields using four surveys: the Team Keck Treasury Redshift Survey \citep[TKRS;][]{wirth:04} in the Great Observatories Origins Deep Survey (GOODS)-North; the European Southern Observatory-GOODS \citep[ESO-GOODS;][]{vanzella:08} spectroscopic program in GOODS-South; DEEP2 \citep[and references therein]{newman:13} (RA = 02h, 14h, and 23h); and the Cosmic Assembly Near-infrared Deep Extragalactic Legacy Survey (CANDELS)-Ultra Deep Survey \citep[UDS;][]{galametz:13}. We target rest-frame \ha{} and \nii{} emission lines in J-band, which corresponds to a redshift range of 0.8 to 1.1. We also target a few $z \sim 1.5$ galaxies, whose \ha{} lines fall in H-band (1.2 \textless z \textless 1.8). Objects are ranked in observational priority based on following criteria: 1) the galaxy must have an accurate spectroscopic redshift; 2) the target's shifted \ha{} line must be located in regions of the J/H-band free from strong OH sky emission lines; 3) filter and atmosphere transmissions need to be high ($\gtrsim$0.7); 4) there must be a nearby TT star with an R-band magnitude below 17 mag within 50'' from the galaxy; and 5) a higher inferred \ha{} flux, and hence SFR, is preferred. To estimate SFR, we infer \ha{} spectroscopic flux from previous \hb{} or \oii{} detections when available. Assuming Case-B recombination, intrinsic flux ratios are estimated as \ha/\hb = 2.8, and \ha/\oii{} = 1.77 \citep{osterbrock:89, mouhcine:05}, not including extinction. Using these relations, we infer the \ha{} fluxes for objects in the TKRS (\hb{} for $z \sim 1$ and \oii{} for $z \sim 1$.5 sources). For ESO-GOODS and DEEP2 targets, information on line fluxes are not available, and we estimate \oii{} line fluxes using their rest-frame B-band magnitude \citep{mostek:12} and then convert to SFR. Objects in the UDS field also do not have line flux information available, and we use their K-band magnitudes, which has been shown to correlate with SFR \citep{reddy:05, erb:06b}, to rank those objects. Lastly, we prioritized sources that have complementary Hubble Space Telescope (HST) imaging. It provides accurate offsets between the galaxies and their TT stars, can aid in morphological comparisons between UV and optical line emissions, and helps choose galaxies that are not too diffuse nor unresolved to increase expected signal detection. Two fields, DEEP2 2d and 23d, are still targeted even though they do not have HST imaging available because they contain key spectroscopic information along with seeing-limited imaging from the Canada France Hawaii Telescope and Sloan Digital Sky Survey.

In total, we observed twenty-five $z\sim1$ and two $z\sim1.5$ systems and successfully detected sixteen $z\sim1$ and one $z\sim1.4$ systems. Table \ref{obs_summary} summarizes IROCKS observations. 
\begin{deluxetable*}{lclcccccccccc}
\tablecolumns{13}
\tablewidth{0pc}
\tablecaption{IROCKS Observation Summary}
\setlength{\tabcolsep}{0.02in}
\tablehead{
\colhead{Survey} & 
\colhead{ID} & 
\colhead{z$_0$\tablenotemark{a}} & 
\colhead{RA} & \colhead{Dec} & 
\colhead{Date} & 
\colhead{$t_{exp}$\tablenotemark{b}} & 
\colhead{Filter} & 
\colhead{$\theta_{TT}$\tablenotemark{c}} & 
\colhead{R$_{TT}$\tablenotemark{d}} & 
\colhead{$\theta_{sm}$\tablenotemark{e}} & 
\colhead{$\theta_{PSF}$\tablenotemark{f}}\\
 & & & J2000.0 & J2000.0 & yy/mm & & & ["] & & [pixel] & ["]
}
\startdata
      &            &        &            & Detected \\
\hline
UDS   & 11655      & 0.8960  & 02 16 58.0 & -05 12 42.6 & 13/08 &  9 & Jn2 & 18.0 & 16.0 & 2.0 & 0.24/0.49\\
UDS   & 10633      & 1.0300  & 02 17 15.6 & -05 13 07.6 & 13/08 &  4 & Jn3 & 21.4 & 16.5 & 2.0 & 0.23/0.48\\
DEEP2 & 42042481   & 0.7934  & 02 31 16.4 & +00 43 50.6 & 14/11 & 10 & Jn1 & 23.2 & 15.4 & 2.0 & 0.26/0.52\\
ESO-G & J033249.73 & 0.9810  & 03 32 49.7 & -27 55 17.4 & 14/09 &  5 & Jn3 & 23.6 & 15.5 & 3.0 & 0.24/0.52\\
TKRS  & 11169      & 1.43249 & 12 36 45.8 & +62 07 54.3 & 13/01 &  6 & Hn2 & 33.7 & 16.4 & 2.0 & 0.37/0.55\\
TKRS  & 7187       & 0.84022 & 12 37 20.6 & +62 16 29.7 & 13/05 &  8 & Jn1 & 48.3 & 14.4 & 2.5 & 0.23/0.48\\
TKRS  & 9727       & 0.90316 & 12 37 05.9 & +62 11 53.6 & 13/05 &  6 & Jn2 & 46.9 & 14.0 & 2.5 & 0.53/0.68\\
TKRS  & 7615       & 1.01268 & 12 37 31.1 & +62 17 14.7 & 13/01 &  6 & Jn3 & 34.3 & 15.4 & 2.5 & 0.48/0.68\\
DEEP2 & 11026194   & 0.9198  & 14 15 43.0 & +52 09 07.6 & 14/06 &  7 & Jn2 & 15.1 & 13.5 & 2.5 & 0.33/0.57\\
DEEP2 & 12008898   & 0.9359  & 14 16 55.5 & +52 27 51.3 & 13/05 & 10 & Jn2 & 20.6 & 16.0 & 1.5 & 0.28/0.39\\
DEEP2 & 12019627   & 0.9040  & 14 18 49.8 & +52 38 08.3 & 13/05 &  9 & Jn2 & 49.7 & 16.4 & 2.0 & 0.23/0.37\\
DEEP2 & 13017973   & 1.0303  & 14 20 13.1 & +52 56 13.7 & 12/06\tablenotemark{g} &  9 & Jn3 & 28.8 & 15.3 & 2.5 & 0.39/0.71\\
DEEP2 & 13043023   & 0.9715  & 14 20 15.8 & +53 06 43.2 & 14/06 &  6 & Jn3 & 35.6 & 13.7 & 2.5 & 0.42/0.59\\
DEEP2 & 32040603   & 1.0327  & 23 28 28.3 & +00 21 55.9 & 14/11 &  5 & Jn3 & 37.9 & 14.7 & 2.5 & 0.23/0.54\\ 
DEEP2 & 32016379   & 0.8335  & 23 29 36.6 & +00 06 12.8 & 13/08 &  9 & Jn1 & 18.8 & 16.8 & 2.0 & 0.27/0.42\\
DEEP2 & 32036760   & 0.8534  & 23 30 32.8 & +00 20 06.9 & 13/08 &  7 & Jn1 & 36.7 & 15.5 & 2.5 & 0.34/0.63\\
DEEP2 & 33009979   & 0.9797  & 23 31 56.3 & -00 02 32.0 & 13/08 &  6 & Jn3 & 41.1 & 13.0 & 2.0 & 0.20/0.42\\
\hline
      &            &        &            & Nondetection \\
\hline
UDS   & 11557      & 0.9180  & 02 17 24.4 & -05 12 52.2 & 14/11 & 4 & Jn2 & 29.3 & 12.6 & \nodata & 0.18 \\
DEEP2 & 42042017   & 0.8070  & 02 28 38.0 & +00 40 14.0 & 14/11 & 3 & Jn1 & 33.1 & 14.2 & \nodata & 0.14 \\ 
TKRS  & 3447       & 0.83457 & 12 36 02.9 & +62 12 01.4 & 12/06\tablenotemark{g} &  5 & Jn1 & 21.3 & 13.6 & \nodata & 0.26 \\
TKRS  & 4512       & 0.84047 & 12 36 08.6 & +62 11 24.4 & 14/05 & 3 & Jn1 & 37.0 & 13.6 & \nodata & 0.24 \\
TKRS  & 9867       & 0.85652 & 12 37 09.0 & +62 12 02.0 & 14/06 & 2 & Jn1 & 31.2 & 14.0 & \nodata & 0.15 \\
TKRS  & 9725       & 1.52079  & 12 37 18.6 & +62 13 15.1 & 13/05 & 2 & Hn3 & 33.2 & 15.8 & \nodata & 0.31 \\
TKRS  & 10137      & 0.90890 & 12 37 19.6 & +62 12 56.2 & 13/05 & 3 & Jn2 & 14.5 & 15.8 & \nodata & 0.31 \\
TKRS  & 3811       & 0.87026 & 12 37 22.6 & +62 20 46.5 & 13/05 & 3 & Jn1 & 17.9 & 13.2 & \nodata & 0.22 \\
TKRS  & 7078       & 0.95492 & 12 37 40.4 & +62 18 53.4 & 14/06 & 3 & Jn2 & 19.8 & 12.8 & \nodata & 0.22 \\
DEEP2 & 12027936   & 1.0385  & 14 19 26.5 & +52 46 09.5 & 13/05 & 3 & Jn4 & 42.3 & 16.7 & \nodata & \nodata
\enddata \label{obs_summary}
\tablenotetext{a}{Spectroscopic redshift from the original selected survey.}
\tablenotetext{b}{Exposure time, multiple of 900 s.}
\tablenotetext{c}{Angular separation to the tip-tilt star.}
\tablenotetext{d}{R magnitude of the tip-tilt star.}
\tablenotetext{e}{FWHM of spatial smoothing Gaussian in pixel unit. 1 pixel = 0.1 arcsecond.}
\tablenotetext{f}{FWHM of PSF during on-axis TT star observation before/after spatial smoothing in arcsecond.}
\tablenotetext{g}{Observation made before OSIRIS grating upgrade.}
\end{deluxetable*}

\subsubsection{TKRS Sample}
TKRS \citep{wirth:04} is a deep spectroscopic survey in GOODS-North undertaken with a visible, multi-slit spectrograph, the DEep Imaging Multi-Object Spectrograph \citep[DEIMOS;][]{faber:03}, on the Keck II telescope. It provides accurate redshift measurements of more than 1500 magnitude-limited objects to $R_{AB}$ = 24.4 mag. To estimate \ha{} fluxes, we use \hb{} and \oii{} emission line fluxes for $z \sim 1$ and $z \sim 1.5$ galaxies, respectively. These \hb{} and \oii{} emission lines were measured from flux-calibrated spectra as described by \citet{juneau:11}, but were not corrected for underlying Balmer absorption. The GOODS-North field has a wealth of optical HST imaging data available. We observed nine $z \sim 1$ and two $z \sim 1.5$ sources from TKRS and successfully detected three $z \sim 1$ (7187, 9727, and 7615) and one $z \sim 1.4$ (11169) sources. 11169 is the only source observed in H-band in our sample.

\subsubsection{ESO-GOODS Sample}
We select our GOODS-South targets from the spectroscopic campaign of \citet{vanzella:08}. The data was taken by the UV FOcal Reducer and low dispersion Spectrograph \citep[FORS2;][]{appenzeller:98} on UT2 at VLT. Their spectroscopic sample was selected by photometric colors and redshifts. The final ESO-GOODS catalog provides more than 850 redshift measurements. We use the relation between \oii{} and rest-frame B-band magnitude shown by \citet{mostek:12} to estimate \ha{} fluxes from rest-frame B-band magnitude. In the GOODS-South field, HST optical observations are available, and we use them to eliminate diffuse sources. We observed and detected only one source, J033249.73, in ESO-G.

\subsubsection{DEEP2 Sample}
DEEP2 is a redshift survey to study the universe at $z \sim 1$ \citep[and references therein]{newman:13} . The observations were done by the visible wavelength Low Resolution Imaging Spectrograph \citep[LRIS;][]{oke:95, rockosi:10} at Keck-I and DEIMOS at Keck-II. It provides more than 38,000 reliable redshift measurements. We select sources in 02h (SDSS deep strip), 14h (EGS -- Extended Groth Strip), and 23h (SDSS deep strip). As GOODS-South sources, we use rest-frame B-band magnitudes to estimate \ha{} fluxes. Multiple optical HST data are available for the EGS field, but not the 02h and 23h fields. We eliminate diffuse sources from EGS using HST images, but sources in the 02h and 23h fields are only selected from their expected \ha{} flux. We observed six sources in EGS, four in 23h, and two in 02h field, and detected five (11026194, 12008898, 12019627, 13017973, 13043023) in EGS, all four (32040603, 32016379, 32036760, 22009979) in 23h, and one (42042481) in 02h.

\subsubsection{UDS Sample}
CANDELS at UDS provides the multi-wavelength (UV to mid-IR) catalog. Among about 36,000 F160W-selected sources, 210 sources have spectroscopic redshift measurements \citep{galametz:13}. At the time of observation, star formation rate estimates were not available, and we use existing K-magnitude measurements to prioritize our samples \citep{erb:06b}. The UDS field has both optical and near infrared HST imaging data. We observed three sources in UDS, and detected two (11655 and 10633) sources.

\subsection{OSIRIS Observations}\label{obs}
IROCKS galaxies were observed with OSIRIS \citep{larkin:06} at the W. M. Keck Observatory in Mauna Kea in Jun 2012, May/Aug 2013, and May/Jun/Sep 2014. OSIRIS is a diffraction-limited IFS with moderate spectral resolution (R $\sim$ 3800). It uses a lenslet array as the sampling element on the sky to achieve low noncommon path error (\textless 30 nm rms). In December 2012, the OSIRIS grating was upgraded, and the final throughput was improved by a factor of 1.83 on average between the old grating at Keck-II and the new grating at Keck-I between 1 and 2.4 $\mu$m \citep{mieda:14}. All IROCKS observations were made after OSIRIS was transferred to Keck-I. Only one target, DEEP2-13017973, was observed with OSIRIS before the grating upgrade.

Our observations use OSIRIS LGS-AO in the coarsest plate scale, 0.1$\arcsec$ per spaxel, corresponding to $\sim$ 800 pc at $z \sim 1$, which gives the highest sensitivity to low surface brightness emission. All observations are made in one of the narrowband J and H filters (5\% bandpasses) in order to observe both \ha{} and \nii{} simultaneously. This combination of plate scale and filter produces a field of view of roughly 4.8'' $\times$ 6.4'', which is sufficient to encompass the entire galaxy and support small $\sim$2'' dithers between exposures on source. For each galaxy we also observe at least one pure sky pointing to ensure proper sky subtraction.

The standard observation procedure is as follows: we acquire a TT star at the optimal position angle (PA) and take a pair of 30 s integrations (center and $\sim$ 1.5'' offset) to check the centering and measure the PSF. Once the telescope pointing matches with the sky, we apply a blind offset and move to the target galaxy. After the AO loop is closed, we take three 900 s exposures in up, down, and center positions. Typically, the up and down positions are separated by 2.2 arcsec. While taking the third frame, the second frame is subtracted from the first. When \ha{} is detected in the first frame, we stay on the target for 1.5h - 2.5h to achieve a high SNR. A different dither offset is used in each exposure to avoid any bad pixel contamination. At the end of each night, we observe an Elias standard star with all filters used that night.

There are two potential problems relating to our target selection that may produce bias in our sample. First, for estimating \ha{} line fluxes to rank our targets, we used B or K broad band magnitudes for all targets, except for TKRS targets, which have existing \hb{} line fluxes (\S \ref{selection}). However, there is no existing empirical data to show a direct correlation between galaxies' B or K broad band magnitudes and \ha{} fluxes, and we only used it because of a lack of alternatives. A more thorough, but also more expensive, approach would be to perform a pre-survey using a near-IR multi-slit spectrograph, such as MOSFIRE \citep{mclean:10, mclean:12}, to measure the galaxies' global \ha{} fluxes. 

Second, during observation, we visually inspected the data after the first 900 seconds exposure to decide whether we continue with a longer exposure. Arguably, 900 seconds may not be sufficient to judge whether the target is a non-detection. Additionally, because our non-TKRS targets have poor \ha{} flux estimates, we spent more time on them, thus potentially generating sample bias. While we do not consider these effects significant, future observations will benefit from a more rigorous methodology.

\subsection{OSIRIS Data Reduction}\label{red}
Data reduction is performed using the OSIRIS data reduction pipeline (DRP) version 3.2 and custom IDL routines. Before we run DRP, we use our own custom IDL code to correct the rectification matrices. The rectification matrices are maps of lenslet point spread functions and are required to extract spectra by DRP. Since the upgrade of the OSIRIS grating and calibration unit, the newly taken matrices have created artificial bad pixels in the reduced cube as they iteratively extract spectra. To resolve this issue, we replace any matrix entry \textgreater 0.8 with its neighbor mean. With the corrected rectification matrices, we first combine several dark frames of that night by DRP to make a master-dark. We then run DRP again to subtract the master-dark; adjust channel levels; remove crosstalk, detector glitches, and cosmic rays; extract spectra using the corrected rectification matrices; assemble data cube; and finally, correct for atmospheric dispersion. After this, we run our own cleaning code on the cube, which, for a given channel, iteratively replaces pixel values with the median of its neighboring pixels, if its original value is more than 15 $\sigma_\lambda$ (standard deviation per channel) away from the spatial median.

After we obtain the cleaned, dark-subtracted cubes, we experiment with two sky subtraction methods, \textit{simple subtraction} and \textit{scale subtraction}, using \textit{pure sky} and \textit{pair sky}. Pair sky is another science frame where the galaxy's location on the detector does not overlap with the current science frame. Simple subtraction, as its name implies, is a simple subtraction of a sky cube from a science cube. Scale sky subtraction, on the other hand, uses an algorithm from \citet{davies:07} that scales OH sky emission lines between adjacent frames to reduce sky subtraction residuals. The final choice of sky and subtraction method is determined by examining the resultant standard deviation in spectral space; a lower standard deviation, (i.e., less noisy), was deemed better. 

76\% of frames are reduced by the scale subtraction method, among which 58\% are with pure sky. Pure sky frames are used more by the scale subtraction method than pair sky frames do. Furthermore, we additionally subtract a channel-dependent constant to the sky-subtracted cube that ensures the median value in the regions away from the source is zero. We then mosaic the reduced cubes using the DRP with the ''meanclip'' combine method with LGS offset. 

\begin{figure}
\centering\includegraphics[width=0.45\textwidth]{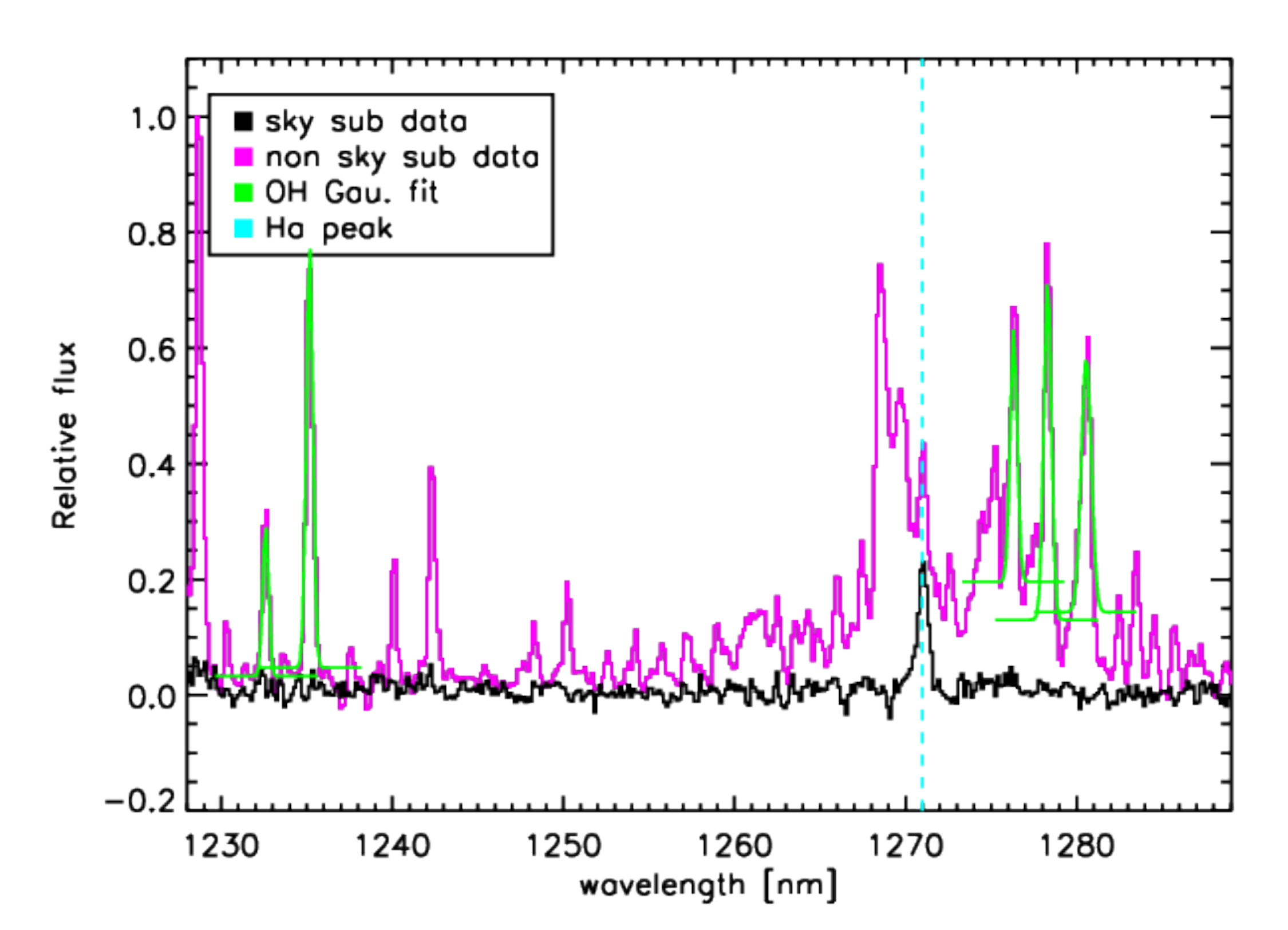}
\caption{An example of unsmoothed non-sky-subtracted (magenta) and fully reduced (black) spectra at a single, bright spaxel of a DEEP2 galaxy, 12008898 ($z=0.936$). The location of \ha{} emission line peak is shown by a cyan vertical line. Brighter OH lines that are well separated were fitted by a Gaussian profile (green) to obtain the instrumental width at that spaxel.}
\label{example_non_sky_spec}
\end{figure}

The effect of sky subtraction can be seen clearly in Figure \ref{example_non_sky_spec}. We are able to largely remove contamination from sky emission lines and recover a well-defined \ha{} emission line from the galaxy. In the end, an additional bad-pixel-removal algorithm is used to replace single, isolated, high-value (6 or 7 $\sigma$ above the spatial median) pixels that are outside of the expected galaxy vicinity, with the spatial median of the given channel. To increase SNR, the cleaned mosaic-ed cube is spatially smoothed by a Gaussian function of FWHM = 1.5 to 3.0 pixel (0.15'' to 0.3''). The smoothing FWHM is chosen by our custom ``adaptive smoothing'' code. The details of this method are documented in Appendix \ref{adaptive}. Finally, flux calibration is done using the Elias telluric standard stars observed on each night.

The error, or uncertainty, in our data is defined by the spatial standard deviation within the region where all mosaic-ed frames are overlapped, for a given channel. Therefore our error is wavelength-dependent, and spatially invariant. The only exception is where not all frames are overlapped. There we scale the error by $\sqrt{n_{\text{max}}/n}$, where $n$ is the number of frames used at that spaxel, and $n_{\text{max}}$ is the maximum number of frames used in the cube.

\subsubsection{\ha{} Maps}
\ha{} flux maps are created by cross-correlating a normalized Gaussian profile of a typical \ha{} width ($\sigma$ = 1.5 channel $\sim$ 50 km s$^{-1}$ at $z$ = 1) with the spectrum at each spaxel to find a correlation peak. We then sum up five channels ($\sim$ 170 km s$^{-1}$) centered on this peak to represent \ha{} flux. The noise map is made by adding the error in the same five channels in quadrature. When the correlation peak does not coincide with the peak of Gaussian fitting (see \S \ref{sec_kine_map} for Gaussian profile fitting to \ha{} lines), we instead use five channels around the redshift from the original surveys, $z_0$ (shown in Table \ref{obs_summary}). In this case, we consider it a non-detection, and the calculated flux reflects the background level. Two \nii{} lines, \nii6549 and \nii6583, are simultaneously observed with the \ha{} line. Both \nii{} lines are well separated ($\sim$ 20 and $\sim$ 28 channels, respectively, from \ha{} line at $z$ = 1), and \ha{} maps are not contaminated by \nii{} line fluxes. Since \nii{} detection is significantly weaker than \ha{}, we do not cross-correlate our spectra to locate it. Rather, we infer its location from the detected \ha{} line, and make its flux map and associated error map by summing up five channels centered on that inferred offset in the spectral dimension. HST images (when available) and the resultant \ha{} flux maps are shown in Figure \ref{kine_11655} on the left and second left panels, respectively.

\subsubsection{1D Spectra}\label{sec_1dspec}
We define an \ha{} segmentation map for each galaxy using the SNR. Spaxels whose SNR$_\text{\ha}$ \textless 3 or Gaussian fitted SNR$_\text{\ha}^G$ (integration of Gaussian parameters with propagated error, see \S \ref{sec_kine_map} for Gaussian fitting) \textless 1.5 are masked out. We then apply final visual inspection to mask out bad spaxels. The integrated 1D spectra of the IROCKS samples (top panel of Figure \ref{1Dspec}) are created by summing up all spaxels in the \ha{} segmentation map. A single Gaussian profile is fitted to the \ha{} emission line in each 1D spectrum to obtain the peak wavelength and integrated width. From the peak wavelength, we measure a systemic redshift ($z_{\text{sys}}$), and from the width, corrected for the instrumental resolution, the global 1D velocity dispersion (\sigmaoned) is obtained. This 1D dispersion \sigmaoned{} \citep[sometimes called $\sigma_{\text{net}}$ or $\sigma_{\text{global}}$;][]{law:09, wisnioski:11, wright:09} is not corrected for terms such as rotation and outflows. In \S \ref{sec_kine_map}, we discuss another velocity dispersion value, \sigmaave{}, which more accurately measures the line-of-sight velocity dispersion.

We note that instrumental resolution varies across the field of view, and for \sigmaoned, we use a spatial average of instrumental width for correction. To calculate the spatially varying instrumental width, we measure the widths of OH lines in non-sky-subtracted data (see example of OH lines in Figure \ref{example_non_sky_spec}). The procedure is as follows: we first smooth the non-sky-subtracted data with a Gaussian function of the same width as the one used for the science data. Using a Gaussian fit, we then measure the widths of bright OH lines that are well separated (\textgreater 5 channels) from other OH lines. This resulted in a few width measurements per spaxial in an individual sky data cube. Since the final science frame is mosaic-ed together at different dither patterns, the instrumental width per spaxel is an average of all the frames combined. We do not see a width trend in wavelengths, and thus we only obtain spatially but not spectrally varying instrumental width. We find that the typical instrumental width corresponds to $\sim$ 45 km s$^{-1}$, and spatial variation is about 10\%.

\subsubsection{Multiple Components in Each Galaxy}\label{section_component}
When there is only a single \ha{} peak in the 1D spectrum, the object is classified as a ''single'' component source: 11655, 10633, 42042481, J033249.73, 9727, 7615, 11026194, 13017973, 13043023, 32040603, 32016379, and 32036760. When there is more than one peak, we spatially separate them and treat them as different components, and the galaxy is classified as a ''multiple'' \ha{} source: 11169 (East and West), 7187 (East and West), 12019627(North, South-East, and South-West), and 33009979 (North and South). There are two special cases: first, the spectrum of 12008898 only has one spectral peak in 1D spectrum, but on both HST and \ha{} maps, its north and south components are spatially separated by $\theta \sim$ 2'' ($\sim$ 3 kpc), so we categorize it as multiple (North and South); and second, the west component of 7187 has more than one spectral peak in 1D spectrum even after it has been separated from the east component, but the peaks cannot be spatially separated, and hence we treat it as a single component. Due to multiple peaks, \sigmaoned{} and other parameters for the west component of 7187 are not well measured. In Appendix \ref{sec_indiv}, each components are separately shown in Figure \ref{11169_comp}.

\subsubsection{Global Fluxes and Star Formation Rates}
Like the top panels of Figure \ref{1Dspec}, the bottom panels are integrated spectra from the segmentation maps, but each \ha{} line spectrum has been shifted to coincide to the same wavelength (i.e., matching each spaxel Gaussian fitting peak to the same systemic redshift, $z_{\text{sys}}$). This procedure removes all large scale velocity trends, such as rotation, from the line width, and is useful for increasing the SNR of \ha{} and boosting the detection of \nii{}. We obtain the global \ha{} and \nii{} fluxes by fitting Gaussian profiles to these shifted integrated 1D spectra, and computing the integral of the fitted Gaussian curves. We also obtain the flux uncertainties using the errors in the fitted parameters.

To convert \ha{} fluxes into luminosities, we use a standard cosmological model (see \S \ref{intro}), and correct for dust extinction, assuming a spatially constant optical depth derived from stellar population models (\S \ref{sec_sed}). These \ha{} luminosities are then converted to SFR using \citet{kennicutt:98} modified by the initial mass function of \citet{chabrier:03}:
\begin{equation}
\label{eq_sfr_ha}
\text{SFR} \text{ }[M_\odot/\text{yr}] = \frac{L_{H\alpha}}{2.23 \times 10^{41} \text{[erg s$^{-1}$]}}.
\end{equation}

The systemic redshift, non-extinction-corrected integrated fluxes of \ha{} and \nii{}, and \nii{} to \ha{} line ratio of each components are summarized in Table \ref{flux_table}. In our $z\sim1$ sample, \ha{} flux spans between 4.1 to 71.8$\times 10^{-17}$ erg s$^{-1}$ cm$^{-2}$, and the average is 21.2$\times 10^{-17}$ erg s$^{-1}$ cm$^{-2}$. In this paper, we report the global \nii/\ha{} ratio, but not its spatial variation. We defer the analysis of spatially resolved \nii/\ha{} to future work. The extinction-corrected/non-corrected \ha{} luminosity and SFR are reported in Table \ref{sed_table}. The following section describes the extinction correction factor.
\begin{deluxetable}{lcccc}
\tabletypesize{\footnotesize}
\tablecolumns{5}
\tablewidth{0pc}
\tablecaption{Emission line fluxes}
\setlength{\tabcolsep}{0.02in}
\tablehead{
\colhead{ID} &
\colhead{z$_{sys}$\tablenotemark{a}} &
\colhead{$f_{\text{\ha}}$\tablenotemark{b}} &
\colhead{$f_{\text{\nii}}$\tablenotemark{c}} &
\colhead{$\log\left(\frac{\text{\nii}}{\text{\ha}}\right)$} 
}
\startdata
11655      & 0.8962 & 20.1 $\pm$  5.0 &  4.9 $\pm$  4.4 & -0.61 $\pm$ 0.40 \\ 
10633      & 1.0318 &  4.1 $\pm$  2.3 & \nodata         & \nodata          \\ 
42042481   & 0.7940 & 43.0 $\pm$  7.8 & 15.2 $\pm$  6.8 & -0.45 $\pm$ 0.21 \\ 
J033249.73 & 0.9813 & 10.8 $\pm$  4.3 &  3.8 $\pm$  6.1 & -0.45 $\pm$ 0.71 \\ 
11169E     & 1.4344 & 14.8 $\pm$  3.4 &  2.3 $\pm$  4.2 & -0.80 $\pm$ 0.79 \\ 
11169W     & 1.4330 & 21.5 $\pm$  3.6 & \nodata       & \nodata            \\ 
7187E      & 0.8404 &  7.1 $\pm$  2.9 &  2.3 $\pm$  4.0 & -0.49 $\pm$ 0.78 \\ 
7187W      & 0.8409 &  6.0 $\pm$  2.9 &  1.4 $\pm$  5.2 & -0.62 $\pm$ 1.59 \\ 
9727       & 0.9038 & 28.2 $\pm$  6.2 & 13.3 $\pm$  6.0 & -0.33 $\pm$ 0.22 \\ 
7615       & 1.0130 & 15.4 $\pm$  5.1 &  3.4 $\pm$  3.6 & -0.66 $\pm$ 0.48 \\ 
11026194   & 0.9205 & 14.3 $\pm$  4.0 &  2.6 $\pm$  3.5 & -0.74 $\pm$ 0.59 \\  
12008898N  & 0.9362 &  5.5 $\pm$  4.3 &  0.6 $\pm$  2.4 & -0.94 $\pm$ 1.68 \\ 
12008898S  & 0.9364 & 55.0 $\pm$  9.4 & 22.2 $\pm$ 44.2 & -0.39 $\pm$ 0.87 \\  
12019627N  & 0.9037 &  8.8 $\pm$  3.5 &  1.8 $\pm$  4.4 & -0.69 $\pm$ 1.08 \\ 
12019627SE & 0.9045 & 15.3 $\pm$  4.8 & \nodata         & \nodata          \\ 
12019627SW & 0.9059 &  9.7 $\pm$  3.5 & \textless 0.2   & \textless -1.70  \\ 
13017973   & 1.0309 & 71.8 $\pm$ 19.6 &  9.9 $\pm$ 16.0 & -0.86 $\pm$ 0.71 \\ 
13043023   & 0.9716 & 27.1 $\pm$  8.2 &  7.6 $\pm$  8.0 & -0.55 $\pm$ 0.47 \\ 
32040603   & 1.0338 & 10.8 $\pm$  3.4 & \textless 0.1   & \textless -2.04  \\ 
32016379   & 0.8339 & 20.0 $\pm$  5.2 &  4.7 $\pm$  4.0 & -0.62 $\pm$ 0.38 \\ 
32036760   & 0.8519 & 16.7 $\pm$  3.7 &  5.0 $\pm$  2.8 & -0.52 $\pm$ 0.26 \\ 
33009979N  & 0.9817 & 12.0 $\pm$  4.2 &  3.5 $\pm$  4.3 & -0.53 $\pm$ 0.55 \\ 
33009979S  & 0.9799 & 44.2 $\pm$  9.5 &  8.3 $\pm$  6.9 & -0.73 $\pm$ 0.37 
\enddata \label{flux_table}
\tablenotetext{a}{Redshift measured from OSIRIS \ha{} detected emission line.}
\tablenotetext{b}{Global \ha{} emission line fluxes obtained by fitting Gaussian profiles to the shifted integrated 1D spectra, in units of 10$^{-17}$ erg/s/cm$^2$.} 
\tablenotetext{c}{Global \nii{} emission line fluxes obtained by fitting Gaussian profiles to the shifted integrated 1D spectra, in units of 10$^{-17}$ erg/s/cm$^2$.}
\end{deluxetable}

\subsection{Stellar Population Modelling}\label{sec_sed}
We make use of publicly available photometric catalogs for each source to construct a consistent spectral energy distribution (SED) and stellar population fit to estimate stellar masses, optical depths, and SFRs. For the four TKRS galaxies in GOODS-North, we use the photometric catalog from version 4.1 3D-HST release \citep{skelton:14}. This catalog contains 22 bands: seven HST, four Spitzer, and nine ground-based, ranging from 0.3 $\mu$m to 8.0 $\mu$m. For our single ESO-GOODS source, we use the GOODS/ISAAC final data release, version 2.0 \citep{retzlaff:10} for J, H, and K photometry, and GOODS/FOR2 final data release version 3.0 for $i-z$, $V-i$, and $B-V$ \citep{vanzella:08}. For the ten DEEP2 sources in our sample, we use the extended photometry catalog of DEEP2 Galaxy Redshift Survey data release 4 \citep{matthews:13}, containing $ugriz$ photometry. For the two UDS sources, we use the CANDELS UDS Multiwavelength catalog \citep{galametz:13}, which contains 19 bands: four HST, four Spitzer, and ten ground-based, ranging from 0.3 to 8.0 $\mu$m. For consistency, our SED fitting uses only ground-based photometry in the 0.3--2.3 $\mu$m range.

The SED fitting method used in this study is further described in \citet{salim:07, salim:09}. In short, the method uses the stellar population synthesis models of \citet{bruzual:03}, with an exponentially declining continuous SFR with random stochastic bursts super-imposed, a range of metallicity (0.1 to 2 $Z_\odot$), and a Chabrier IMF \citep{chabrier:03}. Each model is attenuated according to a two-component prescription of \citet{charlot:00}, whose extinction curve is age-dependent and typically steeper than the \citet{calzetti:01} curve. The model assumes extra attenuation toward HII regions, where young stars are embedded within dense birth clouds as well as the interstellar medium (ISM) in the galaxy at $t < 10^7$ yr. At $t > 10^7$ yr, the birth cloud disappears and only ISM attenuation is considered. We define a total optical depth, $\tau_V$, to indicate attenuation from both HII and ISM, and $\mu\tau_V$ for ISM only attenuation. The coefficient $\mu$ is determined from SED fitting, and in our sample,the average $\mu$ is 0.48.

\begin{figure}
\centering
\includegraphics[width=0.45\textwidth]{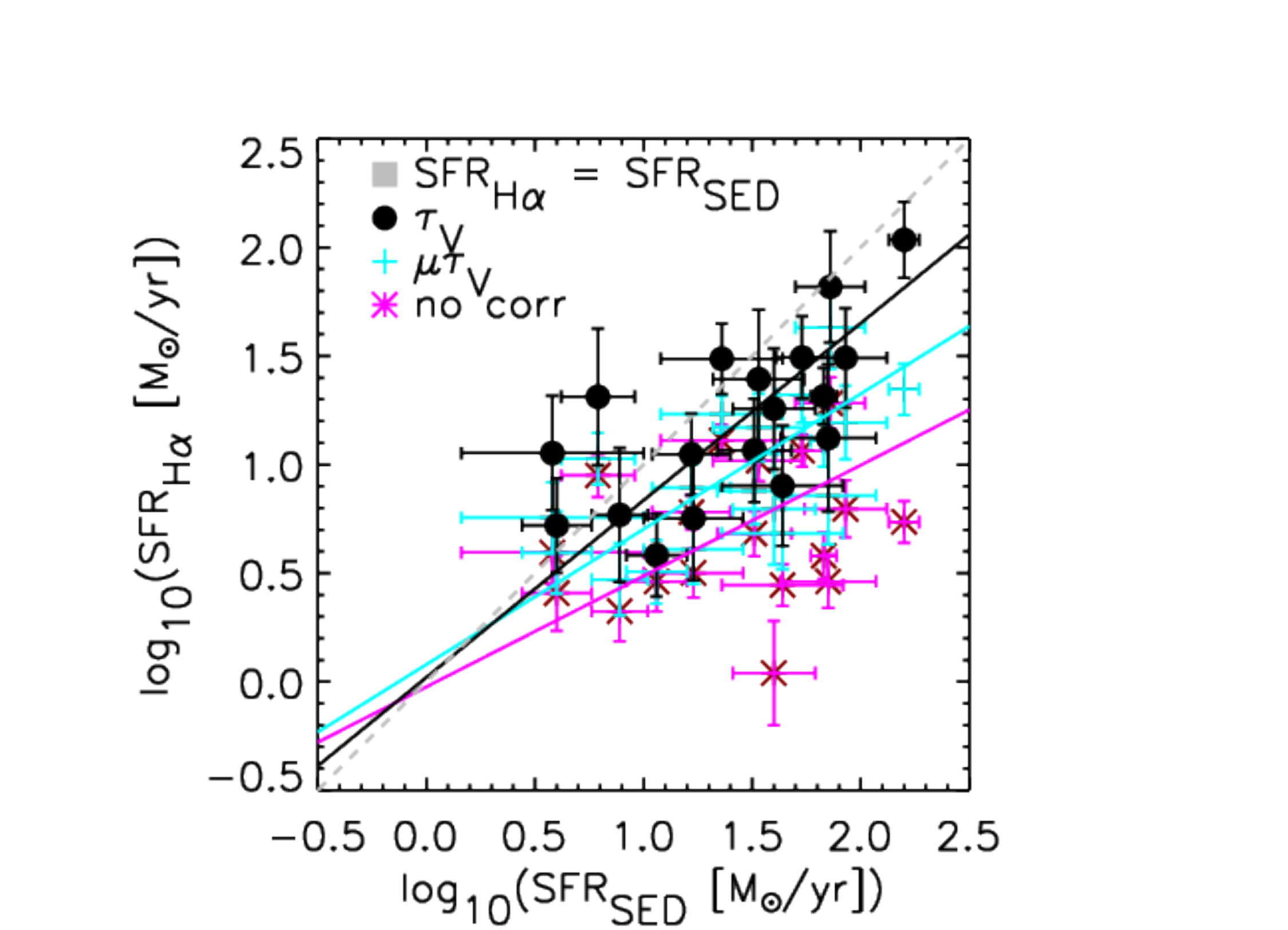}
\caption{Comparison of SFRs with different extinction corrections derived from the SED fits. HII+ISM dust corrected (SFR$_{\text{\ha}}^{00}$, black circle), ISM dust only corrected (SFR$_{\text{\ha}}^0$, cyan plus), and uncorrected (SFR$_{\text{\ha}}$, magenta asterisk) SFR estimated from \ha{} luminosity using \citet{kennicutt:98} and \citet{chabrier:03} vs. SFR estimated from SED fitting. One on one relation (SFR$_{\text{\ha}}$ = SFR$_{\text{SED}}$ is shown in gray dotted line. The average HII+ISM attenuation is $<\tau_V>$ = 1.2, and the average ISM only attenuation is $<\mu\tau_V>$ = 0.6 (SFR$^{00}_{\text{\ha}}$ = SFR$_{\text{\ha}} e^{\tau_{\text{\ha}}}$ and SFR$^{0}_{\text{\ha}}$ = SFR$_{\text{\ha}} e^{\mu\tau_{\text{\ha}}}$). Correcting for the dust attenuation in HII region and ISM yields the best match between the derived SFR$_{\text{\ha}}$ and SFR$_{\text{SED}}$, with a best-fit line of $\log{\text{SFR}_\text{SED}} = 0.02+0.82\log{\text{SFR}_\text{\ha}^{00}}$ and has mean SFR$_{\text{\ha}}^{00}$/SFR$_\text{SED}$ = 0.88.}
\label{sed_sfr_comp}
\end{figure}

Individual values for stellar mass ($M_*$), $\tau_V$, $\mu$, and SFR (SFR$_\text{SED}$) obtained by SED fitting are tabulated in Table \ref{sed_table}. The table also contains uncorrected, ISM only corrected, and HII+ISM corrected \ha{} luminosities ($L_\text{\ha}$, $L_\text{\ha}^0$, and $L_\text{\ha}^{00}$), and the SFRs estimated from these luminosities (SFR$_\text{\ha}$, SFR$_\text{\ha}^0$, and SFR$_\text{\ha}^{00}$). The comparison of these three versions of SFR$_\text{\ha}$ with respect to SFR$_\text{SED}$ is shown in Figure \ref{sed_sfr_comp}.
HII+ISM corrected SFR$_{\text{\ha}}$ best agrees with SFR$_{\text{SED}}$, as shown by the black best-fit line in Figure \ref{sed_sfr_comp}, which has a power of 0.81, mean SFR$_{\text{\ha}}$/SFR$_{\text{SED}}$ = 0.86, and \rchi = 1.24.

\begin{figure}
\centering
\includegraphics[width=0.45\textwidth]{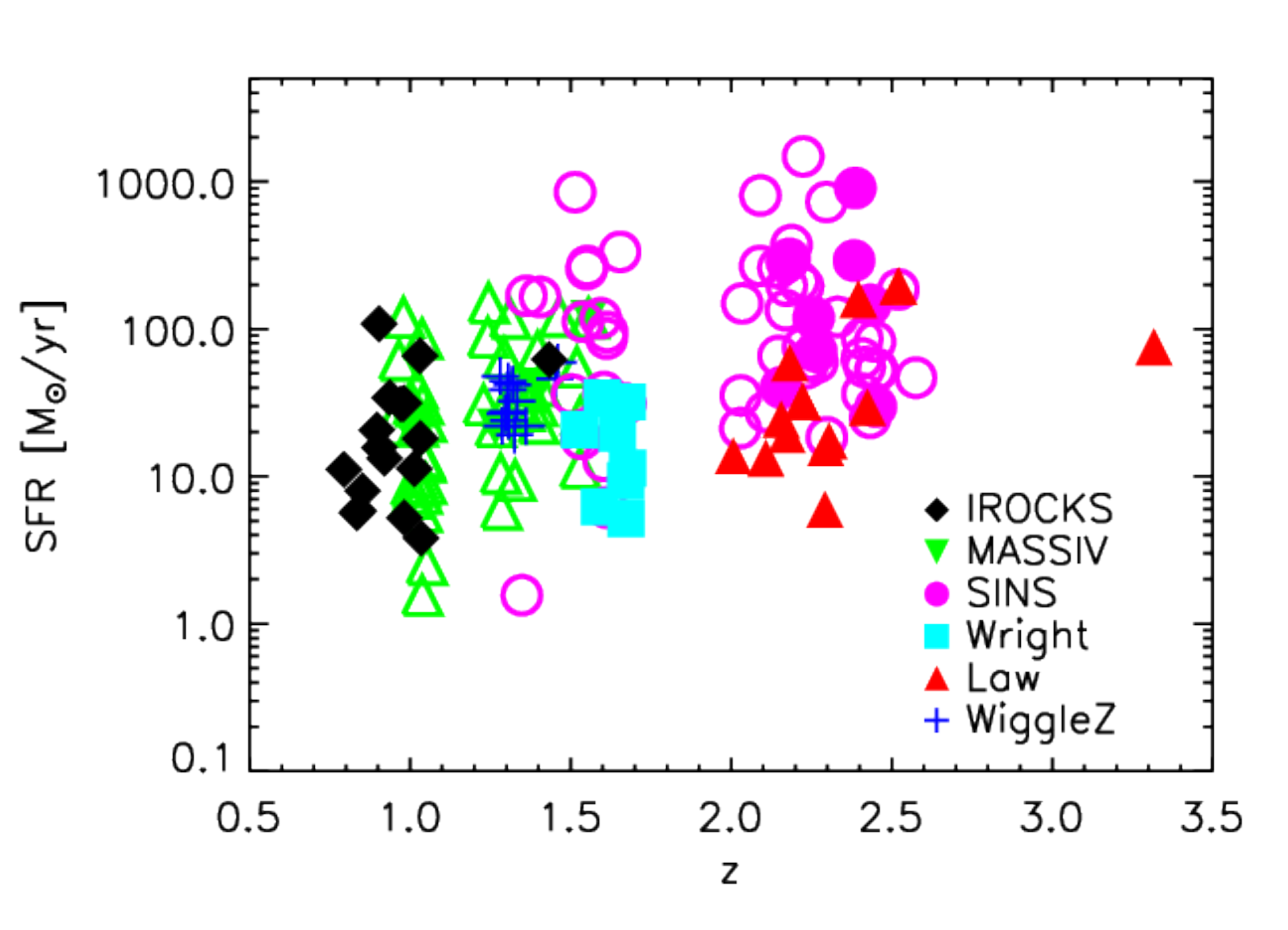}
\caption{Global SFR of individual galaxies in the IROCKS sample compared to other high redshift IFS samples, \citet{wright:09, law:09}; SINS \citep{forster:09}; WiggleZ \citep{wisnioski:11}; and MASSIV \citep{queyrel:12}, as a function of redshift. The SFRs shown here are estimated from \ha{} or \oiii{} fluxes using Plank cosmology (see \S \ref{intro}) and are corrected for ISM-only extinction. Same symbol and color but filled/open are AO/non-AO observation. Because this figure shows global SFR, AO/non-AO is almost irrelevant, but two cases are shown separately to highlight the differences between different surveys.}
\label{sfr_z_cosmo}
\end{figure}

Most IFS studies of high redshift galaxies assume E(B-V)$_{\text{stellar}}$ = E(B-V)$_{\text{nebular}}$ \citep[e.g.,][]{law:09, wright:09, wisnioski:11, queyrel:12}. On the other hand, the SINS survey \citep{forster:09} used a locally found relation, E(B-V)$_{\text{stellar}}$ = 0.44E(B-V)$_{\text{nebular}}$ \citep{calzetti:01}, and found a better agreement between \ha- and UV-continuum-estimated SFR of z $\sim$ 2 galaxies. More recent studies \citep[e.g.,][]{kashino:13, pannella:15} found E(B-V)$_{\text{stellar}} \sim$ 0.75E(B-V)$_{\text{nebular}}$ over the redshift range $0.5 < z < 4$, with more massive galaxies being more dust attenuated. Even though the emission lines are not attenuated by the same amount as the stellar continuum, and extra attenuation toward HII region may be more appropriate, in this paper, we use ISM-only extinction corrected values, otherwise specified, to be consistent with other IFS studies. Figure \ref{sfr_z_cosmo} shows the instantaneous global SFR estimated from \ha{} luminosity as a function of redshift. Major IFS high redshift galaxy observations \citep{law:09, wright:09, forster:09, wisnioski:11, queyrel:12} are over-plotted. On this figure, whether AO is used or not is irrelevant to the global SFR estimate, but we distinguish the two cases to show which survey focuses on what redshift range with what type of observation mode.
\begin{turnpage}[p]
\begin{deluxetable*}{lcccccccccc}
\tabletypesize{\footnotesize}
\tablecolumns{11}
\tablewidth{0pc}
\tablecaption{Stellar Population Parameters}
\setlength{\tabcolsep}{0.02in}
\tablehead{
\colhead{ID} &
\colhead{$\log{(\text{M}_*[\text{\msun}])}$\tablenotemark{a}} & 
\colhead{$\tau_V$\tablenotemark{b}} &
\colhead{$\mu$\tablenotemark{c}} & 
\colhead{L$_{\text{\ha}}$\tablenotemark{d}} &
\colhead{L$_{\text{\ha}}^{0}$\tablenotemark{e}} & 
\colhead{L$_{\text{\ha}}^{00}$\tablenotemark{f}} &
\colhead{SFR$_{\text{\ha}}$\tablenotemark{g}} &
\colhead{SFR$_{\text{\ha}}^{0}$\tablenotemark{h}} & 
\colhead{SFR$_{\text{\ha}}^{00}$\tablenotemark{i}} &
\colhead{SFR$_{\text{SED}}$\tablenotemark{j}} \\
 & & & & [10$^{41}$ erg/s] & [10$^{41}$ erg/s] & [10$^{41}$ erg/s] & [\msun/yr] & [\msun/yr] & [\msun/yr] & [\msun/yr]
}
\startdata
11655      & 10.2 $\pm$ 0.1 & 2.07 $\pm$ 0.21 & 0.72 & 8.5 $\pm$ 2.1 & 28.7 $\pm$ 7.9 & 46.0 $\pm$ 13.8 & 3.8 & 12.9 & 20.6 & 67.6 \\
10633      & 11.2 $\pm$ 0.0 & 3.43 $\pm$ 0.40 & 0.62 & 2.4 $\pm$ 1.3 & 13.9 $\pm$ 8.2 & 40.3 $\pm$ 25.8 & 1.1 & 6.2 & 18.1 & 39.8 \\
42042481   & 10.6 $\pm$ 0.2 & 0.75 $\pm$ 0.48 & 0.42 & 13.5 $\pm$ 2.5 & 17.4 $\pm$ 4.3 & 24.9 $\pm$ 10.8 & 6.0 & 7.8 & 11.1 & 16.6 \\
J033249.73 & 10.5 $\pm$ 0.1 & 0.88 $\pm$ 0.37 & 0.60 & 5.7 $\pm$ 2.3 & 8.8 $\pm$ 3.8 & 11.7 $\pm$ 5.8 & 2.6 & 3.9 & 5.2 & 4.0 \\
11169E     & 10.8 $\pm$ 0.1 & 1.02 $\pm$ 0.84 & 0.21 & 19.9 $\pm$ 4.6 & 23.7 $\pm$ 6.4 & 45.8 $\pm$ 33.1 & 8.9 & 10.6 & 20.5 & 6.2 \\
11169W     & 10.1 $\pm$ 0.0 & 1.06 $\pm$ 0.41 & 0.32 & 28.8 $\pm$ 4.8 & 38.0 $\pm$ 7.5 & 68.5 $\pm$ 25.6 & 12.9 & 17.0 & 30.7 & 22.9 \\
7187       & 10.3 $\pm$ 0.1 & 1.25 $\pm$ 0.78 & 0.33 & 4.7 $\pm$ 1.5 & 6.6 $\pm$ 2.5 & 13.1 $\pm$ 9.3 & 2.1 & 3.0 & 5.9 & 7.8 \\
\hspace{1em} 7187E      & \nodata & \nodata & \nodata & 2.5 $\pm$ 1.1 & 3.6 $\pm$ 1.7 & 7.1 $\pm$ 5.4 & 1.1 & 1.6 & 3.2 & \nodata \\
\hspace{1em} 7187W      & \nodata & \nodata & \nodata & 2.2 $\pm$ 1.0 & 3.0 $\pm$ 1.6 & 6.0 $\pm$ 4.8 & 1.0 & 1.4 & 2.7 & \nodata \\
9727       & 11.0 $\pm$ 0.0 & 3.66 $\pm$ 0.41 & 0.47 & 12.1 $\pm$ 2.7 & 49.5 $\pm$ 13.4 & 241.7 $\pm$ 96.9 & 5.4 & 22.2 & 108.4 & 158.5 \\
7615       & 10.7 $\pm$ 0.1 & 1.29 $\pm$ 0.62 & 0.35 & 8.8 $\pm$ 2.9 & 12.7 $\pm$ 4.8 & 25.2 $\pm$ 15.3 & 3.9 & 5.7 & 11.3 & 3.8 \\
11026194   & 10.2 $\pm$ 0.2 & 1.86 $\pm$ 0.89 & 0.60 & 6.5 $\pm$ 1.8 & 16.1 $\pm$ 8.3 & 29.5 $\pm$ 23.0 & 2.9 & 7.2 & 13.2 & 70.8 \\
12008898N  & \nodata & \nodata & \nodata & 2.6 $\pm$ 2.0 & 4.7 $\pm$ 3.9 & 7.0 $\pm$ 6.1 & 1.2 & 2.1 & 3.1 & \nodata \\
12008898S  & 9.9 $\pm$ 0.1 & 1.21 $\pm$ 0.49 & 0.60 & 25.8 $\pm$ 4.4 & 46.7 $\pm$ 13.8 & 69.4 $\pm$ 30.2 & 11.6 & 21.0 & 31.1 & 53.7 \\
12019627N  & \nodata & \nodata & \nodata & 3.8 $\pm$ 1.5 & 5.9 $\pm$ 2.8 & 9.1 $\pm$ 5.8 & 1.7 & 2.7 & 4.1 & \nodata \\
12019627S  & 10.0 $\pm$ 0.1 & 1.08 $\pm$ 0.60 & 0.51 & 10.7 $\pm$ 2.6 & 16.8 $\pm$ 5.8 & 25.9 $\pm$ 14.2 & 4.8 & 7.5 & 11.6 & 32.4 \\
\hspace{1em} 12019627SE & \nodata & \nodata & \nodata & 6.6 $\pm$ 2.1 & 10.4 $\pm$ 4.2 & 16.0 $\pm$ 9.3 & 3.0 & 4.7 & 7.2 & \nodata \\
\hspace{1em} 12019627SW & \nodata & \nodata & \nodata & 4.2 $\pm$ 1.5 & 6.6 $\pm$ 2.9 & 10.1 $\pm$ 6.2 & 1.9 & 2.9 & 4.5 & \nodata \\
13017973   & 10.6 $\pm$ 0.2 & 1.51 $\pm$ 0.64 & 0.65 & 42.7 $\pm$ 11.7 & 95.3 $\pm$ 41.5 & 146.8 $\pm$ 86.6 & 19.2 & 42.7 & 65.8 & 72.4 \\
13043023   & 10.4 $\pm$ 0.1 & 1.96 $\pm$ 0.53 & 0.57 & 13.9 $\pm$ 4.2 & 34.8 $\pm$ 13.6 & 69.2 $\pm$ 36.6 & 6.3 & 15.6 & 31.0 & 85.1 \\
32040603   & 9.6 $\pm$ 0.3 & 0.34 $\pm$ 0.37 & 0.37 & 6.5 $\pm$ 2.0 & 7.2 $\pm$ 2.4 & 8.5 $\pm$ 3.7 & 2.9 & 3.2 & 3.8 & 11.5 \\
32016379   & 10.4 $\pm$ 0.2 & 0.71 $\pm$ 0.74 & 0.43 & 7.1 $\pm$ 1.8 & 9.1 $\pm$ 3.3 & 12.6 $\pm$ 8.3 & 3.2 & 4.1 & 5.7 & 17.0 \\
32036760   & 10.7 $\pm$ 0.2 & 1.29 $\pm$ 0.73 & 0.52 & 6.2 $\pm$ 1.4 & 10.7 $\pm$ 4.1 & 17.8 $\pm$ 11.3 & 2.8 & 4.8 & 8.0 & 43.7 \\
33009979N  & \nodata & \nodata & \nodata & 6.3 $\pm$ 2.2 & 9.0 $\pm$ 4.1 & 15.0 $\pm$ 11.8 & 2.8 & 4.0 & 6.7 & \nodata \\
33009979S  & 10.3 $\pm$ 0.2 & 1.06 $\pm$ 0.86 & 0.41 & 23.2 $\pm$ 5.0 & 33.1 $\pm$ 11.9 & 55.1 $\pm$ 40.5 & 10.4 & 14.8 & 24.7 & 33.9
\enddata \label{sed_table}
\tablenotetext{a}{Stellar mass derived from the SED fits.}
\tablenotetext{b}{Total optical depth for HII+ISM extinction} 
\tablenotetext{c}{Correction to the optical depth for ISM only extinction}
\tablenotetext{d}{\ha{} luminosity not corrected for extinction.}
\tablenotetext{e}{\ha{} luminosity corrected for ISM only extinction ($\mu \tau_V$).}
\tablenotetext{f}{\ha{} luminosity corrected for HII+ISM extinction ($\tau_V$).}
\tablenotetext{g}{SFR estimated from uncorrected \ha{} luminosity.}
\tablenotetext{h}{SFR estimated from ISM only ($\mu \tau_V$) extinction corrected \ha{}.}
\tablenotetext{i}{SFR estimated from HII+ISM ($\tau_V$) extinction corrected \ha{}.}
\tablenotetext{j}{SFR estimated from SED fitting.}
\end{deluxetable*}
\end{turnpage}

\section{Morphologies}\label{morph}
We quantify morphologies of star-forming regions by examining \ha{} maps with the same segmentation criteria in \S \ref{obs} applied. \ha{} flux distribution and its segmentation map are best described by Figure \ref{kine_11655} and Figure \ref{clump_im}. We measure a size scale and three morphological parameters for each galaxy. We define a radius of gyration, $r_g$ as a size scale. It yields a typical distance from a given origin using the second moment of flux:
\begin{equation}
r_g = \sqrt{\frac{\sum\limits_{i}{d_i^2 f_i}}{\sum\limits_{i}{f_i}}},
\end{equation}
where $d_i$ is the distance between the given origin to the $i^{\rm th}$ pixel whose flux value is $f_i$. Our choice of origin is the flux-weighted centroid. This is a mathematically robust way to define a galaxy size, especially for systems with asymmetric and clumpy flux distributions since it does not assume a specific galaxy model (e.g., Sersic index). Many of our \ha{} maps exhibit clumpy morphologies, and $r_g$ has the additional advantage of being largely insensitive to PSF and spatial smoothing because it gives the typical distance between each clump center (a galaxy with a single concentrated nucleus has a small $r_g$ while a galaxy with multiple nuclei has a $r_g$ that is roughly the distance between nuclei). Compared to typical size measurements, such as a half light radius, $r_g$ is always smaller and more sensitive to the distribution of the light. The values of $r_g$ are reported in Table \ref{mor_table}. In our sample, $r_g$ ranges from 1.0 to 7.6 kpc, and the average is 3.5 kpc. When the source has more than one distinct component (TKRS11169, DEEP2-12008898, DEEP2-12019627, and DEEP2-33009979), we also report their separation in Table \ref{mor_table}. In our sample, the smallest source, UDS 10633, is smaller than the smoothing width and hence not resolved. However, its spectrum has good signal at the expected redshift, so we consider this source as a real detection (not noise spike) and keep it in our analysis. The biggest galaxy, DEEP2 13017973, has $r_g$ =7.6 kpc, but the most extended one is DEEP2 12019627, whose separation between the different components spans 24 kpc. 

In \S \ref{sec_clump}, we measure the individual sizes (half light radii) of clumps in galaxies. While the radius of gyration and the component separation distance describe the whole extent of the galaxy, the clump size describes the scale of local star-forming regions.

We also calculate three morphological parameters for our \ha{} maps: the Gini coefficient \citep[$G$;][]{abraham:03}, $M_{20}$ \citep{lotz:04}, and multiplicity \citep[$\Psi$;][]{law:07a}. The Gini coefficient is commonly used in econometrics, and when applied to galaxy morphologies it quantifies the relative distribution of galaxy flux among its constituent pixels. $G$ is one when all light is concentrated in one pixel while $G$ is zero when every pixel has the same value. $M_{20}$ is the normalized second-order moment of the brightest 20\% of the galaxy's flux and has low negative value when galaxies are extended with multiple nuclei and high negative value when galaxies are smooth with bright nucleus. $\Psi$ is designed to measure how multiple the source appears by measuring the projected potential energy of the light distribution, normalized by the most compact arrangement of the flux pixels. Low $\Psi$ means compact single nuclei galaxies, while high $\Psi$ means clumpy multiple nuclei galaxies. For example, see Figure 3 of \citet{lotz:04} and Figure 10 of \citet{law:07a} for how $G$, $M_{20}$, and $\Psi$ change with different HST morphologies. $G$, $M_{20}$, and $\Psi$ are listed in Table \ref{mor_table}.
\begin{deluxetable}{lccccc}
\tablecolumns{6}
\tablewidth{0pc}
\tablecaption{\ha{} Morphology Parameters}
\setlength{\tabcolsep}{0.02in}
\tablehead{
\colhead{ID} & 
\colhead{$r_g$\tablenotemark{a}} &
\colhead{$d$\tablenotemark{b}} &
\colhead{$G$\tablenotemark{c}} &
\colhead{$M_{20}$\tablenotemark{d}} & 
\colhead{$\Psi$\tablenotemark{e}} \\
 & [kpc] & [kpc]
}
\startdata
11655                   & 2.79    & \nodata & 0.22 & -1.29 & 2.25 \\
10633                   & $<$0.88    & \nodata & 0.14 & -0.88 & 0.32 \\
42042481                & 5.88    & \nodata & 0.19 & -1.18 & 5.16 \\
J033249.73              & 4.68    & \nodata & 0.11 & -0.74 & 13.78 \\
11169                   & \nodata & 8.97    & 0.18 & -0.86 & 10.11 \\
\hspace{1em} 11169E     & 2.60    & \nodata & 0.11  & -0.86 & 3.34 \\
\hspace{1em} 11169W     & 2.55    & \nodata & 0.21  & -1.30 & 2.30 \\
7187                    & \nodata & 9.19    & 0.14  & -0.90 & 12.98 \\
\hspace{1em} 7187E      & 2.33    & \nodata & 0.16  & -1.42 & 3.71 \\
\hspace{1em} 7187W      & 3.55    & \nodata & 0.10  & -0.61 & 13.34 \\
9727                    & 4.82    & \nodata & 0.12 & -0.96 & 6.10 \\
7615                    & 5.00    & \nodata & 0.11 & -0.69 & 13.11 \\
11026194                & 3.07    & \nodata & 0.14 & -0.85 & 4.95 \\
12008898                & \nodata & 17.35   & 0.29 & -1.47 & 7.50 \\
\hspace{1em} 12008898N  & 1.04    & \nodata & 0.22  & -1.07 & 1.05 \\
\hspace{1em} 12008898S  & 2.88    & \nodata & 0.29  & -1.23 & 3.61 \\
12019627                & \nodata & 24.20   & 0.18 & -1.08 & 17.42 \\
\hspace{1em} 12019627N  & 4.41    & \nodata & 0.17  & -1.15 & 14.61 \\
\hspace{1em} 12019627SE & 2.66    & \nodata & 0.18  & -1.29 & 4.56 \\
\hspace{1em} 12019627SW & 1.87    & \nodata & 0.15  & -0.94 & 1.72 \\
13017973                & 7.59    & \nodata & 0.09 & -0.67 & 16.31 \\
13043023                & 4.72    & \nodata & 0.11 & -0.87 & 9.80 \\
32040603                & 1.96    & \nodata & 0.21 & -1.33 & 0.53 \\
32016379                & 4.32    & \nodata & 0.17 & -0.76 & 8.62 \\
32036760                & 3.09    & \nodata & 0.16 & -1.21 & 0.79 \\
33009979                & \nodata & 16.37   & 0.31 & -1.37 & 7.66 \\
\hspace{1em} 33009979N  & 2.04    & \nodata & 0.16  & -1.06 & 2.80 \\
\hspace{1em} 33009979S  & 2.83    & \nodata & 0.31  & -1.58 & 1.58 
\enddata \label{mor_table}
\tablenotetext{a}{Radius of gyration by \ha{} flux with respect to the flux weighted centroid.}
\tablenotetext{b}{Distance between two components. When there are more than two components, it is the distance between the two farthest components.}
\tablenotetext{c}{Gini parameter on a segmentation map.}
\tablenotetext{d}{Second-order moment on a segmentation map.}
\tablenotetext{e}{Multiplicity parameter on a segmentation map.}
\end{deluxetable}

As discussed by \citet{law:09}, OSIRIS \ha{} morphologies are difficult to compare to high resolution rest-UV HST morphologies. IFS data typically have high background levels, and the special background reduction techniques employed by the OSIRIS pipeline results in highly customized segmentation maps (see \S \ref{red}). These segmentation maps are different from the ones commonly used for imaging data, such as a quasi-Petrosian isophotal cut \citep{abraham:07}. Even with all these techniques, we still are unable to achieve the same level of low brightness sensitivity as narrow band data, and as a result, our $G$ values are systematically lower than the rest-frame UV imaging data \citep[e.g.,][]{lotz:04, law:07b, law:09}. 

Because of the extremely narrow field of view of OSIRIS, there are no reference stars that can be used for astrometric calibration between HST and OSIRIS data. This is another uncertainty for morphological comparisons, but we included HST images in Figure \ref{kine_11655} when available, and we align the images by visual inspection. Exactly how HST-\ha{} alignments are done changes from source to source, and the alignment details can be found in Appendix \ref{sec_indiv}.

\section{Kinematics}
\subsection{Kinematic Maps}\label{sec_kine_map}
\begin{deluxetable*}{lccccc}
\tablecolumns{6}
\tablewidth{0pc}
\tablecaption{Kinematics Parameters}
\setlength{\tabcolsep}{0.02in}
\tablehead{
\colhead{ID} & 
\colhead{\sigmaoned\tablenotemark{a}} & 
\colhead{\sigmaave\tablenotemark{b}} & 
\colhead{\vshear\tablenotemark{c}} &
\colhead{\vshear/\sigmaave} &
\colhead{\s\tablenotemark{d}}\\
 & [km/s] & [km/s] & [km/s] 
}
\startdata
11655      & 100.8 $\pm$ 25.5 & 54.7 $\pm$ 3.0 (14.2) & 125.9 $\pm$  8.6 (19.6) & 2.30 $\pm$ 0.20 & 104.5 $\pm$ 5.4 \\
10633      &  58.4 $\pm$ 36.5 & 54.5 $\pm$ 4.0 (11.2) &   7.8 $\pm$  5.7 (14.3) & 0.14 $\pm$ 0.11 &  54.8 $\pm$ 4.0 \\
42042481   &  86.9 $\pm$ 15.6 & 66.6 $\pm$ 1.4 (14.8) & 179.4 $\pm$ 15.7 (25.7) & 2.70 $\pm$ 0.24 & 143.2 $\pm$ 9.9 \\
J033249.73 &  88.0 $\pm$ 33.2 & 71.0 $\pm$ 2.9 (15.1) &  97.0 $\pm$ 14.1 (26.6) & 1.37 $\pm$ 0.21 &  98.7 $\pm$ 7.2 \\
11169E     & 140.5 $\pm$ 23.0 & 96.5 $\pm$ 3.4 (13.6) & 125.5 $\pm$ 14.0 (25.9) & 1.30 $\pm$ 0.15 & 131.1 $\pm$ 7.1 \\
11169W     & 110.4 $\pm$ 13.7 & 88.0 $\pm$ 2.1 (10.1) &  57.4 $\pm$  9.8 (20.8) & 0.65 $\pm$ 0.11 &  96.9 $\pm$ 3.5 \\
7187E      &  85.9 $\pm$ 35.2 & 80.5 $\pm$ 2.5 (12.5) & 130.3 $\pm$ 12.6 (25.5) & 1.62 $\pm$ 0.16 & 122.3 $\pm$ 6.9 \\
7187W      & 190.6\tablenotemark{e} $\pm$ 107.5 & 62.0 $\pm$ 3.5 (18.9) & 239.8 $\pm$ 13.4 (29.9) & 3.87 $\pm$ 0.31 & 180.5 $\pm$ 9.0 \\
9727       &  65.6 $\pm$ 17.4 & 64.8 $\pm$ 2.5 (14.7) &  89.1 $\pm$ 11.3 (22.6) & 1.37 $\pm$ 0.18 &  90.4 $\pm$ 5.9 \\
7615       &  75.1 $\pm$ 24.6 & 66.1 $\pm$ 2.3 (16.6) &  89.0 $\pm$ 12.7 (25.0) & 1.35 $\pm$ 0.20 &  91.3 $\pm$ 6.4 \\
11026194   &  72.1 $\pm$ 21.7 & 64.0 $\pm$ 2.9 (15.4) &  71.9 $\pm$  9.7 (20.3) & 1.12 $\pm$ 0.16 &  81.7 $\pm$ 4.8 \\
12008898N  &  65.1 $\pm$ 52.0 & 61.5 $\pm$ 6.8 (14.4) &  62.6 $\pm$ 12.6 (20.6) & 1.02 $\pm$ 0.23 &  75.8 $\pm$ 7.5 \\
12008898S  &  68.2 $\pm$ 11.4 & 61.6 $\pm$ 1.2 ( 8.6) &  73.6 $\pm$  7.6 (14.3) & 1.19 $\pm$ 0.12 &  80.7 $\pm$ 3.6 \\
12019627N  &  72.8 $\pm$ 44.2 & 55.8 $\pm$ 5.0 (12.0) & 191.9 $\pm$ 14.3 (19.5) & 3.44 $\pm$ 0.40 & 146.7 $\pm$ 9.6 \\
12019627SE &  65.4 $\pm$ 23.4 & 48.0 $\pm$ 2.9 (18.1) &  71.2 $\pm$ 14.6 (23.7) & 1.48 $\pm$ 0.32 &  69.5 $\pm$ 7.7 \\
12019627SW &  51.9 $\pm$ 27.3 & 59.5 $\pm$ 4.0 (15.3) &  68.3 $\pm$ 11.9 (18.0) & 1.15 $\pm$ 0.21 &  76.6 $\pm$ 6.1 \\
13017973   &  62.6 $\pm$ 19.1 & 65.3 $\pm$ 1.9 (17.8) & 117.7 $\pm$ 15.6 (27.0) & 1.80 $\pm$ 0.25 & 105.8 $\pm$ 8.8 \\
13043023   &  59.5 $\pm$ 20.4 & 63.9 $\pm$ 1.9 (15.8) &  65.4 $\pm$  9.2 (24.7) & 1.02 $\pm$ 0.15 &  78.8 $\pm$ 4.1 \\
32040603   &  49.7 $\pm$ 19.4 & 55.1 $\pm$ 1.9 ( 8.1) &  40.5 $\pm$  8.4 (15.7) & 0.73 $\pm$ 0.15 &  62.1 $\pm$ 3.2 \\
32016379   &  54.0 $\pm$ 18.3 & 63.3 $\pm$ 2.0 (16.8) &  64.0 $\pm$ 12.7 (23.0) & 1.01 $\pm$ 0.20 &  77.8 $\pm$ 5.5 \\
32036760   &  53.1 $\pm$ 13.2 & 55.0 $\pm$ 1.9 (11.4) &  60.3 $\pm$ 12.7 (16.8) & 1.10 $\pm$ 0.23 &  69.6 $\pm$ 5.7 \\
33009979N  &  49.7 $\pm$ 19.2 & 49.9 $\pm$ 2.3 (11.4) &  43.0 $\pm$  8.2 (16.1) & 0.86 $\pm$ 0.17 &  58.4 $\pm$ 3.6 \\
33009979S  &  79.8 $\pm$ 16.8 & 61.0 $\pm$ 2.1 (13.2) & 128.9 $\pm$ 15.3 (22.6) & 2.11 $\pm$ 0.26 & 109.7 $\pm$ 9.1 
\enddata \label{kine_table}
\tablenotetext{a}{Gaussian width of 1D spectrum.}
\tablenotetext{b}{SNR weighted average of dispersion map. The errors in the weighted average are reported. We also report the median errors of the dispersion maps in parentheses. They represent typical error per spaxel.}
\tablenotetext{c}{\vshear = $1/2(v_{max}-v_{min})$. The error is the error in \vshear. For a reference, the median errors of the rotation maps are reported in parentheses to represent typical error per spaxel.}
\tablenotetext{d}{\s = $\sqrt{0.5v_{\text{shear}}^2+\sigma_{\text{ave}}^2}$. Note that \s{} is uncorrected for an inclination while $S_{0.5}$ is corrected for an inclination \citep{kassin:12}.}
\tablenotetext{e}{This component has double peak that cannot be separated spatially. See \S \ref{red} and Appendix.}
\end{deluxetable*}
We create kinematic velocity maps of star-forming regions by fitting a Gaussian profile to the \ha{} emission line in each spaxel. Intensity, width, center position, and constant offset are fitted, and these parameters are then converted to physical quantities of interest. The radial velocity map is obtained from the peak position with respect to \ha{} at the systemic redshift ($z_{\text{sys}}$ in Table \ref{flux_table}). The velocity dispersion map is calculated from the width of the Gaussian function, corrected for the spatially varying instrumental resolution (see \S \ref{red} for instrumental width). The third and last panels of Figure \ref{kine_11655} in Appendix \ref{sec:kinemapfigure} show our radial velocity and velocity dispersion maps. For these kinematics maps, we apply the same segmentation criteria as those specified in \S \ref{obs}.

We measure the SNR weighted averages of velocity dispersion, \sigmaave{} \citep[sometimes referred as \sigmamean;][]{law:09, wisnioski:11}, in our segmented kinematics maps. Since it excludes the global velocity gradient, it represents a more accurate measurement of the line-of-sight velocity dispersion compared to \sigmaoned. However, the gradient within a pixel, 0.1'' per spaxel, beam smearing, and weighting method can still potentially bias the value.

In addition to velocity dispersion, we also measure the velocity shear, \vshear, which is defined as a half of the maximum difference in rotational velocity, 0.5($v_{\text{max}}$ - $v_{\text{min}}$), in a galaxy. Because the axis of rotation is not well defined in most of our galaxies, instead of $v_{max}$ and $v_{min}$ being maximum and minimum velocities along the kinematic major axis \citep[e.g,][]{forster:06, law:09}, we use velocities in the main bodies of the galaxies. In order to avoid possible outliers and artifacts, we use a modified version of the method by \citet{goncalves:10}. We calculate $v_{max}$ and $v_{min}$ as the mean of the highest and lowest 3 values. Given that the inclinations of the galaxies are not well constrained, and that the depth of observation is not sufficient to detect the full spatial extent, some galaxies do not show obvious disk-like velocity gradients. For these galaxies, \vshear{} represents the best possible unbiased rotation measurement. We discuss the effect of smoothing on the kinematics in Appendix \ref{sec:smooth}. \vshear, \sigmaave, the ratio \vshear/\sigmaave, and \sigmaoned{} are listed in Table \ref{kine_table}.

In Table \ref{kine_table}, we also report a combined velocity scale, $S_K$. This is a velocity indicator for tracing galaxy potential well depths proposed by \citet{weiner:06}, and defined as $S_K \equiv \sqrt{Kv^2+\sigma^2}$. We adopted $K$ = 0.5 for a flat rotation curve whose density profile is $\propto r^{-2}$. We use the notation of \s{} $= \sqrt{v_\text{shear}^2 + \sigma_\text{ave}^2}$ to emphasize the difference between inclination uncorrected \vshear{} and inclination corrected $V_{rot}$ for $S_{0.5}$ \citep{kassin:12}. Both \s{} and \sigmaoned{} describe the total kinematic/potential energy of the galaxy and should have similar values, and they can serve as a consistency check. Most sources have similar values between \s{} and \sigmaoned. For a few cases when they are significantly different, those sources with high \vshear{} are likely interacting or dominated by low SNR regions in the data.

\begin{figure*}
\centering
\includegraphics[width=0.99\textwidth]{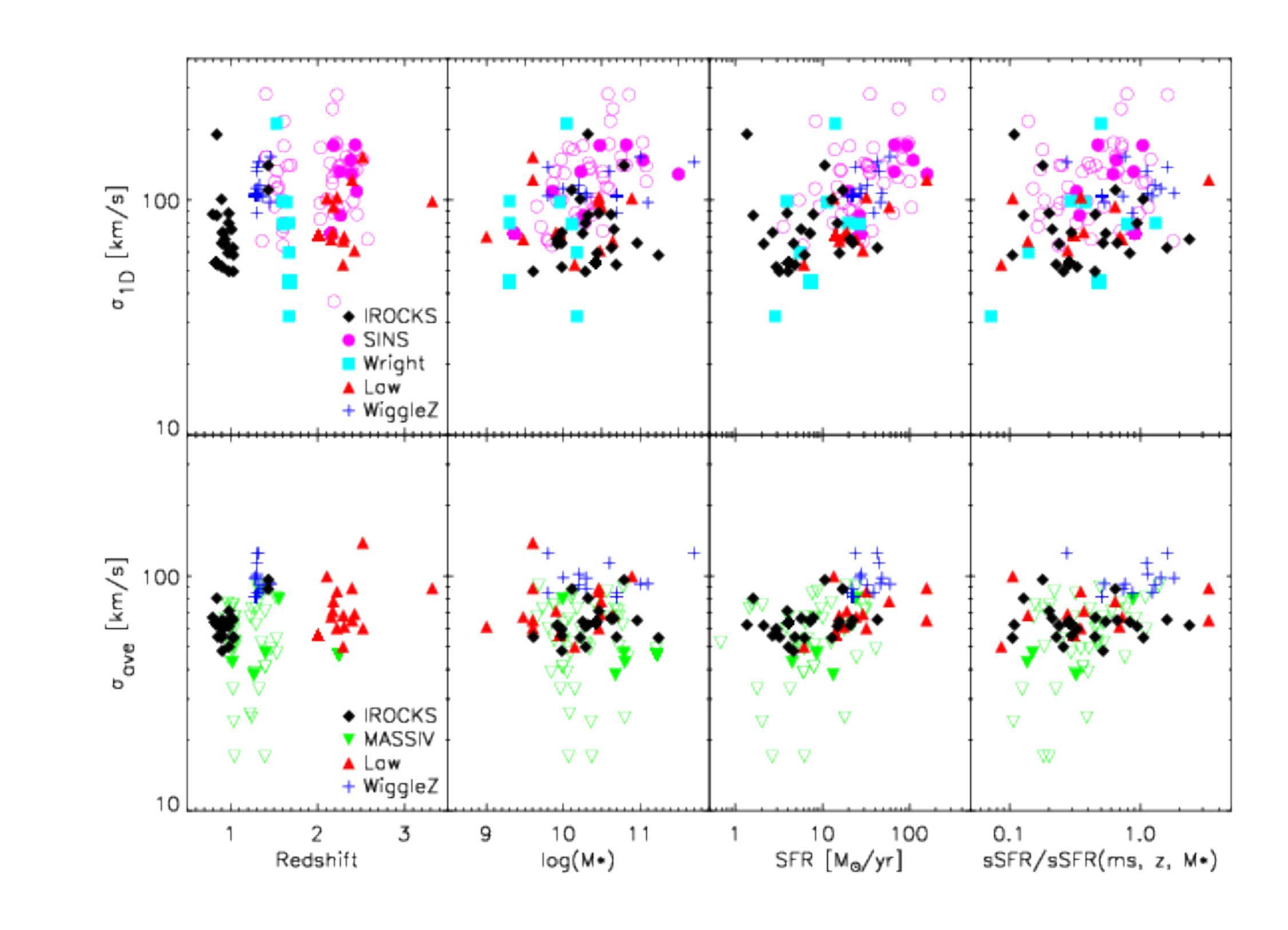}
\caption{Dispersions \sigmaoned{} (top) and \sigmaave{} (bottom) measured by IROCKS and other IFS studies, SINS \citep{forster:09}; \citet{wright:09, law:09}; WiggleZ \citep{wisnioski:11}; and MASSIV \citep{epinat:12, queyrel:12} as a function of redshift, stellar mass, SFR, and specific SFR normalized to the \citet{genzel:15} version of the star-formation main sequence of \citet{whitaker:12} (see Equation \ref{eq_ms}). The symbols whose colors and shapes are the same but are open/filled are the difference between non-AO/AO within the same survey. The average error of IROCKS \sigmaoned{} is 28.2 km s$^{-1}$, and \sigmaave{} is 2.5 km s$^{-1}$.}
\label{other_para_comp}
\end{figure*}
Figure \ref{other_para_comp} shows how \sigmaoned{} and \sigmaave{} change in redshift, stellar mass, star formation rate, and normalized specific star formation rate \citep[see Equation \ref{eq_ms},][]{whitaker:12, genzel:15}. Measurements of \citet{wright:09}, \citet{law:09}, \citet{forster:09}, \citet{wisnioski:11}, and \citet{epinat:12} are also shown. While our \sigmaoned{} spans a similar, wide range of 49 \textless \sigmaoned{} \textless 150 km s$^{-1}$, as other surveys, our \sigmaave{} spans very narrow range at lower values than other surveys. The narrow range may be due to the observational limitation. 50 km s$^{-1}$ corresponds to $\sigma$ = 1.5 channel. In the low SNR regime we work in, a line narrower than this width is difficult to distinguish from noise spikes, and what we see in IROCKS sample at $z \sim 1$ might be an upper limit. Interestingly, our only z $\sim$ 1.4 source (TKRS11169) shows a higher dispersion, \sigmaave{} $\sim$ 90 km s$^{-1}$, on both east and west components. A previous X-ray observation \citep{alexander:03} identified this source as AGN, and high dispersion is consistent with AGN narrow line region kinematics. However, it is unlikely that both components each host an ANG. Thus, the high dispersion we observed is most likely the kinematic evolution (higher dispersion at higher redshift) seen in the other surveys.

\citet{weiner:06, weiner:06b} show that \sigmaoned{} is relatively robust against observational effects to measure internal kinematics of galaxies, and can be used to study the Tully-Fisher (TF) relation with a large scatter. The second and third panels of Figure \ref{other_para_comp} on the top row is a representative of the TF relation.  Our sample shows an increasing \sigmaoned{} with both an increasing stellar mass and SFR. However, to properly conduct an investigation of the TF relation, we need to greatly increase the number of disks in our sample (of order hundreds to thousands) to overcome the intrinsic scatter.

The right most panels of Figure \ref{other_para_comp} show that most samples have specific SFR below the main sequence. This is because we only apply ISM only extinction to estimate SFR to be consistent with other surveys (see \S \ref{sec_sed}). By applying extra-attenuation, the SFR of IROCKS sources increase by, on average, a factor of 1.8. Assuming galaxies from other surveys also get a factor of $\sim$ 2 increase in SFRs, the center of normalized sSFR is shifted to around 1. After the correction, IROCKS and other high-z samples are near the main sequence within an order of magnitude.  Samples in KMOS$^{3D}$ \citep{wisnioski:15} and KROSS \citep{stott:16} surveys also span similar range. We discuss the comparison between the IROCKS, KMOS$^{3D}$, and KROSS kinematics in \S \ref{kine_discussion}.

The $z \sim 1$ sample spans line-of-sight velocity dispersions of 48 $\lesssim$ \sigmaave{} $\lesssim$ 80 km s$^{-1}$, velocity shears of 40 $\lesssim$ \vshear{} $\lesssim$ 192 km s$^{-1}$, and combined velocity scales of 58 $\lesssim$ \s{} $\lesssim$ 147 km s$^{-1}$ (excluding 10633 and 7187W, see \S \ref{red} and \S \ref{morph}). We will further discuss kinematic properties, in particular, disk settling using \vshear/\sigmaave{} values in \S \ref{kine_discussion}.

\subsection{Disk Fits}\label{kine_model}
\begin{figure*}[p]
\centering
\includegraphics[width=0.60\textwidth]{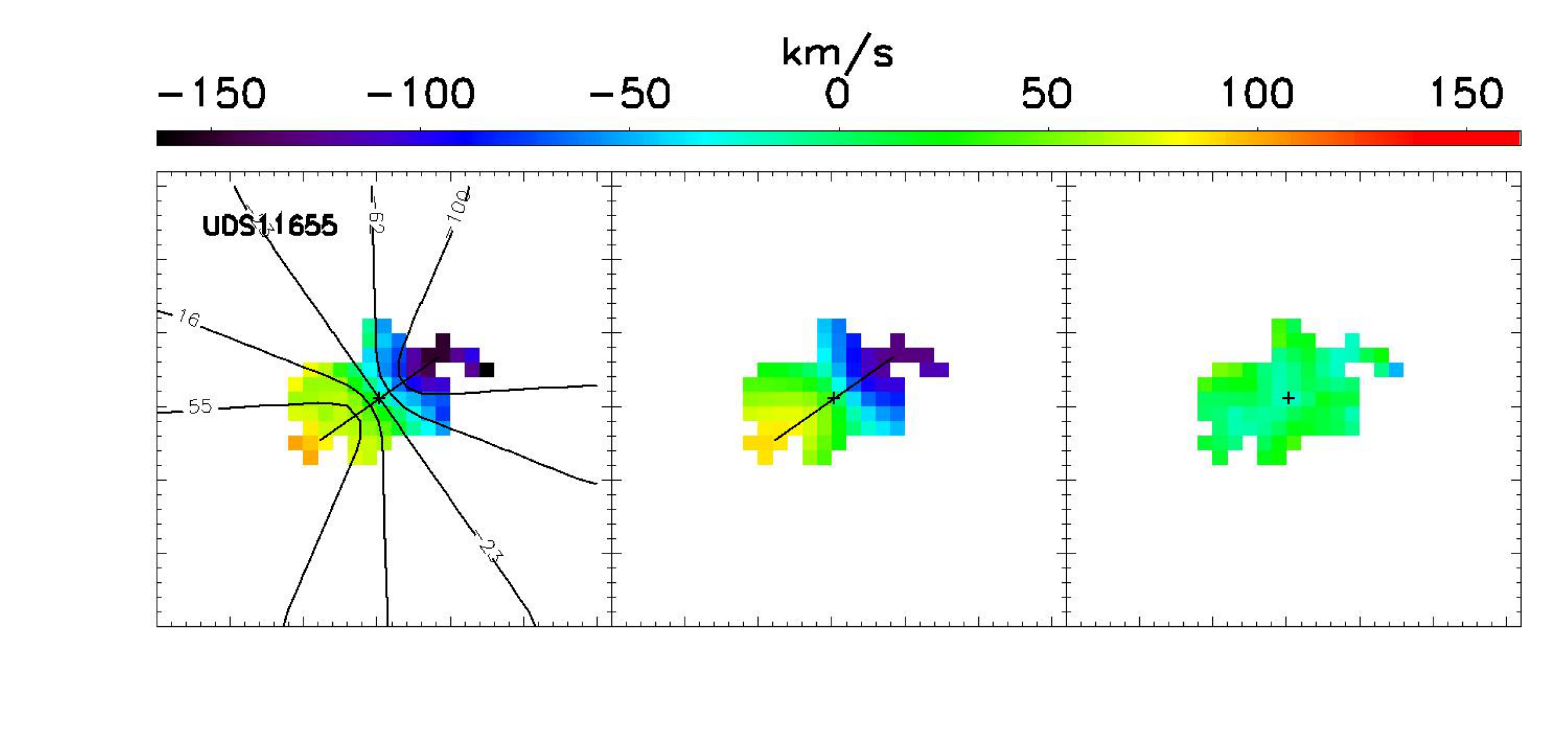}
\includegraphics[width=0.35\textwidth]{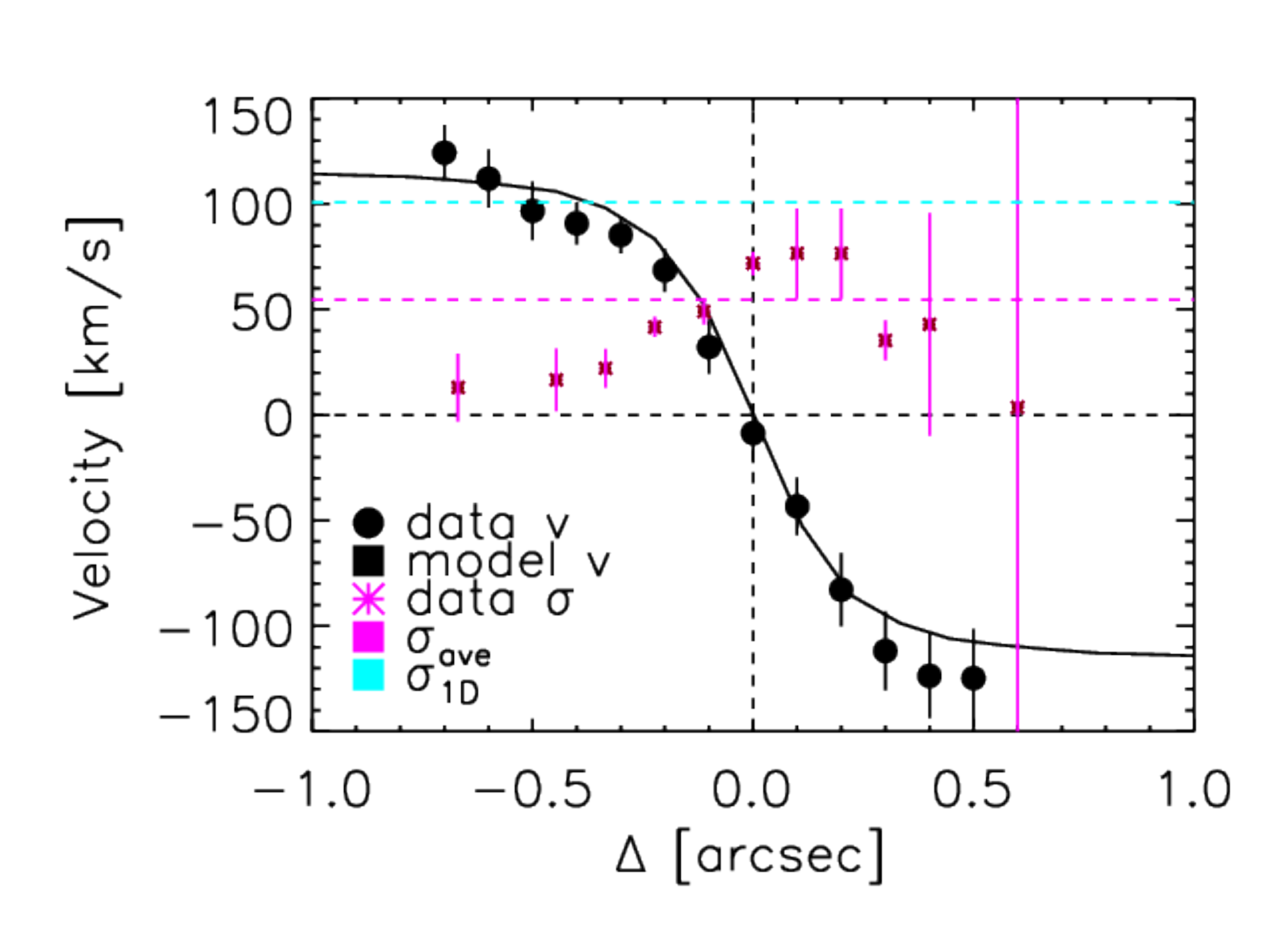}
\includegraphics[width=0.60\textwidth]{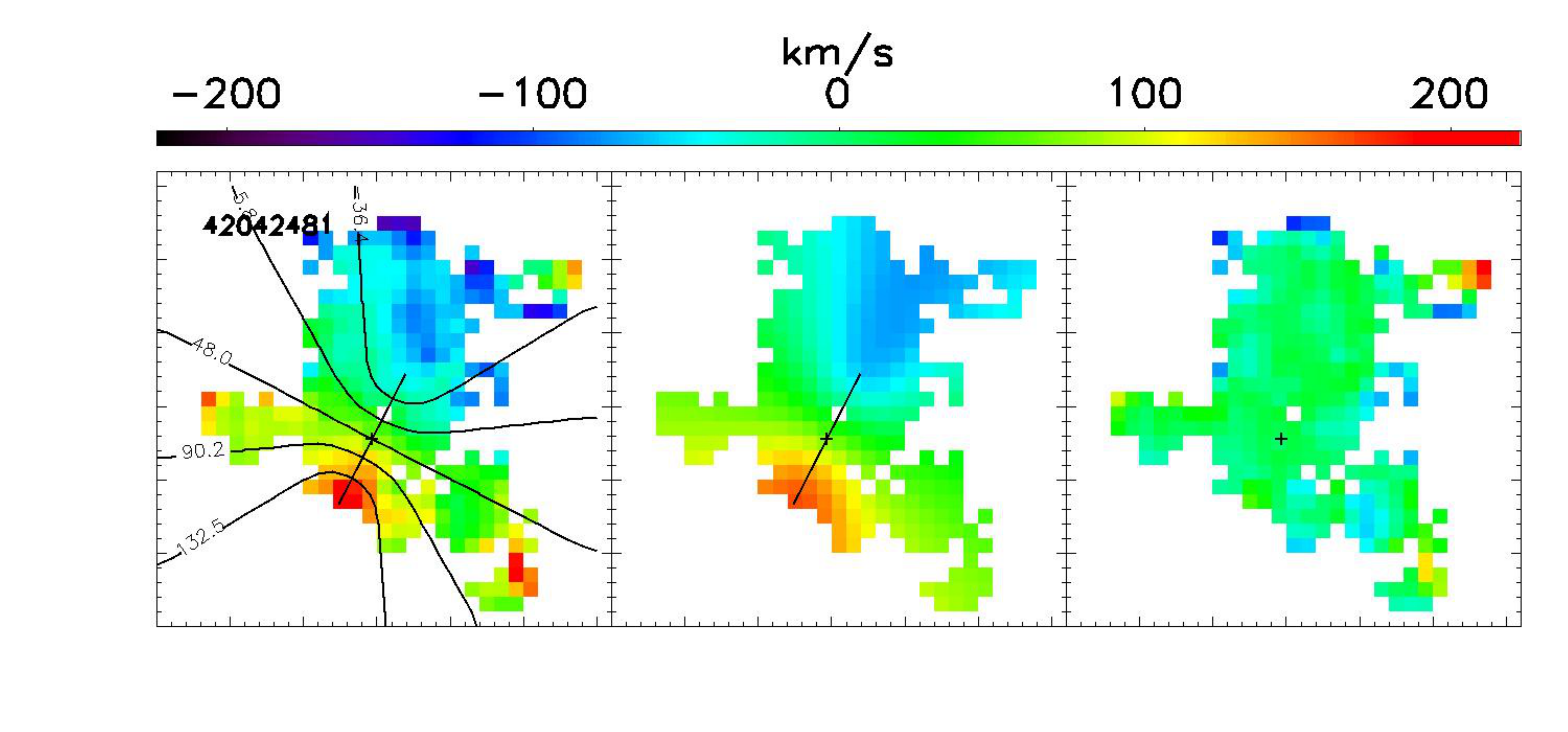}
\includegraphics[width=0.35\textwidth]{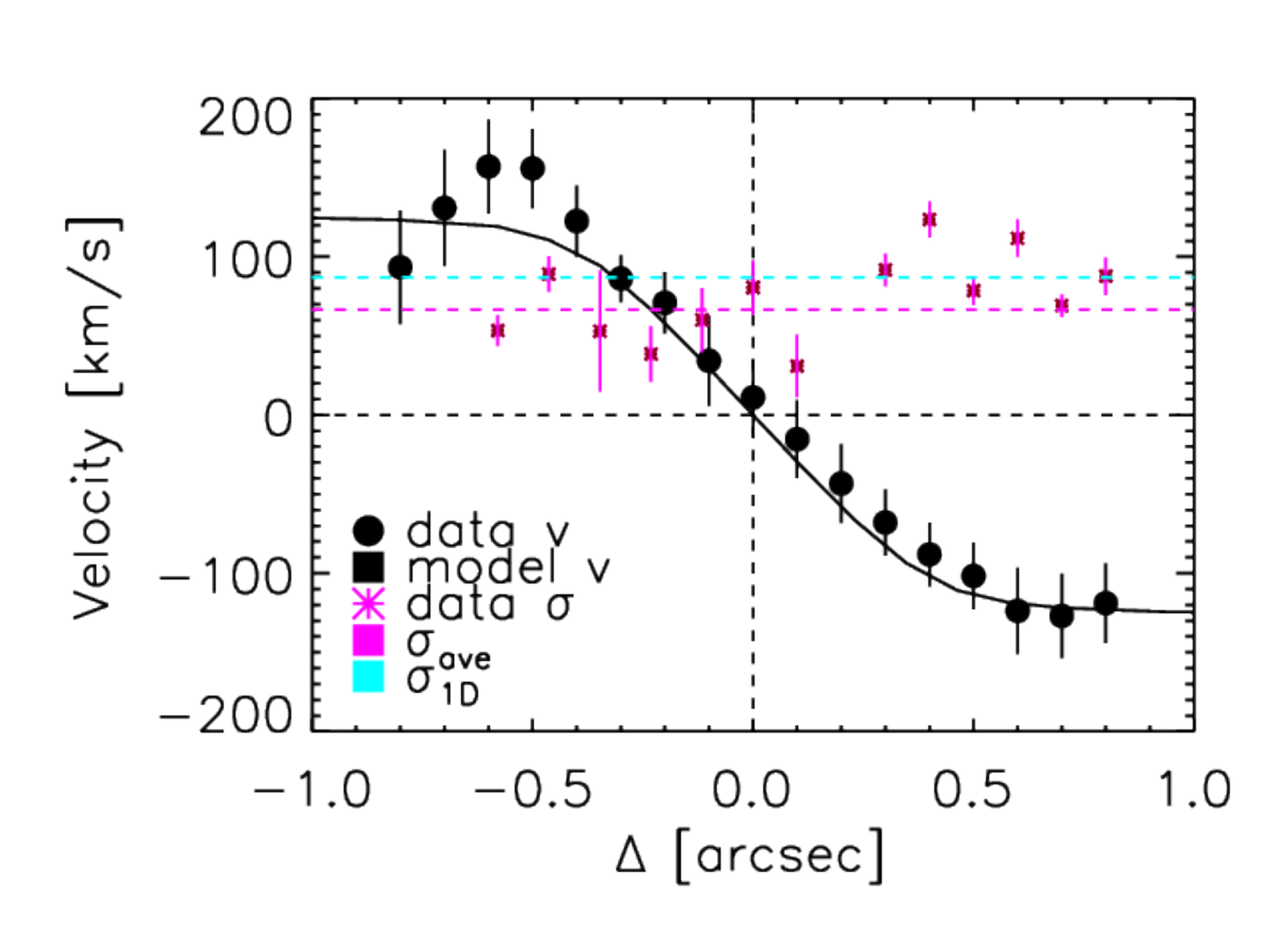}
\includegraphics[width=0.60\textwidth]{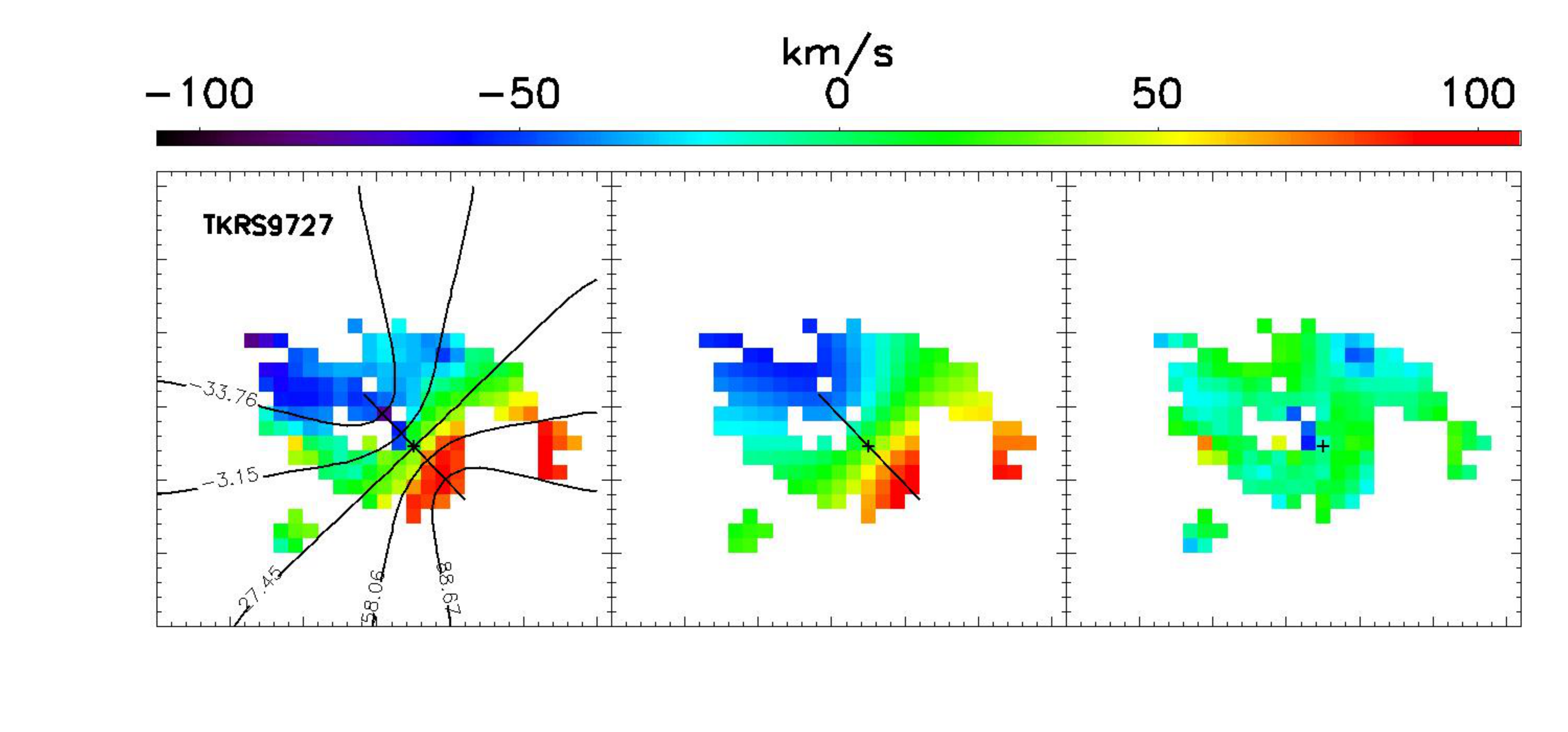}
\includegraphics[width=0.35\textwidth]{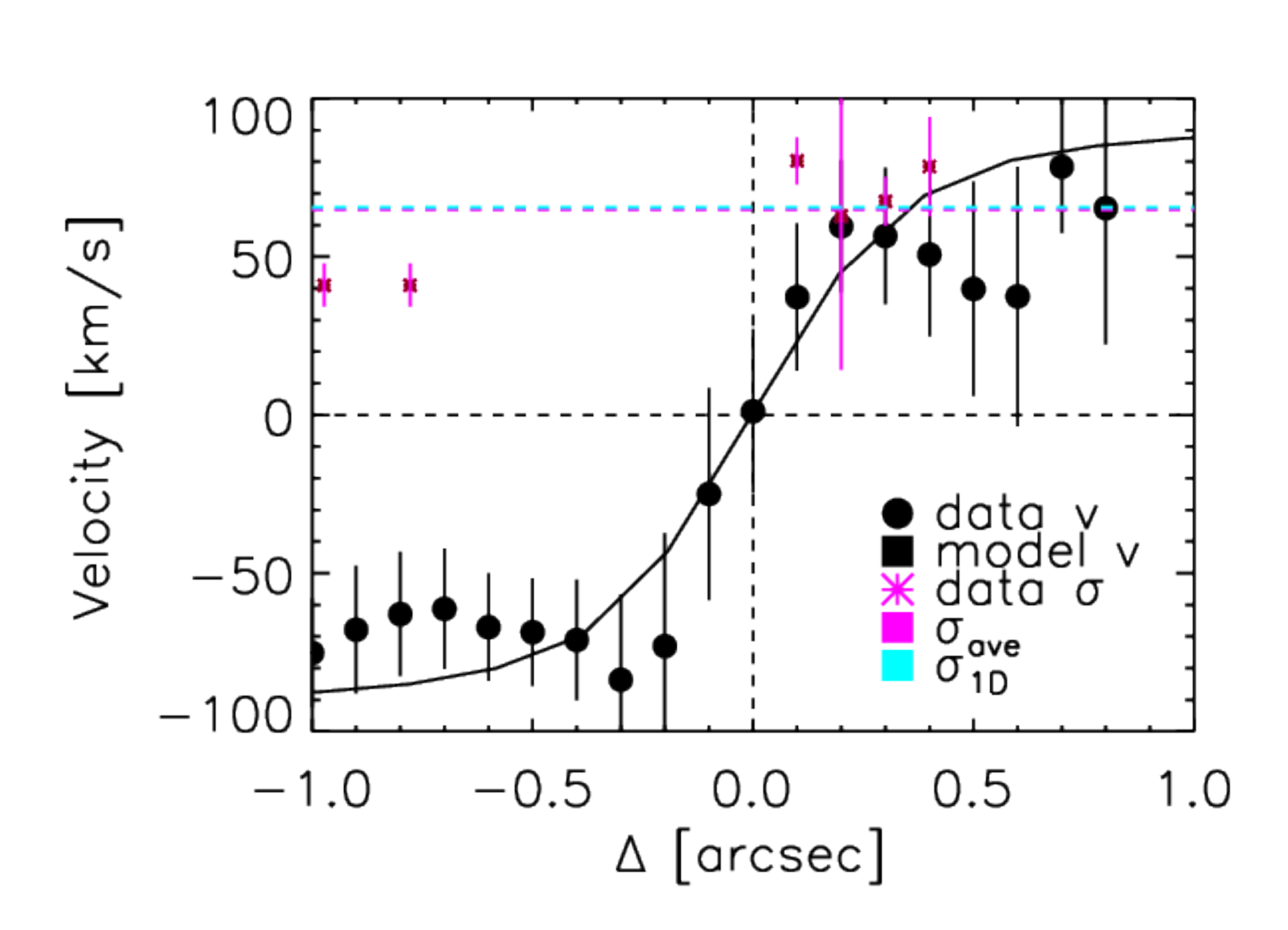}
\includegraphics[width=0.60\textwidth]{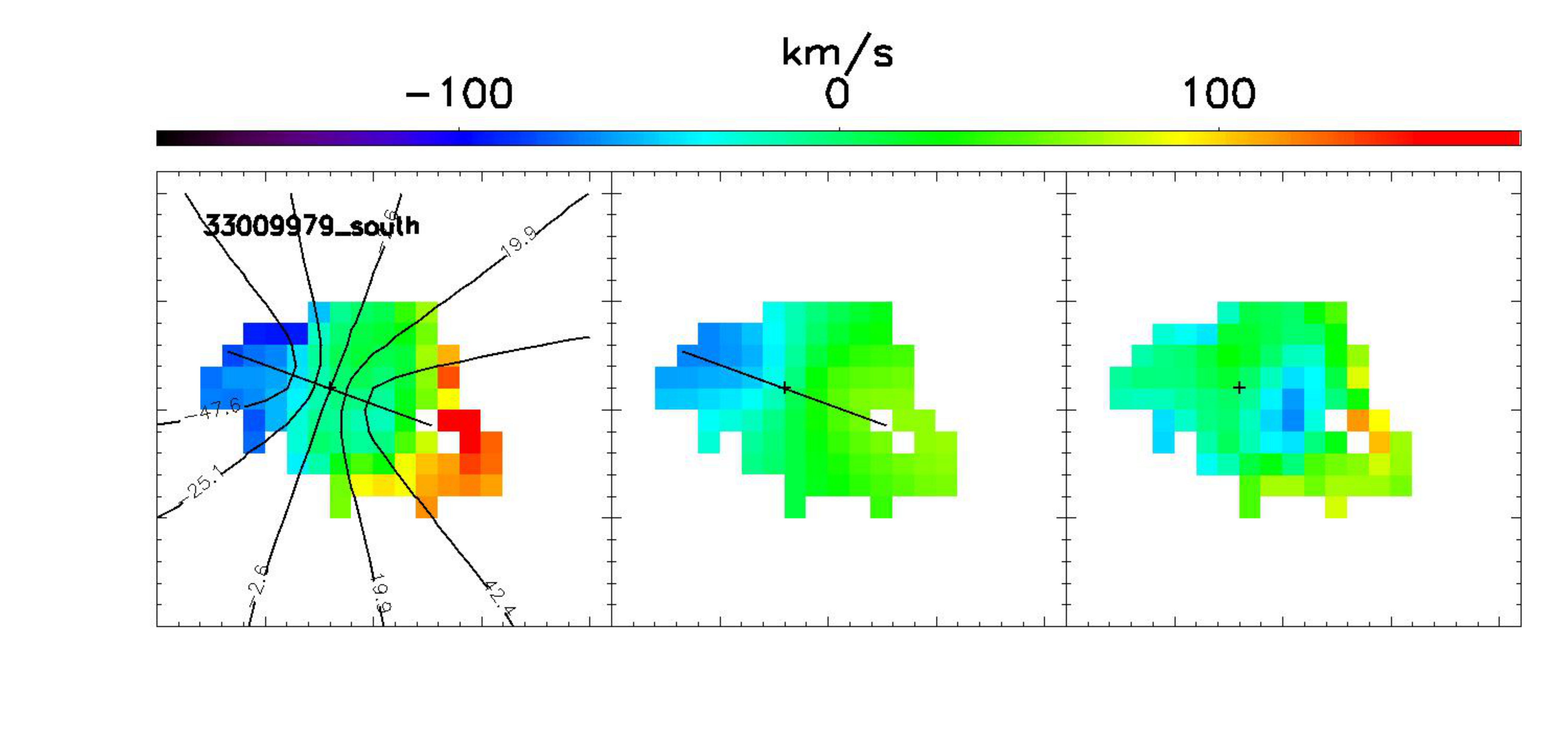}
\includegraphics[width=0.35\textwidth]{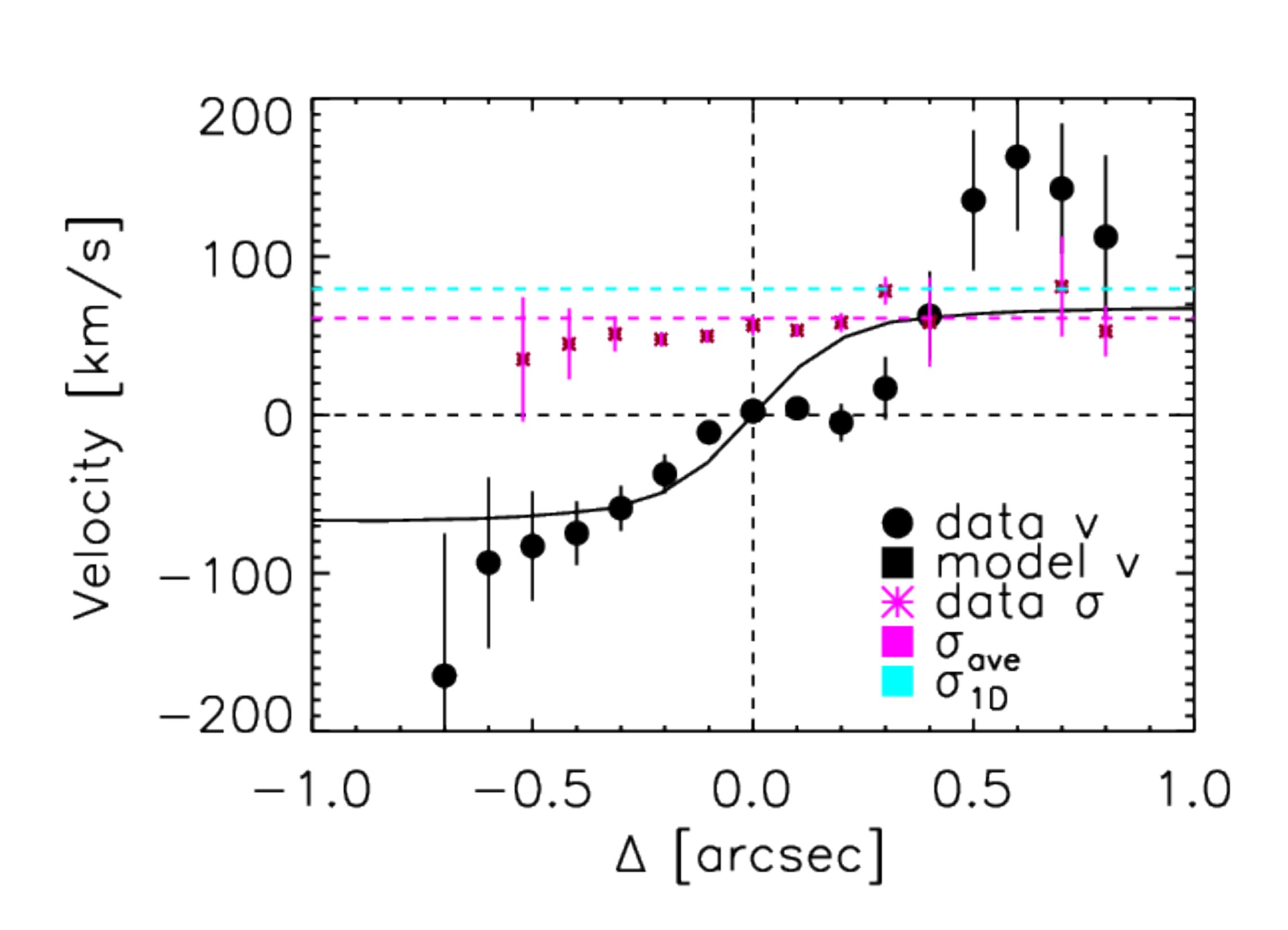}
\caption{Kinematic inclined-disk best fit to four $z \sim 1$ galaxies in our sample (UDS11655, DEEP2-42042481, TKRS9727, and DEEP2-33009979S). Shown on the left panels are the observed radial velocity (left), fitted inclined disk model (middle), and the residual between observed and model radial velocities (right). Plus sign ($+$) shows the dynamical center, and the black straight line shows the direction of velocity gradient. The figure on the right shows the observed (black filled circle) and fitted model (black line) rotation curve. The line-of-sight dispersion (magenta asterisks), \sigmaave{} (magenta dashed line), and \sigmaoned{} (cyan dashed line) are overplotted. The velocity field of DEEP2-33009979S near the center behaves differently compared to the rest of the main body. Because our fitting algorithm cannot capture such complicated structure, we enforce the dynamical center to be at the \ha{} flux peak and the plateau radius to be within the detected area.}
\label{kine_model_figure}
\end{figure*}
Following a disk fitting analysis by \citet{wright:09}, we fit an inclined disk model to all galaxy's radial velocity map to determine if it is consistent with a disk galaxy. The disk model we use is a tilted ring algorithm for a symmetrically rotating disk \citep{begeman:87}, which contains seven parameters; the center of rotation in the sky coordinates ($x_0$, $y_0$), position angle (PA) of the major axis ($\phi$), inclination angle ($i$), velocity slope ($m_v$), radius at which the plateau velocity is achieved in the plane of the disk ($R_p$), and systemic velocity offset ($v_0$). The observed radial velocity in the sky coordinates is described by:
\begin{equation}
v(x, y) = v_0 + V_c(R)\sin(i)\cos(\Theta), 
\end{equation}
where $R$ and $\Theta$ are the polar coordinates in the plane of the galaxy, and $V_c$ is the azimuthally symmetric circular velocity. $\Theta$ is related to the other parameters as follows:
\begin{equation}
\cos(\Theta) = \frac{-(x-x_0)\sin(\phi)+(y-y_0)\cos(\phi)}{R}
\end{equation}
\begin{equation}
\sin(\Theta) = \frac{-(x-x_0)\cos(\phi)-(y-y_0)\sin(\phi)}{R\cos(i)}.
\end{equation}
This model defines for a given radius, $R$, from the center in the plane, the velocity profile is increasing linearly, until it reaches the plateau velocity, $V_p$, at a plateau radius,$R_p$:
\begin{equation}
V_c =
\begin{cases}
~m_vR & \quad \text{if } R < R_p, \\
~ V_p = m_vR_p & \quad \text{if } R \geq R_p. \\
\end{cases}
\end{equation}
Since the observed velocity map is a velocity field convolved with a PSF, we also convolve our model with a Gaussian profile whose FWHM is the summation in quadrature of the un-smoothed TT star FWHM and smoothing FWHM used in the science data (Table \ref{obs_summary}). 
\begin{deluxetable}{lccccc}
\tablecolumns{6}
\tablewidth{0pc}
\tablecaption{Kinematic Model Parameters}
\setlength{\tabcolsep}{0.02in}
\tablehead{
\colhead{ID} & 
\colhead{$P.A.$\tablenotemark{a}} &
\colhead{$R_{peak}$\tablenotemark{b}} &
\colhead{$V_p$\tablenotemark{c}} &
\colhead{$<\Delta>$\tablenotemark{d}} &
\colhead{\rchi\tablenotemark{e}}\\
 & [deg] & [kpc] & [km/s] & [km/s]
}
\startdata
11655    & 125.1 & 1.2 & 140.7 & 13.4 & 0.1 \\
42042481 & 152.6 & 3.2 & 151.7 & 23.6 & 0.4 \\
9727     & 223.8 & 0.5 & 109.9 & 13.2 & 0.1 \\
33009979S\tablenotemark{f} & 249.9 & 0.5 & 81.7 & 30.8 & 0.7
\enddata \label{kine_model_table}
\tablenotetext{a}{Position angle}
\tablenotetext{b}{Radius where the rotational velocity reaches its peak}
\tablenotetext{c}{Plateau velocity, $V_p = m_vR_p$.}
\tablenotetext{d}{Average residual of $|$observed - model$|$ kinematics.}
\tablenotetext{e}{Reduced $\chi^2$ between observed and model velocity field.}
\tablenotetext{f}{Dynamical center is forced to be the \ha{} peak.}
\end{deluxetable}
Since \ha{} detections only represent the regions of on-going star formation, which is not necessarily distributed uniformly in the disk, we cannot satisfactorily set a constraint on the inclination angle from the \ha{} morphology alone. Use of deep HST images to determine the inclination angle may be more robust, but we do not have HST images for all of our sample, and instead, we fix the inclination angle to be an expectation value, $<i>$, of 57.3$^\circ$ \citep[e.g.,][]{law:09} to be consistent throughout our sample. This reduces the number of final fitted parameters to be six. The best fit model is determined by the least-square method, weighted by error. Among the 23 components in our 17 IROCKS sources, four (11655, 42042481, 9727, and 33009979S) are well fitted by a disk model. We note that one of the four, 33009979S, has a velocity field that behaves differently near the center of the system compared to the rest of the main body. Our simple model does not fully capture its complex velocity pattern, and our fitting algorithm does not easily converge. To aid with numerical convergence, we enforce the dynamical center to be at the \ha{} flux peak. Additionally, we enforce the plateau radius to be within the detected area for our later analysis in \S \ref{sec:diskmass}. Given priors on these values, these constraints yield the lowest converged $\chi^2$ parameter space with the most realistic values for this source. The resultant disk parameters, average residuals, and reduced $\chi^2$ values are listed in Table \ref{kine_model_table}.

In Figure \ref{kine_model_figure}, the observed velocity maps, best fit models, and residuals are shown on the left with the projected major-axis rotation curves on the right. On the rotation curve, the line-of-sight dispersion, \sigmaave{}, and \sigmaoned{} are overplotted. The line-of-sight dispersion of UDS 11655 peaks at the center and flattens at the large radius. This dispersion profile is similarly seen in the majority of the KMOS$^{3D}$. One possible reason for this dispersion profile is beam smearing \citep{newmans:13}, but our other three disk candidates show essentially flat dispersion profiles. In fact, regardless of their kinematic classification, the majority of our sample show a flat dispersion across the spatial extent of the galaxies (see the rightmost panels in Figure \ref{kine_11655}). This most likely indicates that our kinematics are not too affected by beam smearing, which may confuse the kinematic classification.

\subsection{Disk Settling}\label{kine_discussion}
\begin{figure*}[t]
\centering
\includegraphics[width=0.9\textwidth]{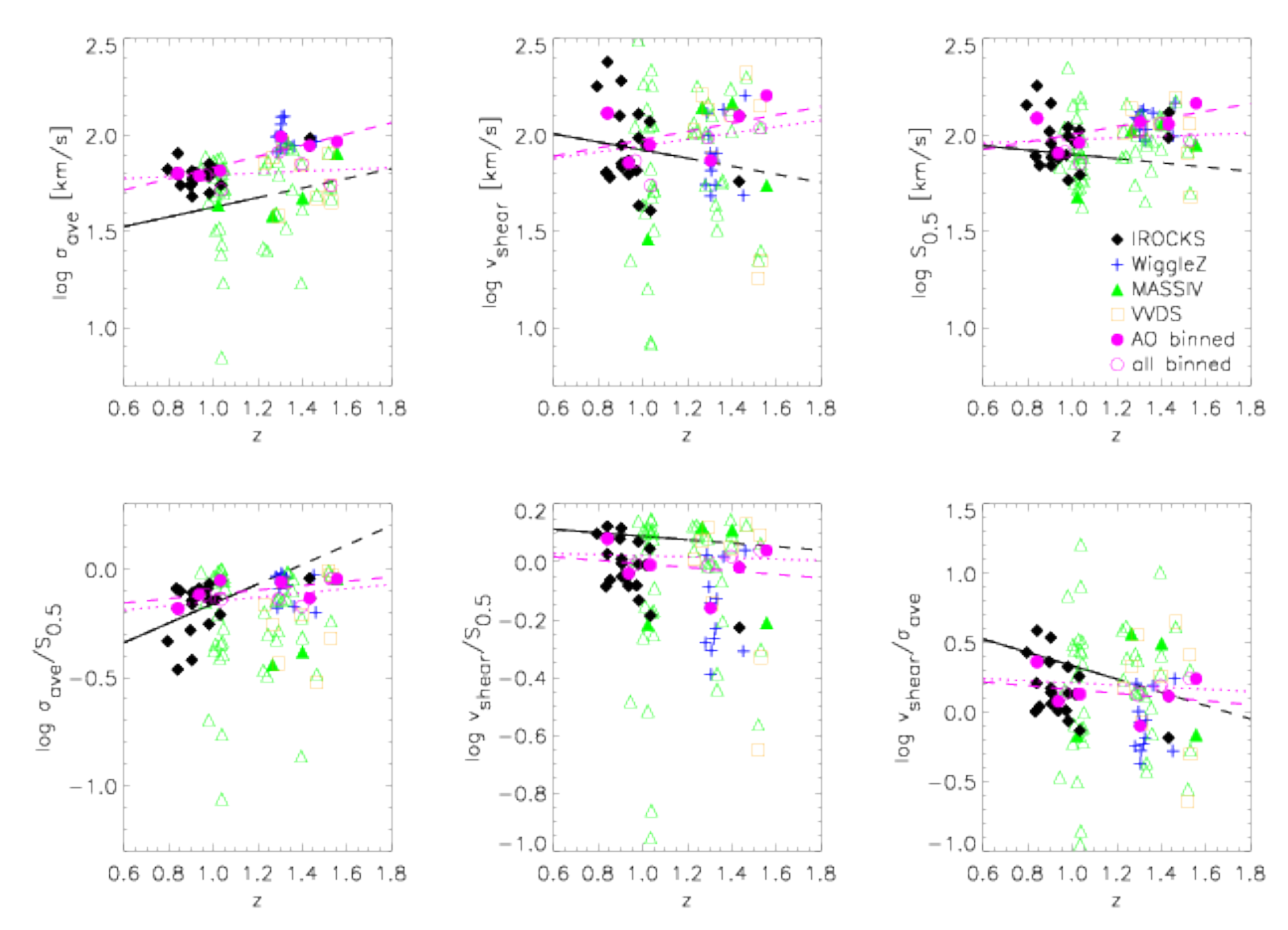}
\caption{Evolution of \sigmaave, \vshear, \s, and their ratios measured by IROCKS and other IFS high redshift galaxy studies, VVDS \citep{epinat:09}; WiggleZ \citep{wisnioski:11}; and MASSIV \citep{epinat:12}. The symbols whose colors and shapes are the same but are open/filled denote non-AO/AO observations within the same survey. Relationships found by the 1D spectrum study of \citet{kassin:12} at $0.2 < z < 1.2$ are over-plotted as a black line. Black dashed lines are extrapolations of \citet{kassin:12} beyond $z>1.2$. Binned median of all AO data combined are shown as magenta filled circles. Linear fits to the binned AO medians are shown as magenta dashed lines. Binned median of all data, both AO and non-AO, are shown as magenta open circles, and linear fits are shown as magenta dotted lines.}
\label{kine_flux_mass}
\end{figure*}

Resolved measurements of kinematics allow one to probe the process of disk settling that leads to present-day spiral galaxies. In Figure \ref{kine_flux_mass}, we compare the IROCKS sample's \sigmaave, \vshear, and \s{} values as a function of redshift with those reported by other high-z surveys \citep{epinat:09, epinat:12, wisnioski:11}. Note that \vshear{} values plotted here for \citet{epinat:09, epinat:12} are the plateau velocities obtained from their kinematic fitting. For comparison, we removed the inclination correction in their calculated values to be consistent with both our sample and \citet{wisnioski:11} data points. We also plot the relationship found by the 1D long-slit study of \citet{kassin:12} at $0.2 < z < 1.2$, for mass limited sample ($9.8 < \log{M}($\msun$) < 10.7$), as black lines for comparison. For a galactic disk to be considered settled, one expects its organized motion in rotation to dominate over random motion, hence \vshear/\sigmaave{} $>>$ 1. This quantity as a function of redshift is shown in the lower right panel of Figure \ref{kine_flux_mass}.

Our measurements deviate from the kinematic relationships found by \citet{kassin:12}: we generally find a higher velocity dispersion, and lower \vshear/\sigmaave{} ratio. Most components in the IROCKS sample have \vshear/\sigmaave{} $\sim$ 1, and only five have \vshear/\sigmaave{} \textgreater 2. If we apply the definition of settled fraction proposed by \citet{kassin:12} (\vshear/\sigmaave{} \textgreater 3), this fraction in our sample would be 2/21, or $\sim10\%$, which is lower than the disk fraction expected. Some of this discrepancy may be reconciled by a difference in the \vshear{} definition: \citet{kassin:12} correct their \vshear{} values for inclinations between $30^\circ < i < 70^\circ$, using axis ratios of $V+I$ band HST images, while we do not include any inclination dependence in ours. This difference accounts for at most a factor of two increase in \vshear{} values, which may be one of the reasons why our settled fraction appears to be low. Besides inclination effects, our \vshear{} measurements are similar to those of \citet{kassin:12}, implying that the velocities in our sample, both in rotation and dispersion, are higher than their sample.

While \vshear{} measurements may be ambiguous due to the lack of inclination information, the elevated dispersion we measure is robust and consistent with previous IFS+AO studies that observed elevated dispersions compared to local ($z=0$) galaxies. In fact, looking at IFS binned median for AO only data (magenta filled circles with linear fit by dashed lines, see upper-left panel in Figure \ref{kine_flux_mass}) and for all combined data (magenta open circles with linear fit by dotted lines), our combined results show a steady decrease of \sigmaave{} with decreasing redshift and increase of \vshear/\sigmaave, consistent with the picture of disk settling, except for the MASSIV survey non-AO data. The median linear fit used on IFS binned data shows a shallower slope, but in general the same trend matches that of \citet{kassin:12}, except for \vshear{} and \s. 

Possible sources of discrepancies in kinematic parameters and their trends between \citet{kassin:12} and IFS surveys may due to differing sample selections and definitions of derived values. While \citet{kassin:12} uses a mass limited sample ($9.8 < \log{M}($\msun$) < 10.7$), we use all available IFS data points in which roughly 20\% are outside of this mass range (see the second panels of Figure \ref{other_para_comp} for rough estimate). As shown in \citet{kassin:12}, more massive galaxies tend to settle earlier than less massive ones, and mixing different populations in IFS studies may lead to redshift trends being washed out. Furthermore, different definitions of kinematic parameters in IFS surveys may introduce varying systematics among surveys. Commonly among IFS studies, \sigmaave{} is an average of the velocity dispersion map that is derived from an emission line-width corrected for the instrumental dispersion; however, the definition of $v$ differs among IFS surveys. For instance, \vshear{} in our sample and \citet{wisnioski:11} are derived from the difference between the maximum and minimum in the velocity map, but \citet{epinat:09, epinat:12} use plateau velocities obtained from the kinematic model fitting. Also, usually IFS observations are less sensitive to galaxies at large radii than traditional seeing-limited spectrograph, and IFS \vshear{} values may be lower regardless of the calculating methods due to sensitivity differences.

A second method of determining a disk fraction is through disk fitting. Re-enforcing our conclusions from \S \ref{kine_model}, we found four components well-fitted by an inclined disk model. Indeed, three of these disk candidates have some of the highest \vshear/\sigmaave{} (\textgreater 2) in our sample, while the last one is a nearly face-on disk. Additionally, there are some components, such as DEEP12008898N and 33009979N, that show velocity gradients consistent with rotation by visual inspection, but their small sizes prevent reliable fitting. Overall, it is likely that the common notion that about one-third of the galaxies in high redshift samples are disk-like also applies in our $z\sim1$ sample, but we need finer sampling and deeper observations to confirm this.

The recent large seeing-limited IFS kinematic surveys KMOS$^{3D}$ \citep{wisnioski:15} and KROSS \citep{stott:16} have observed 90 and 584 $z\sim1$ galaxies, respectively. In the KMOS$^{3D}$ survey, the \sigmaave{} equivalent dispersion is measured from the outer region of galaxies to avoid rotation and beam smearing effects, and their average dispersion at $z\sim1$ is found to be 25 \kms. Their global rotation is corrected for the inclination and is defined as half of the difference between maximum and minimum, which is similar to our \vshear. Under these definitions, they find 70 to 93 \% of galaxies are rotation dominated. In the KROSS survey, the definition of dispersion is similar to our $\sigma^{\text{corr}}$ discussed in Appendix \ref{sec:smooth}, where the local velocity gradient is removed from the dispersion, and their sample average is 60 \kms. The global rotation is the average of velocities in the model velocity map at a radius 2.2 times the effective radius along the semimajor axis that is corrected for inclination. They find 83 \% of their sample as rotation dominated.

While our definition and sample average of dispersion is consistent with KROSS, the KMOS$^{3D}$ average dispersion definition is different from ours, and their final average value is a factor of two lower. Even if we apply the same method as KMOS$^{3D}$, the majority of the IROCKS sample has a flat dispersion and the average value would not be as low as their measured value. Also, in both studies, they find significantly higher disk fractions than IROCKS. In general, seeing-limited IFS observations are more sensitive to low surface brightness regions, and deeper observations by AO+IFS is warranted to fully compare the measurements in the outer regions of galaxies.

\section{Derived Masses}\label{mass}
In this section, we estimate the gas masses of our galaxies using the measured \ha{} fluxes. We then derive their virial masses using their kinematics. At last, for the four galaxies well fitted by disk models, we calculate their dark matter halo and enclosed masses using their fitted disk parameters.
\subsection{Gas Mass}
The gas mass ($M_\text{gas}$) of a galaxy can be expressed with respect to its gas depletion timescale ($t_{dep}$) as:
\begin{equation}
\label{gas_mass_to_t_dep}
M_{\text{gas}} = t_{\text{dep}}/\text{SFR} \, .
\end{equation}
We can obtain estimates for the gas masses of our galaxies by inferring their $t_{dep}$ from an empirical relationship between $t_{dep}$ and the specific star formation rate (sSFR) normalized to the star-formation main sequence (SFMS) \citep{genzel:15}:
\begin{equation}
\label{genzel_gas_mass}
\begin{split}
\lefteqn{\log(t_{\text{dep}}(z, \text{sSFR}, M_*)|_{\alpha=\alpha_{\text{MW}}})} \\
& = \alpha_{f} + \xi_{f}\log(1+z) + \xi_{g}\log(\text{sSFR}/\text{sSFR}(\text{ms}, z, M_*)) \\
& + \xi_{h}(\log(M_*) - 10.5) \, ,
\end{split}
\end{equation}
where $t_{dep}$ is in the units of [Gyr$^{-1}$], $M_*$ is the stellar mass in [$M_\odot$], and $\{\alpha_{f1}, \xi_{f1}, \xi_{g1}, \xi_{h1}\} = \{+0.1, -0.34, -0.49, +0.01\}$ are fit parameters. $\text{sSFR}(\text{ms}, z, M_*)$ is the specific star formation rate in the star-formation main sequence, which follows a fitted function \citep{whitaker:12}\footnote{The coefficients in this equation are different from the ones in the original equation in \citet{whitaker:12}: $\log(\text{sSFR}(\text{ms}, z, M_*)) = 0.38 + 1.14z - 0.19z^2 - (-0.7 + 0.13z)(\log{M_*} - 10.5)$. We use \citet{genzel:15} version to follow their method to estimate gas mass.}:
\begin{equation}
\label{eq_ms}
\begin{split}
\lefteqn{\log(\text{sSFR}(\text{ms}, z, M_*))} \\
& = -1.12 + 1.14z - 0.19z^2 - (0.3 + 0.13z)(\log{M_*} - 10.5),
\end{split}
\end{equation}
where sSFR is in the units of [Gyr$^{-1}$]. Combing Equations \ref{gas_mass_to_t_dep} to \ref{eq_ms}, we obtain our first gas mass estimates, which we denote $M_{gas, 1}$, and they are listed in Table \ref{mass_table}.

For comparison, we use an independent method to calculate a second gas mass estimate, which we denote $M_{gas, 2}$. The gas surface density ($\Sigma_{\text{gas}}$) is related to the SFR per area ($\Sigma_{\text{SFR}}$) by an empirical relation \citep{kennicutt:07}. Modified for a Chabrier IMF, it is:
\begin{equation}
\label{gas_mass_eq_genzel}
\log{\left(\frac{\Sigma_{gas}}{\text{M}_\odot \text{pc}^{-2}}\right)} = 0.73\log{\left(\frac{\Sigma_{SFR}}{\text{M}_\odot \text{yr}^{-1} \text{kpc}^{-2}}\right)}+2.91 \, .
\end{equation}
Replacing SFR by the observed \ha{} luminosity using Equation \ref{eq_sfr_ha}, the gas mass is:
\begin{equation}
\label{gas_mass_eq}
M_{gas, 2} = 1.27 \times 10^{-23}L_{\text{\ha}}^{0.73}A_{pc}^{0.27},
\end{equation}
where $A_{pc}$ is the area of a pixel in [parsec$^2$]. The values of $M_{gas, 2}$ are listed in Table \ref{mass_table}. This second method has the additional advantage of allowing us to convert a spatial SFR distribution to a gas distribution using Equation \ref{gas_mass_eq_genzel}. This enables us to investigate local properties of the galaxies, such as their gravitational stability (\S \ref{sec_toomre}). This is not possible with the first method, because we only have the global value for $M_*$. 

We note that we have elected to carry out the \citet{genzel:15} empirical estimate of gas mass to z$\sim$1 sample, since our group has verified that OSIRIS \ha{} emission of z$\sim$ 1.5 galaxies matches the estimated gas mass directly from Plateau de Bure Interferometer CO 3-2 observations. This gas mass estimate was in better agreement than the standard Kennicutt law used above. We thus show the gas fraction using the first method, $M_\text{gas, 1}/(M_*+M_{\text{gas, 1}})$, in Table \ref{mass_table}.

\subsection{Virial Mass}\label{sec:virial}
For a virialized system, the virial mass within a given radius, $r_{vir}$, can be estimated by assuming a symmetric gravitational potential. We use \sigmaoned{} to represent the kinetic energy of the system, which includes both global rotation and line-of-sight velocity dispersion. Then the virial mass can be written as:
\begin{equation}\label{eq:virial}
M_{vir} = \frac{C\sigma_{1D}^2r_{vir}}{G},
\end{equation}
where $G$ is the gravitational constant, and $C$ is a constant factor that represents the shape of the potential with respect to our viewing angle. For example, $C$ = 5 if the mass is uniformly distributed in a sphere, and $C$ = 3.4 if it is a uniform thin disk with an average inclination \citep[e.g.][]{erb:06a}. We use $C$ = 3.4 for our four disk candidate galaxies (UDS11655, 42042481, TKRS9727, and 33009979S), and $C$ = 5 for the rest of the sample. Since our galaxies have no clear boundaries (see \S \ref{morph}), we use the radius of gyration, $r_g$, as $r_{vir}$, although it is most likely an underestimate because $r_g$ decreases for more centrally concentrated galaxies.

\subsection{Masses for Disk Galaxies}
\label{sec:diskmass}
\begin{deluxetable*}{lccccccc}
\tablecolumns{8}
\tablewidth{0pc}
\tablecaption{Masses}
\setlength{\tabcolsep}{0.02in}
\tablehead{
\colhead{ID} & 
\colhead{$\log{\text{M}_*}$\tablenotemark{a}} &
\colhead{$\log{\text{M}_{\text{gas, 1}}}$\tablenotemark{b}} &
\colhead{$f_{\text{mol gas}}$\tablenotemark{c}} & 
\colhead{$\log{\text{M}_{\text{gas, 2}}}$\tablenotemark{d}} & 
\colhead{$\log{\text{M}_{\text{vir}}}$\tablenotemark{e}} & 
\colhead{$\log{\text{M}_{\text{halo}}}$\tablenotemark{f}} &
\colhead{$\log{\text{M}_{\text{enc}}}$\tablenotemark{g}}
}
\startdata
11655      & 10.22 & 10.32 & 0.56 & 9.82 & 10.35 & 11.80 & 10.49 \\
10633      & 11.24 & 10.26 & 0.09 & 9.78 &  9.54 & \nodata & \nodata \\
42042481   & 10.62 & 10.06 & 0.22 & 9.61 & 10.55 & 11.95 & 10.79 \\
J033249.73 & 10.46 &  9.72 & 0.15 & 9.39 & 10.62 & \nodata & \nodata \\
11169E     & 10.79 & 10.28 & 0.24 & 9.83 & 10.78 & \nodata & \nodata \\
11169W     & 10.11 & 10.45 & 0.69 & 9.96 & 10.56 & \nodata & \nodata \\
7187       & 10.32 &  9.78 & 0.22 & 9.41 & \nodata & \nodata & \nodata \\
\hspace{1em} 7187E      & \nodata & \nodata & \nodata & 9.22 & 10.30 & \nodata & \nodata \\
\hspace{1em} 7187W      & \nodata & \nodata & \nodata & 9.17 & 11.18 & \nodata & \nodata \\
9727       & 10.96 & 11.04 & 0.55 & 10.34 & 10.22 & 11.50 & 10.28 \\
7615       & 10.66 & 10.05 & 0.20 & 9.63 & 10.52 & \nodata & \nodata \\
11026194   & 10.25 & 10.12 & 0.43 & 9.68 & 10.27 & \nodata & \nodata \\
12008898N  & \nodata & \nodata & \nodata  & 9.22 & 9.71 & \nodata & \nodata \\
12008898S  &  9.92 & 10.49 & 0.79 & 9.95 & 10.19 & \nodata & \nodata \\
12019627   &  9.98 & 11.36 & 0.96 & 9.23 & \nodata & \nodata & \nodata \\
\hspace{1em} 12019627N  & \nodata & \nodata & \nodata & 9.31 & 10.43 & \nodata & \nodata \\
\hspace{1em} 12019627SE & \nodata & \nodata & \nodata & 9.23 & 10.12 & \nodata & \nodata \\
\hspace{1em} 12019627SW & \nodata & \nodata & \nodata & 9.34 & 9.77 & \nodata & \nodata \\
13017973   & 10.63 & 10.82 & 0.60 & 10.19 & 10.54 & \nodata & \nodata \\
13043023   & 10.44 & 10.49 & 0.53 & 9.95 & 10.29 & \nodata & \nodata \\
32040603   &  9.61 &  9.10 & 0.24 & 9.40 & 9.75 & \nodata & \nodata \\
32016379   & 10.42 &  9.76 & 0.18 & 9.51 & 10.17 & \nodata & \nodata \\
32036760   & 10.69 &  9.91 & 0.14 & 8.93 & 10.01 & \nodata & \nodata \\
33009979N  & \nodata & \nodata & \nodata & 9.47 & 9.77 & \nodata & \nodata \\
33009979S  & 10.29 & 10.39 & 0.56 & 9.88 & 10.15 & 11.08 & 10.40
\enddata \label{mass_table}
\tablenotetext{a}{Stellar mass from SED model}
\tablenotetext{b}{Total gas mass derived by the method of \citet{genzel:15}}
\tablenotetext{c}{Gas mass fraction by the method of \citet{genzel:15}}
\tablenotetext{d}{Total gas mass derived by the method of \citet{kennicutt:98}}
\tablenotetext{e}{Virial mass estimate, $C$ = 3.4 for disk candidates and $C$ = 5 for non disks.}
\tablenotetext{f}{Dark matter halo mass}
\tablenotetext{g}{Enclosed (dynamical) mass}
\end{deluxetable*}

\subsubsection{Dark Matter Halo Mass}
We assume that for spherical and virialized dark matter halos, the circular velocity is $V_c = [GM(r)/r]^{1/2}$, where $M(r)$ is the total mass enclosed within $r$. Following common practice, we consider a dark halo within a radius $r_{200}$, defined as where the mean enclosed density is 200 times the mean cosmic value $\overline{\rho}$:
\begin{equation}
r_{200} = \left[\frac{GM(r_{200})}{100\Omega_m(z)H^2(z)}\right]^{1/3},
\end{equation}
where the Hubble's parameter $H$ and matter density parameter $\Omega_m$ are related to their present values by $H(z) = H_0E(z)$, $\Omega_m(z) = \Omega_{m, 0}(1+z)^3/E^2(z)$, and $E(z) = [\Omega_{\Lambda, 0} + (1-\Omega_0)(1+z)^2+\Omega_{m, 0}(1+z)^3]^{1/2}$. The halo mass is then written as:
\begin{equation}
\label{halo_mass_eq}
M_{halo} = \frac{0.1 V_c^3}{H_0 G \Omega_m^{0.5}(1+z)^{1.5}}.
\end{equation}
We use the plateau velocity $V_p$ found in \S \ref{kine_model} for $V_c$ and report $M_{halo}$ in Table \ref{mass_table}.

\subsubsection{Enclosed Mass}
The enclosed mass, which is often called the dynamical mass, refers to the mass residing in the disk-like component of the galaxy. It is calculated by assuming circular motion in a highly flattened spheroid described by the following equation:
\begin{equation}
M_{enclosed} = \frac{2V^2_cr}{\pi G}.
\end{equation}
Again, we use the plateau velocity, $V_p$ for $V_c$. For $r$, we use the farthest distance from the dynamical center to the edge of the galaxy, as seen in the segmentation maps. The resultant enclosed masses are listed in Table \ref{mass_table}.

\subsection{Mass Summary}
In this section, we have estimated the gas masses, by two independent methods \citep{genzel:15, kennicutt:07}, and virial masses for our sources. For our four disk candidates, we have also estimated their halo masses and enclosed (dynamical) masses. While the stellar masses in our sample range from $\log{M_*/\text{\msun}}$ = 9.61 to $\log{M_*/\text{\msun}}$ = 11.24, the gas masses estimated with the \citet{genzel:15} method span 9.10 $\lesssim \log{M_{\text{gas, 1}}/\text{\msun}} \lesssim$ 11.36, and the gas fractions, $f_{\text{mol gas}}=M_{\text{gas}}/(M_{\text{gas}}+M_*)$, span 0.14 $\lesssim f_{\text{mol gas}} \lesssim$ 0.80.

The virial masses span 9.54 $\lesssim \log{M_{vir}/\text{\msun}} \lesssim$ 10.62 (excluding 7187W; see \S \ref{red}), and are overall in order-of-magnitude agreement with $M_*$ and $M_{\text{gas,1}}$. However, one particular case, the source 10633, shows notable disagreement in its mass estimates. Specifically, its virial mass, $\log{\text{M}_{\text{vir}}/\text{\msun}}$ = 9.54, is nearly two orders of magnitude lower than the sum of its stellar ($\log{\text{M}_{*}/\text{\msun}}$ = 11.24) and gas ($\log{\text{M}_{\text{gas}}/\text{\msun}}$ = 10.26) masses. This discrepancy is most likely due to the result of incomplete detection: while the HST image shows three separate components, the \ha{} map only has one component (see Appendix \ref{sec_indiv} for detail). 

For the four disk candidates, we additionally calculated enclosed masses, which are in good agreement with their virial, and halo masses, which span 11.08 $\lesssim \log{\text{M}_{\text{halo}}/\text{\msun}} \lesssim$ 11.95. In order to obtain a rotation curve with a plateau velocity ($V_p$), we require the model to fit the plateau radius ($R_p$) within the detected area (see \S \ref{kine_model}). The assumption of $V_p = V_c$ may be too simplified to model disks since even in well-ordered (high $v/\sigma$) local disks, $V_{opt}/V_{200c}$ (optical-to-virial velocity ratio) is found to differ by 30 to 40 \% \citep[e.g.][]{reyes:12}. Taking into account these considerations, $M_\text{enc}$ and $M_\text{halo}$ are order-of-magnitude estimates.

Similarly, the assumption in the constant factor $C$ in the virial mass calculation (Equation \ref{eq:virial}) has a high uncertainty. We only assume two cases: $C=3.4$ for the four disk candidates and $C=5$ for the other galaxies. Between the two, there is a factor of 1.5 difference if we mis-classify galaxies. Moreover, these two cases are assuming uniform thin disks (C=3.4) and uniform spheres (C=5), which are simplifications in themselves. Additionally, as mentioned in \S \ref{sec:virial}, the use of $r_g$ as $r_{vir}$ also adds uncertainty in$ M_{vir}$. Therefore combining these factors, we expect uncertainties of order unity in $M_{vir}$.

\section{Clumps}\label{sec_clump}
Observations of star-forming galaxies at high redshift show irregular morphologies, dominated by kpc-scale star-forming clumps \citep[e.g.,][]{elmegreen:09, forster:09, livermore:12}. These clumps are likely a result of gravitational instability in the disk. They are speculated to migrate toward the galactic center through dynamical friction and form the galactic bulge \citep[][and reference therein]{bournaud:15}; and/or be disrupted by stellar feedback and recycle its gas back to the ISM \citep{hopkins:12, oklopcic:16}. In this section, we explain how we define the observed z$\sim$1 clumps and present their properties. 

\subsection{Clump Definition}
\begin{deluxetable}{cccccc}
\tabletypesize{\footnotesize}
\tablecolumns{6}
\tablewidth{0pc}
\tablecaption{Clump Parameters}
\setlength{\tabcolsep}{0.02in}
\tablehead{
\colhead{ID} &
\colhead{Clump} &
\colhead{$r_{1/2}$\tablenotemark{a}} &
\colhead{$r_{\text{ap}}$\tablenotemark{b}} &
\colhead{SFR\tablenotemark{c}} &
\colhead{$\sigma_{1D}$\tablenotemark{d}} \\
 & & [kpc] & [kpc] & [M$_\odot$/yr] & [km/s]
}
\startdata
11655      & A & 2.50 & 2.63 & 8.43 & 48.0 \\
           & B & 0.58 & 0.99 & 0.56 & 67.8 \\
10633      & A & 0.91 & 1.23 & 7.51 & 57.3 \\
42042481   & A & 3.04 & 3.13 & 2.40 & 61.0 \\
           & B & 0.79 & 1.10 & 0.21 & 43.6 \\
           & C & 0.85 & 1.15 & 0.19 & 51.2 \\
           & D & 0.63 & 1.00 & 0.16 & 50.1 \\
           & E & 0.61 & 0.98 & 0.12 & 32.6 \\
           & F & \nodata & 0.66 & 0.10 & 131.8 \\
           & G & \nodata & 0.70 & 0.06 & 79.3 \\
J033249.73 & A & 1.25 & 1.75 & 0.84 & 77.2 \\
           & B & 0.98 & 1.57 & 0.78 & 61.8 \\
           & C & \nodata & 1.22 & 0.31 & 60.4 \\
           & D & \nodata & 0.76 & 0.19 & 93.4 \\
11169      & A & 2.47 & 2.62 & 14.12 & 96.8 \\
           & B & 3.02 & 3.14 & 9.49 & 113.9 \\
           & C & 0.42 & 0.96 & 0.94 & 66.6 \\
           & D & 0.65 & 1.08 & 0.88 & 64.4 \\
           & E & 0.19 & 0.89 & 0.71 & 70.8 \\
7187       & A & 3.01 & 3.17 & 1.88 & 87.7 \\
           & B & 1.52 & 1.81 & 0.50 & 99.5 \\
           & C & \nodata & 0.93 & 0.17 & 71.7 \\
           & D & 0.45 & 1.08 & 0.17 & 56.6 \\
           & E & 0.33 & 1.03 & 0.14 & 24.5 \\
           & F & 0.53 & 1.11 & 0.11 & 73.5 \\
9727       & A & 4.45 & 4.56 & 26.74 & 89.7 \\
           & B & 1.14 & 1.52 & 3.37 & 86.6 \\
           & C & 0.51 & 1.13 & 1.69 & 45.4 \\
           & D & 0.80 & 1.28 & 0.99 & 13.3 \\
           & E & 0.42 & 1.09 & 0.78 & 40.6 \\
           & F & \nodata & 0.82 & 0.66 & 45.9 \\
7615       & A & 2.22 & 2.45 & 1.77 & 79.4 \\
           & B & 1.28 & 1.64 & 1.17 & 79.6 \\
           & C & 1.63 & 1.93 & 1.15 & 60.6 \\
           & D & 1.89 & 2.16 & 0.94 & 64.3 \\
           & E & 1.13 & 1.53 & 0.60 & 70.8 \\
           & F & 0.69 & 1.24 & 0.44 & 59.8 \\
11026194   & A & 2.63 & 2.82 & 4.56 & 65.0 \\
           & B & 1.20 & 1.56 & 2.25 & 76.4 \\
12008898   & A & 2.84 & 2.91 & 13.87 & 61.4 \\
           & B & 2.74 & 2.81 & 11.92 & 59.6 \\
           & C & 1.35 & 1.48 & 2.07 & 54.0 \\
12019627   & A & 1.82 & 1.99 & 1.93 & 45.6 \\
           & B & 2.34 & 2.47 & 1.66 & 42.6 \\
           & C & 1.10 & 1.36 & 0.92 & 58.5 \\
           & D & 0.82 & 1.15 & 0.58 & 71.1 \\
           & E & 0.46 & 0.92 & 0.39 & 54.4 \\
           & F & 0.73 & 1.08 & 0.34 & 25.3 \\
13017973   & A & 2.77 & 2.96 & 13.12 & 36.2 \\
           & B & 1.88 & 2.14 & 12.03 & 161.5 \\
           & C & 1.69 & 1.98 & 4.83 & 39.6 \\
           & D & 1.37 & 1.71 & 4.17 & 60.4 \\
           & E & 0.82 & 1.32 & 3.75 & 62.0 \\
           & F & \nodata & 0.69 & 1.31 & 117.6 \\
           & G & \nodata & 0.90 & 1.23 & 76.6 \\
           & H & \nodata & 1.03 & 0.99 & 46.6 \\
13043023   & A & 1.08 & 1.49 & 2.31 & 104.5 \\
           & B & 0.52 & 1.15 & 1.39 & 69.3 \\
           & C & 0.51 & 1.14 & 0.96 & 61.1 \\
           & D & \nodata & 0.69 & 0.66 & 58.6 \\
32040603   & A & 1.53 & 1.85 & 1.75 & 52.5 \\    
32016379   & A & 1.94 & 2.09 & 1.20 & 64.9 \\
           & B & 1.48 & 1.68 & 0.63 & 27.2 \\
           & C & 0.72 & 1.06 & 0.26 & 58.6 \\
32036760   & A & 2.78 & 2.95 & 2.60 & 55.1 \\
33009979   & A & 2.15 & 2.30 & 7.45 & 60.8 \\
           & B & 1.92 & 2.09 & 2.43 & 42.8 \\
           & C & 0.74 & 1.10 & 0.61 & 56.5 
\enddata \label{clump_table}
\tablenotetext{a}{Half-light radius of clump.}
\tablenotetext{b}{Aperture size (i.e., non corrected size).}
\tablenotetext{c}{ISM corrected SFR inside the half-light radius.}
\tablenotetext{d}{Integrated velocity dispersion inside the half-light radius.}
\end{deluxetable}
There have been many definitions of ''clumps'' in the literature. For imaging studies, the definition ranges from visual inspection \citep[e.g.,][]{cowie:95, elmegreen:07}, which is difficult to reproduce, to automated definitions based on the intensity contrast between the peak and the local background in galaxy images \citep{guo:12, wuyts:12}. For example, \citet{guo:15} suggested UV-bright clumps as discrete regions that individually contribute more than 8 \% of the rest frame UV light of their galaxies. In IFS studies, \citet{genzel:11} required a clump to be a local maximum in at least two separate velocity channels; while \citet{wisnioski:12} identified their clumps solely from local \ha{} peaks in 2D \ha{} maps. 

\begin{figure*}[t]
\centering
\includegraphics[width=0.32\textwidth]{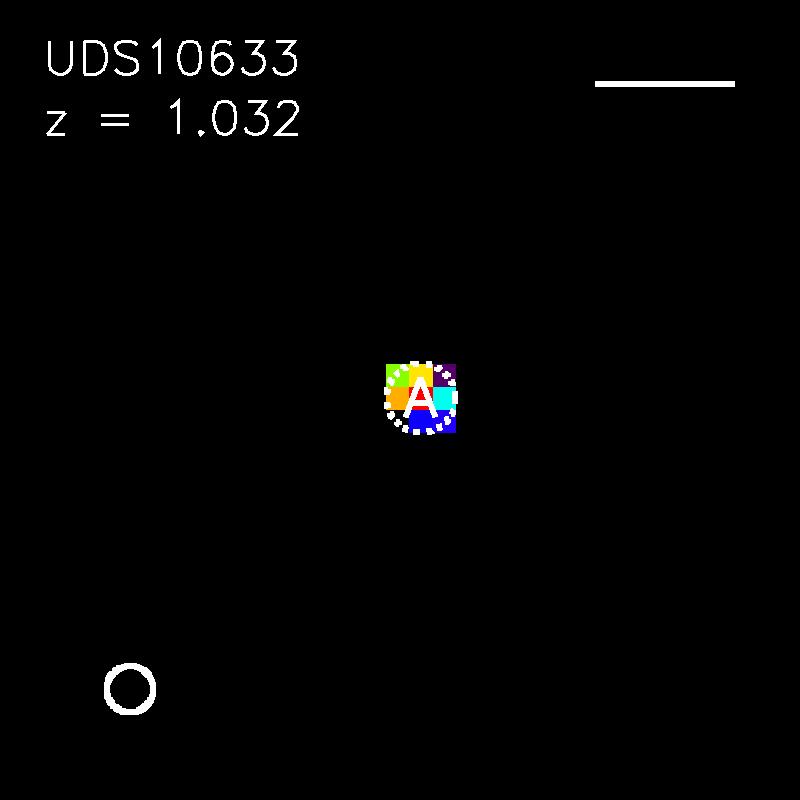}
\includegraphics[width=0.32\textwidth]{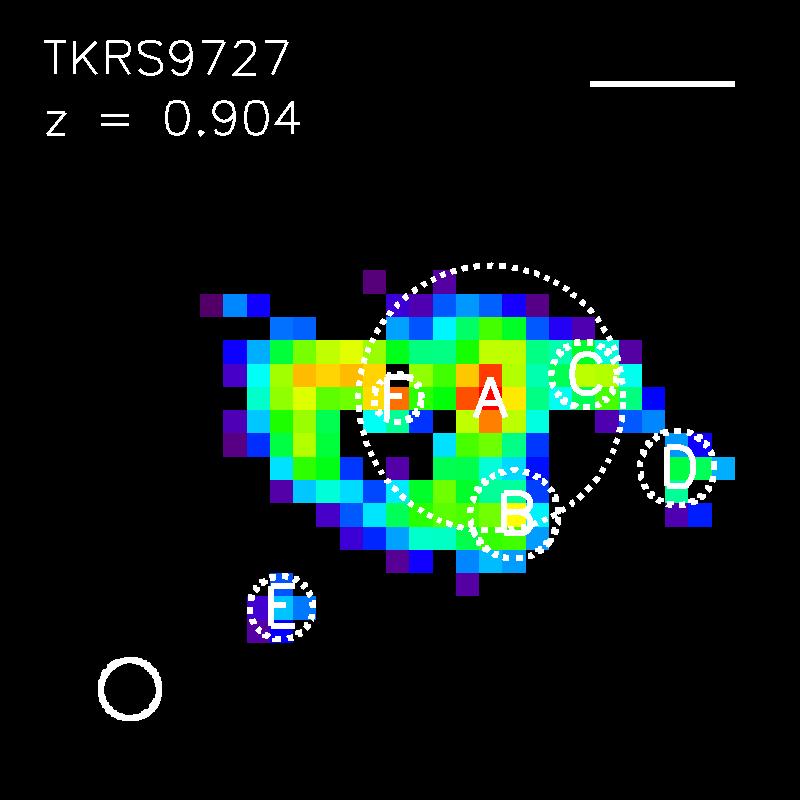}
\includegraphics[width=0.32\textwidth]{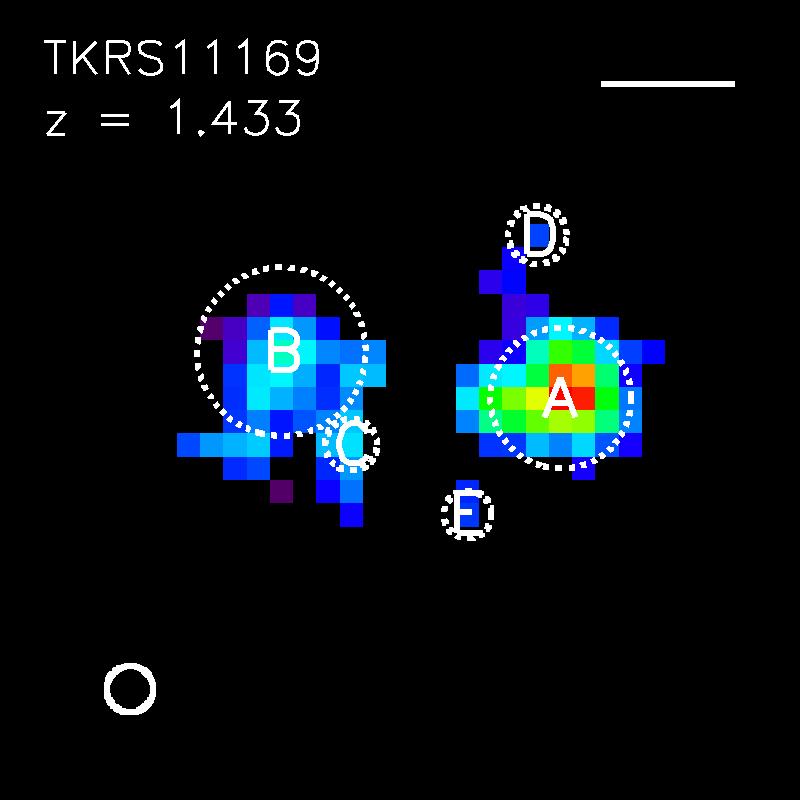}
\includegraphics[width=0.32\textwidth]{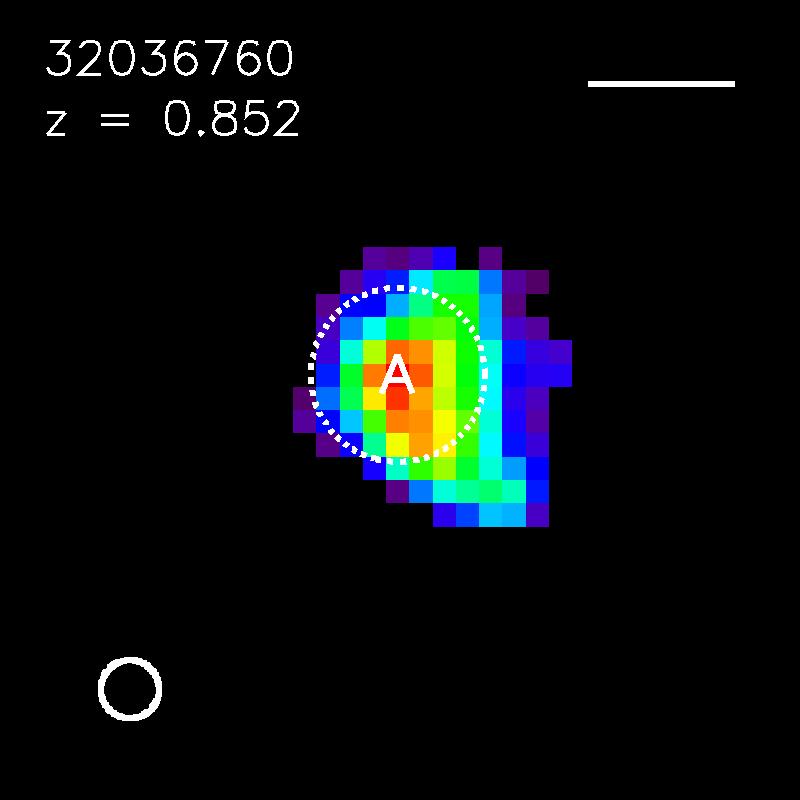}
\includegraphics[width=0.32\textwidth]{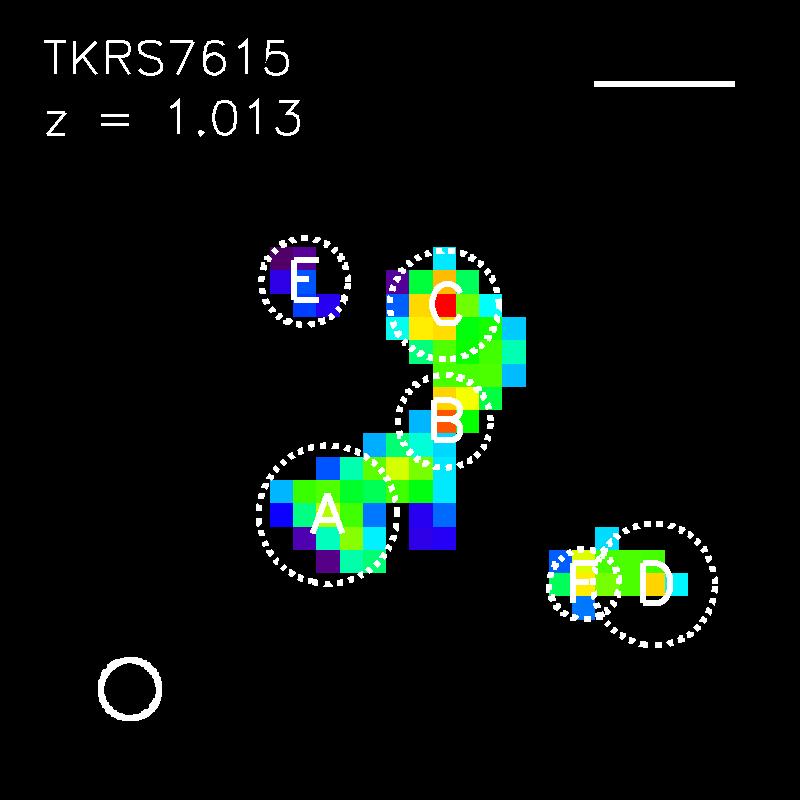}
\includegraphics[width=0.32\textwidth]{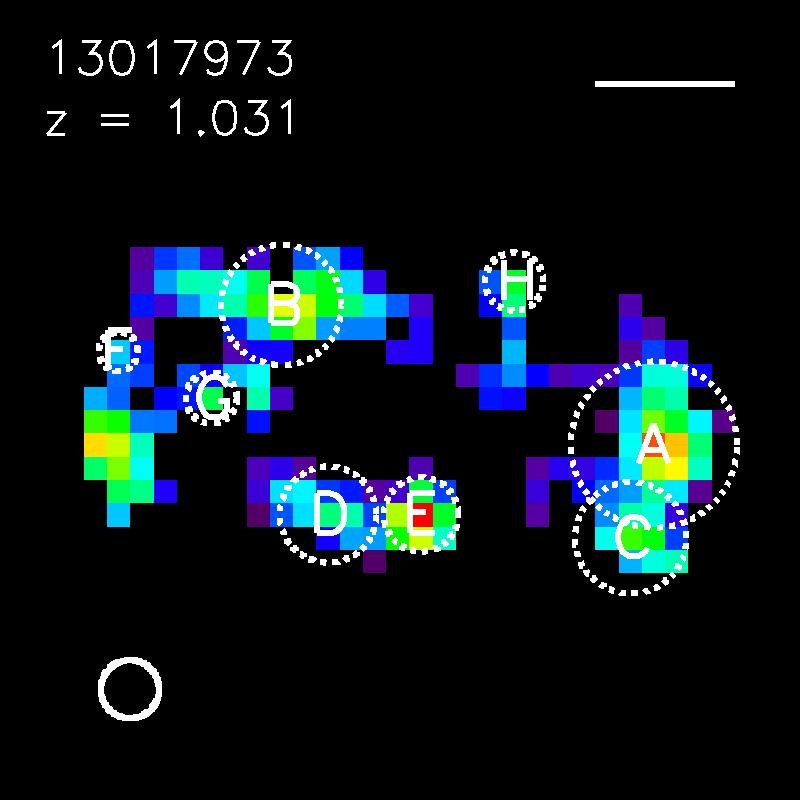}
\includegraphics[width=0.32\textwidth]{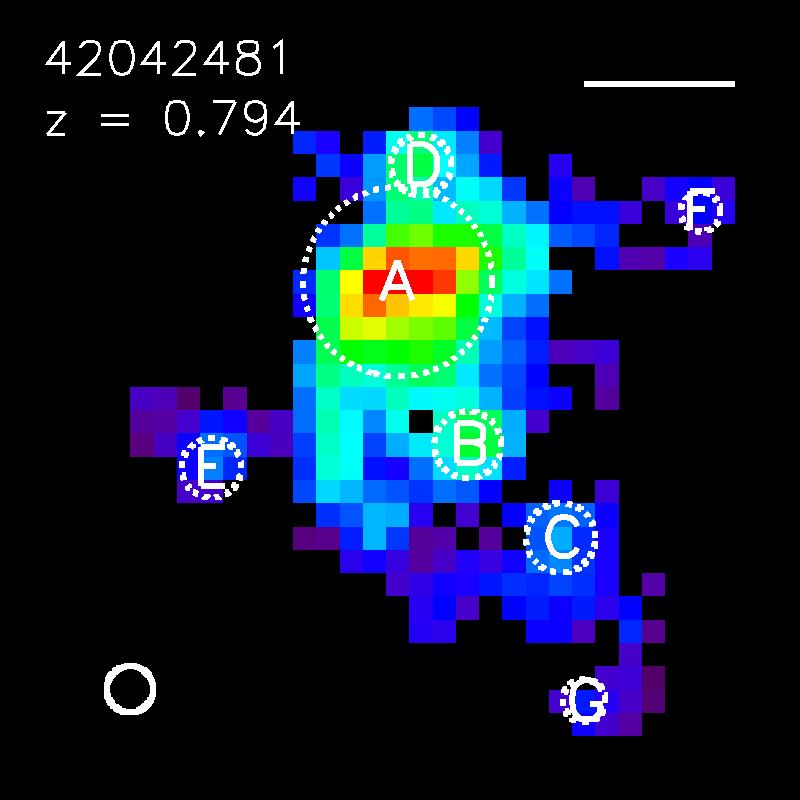}
\includegraphics[width=0.32\textwidth]{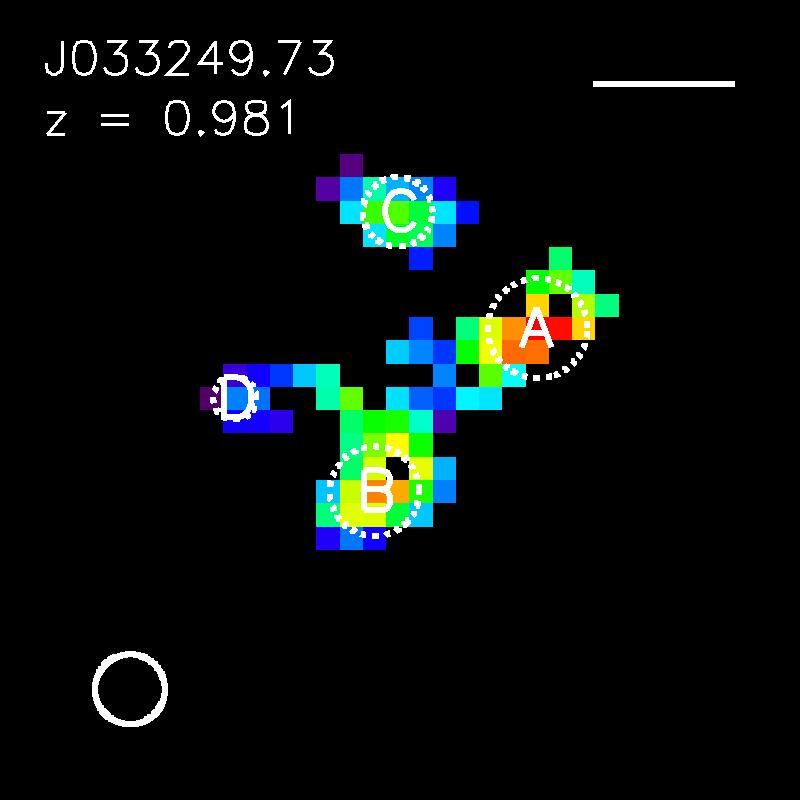}
\includegraphics[width=0.32\textwidth]{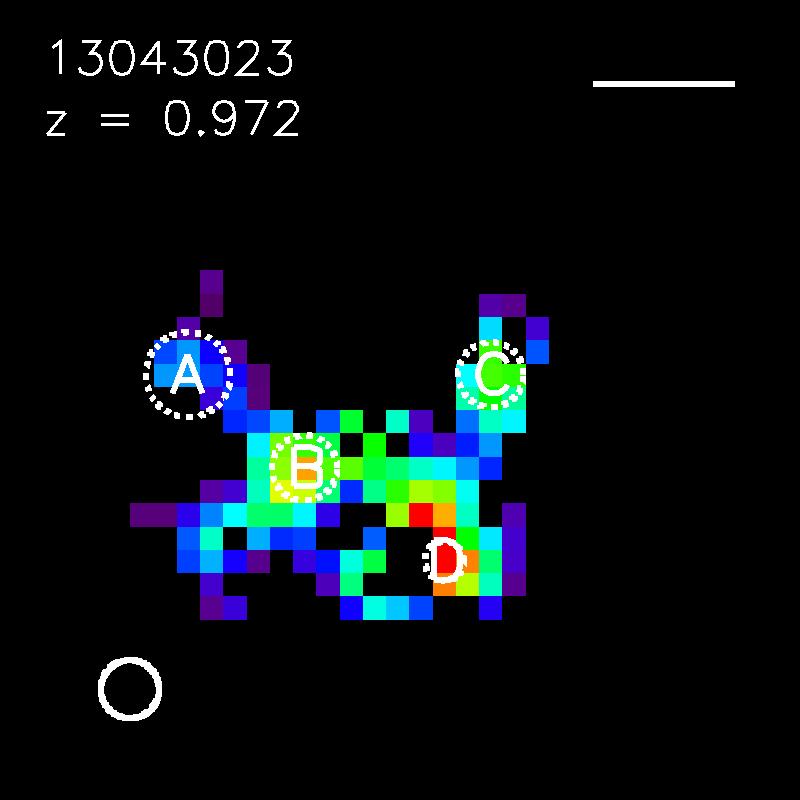}
\includegraphics[width=0.32\textwidth]{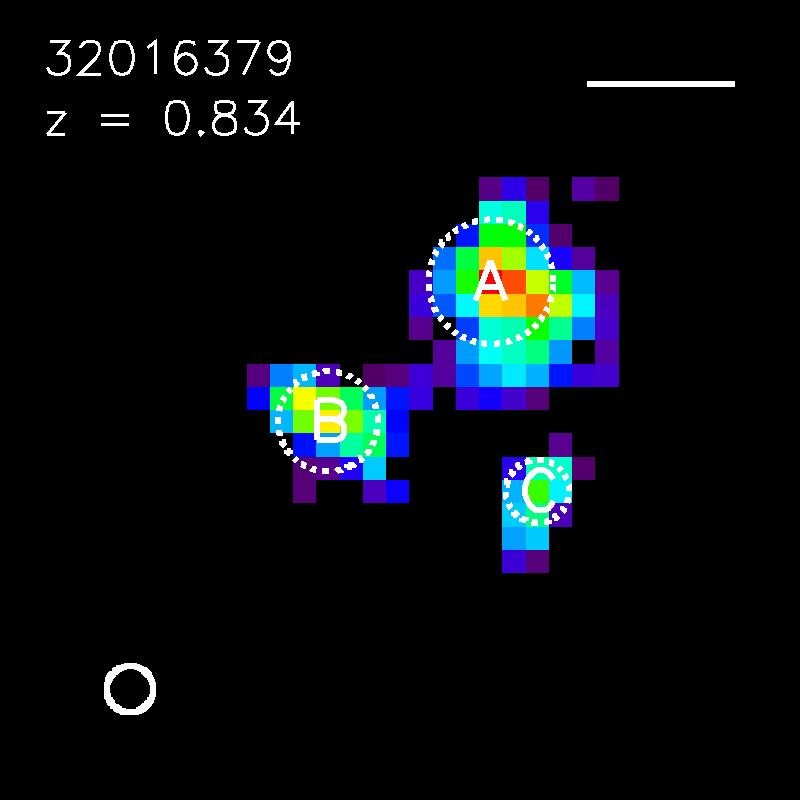}
\includegraphics[width=0.32\textwidth]{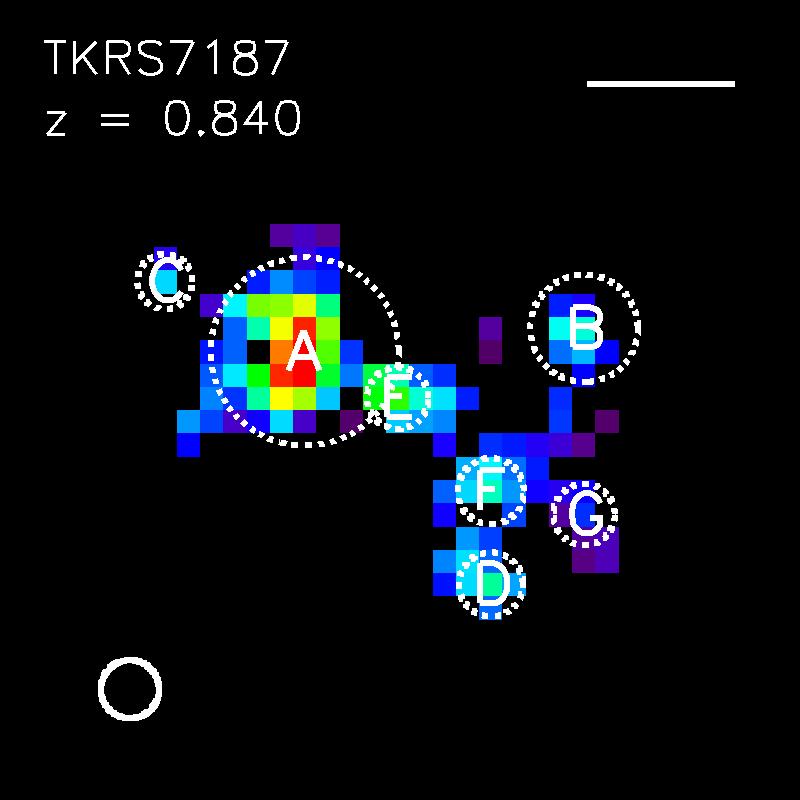}
\includegraphics[width=0.32\textwidth]{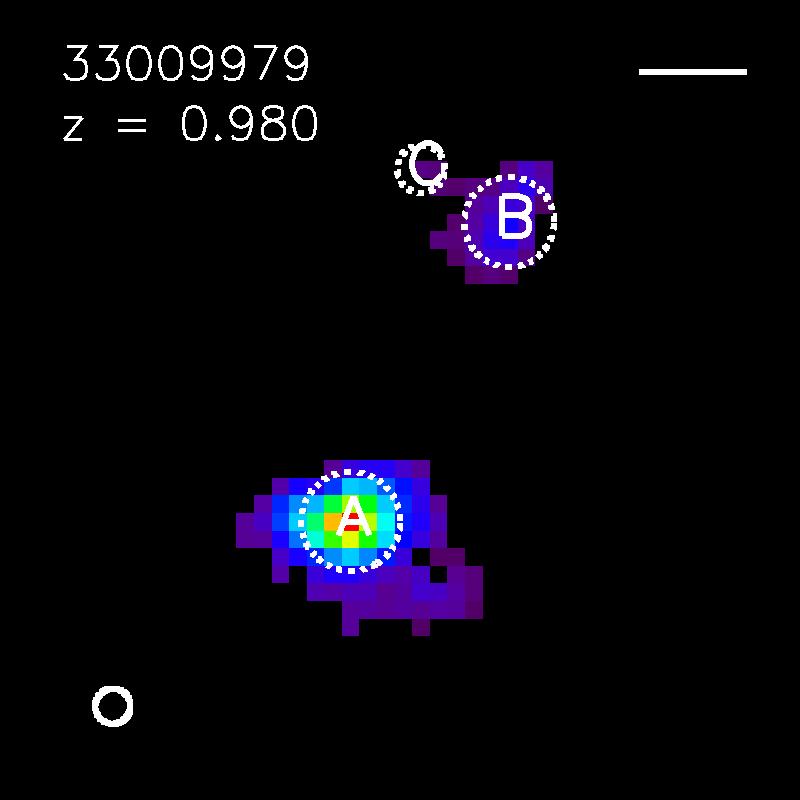}
\caption{Identified clump locations and sizes. The radii of dashed circles are the size used to obtain the total clump \ha{} flux ($r_{\text{ap}}$). In our definition, a clump is a local \ha{} peak that is separated by more than two pixels from neighbor peaks in \ha{} maps (second panels in Figure \ref{kine_11655}). The clumps are marked as A, B, and so forth in a descending order of brightness. Panels are organized from the highest to lowest stellar mass estimated by SED fitting. The name and redshift of the galaxy are listed at the top left corner. The length of top right line presents 5 kpc at that redshift. The solid circle at the bottom left presents the size of smoothing FWHM.}
\label{clump_im}
\end{figure*}
\begin{Contfigure}[t]
\centering
\includegraphics[width=0.32\textwidth]{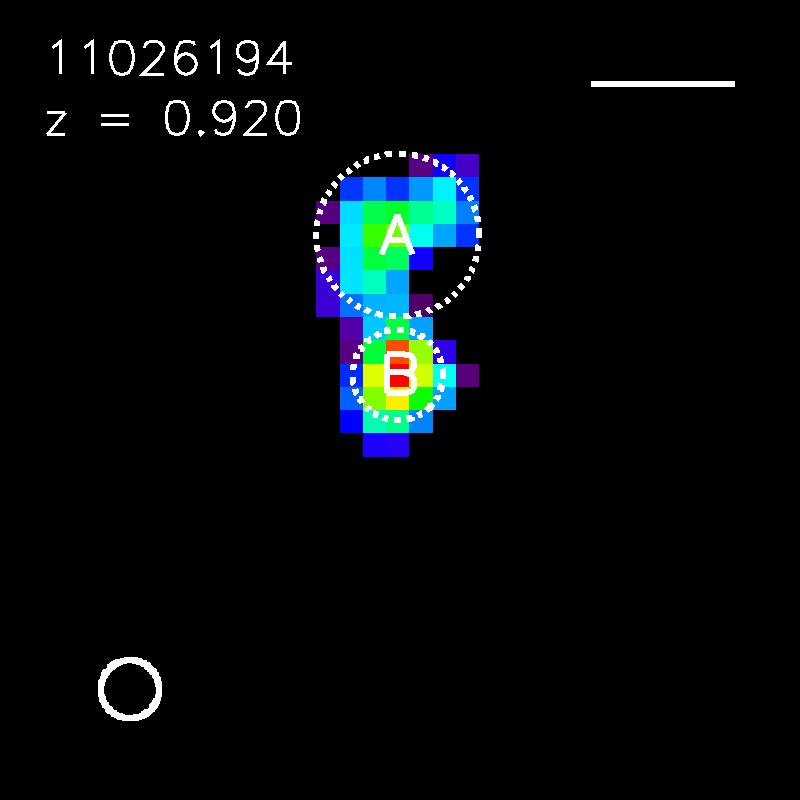}
\includegraphics[width=0.32\textwidth]{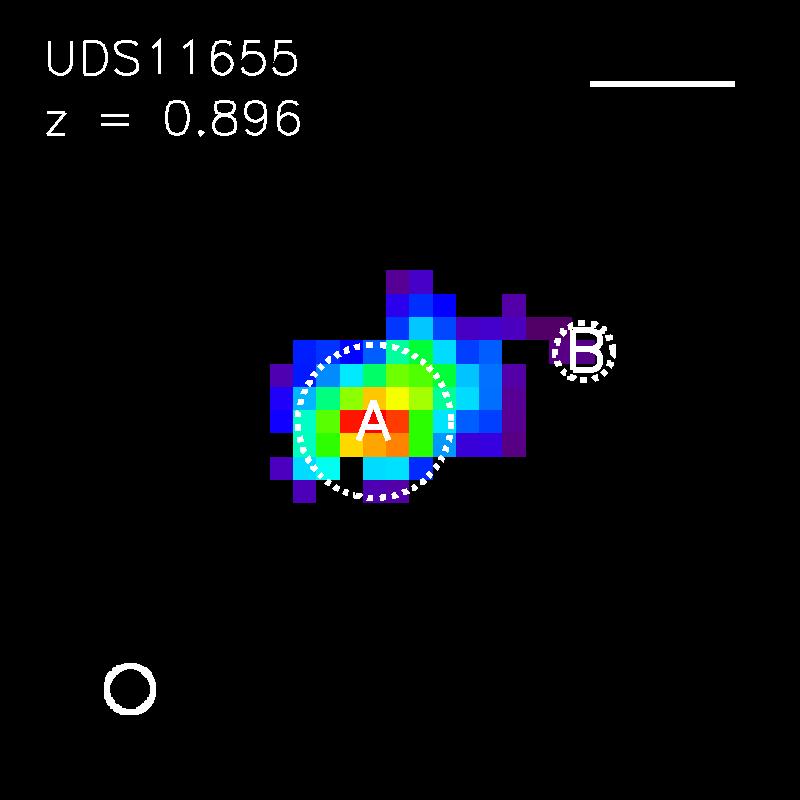}
\includegraphics[width=0.32\textwidth]{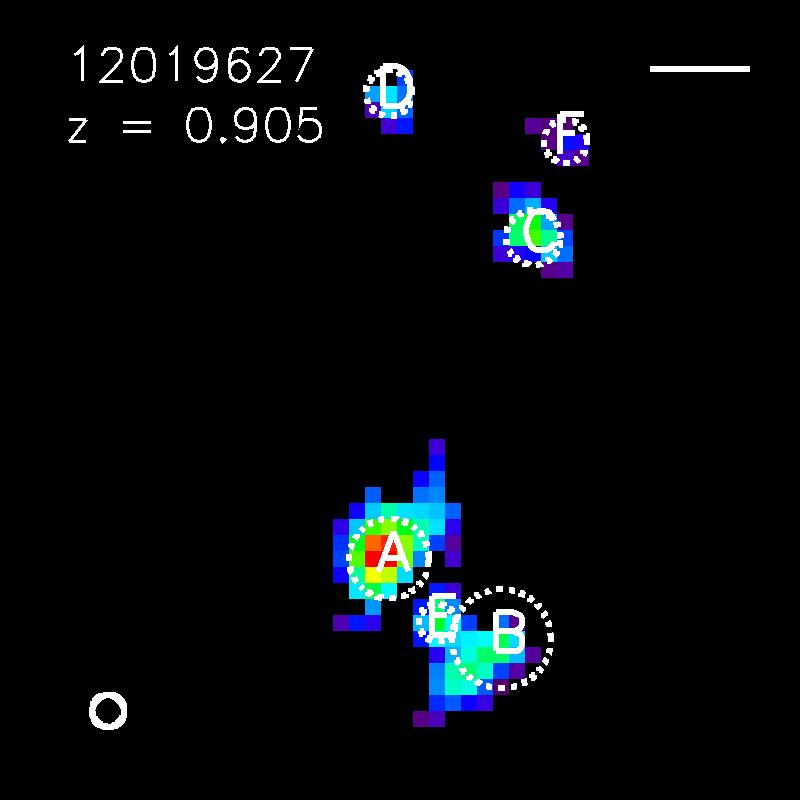}
\includegraphics[width=0.32\textwidth]{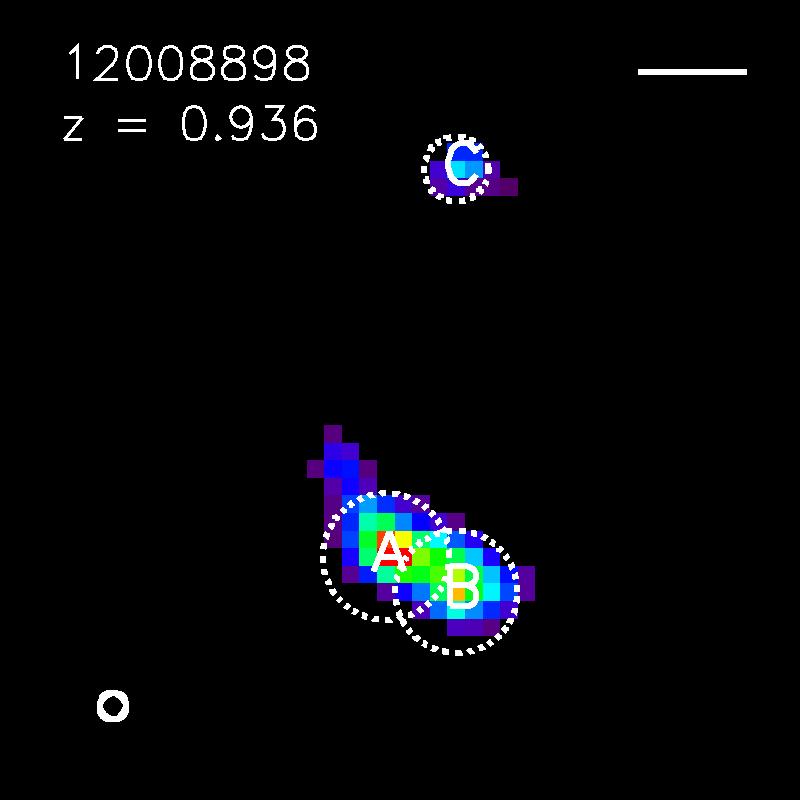}
\includegraphics[width=0.32\textwidth]{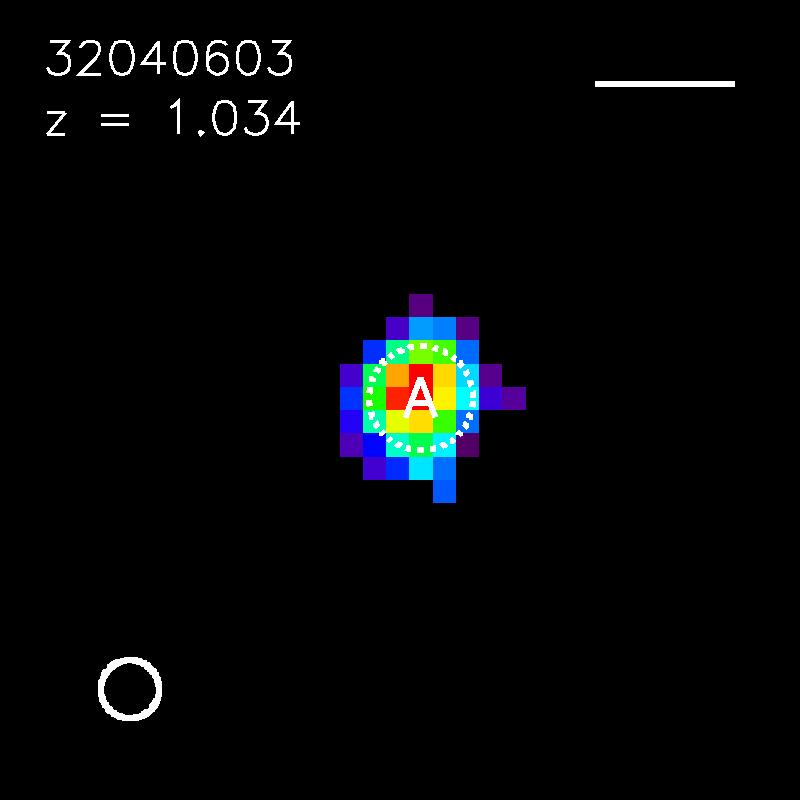}
\caption{}
\end{Contfigure}

We define a clump as a local \ha{} flux peak that is separated by more than two pixels from other peaks in \ha{} maps (second panels in Figure \ref{kine_11655}). We apply this definition to the smoothed \ha{} maps. When this definition is applied to a compact, single nuclei galaxy, the whole galaxy itself is classified as a ''clump'' (e.g., UDS 10633). It is technically \emph{not} a clump, but we include them in our analysis for completeness. Under this definition, we identify 68 isolated \ha{} peaks among 17 sources. We use the 68 isolated \ha{} peaks to investigate their \ha{} flux and velocity dispersion, and where we resolve the clumps we are able to measure their physical size.

We measure clump sizes through the following procedure: 1) we make an azimuthally averaged surface brightness profile centered at the peak, 2) compute the derivative of the surface brightness profile with respect to radius, 3) set the background to be the radius ($r_{\text{back}}$) at which the derivative crosses 0 or reaches less than a cut off value (in our case $3^{-18}$erg s$^{-1}$ cm$^{-2}$ arcsec$^{-1}$), 4) subtract the background from the \ha{} map, and 5) calculate the radius at which half of the total flux within $r_{\text{back}}$ is included. The size obtained by this method is denoted as $r_{\text{ap}}$. This method is robust when the surface brightness profile is steep. When the profile is shallow (i.e., size is large), the derivative slowly plateaus to 0, and our choice of the cut-off value is not necessarily the best; however, a shallow profile also means the background value is not sensitive to the choice of the background location, so we do not expect this uncertainty to have a significant effect on our measurements.

When the surface brightness profile is approximated by a Gaussian function, using its standard deviation ($\sigma_{\text{G}}$), the half light radius ($r_{1/2}^{\text{G}}$) and FWHM can be written as $r_{1/2}^{\text{G}} = \sqrt{-2\ln{0.5}}\sigma_{\text{G}}$ and FWHM $ = 2\sqrt{2\ln{2}}\sigma_{\text{G}}$, respectively. Using these relationships, the final clump sizes, denoted as $r_{1/2}$, are expressed as follows:
\begin{equation}
    r_{1/2} = \sqrt{r_{\text{ap}}^2 + \frac{2\ln{0.5}}{4\times2\ln{2}}\text{FWHM}^2}.
\end{equation}
The final values of $r_{\text{ap}}$ and $r_{1/2}$ are listed in Table \ref{clump_table}. Some clumps are smaller than beam sizes, and are considered unresolved. Among 68 isolated peaks, 58 are resolved clumps. The uncorrected sizes ($r_{\text{ap}}$) of identified clumps are shown in Figure \ref{clump_im} as the size of dashed circles centered at the peaks. Figures are ordered from the highest to the lowest stellar mass estimated by SED fitting \S \ref{sec_sed}.

The total \ha{} flux for each clump is measured by summing up the spectra inside the uncorrected aperture radius ($r_{\text{ap}}$), and fitting a Gaussian profile to the \ha{} emission line in a total spectrum. To compare with other surveys, we assume a spatially uniform, ISM-only extinction to convert \ha{} fluxes into SFRs (see \S \ref{sec_sed} for HII and ISM extinction). We also obtain each clump's \sigmaoned{}, measured from the width of the Gaussian function, and corrected for an average instrumental width within the aperture radius. The values of SFR and \sigmaoned{} are listed in Table \ref{clump_table}. When the clump is unresolved, its SFR and dispersion values are still valid within the aperture, and we include them in our analysis. The clumps are marked as A, B, and so forth in a descending order of brightness in Figure \ref{clump_im}.

We find that among the $z \sim 1$ sample, star-forming clumps have a half-light radius between 0.17 to 4.5 kpc, \sigmaoned{} between 13 to 160 km s$^{-1}$, and SFR between 0.1 to 27 \myr. 

\subsection{Disk Stability}\label{sec_toomre}
\begin{figure*}
\centering
\includegraphics[width=0.24\textwidth]{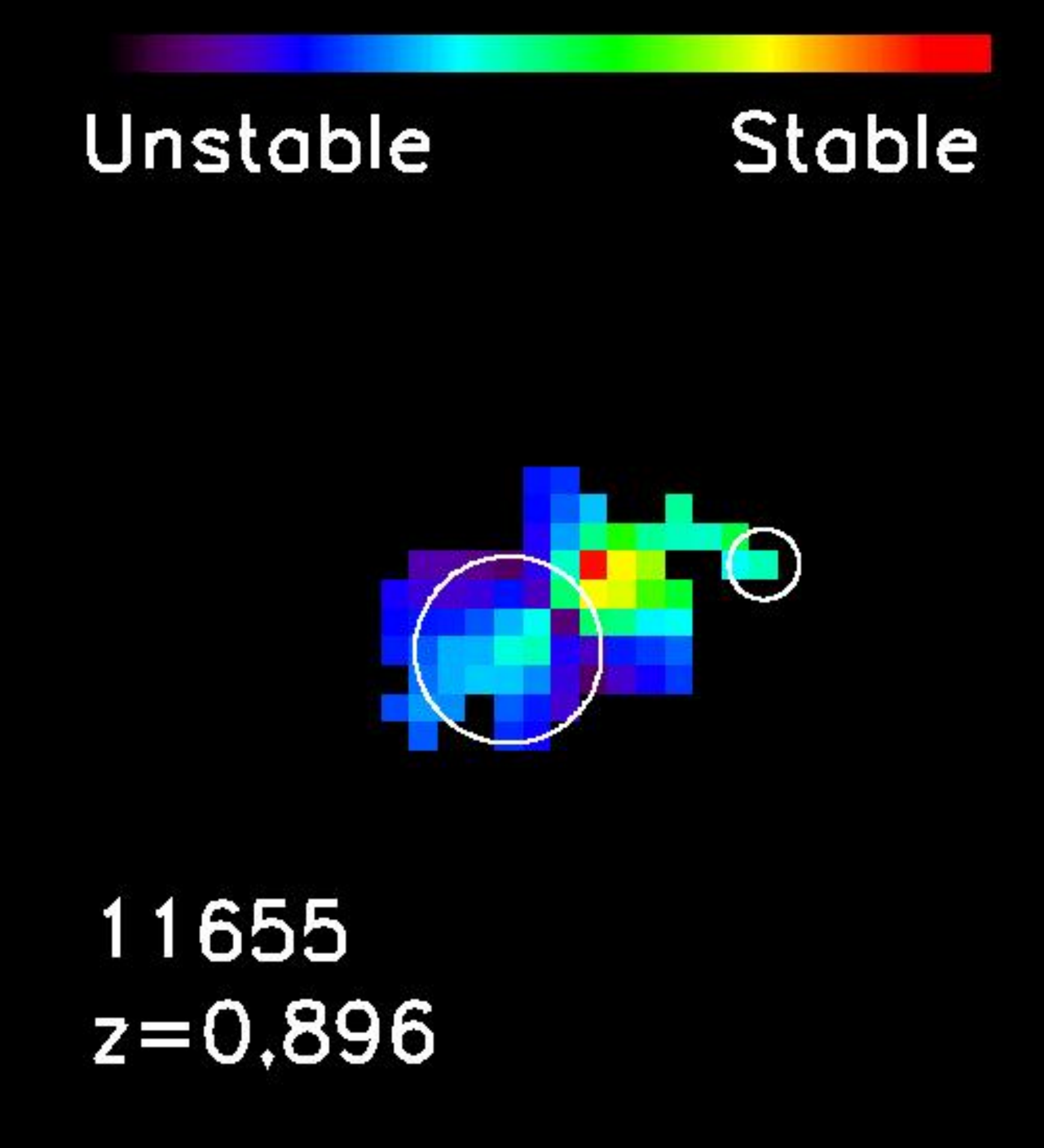}
\includegraphics[width=0.24\textwidth]{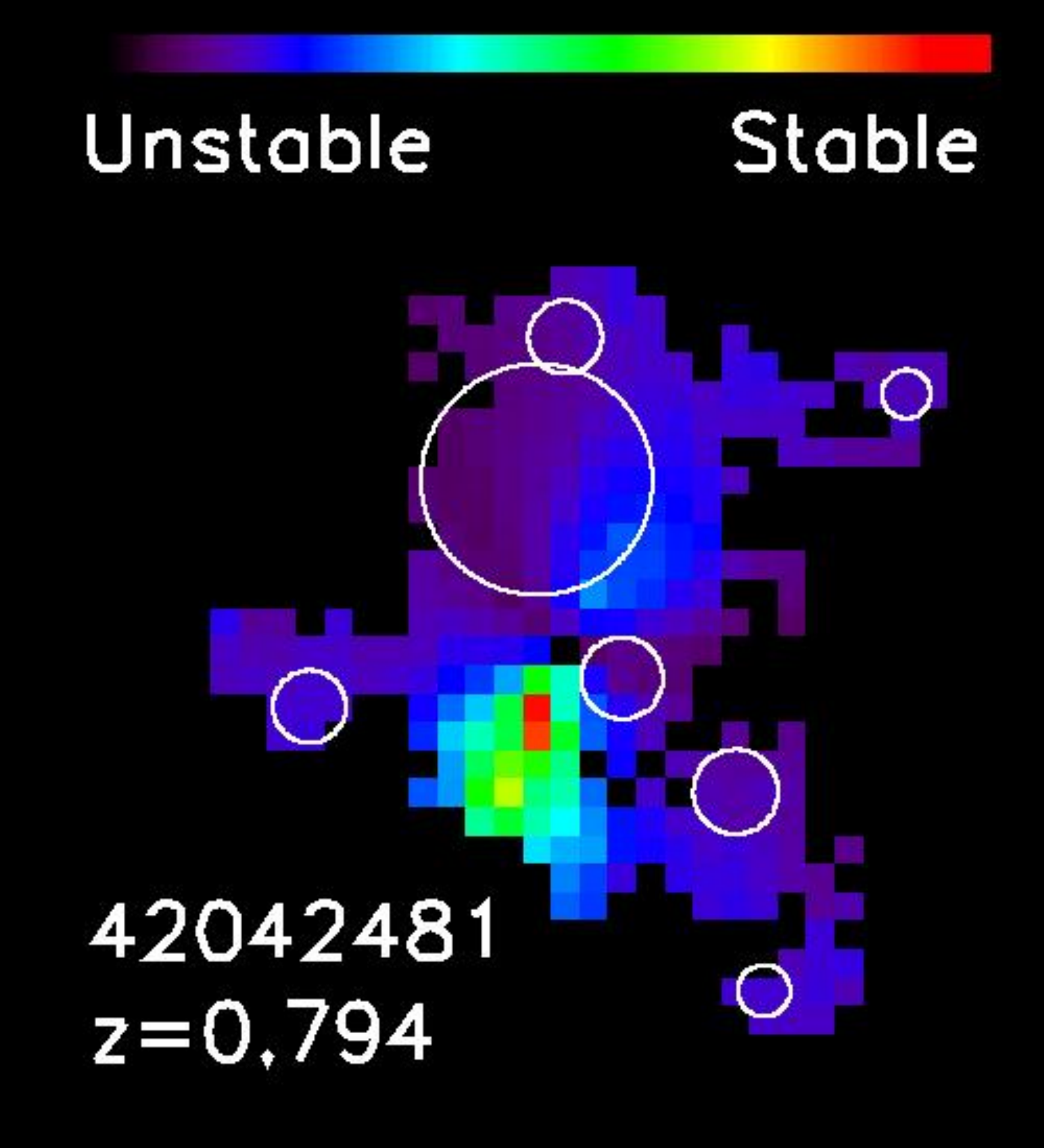}
\includegraphics[width=0.24\textwidth]{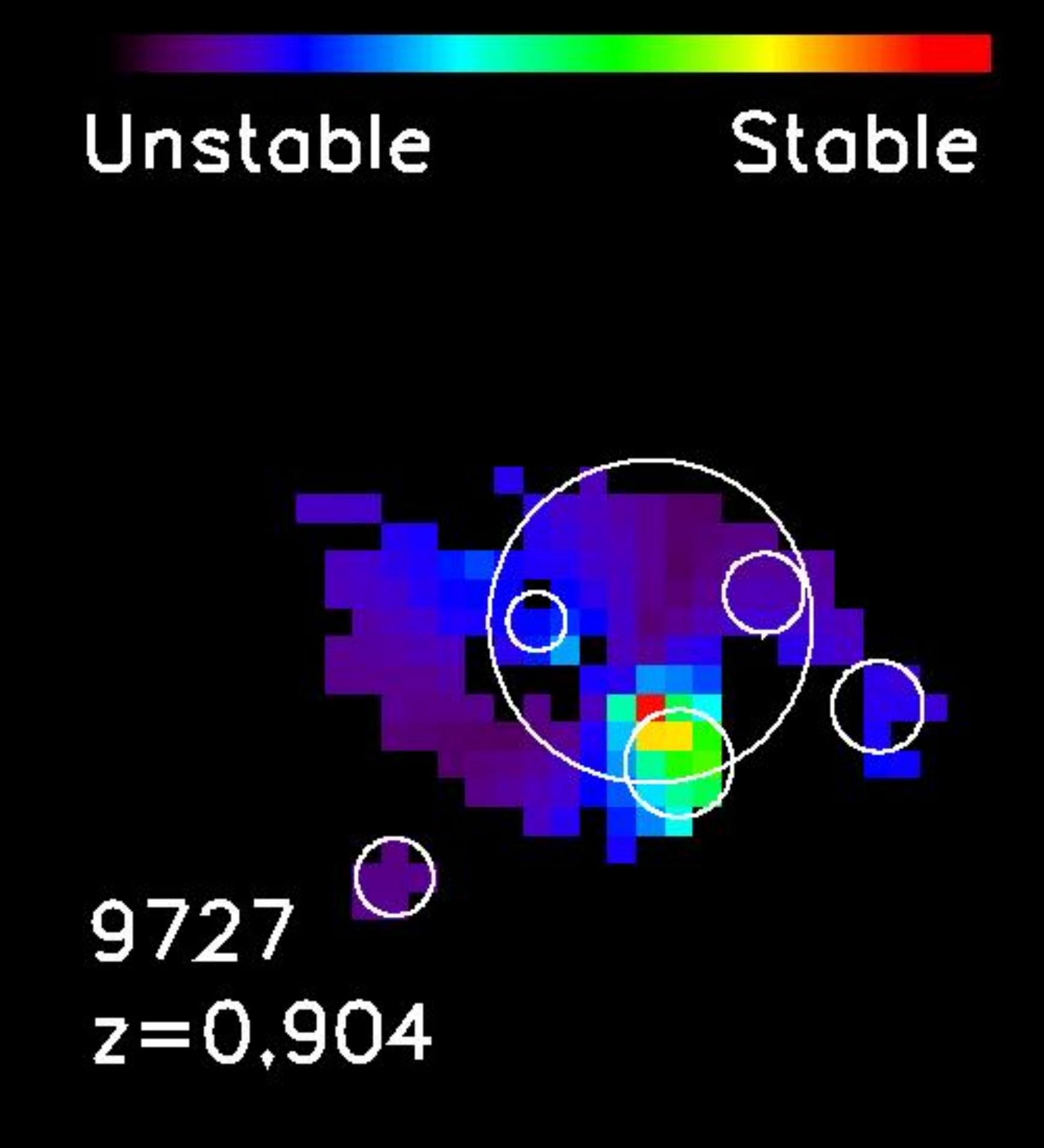}
\includegraphics[width=0.24\textwidth]{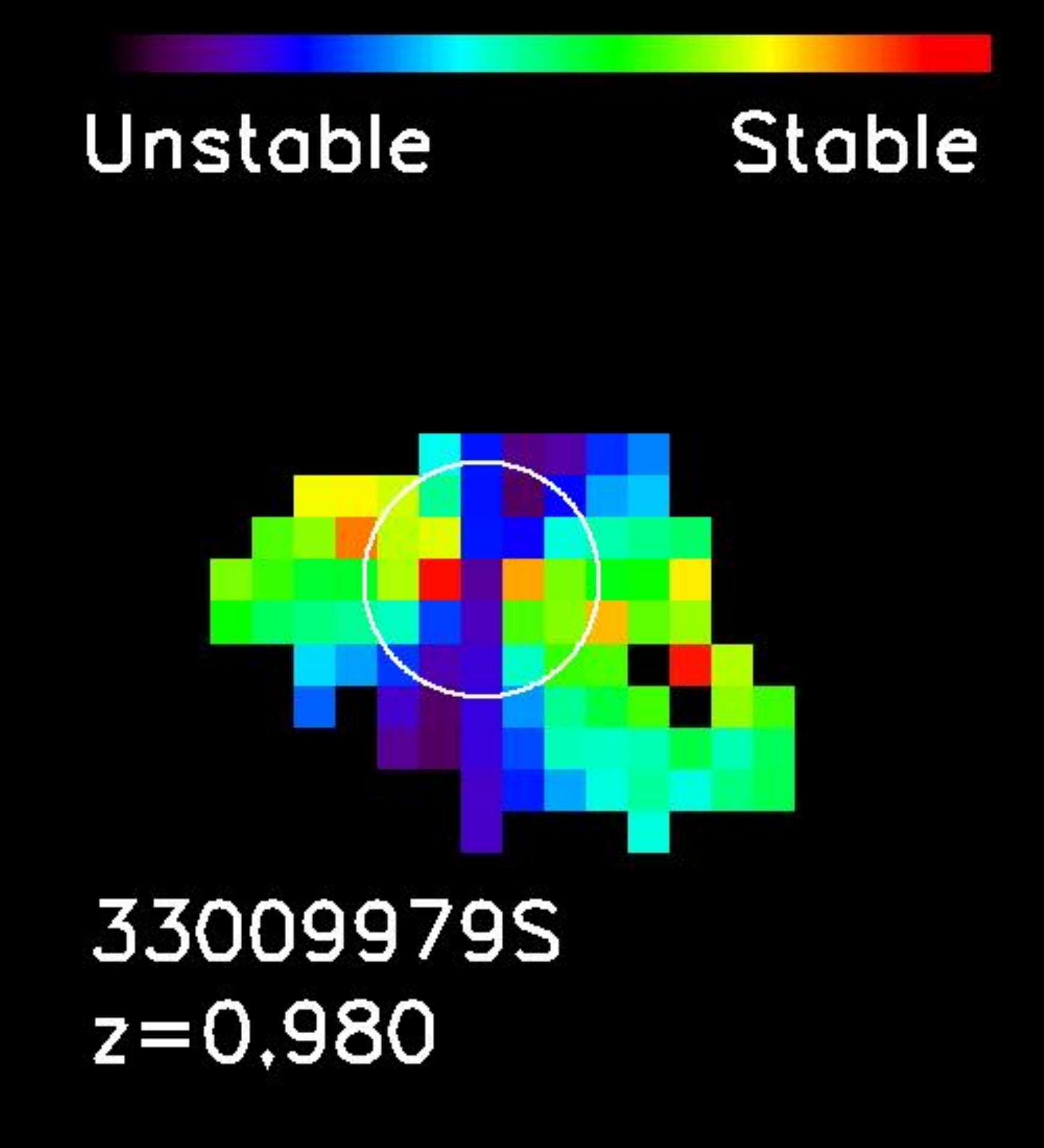}
\caption{Relative Toomre parameter ($Q$, Equation \ref{toomre_eq}) maps of four disk candidates (UDS11655, DEEP2-42042481, TKRS9727, and DEEP2-33009979S) in IROCKS sample. Circles are centered at the peaks of the clumps, and their radii represent the sizes of the clumps, $r_{\text{ap}}$. Most clumps are located where $Q$ is low (unstable), which is seen by high-z observations \citep{genzel:11, wisnioski:12}.}
\label{toomre_map}
\end{figure*}

We investigate the dynamical stability of the candidate disks using \ha{} flux maps and fitted disk models. The Toomre parameter, $Q_{\text{gas}}$, describes the gravitational stability of a gaseous disk by using the local velocity shear and random motion and is expressed as:
\begin{equation}
Q_{\text{gas}} = \frac{\sigma \kappa}{\pi G \Sigma_{\text{gas}}} 
\end{equation}
where $\sigma$ is the local velocity dispersion, $G$ is the gravitational constant, $\Sigma_{\text{gas}}$ is the gas surface density (evaluated from Equation \ref{gas_mass_eq}), and $\kappa$ is the epicyclic frequency of the disk. $\kappa$ can be replaced by the orbital frequency, $\Omega$, if the system is Keplerian. $Q_{\text{gas}} \lesssim$ 1 to 2 can cause instability-driven large scale turbulence. Following \citet{thompson:05} and \citet{genzel:11}, if we assume the total mass $M_T \propto v^2r/G$ and total gas mass $M_g$ inside the radius $r$, then the Toomre parameter can be written as follows:
\begin{equation}
\label{toomre_eq}
Q = a\frac{\sigma}{v}\left(\frac{M_T}{M_g}\right) = \frac{\sigma}{v}\frac{a}{f_g},
\end{equation}
where $f_g$ is the gas fraction within radius $r$, and the constant $a$ represents different potentials. We apply $a = \sqrt{2}$ for a flat rotation curve for a disk. In our sample, four galaxies (11655, 42042481, 9727, and 33009979S) are well fit to a disk model, and $Q_{\text{gas}}$ can be computed spatially using the locally measured gas surface density, velocity dispersion, and modeled rotation. The inclinations of these galaxies are not well constrained with \ha{} detection, and hence we use the expectation value $<i> = 57.3^\circ$ for all four disk fitting (\S \ref{kine_model}), and therefore this adds uncertainties into derived Toomre values. 

Our model assumes rotation-supported disks with $v = v_\text{circ}$. This may not be appropriate for disks that are partially supported by turbulence, and in those cases, their potential would likely be better traced by \s{} (or $S_{0.5}$) instead of $v_\text{circ}$. Also, we use $a = \sqrt{2}$ for a flat rotation curve, while inner part of the disks’ rotation may better resemble solid body rotation, which would give $a=2$. Despite these shortcomings in our model, we keep these assumptions for a one-to-one comparison with other IFS studies.

Instead of showing absolute values, we show relative Toomre maps in Figure \ref{toomre_map} \citep{wisnioski:12} for our four disk candidates (from the left, 11655, 42042481, 9727, and 33009979S). We overplot circles with centers located at the peaks and radii representing the aperture radius ($r_{ap}$) of the clumps. Most clumps reside where $Q_{\text{gas}}$ is low (unstable), as seen in higher redshift observations \citep{genzel:11, wisnioski:12}.

\subsection{Clump Evolution}
\label{clump_discussion}
The empirical properties of star-forming clumps can provide clues to the physical mechanisms that drive their formation and evolution, and it is interesting to compare them to local HII regions. \citet{wisnioski:12} compared their observations on $z\sim1.3$ star-forming clumps with data on local HII regions and found tight scaling relations between the clump size, luminosity, and velocity dispersion regardless of clump redshifts. This led them to conclude that clumps at $z\sim1.3$ are likely larger analogs of local HII regions, and turbulence sets the scaling relation. On the other hand, \citet{livermore:15} using observations on gravitationally lensed galaxies, combined with previous lensed and non-lensed galaxies, found that the mean surface brightness and characteristic luminosity of clumps evolves with redshift, becoming brighter as redshift increases. They argued that this can be explained by an evolving gas mass fraction that increases with redshift, which translates to a higher SFR density if the clumps are results of disk fragmentation via gravitational instability. These two results imply two distinct mechanisms that set the characteristics of star-forming clumps. We will compare our IROCKS measurements with these results, and attempt to reconcile the differences.

\label{clump_discussion}
\begin{figure}[b]
\centering
\includegraphics[width=0.45\textwidth]{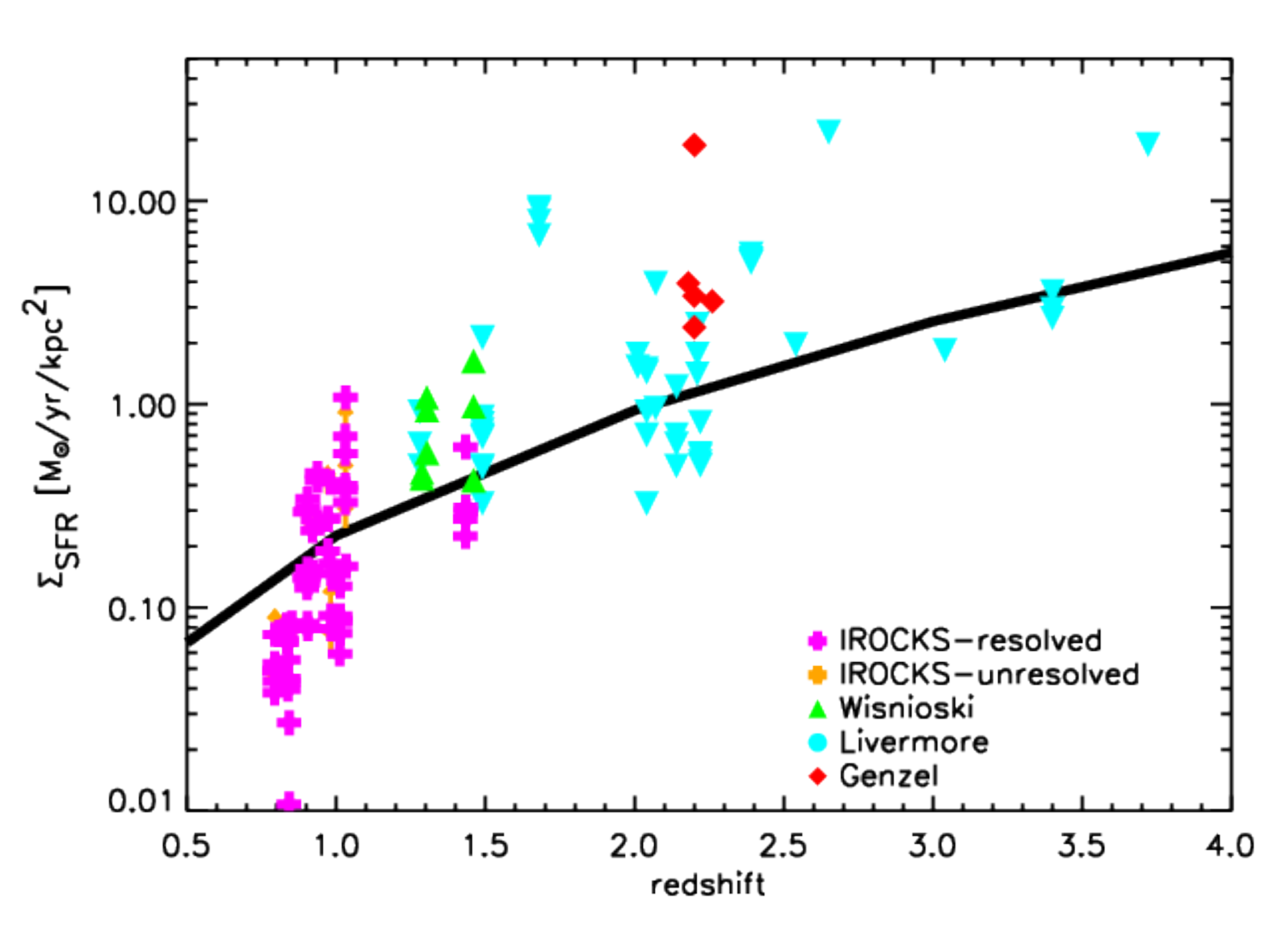}
\caption{Star formation rate surface density of clumps as a function of redshift. IROCKS and previous survey \citep{genzel:11, wisnioski:12, livermore:15} measurements are plotted with an empirical fit by \citet{livermore:15}. IROCKS data points are separated between resolved (magenta) and unresolved (orange) (see \S \ref{sec_clump}).}
\label{clump_surfaceSFR}
\end{figure}

\begin{figure*}
\centering
\includegraphics[width=0.9\textwidth]{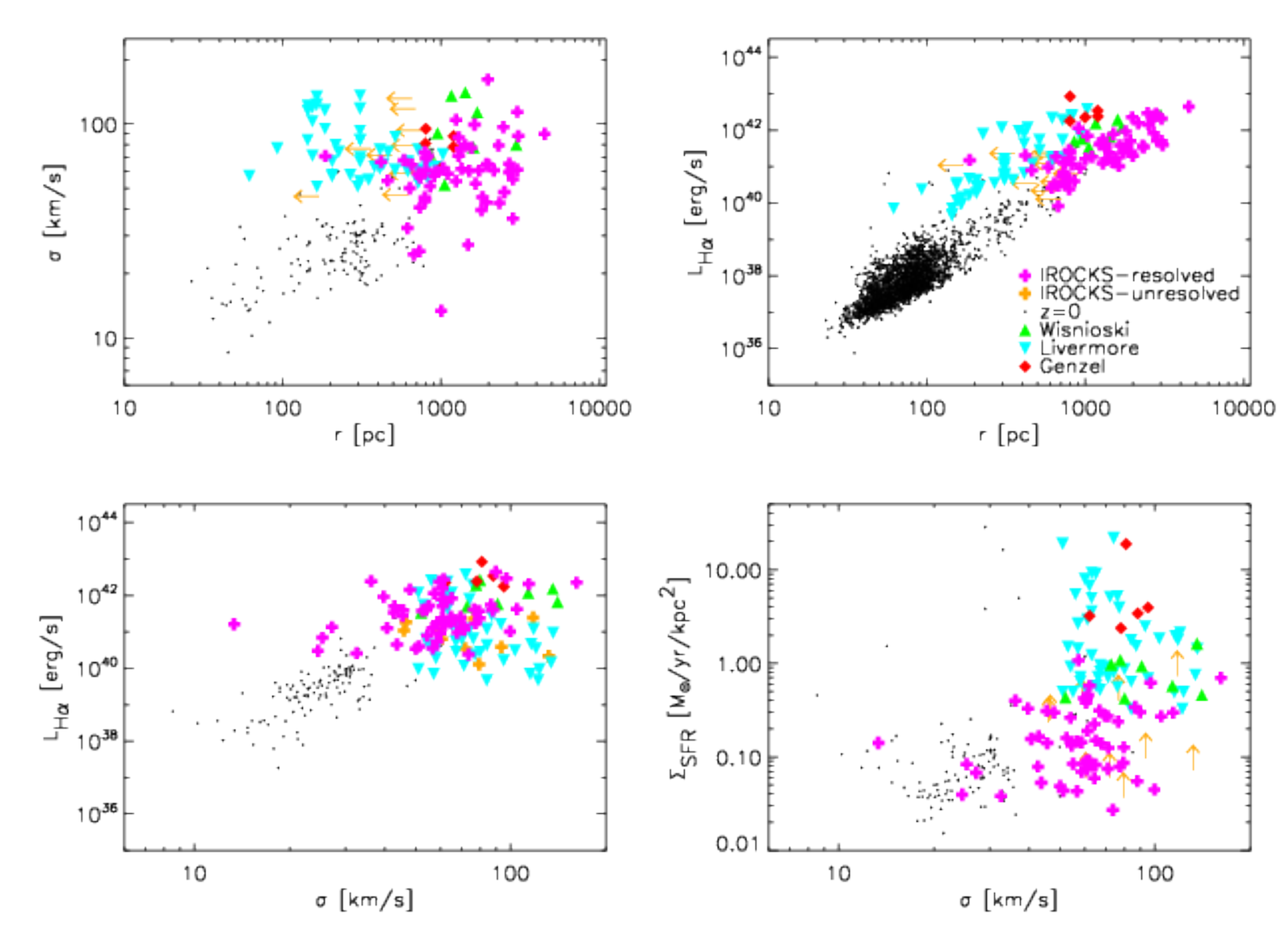}
\caption{Clump size, velocity dispersion, luminosity, and SFR surface density relations. IROCKS and previous surveys \citep{genzel:11, wisnioski:12, livermore:15} are shown. $z=0$ data points are described in \citet{wisnioski:12}. IROCKS data points are separated between resolved (magenta) and unresolved (orange) clumps or regions. For the two top panels, where the x-axis is in units of radii [pc], unresolved points are shown as left point arrows to emphasize these size measurements are upper limits. On the bottom right panel, the SFR densities for unresolved clumps are shown as up point arrows as they are the lower limits.}
\label{clump_relation}
\end{figure*}

Figure \ref{clump_surfaceSFR} shows the clump SFR surface density, $\Sigma_{\rm SFR}$, as a function of redshift, of IROCKS and data points from other surveys \citep{genzel:11,wisnioski:12,livermore:15}. Also shown in the figure is Equation 5 of \citet{livermore:15}, the empirical relation they found. We find excellent agreement with their relation, which we consider to be one of the supporting evidences for the disk fragmentation scenario. Figure \ref{clump_relation} shows the relations between our clump size, luminosity, velocity dispersion, and $\Sigma_{\rm SFR}$, together with data points from the same surveys as Figure \ref{clump_surfaceSFR}. \citet{wisnioski:12} found that assuming equal weighting for all points, combining local HII regions and $z>1$ clumps, luminosity scales with size by the relation, $L\propto r^{2.72\pm0.04}$. When only $z>1$ clumps are considered (eight clumps), this relation becomes $L\propto r^{1.42\pm0.45}$. Using IROCKS resolved clumps (58 clumps), we find $L\propto r_{1/2}^{1.47\pm0.15}$, and this is consistent with \citet{wisnioski:12}. In fact, like \citet{wisnioski:12}, we find our relation can be reasonably extended to HII regions at $z\sim0$. However, as already shown by Figure \ref{clump_surfaceSFR}, this does not imply a lack of time evolution in clump properties. Interestingly, we find that even though we have similar velocity dispersions as the other IFS studies, the SFR surface density is lower in our sample. Clumps with a given velocity dispersion are able to occupy a range of SFR surface density conditions. This probably indicates that clumps are not necessarily virialized, and gravitational instability contributes to the high dispersion observed \citep{livermore:15}. 

The $z\sim1$ clumps agree well with the slightly higher redshift IFS samples from \citet{wisnioski:12} on $\sigma-r$, $L-r$, and $L-r$ relations, but have some deviation on $\Sigma-\sigma$ relation. On $\Sigma-z$ relation, the $z\sim1$ clumps agree well with the IFS lensed galaxy samples from \citet{livermore:15}. Our clump SFR surface density measurements support the hypothesis of clumps forming from disk fragmentation. We find similarities between local HII regions and high-z star-forming clumps. Yet a larger statistical sample is still needed to explore redshift, stellar mass, and gas fraction trends that could point to some environmental impact on clump properties. Also, a better understanding between observational and analysis differences between IFS lensed and un-lensed population is still warranted.

\section{Conclusion}\label{dis}
In this paper, we have presented the first results of the IROCKS survey, which is currently the largest sample of IFS+AO observations of star-forming galaxies at $z \sim 1$. The sample consists of sixteen $z \sim 1$ and one $z \sim 1.4$ star-forming galaxies, selected from the four well studied fields, GOODS-North, GOODS-South, DEEP2, and UDS. All of our targets, but one, were observed with the upgraded OSIRIS grating at the Keck I telescope, with the assistance of a newly upgraded AO system. We focused on the kinematics and morphological properties of star-forming galaxies at $z \sim 1$ by using \ha{} emission line as a star formation tracer. The results of our survey are summarized as follows:

\begin{itemize}

\item[1] In our sample of sixteen star-forming galaxies with 0.794 $\leq z \leq$ 1.03 (median $z$ = 0.936), twelve are classified as single and four as multiple systems, based on the number of spectrally and/or spatially separated components observed. Our seventeenth source 11169 has $z = 1.43$, and is classified as a multiple system.

\item[2] We computed the SFR for each galaxy. Taking into account only extinction by the ISM (SFR$_{\text{\ha}}^0$) spans 0.2 $\leq \text{SFR}_{\text{\ha}}^0 \leq$ 42.7 \myr. Applying extra attenuation from HII regions, it increases by a factor of $\sim$ 2 to 5 and becomes 0.3 $\leq \text{SFR}_{\text{\ha}}^{00} \leq$ 108.4 \myr. We find that applying both ISM and HII extinction provides better agreement with the SFR esitmated from SED fitting.

\item[3] Using line width measurements, we find all $z \sim 1$ components to have line-of-sight velocity dispersions of \sigmaave{} $\gtrsim$ 48 km s$^{-1}$, with a median value of 61.6 km s$^{-1}$. In comparison, both components in 11169 ($z \sim 1.4$) have even higher dispersion, \sigmaave{} $\sim$ 90 km s$^{-1}$. Considering disk fraction using both disk model fitting and \vshear/\sigmaave criteria, $z \sim 1$ galaxies resemble $z > 1$ galaxies in that about one-third are disk-like. 

\item[4] The stellar mass of each galaxy is estimated using SED fitting, and it ranges between 9.6 $\leq \log{\text{M}_*/\text{\msun}} \leq$ 11.2. Gas mass and virial mass are given through SFR and kinematics arguments, and they are between 9.10 $\lesssim \log{\text{M}_{\text{gas, 1}}/\text{\msun}} \lesssim$ 11.04 and 9.54 $\lesssim \log{M_{vir}} \lesssim$ 10.62, respectively. Using both stellar and gas mass, we find the gas fraction in these galaxies ranges between 0.14 $< f_{\text{gas}} <$ 0.80.

\item[5] Clump properties in the $z \sim 1$ galaxies were explored for the first time by IFS, and we identified 68 star-forming clumps, among which 58 are resolved. The sizes of resolved clumps are 0.3 $ \lesssim r_{1/2} \lesssim$ 4.5 kpc, their SFRs are 0.1 $\lesssim$ SFR $\lesssim$ 26.7 \myr, and integrated dispersions are 13 $\lesssim$ \sigmaoned{} $\lesssim$ 132 km s$^{-1}$.

\item[6] Compared to the other high-z clump sample, they support the disk fragmentation model as the clump formation mechanism while the $z \sim 1$ clumps follow a similar size-luminosity clump relation as local HII regions even though they are orders of magnitude larger in SFR and size. 

\end{itemize}

The high spatial resolution that IFS+AO provides comes with a sacrifice of SNR that impacts our measurements of galaxy rotation. Compared to observations without AO, our observations are less sensitive to low surface brightness regions of the galaxies, which is where the plateau velocity should be measured. Consequently, our rotation measurements are biased toward the more dispersed, central portions of the galaxies, and should not be used as a direct comparison to 1D slit-based spectroscopy observations that probe the fainter outskirts of galaxies. Indeed, IFS+AO observations find more dispersion dominated galaxies while non-AO find more rotationally-dominated systems because of this effect \citep{newman:12b}.

In order to boost SNR in the low surface brightness regions of the galaxies, where plateau velocities are reached, we apply a smoothing to each galaxy data cube. This smoothing does not have a significant impact on the global dispersions (Appendix \ref{sec:smooth}) and dispersion profiles (\S \ref{kine_model}). However, it may soften the velocity gradient, resulting in a lower estimate underestimate of \vshear{} (Appendix \ref{sec:smooth}).

High-redshift kinematic studies systematically classify their kinematic types by using disk model fitting and the global $v/\sigma$ parameter. However, interacting pairs and late-stage merger remnants have been shown that they can sometimes produce similar kinematic fields to high-redshift disk systems. Using both kinematic- and morphological-analysis is suggested to help distinguish between late-stage mergers and rotating disks \citep{hung:15}. It is also important to note that when comparing kinematic properties (such as the disk or merger fraction), we need to use more uniform kinematic distinction criteria between different spectroscopy studies. This has been challenging since each group has been re-defining their kinematic distinction criteria, and their disk modeling procedures varies. The community should be careful when combining data sets, and we need to push more for unified data samples and analysis techniques, especially between differing instruments.

In the last few years, more physically realistic high resolution simulations have become available, and galaxy formation and evolution are now studied at individual galaxy structure size scales ($\sim$ kpc). Comparing our kinematic results against zoomed-in hydrodynamics simulation data points of \citet{kassin:14}, our results fall between the cold (without feedback) and warm (with stellar feedback) models, suggesting at least a moderate amount of feedback is needed to reproduce our results (see Figure 1 of \citet{kassin:14}). However, \citet{kassin:14} have commented on their results' possible dependencies on poorly constrained quantities such as the average stellar mass of galaxies and the spatial variations of gas density and temperature. Simulations which probe parameters such as stellar mass, feedback mechanism, and metallicity would certainly be helpful for pinpointing the physics that dictate ''feedback'' in galaxy evolution.

Our $z \sim 1$ clumps are consistent with the SFR surface density and redshift relation found by \citet{livermore:15}, who argued that their relation suggests gravitational instability as the clump formation mechanism. We can further test this theory by measuring the gas fractions of individual clumps using, for example, molecular line emissions from ALMA, and comparing them to their luminosities. The luminosity of a clump is related to its mass, which, if formed from gravitational instability, is higher for larger gas fractions.

In this study, we extended the IFS study of high-z kinematics and morphologies to $z \sim 1$ regime with sixteen additional sources. However the number of IFS high-z samples are still limited and currently only able to probe the most massive and luminous star forming galaxies. Extremely large telescopes coming in a few years combined with IFS+AO will enable us to see high-z galaxies at the scale of a giant molecular cloud, and will provide us key information to understand galaxy evolution. 

This is the first paper from the IROCKS study, and a second paper on the nebular diagnostic of these galaxies is forthcoming, where we will focus on resolved metallicity gradients on ten galaxies and explore ionization and feedback mechanisms like shocks and AGN.

\acknowledgements
The Dunlap Institute is funded through an endowment established by the David Dunlap family and the University of Toronto. This research was partly supported by the Natural Sciences and Engineering Research Council (NSERC) of Canada Discovery Grant. We extend our gratitude to Nick Mostek, Alison Coil, and Bahram Mobasher for generously sharing their SED parameters for these sources. We also thank Emily Wisnioski for providing local HII measurements used in our clump analysis. We are grateful to the Keck Observatory staff, Jim Lyke and Randy Campbell, for helping with many of the observations and the OSIRIS data reduction pipeline. We thank the referee for a thorough reading and valuable comments. The data presented herein were obtained at the W.M. Keck Observatory, which is operated as a scientific partnership among the California Institute of Technology, the University of California, and the National Aeronautics and Space Administration. The Observatory was made possible by the generous financial support of the W.M. Keck Foundation. The authors wish to recognize and acknowledge the very significant cultural role and reverence that the summit of Mauna Kea has always had within the indigenous Hawaiian community. We are most fortunate to have the opportunity to conduct observations from this mountain.

\bibliography{reference}
\bibliographystyle{apj}

\clearpage
\appendix
\section{Adaptive smoothing}\label{adaptive}
\begin{figure}[b]
\centering
\includegraphics[width=0.5\textwidth]{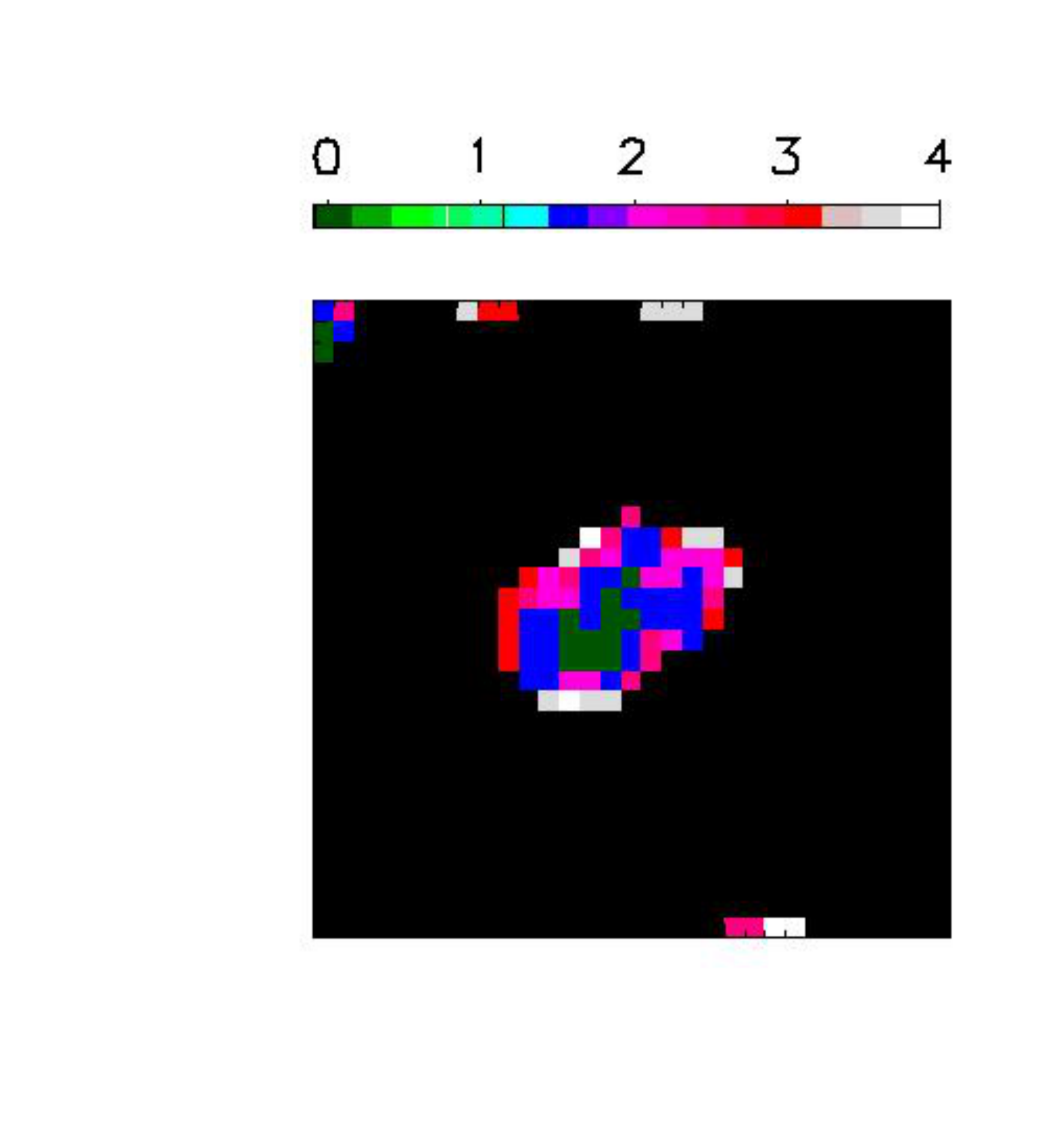}
\caption{FWHM map of UDS11655. The color presents the smallest value of FWHM that makes the particular spaxel reach a desired SNR. The final single smoothing width for each data cube is the mean of FWHM map. For UDS11655, FWHM = 2.0 pixel is chosen as a final FWHM.}
\label{adaptive_ex}
\end{figure}
In IFS studies of high redshift galaxies, very often data cubes are spatially smoothed by a Gaussian function of FWHM $\sim$ 2 pixels to increase the signal-to-noise ratio \citep[e.g.][]{law:09, wright:09, forster:09, genzel:11, wisnioski:11, epinat:12}. Some properties (e.g., \sigmaoned, \sigmaave, and SFR) are not significantly affected by smoothing, but other parameters need careful treatment. For example, when we study the spatially resolved quantities, such as the metallicity gradient across the galaxy and resolved clumps, the smoothing process distributes the flux to neighbour pixels and as a result smears out the information. In particular, observations with AO, where diffraction limited observation is potentially achievable, lowering the spatial resolution in the data reduction process is detrimental. In order to increase SNR while preserving as high spatial resolution as possible, the choice of optimum width is crucial. We develop an adaptive smoothing code to find the best choice of smoothing width. 

In short, the code iteratively applies smoothing of increasing FWHM to a data cube until spaxels reach a desired or optimal SNR. In each iteration, the entire reduced, un-smoothed cube is smoothed by a single FWHM, and the SNR of each spaxel in an \ha{} flux map is calculated using the method described in \S \ref{red}. For the next iteration, the same original, reduced, un-smoothed cube is then smoothed by a wider FWHM, usually increasing by 0.5 pixel for each iteration, and we repeat the process until the maximum FWHM is reached, or most spaxels achieve a high SNR. The smallest smoothing FWHM that allows the spaxel at $[i,j]$ to reach the desired SNR is then recorded as FWHM$_{i, j}$. The most optimized, final smoothing width for the particular data cube is the mean FWHM$_{i,j}$ within the region of interest. Figure \ref{adaptive_ex} shows a FWHM$_{i, j}$ map of UDS11655 as an example. The color presents the value of FWHM, and is illustrative how adaptive smoothing can potentially be powerful at increasing the SNR of low surface brightness emission.

In the analysis, we use this code only to find the most optimized smoothing width. However, this code has the potential to produce an adaptively smoothed data cube, where spaxels of higher signal would be smoothed by a narrower FWHM. Such a method is suitable for morphology related analysis (e.g., morphology parameter, size, peak location), and particularly beneficial when (1) the galaxy contains an AGN with a high single \nii/\ha{} peak, which would allow for a more accurate measurement of the location of the AGN; also, when (2) multiple star forming clumps are located close to each other, which would prevent excess smoothing to smear the boundaries between them. On the other hand, a spatial varying smoothing length makes it difficult to model the beam size correctly. The potential of this method and its numerous merits will be explored in future studies.

\section{Effect of smoothing on kinematics}\label{sec:smooth}
\begin{figure*}[t]
\centering
\includegraphics[width=0.8\textwidth]{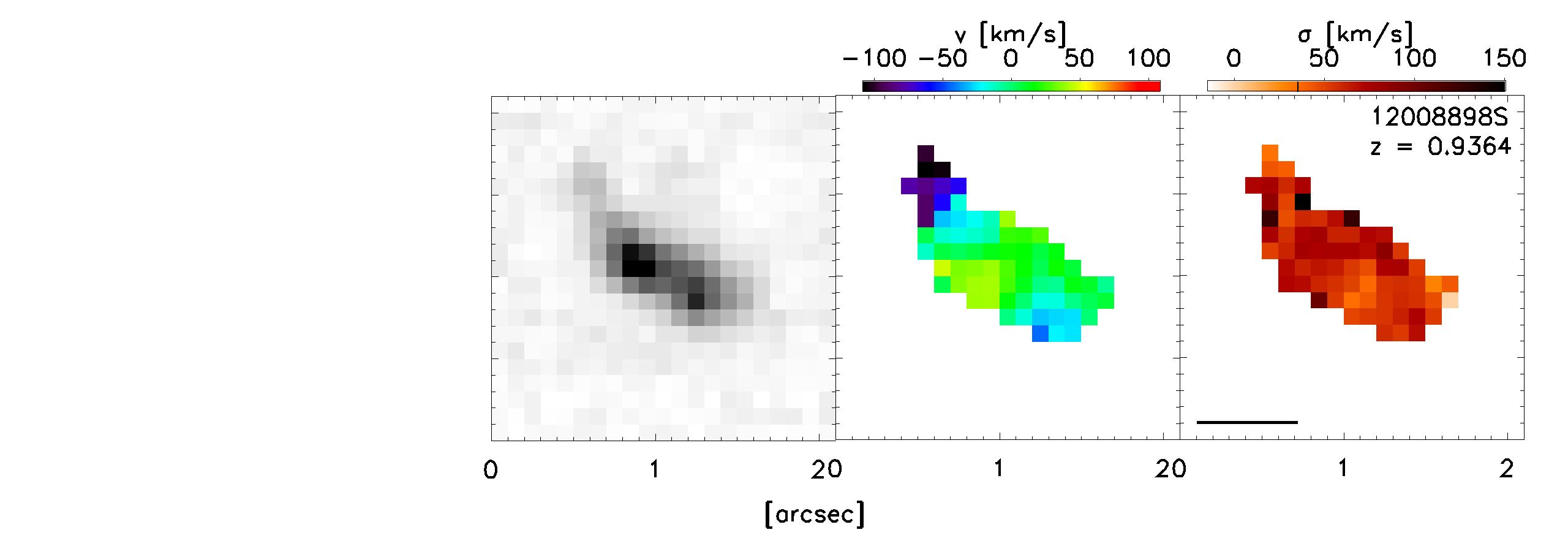}
\includegraphics[width=0.8\textwidth]{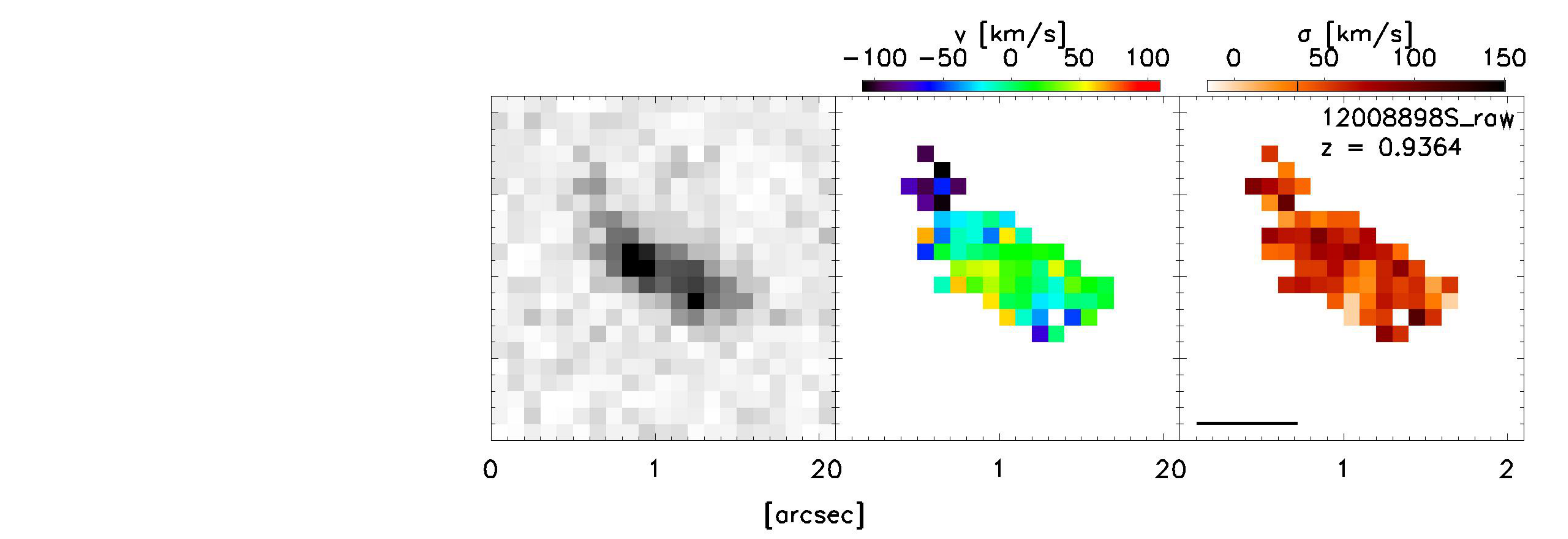}
\includegraphics[width=0.8\textwidth]{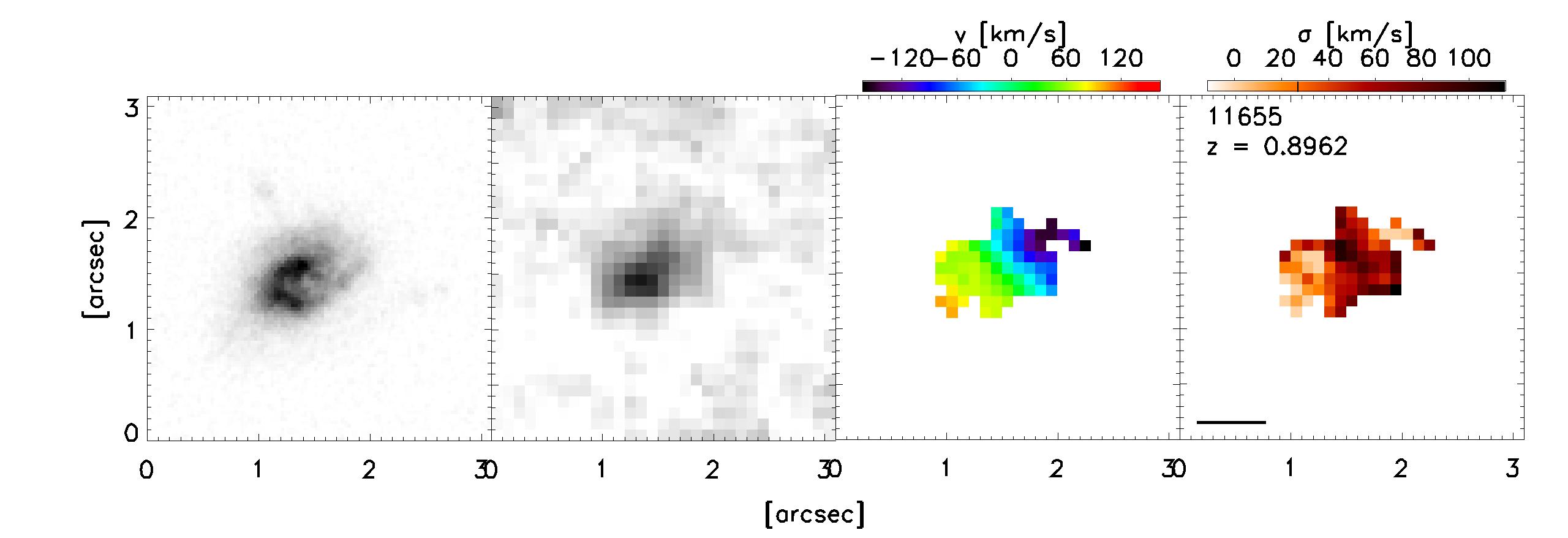}
\includegraphics[width=0.8\textwidth]{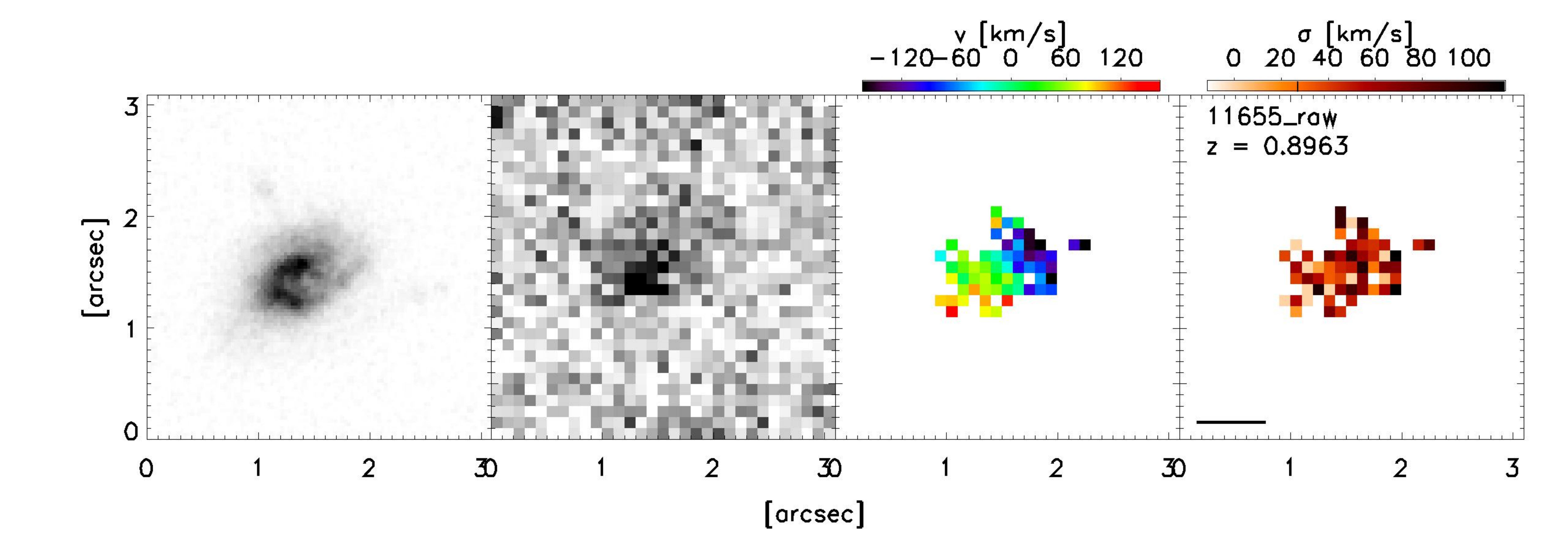}
\caption{From the left, HST (when available), \ha, rotation velocity, and velocity dispersion maps of DEEP2-12008898S smoothed (first row) and unsmoothed (second row) as an example of irregular galaxy, and UDS-11655 smoothed (third row) and unsmoothed (bottom) as an example of disk galaxy. The smoothing FWHM of DEEP2-12008898S is 1.5 pixel, and that of UDS-11655 is 2.0 pixel.}
\label{irregular_comp}
\end{figure*}

Side-by-side comparisons of the AO and AO+artificial smoothing resolution kinematic maps for an irregular galaxy (DEEP2-12008898S smoothed by FWHM = 1.5pixel) and a disk candidate (UDS11655 smoothed by FWHM = 2.0pixel) are shown in Figure \ref{irregular_comp}.

We apply a local velocity gradient correction to the dispersion. Half of the biggest velocity difference between vertical or horizontal immediate neighbor pixels, $\Delta v = 0.5 \times \text{max}(|v_{i+1, j} - v_{i-1, j}|,|v_{i, j+1} - v_{i, j-1}|)$, is subtracted from the local dispersion in quadrature, $\sigma^{\text{corr}} = \sqrt{\sigma^2 - \Delta v^2}$. The SNR weighted average of $\sigma^{\text{corr}}$ in our sample is typically $\sim60$ km s$^{-1}$, compared to $\sim64$ km s$^{-1}$ for the non-corrected \sigmaave, which indicates the local velocity gradient within a pixel is small compared to the line-of-sight dispersion.

We also investigate the effects of beam smearing on the observed velocities. Using one of the highest SNR sources in the sample, we find the un-smoothed data to have a dispersion lower by $\sim$4 km s$^{-1}$ compared to the smoothed data set. When the additional local gradient correction is applied to the un-smoothed data, the dispersion is lowered further by $\sim$5 km s$^{-1}$. This confirms our local velocity gradient correction analysis with the smoothed data sets.

Overall the line-of-sight velocity dispersion measurements are resolved (i.e., measured widths are not the widths of smoothing nor local rotation), and after the local gradient corrections have been applied, they are found to be $\gtrsim$ 55 km s$^{-1}$ across our sample. As shown in previous studies, this is significantly higher than velocity dispersions found in local galaxies. We note that our method for removing the local velocity gradient is not rigorous: we have included it to provide a rough quantitative estimate of the contribution of our finite spatial resolution to the line-of-sight dispersion. For the rest of our analysis, we will not apply this correction, which as we have shown has a $\lesssim 10\%$ effect on our results.

For the two galaxies shown in Figure \ref{irregular_comp}, \vshear{} is reduced by 16\% (irregular) and by 22\% (disk). This is consistent with the expectation that smoothing softens velocity gradients, and may result in an underestimate of \vshear{} in the disk candidates. Smoothing is nonetheless necessary, since, as shown by Figure \ref{irregular_comp}, the increased SNR of the observations allows more robust kinematics measurements on each of the sources.
\clearpage
\section{Kinematic maps}\label{sec:kinemapfigure}
In this section, we show IROCKS \ha{} flux, radial velocity, and velocity dispersion maps, extracted from OSIRIS data cube. When available, HST images are also shown.
\begin{figure*}[b]
\centering
\includegraphics[width=0.82\textwidth]{convert_image/UDS11655_combine.jpg}
\includegraphics[width=0.82\textwidth]{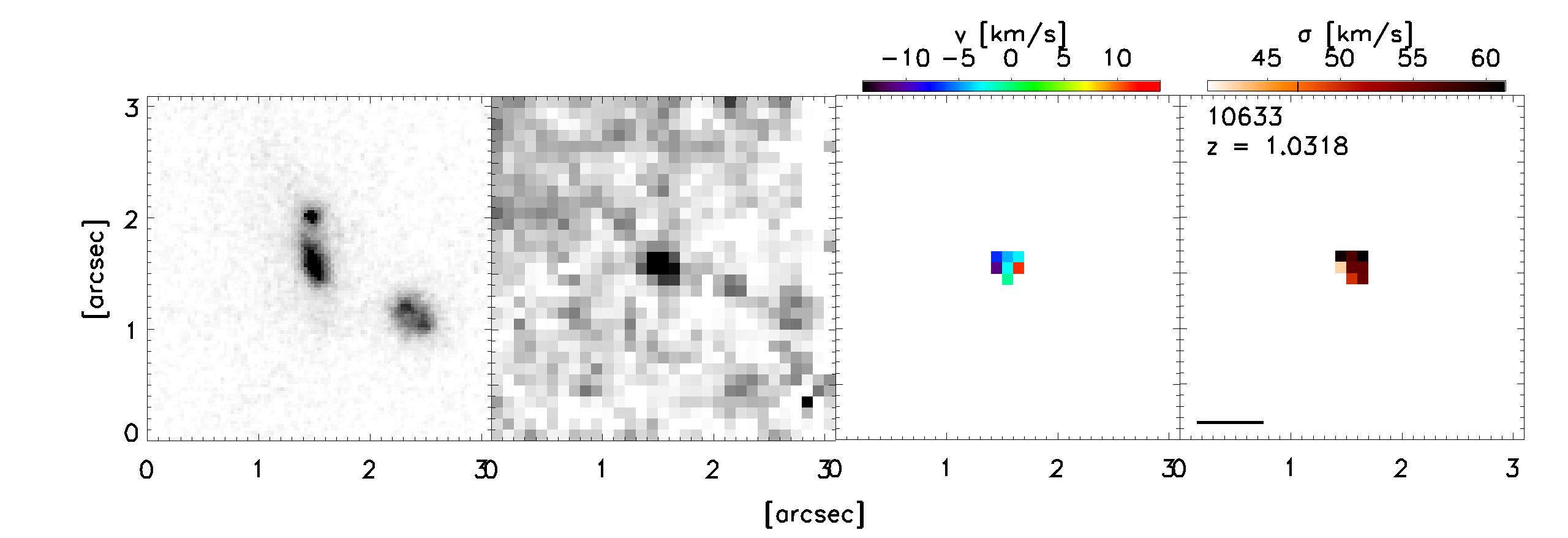}
\includegraphics[width=0.82\textwidth]{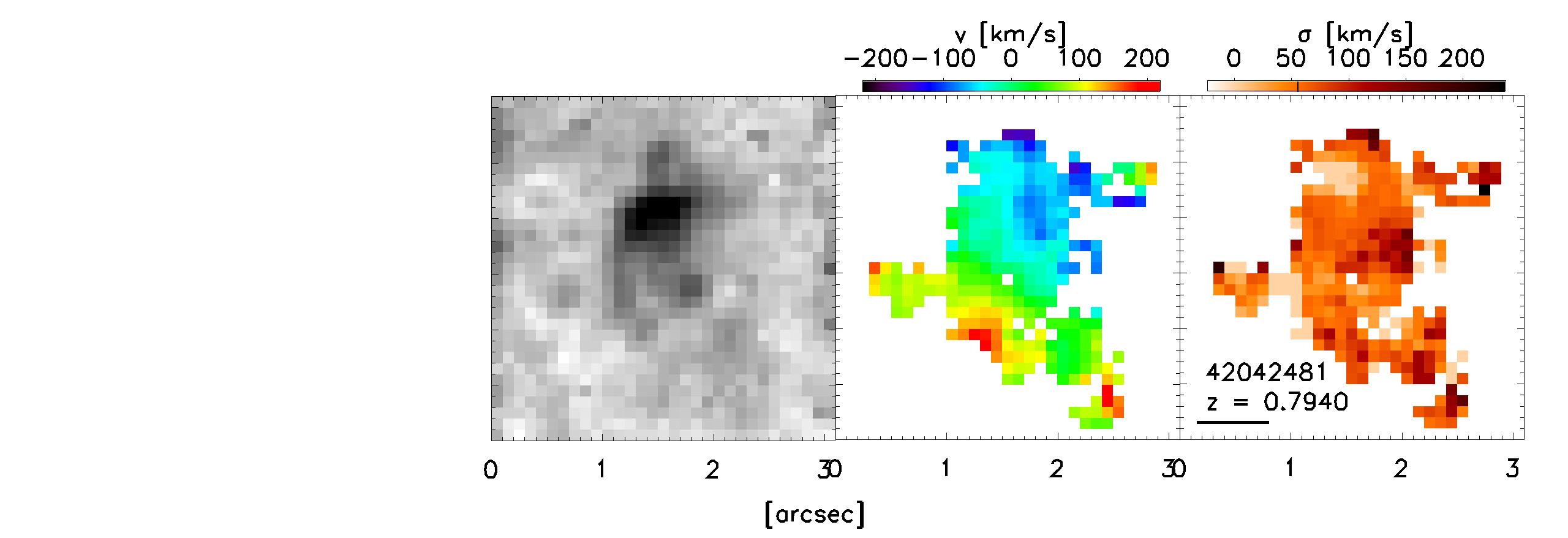}
\includegraphics[width=0.82\textwidth]{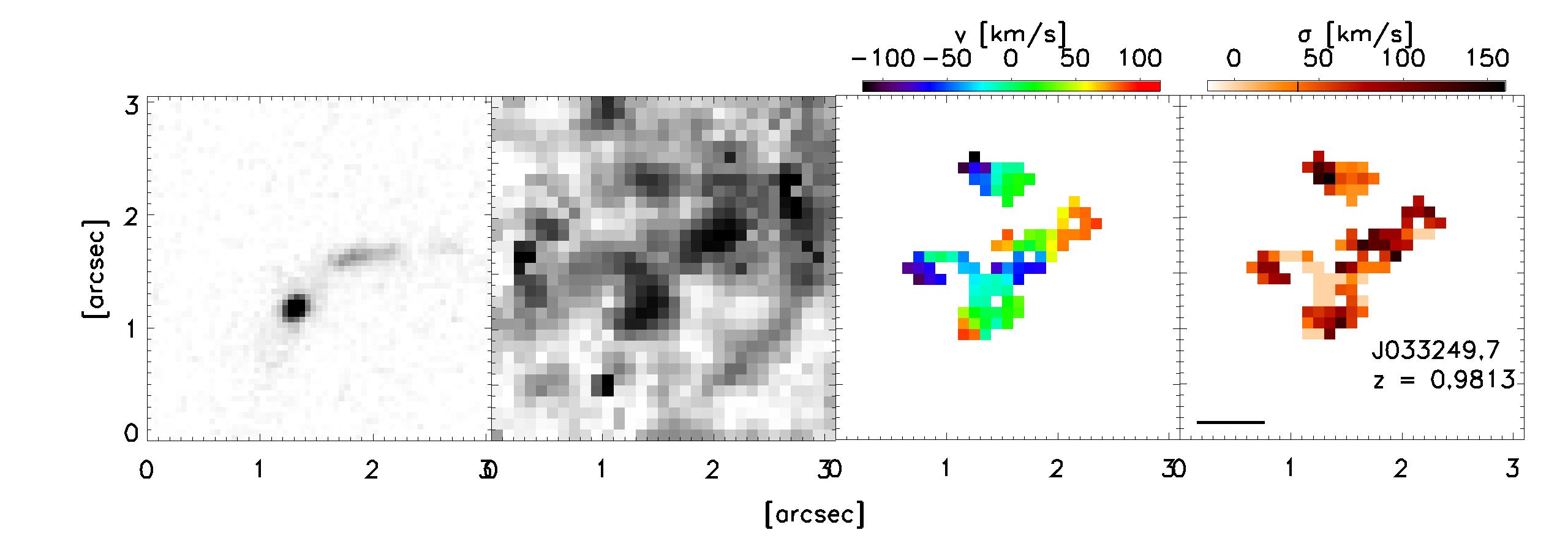}
\caption{From the left, HST (when available), \ha{} flux, radial velocity, and velocity dispersion maps. The orientation of the images are fixed to be North up and East to the left. On the right panel, the name of the source and its redshift are shown in the top (or other location when the text overlaps with the map), and the length of the black line on the left bottom corner represents a projected size of 5 kpc at the redshift of the galaxy. All HST images are taken by F814W filter, except for J033249.7 (F606W), TKRS11169 (F850LP), TKRS9727 (F850LP), and TKRS7615 (F850LP).}
\label{kine_11655}
\end{figure*}
\begin{Contfigure}[p]
\centering
\includegraphics[width=0.82\textwidth]{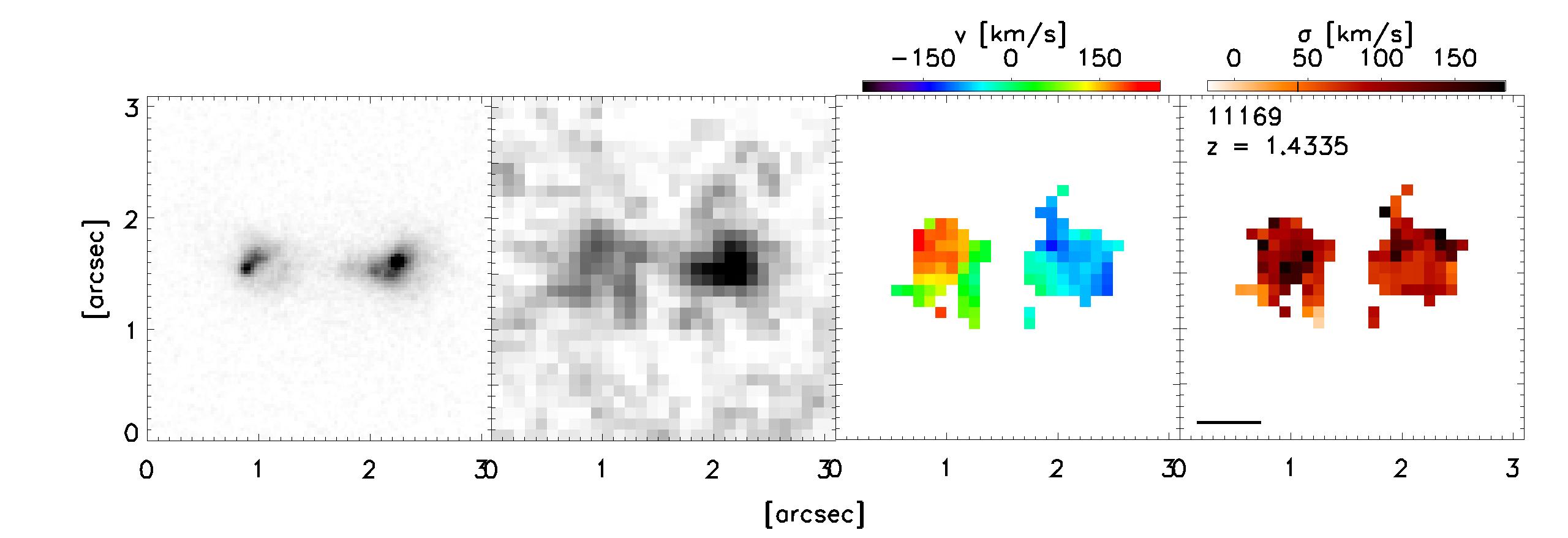}
\includegraphics[width=0.82\textwidth]{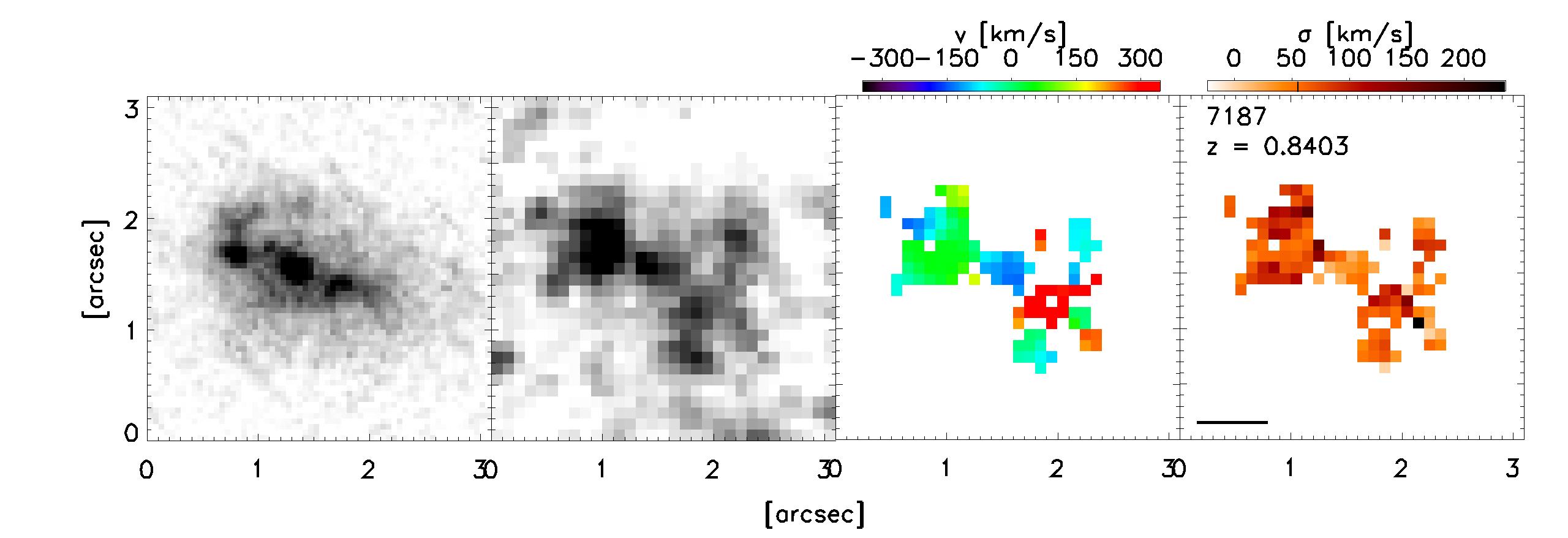}
\includegraphics[width=0.82\textwidth]{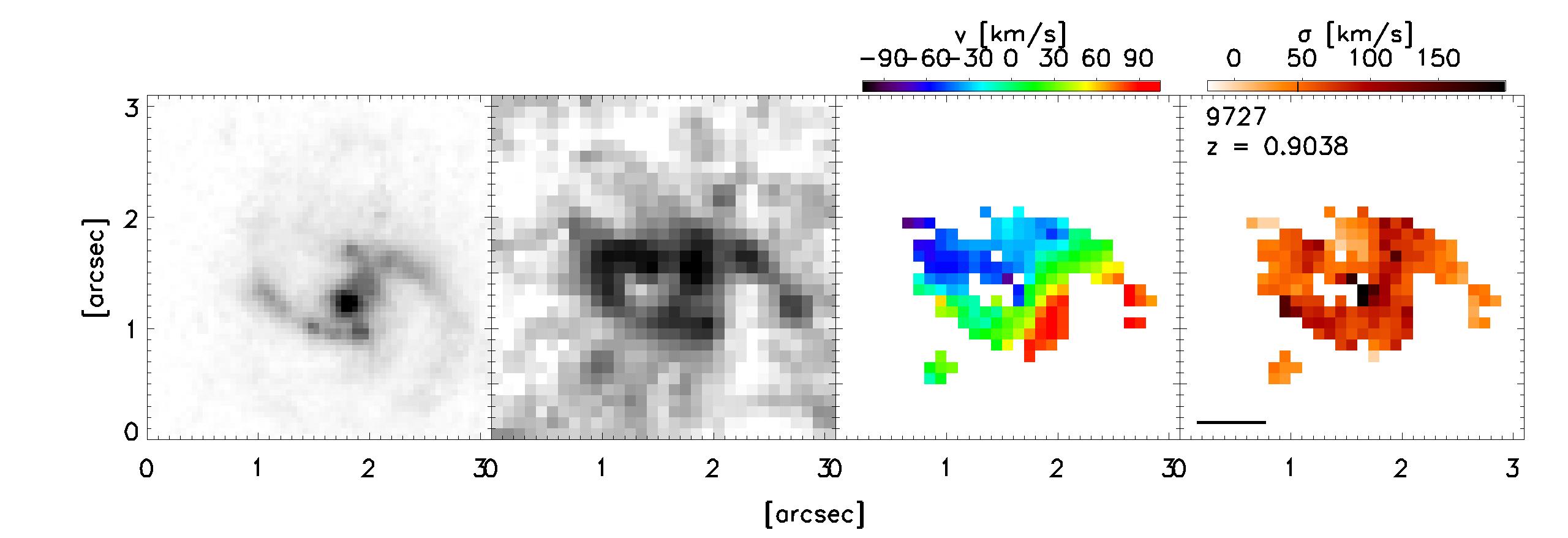}
\includegraphics[width=0.82\textwidth]{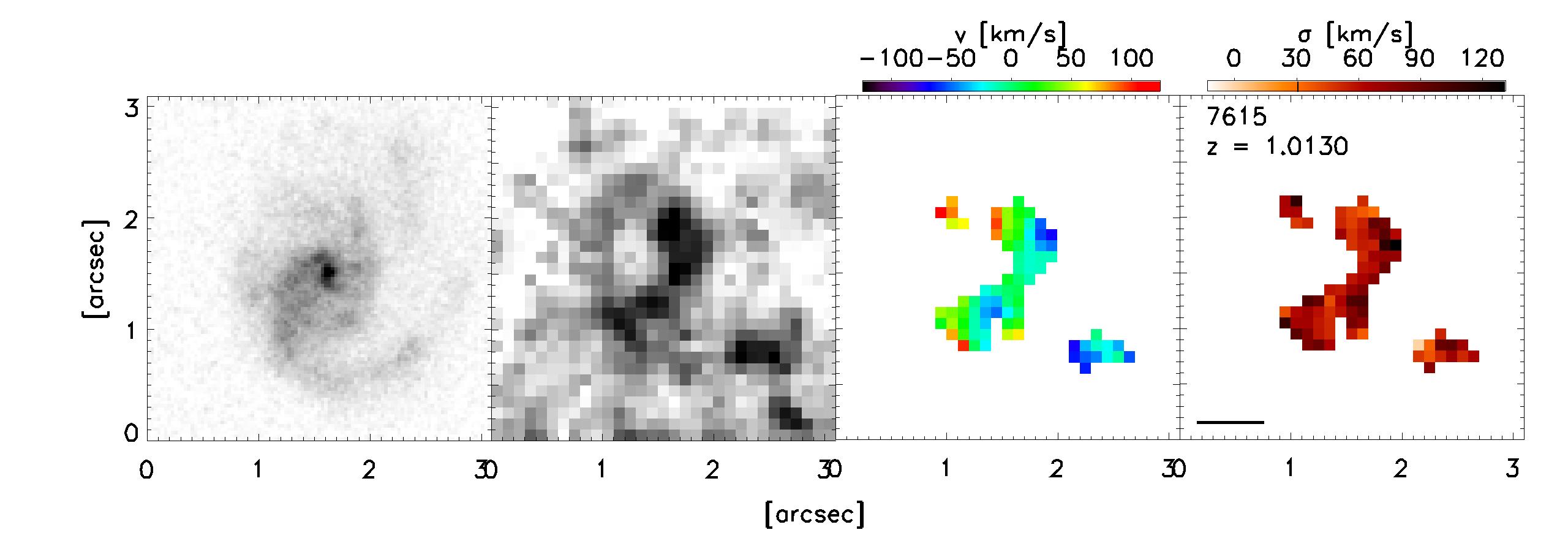}
\caption{}
\end{Contfigure}
\begin{Contfigure}[p]
\centering
\includegraphics[width=0.82\textwidth]{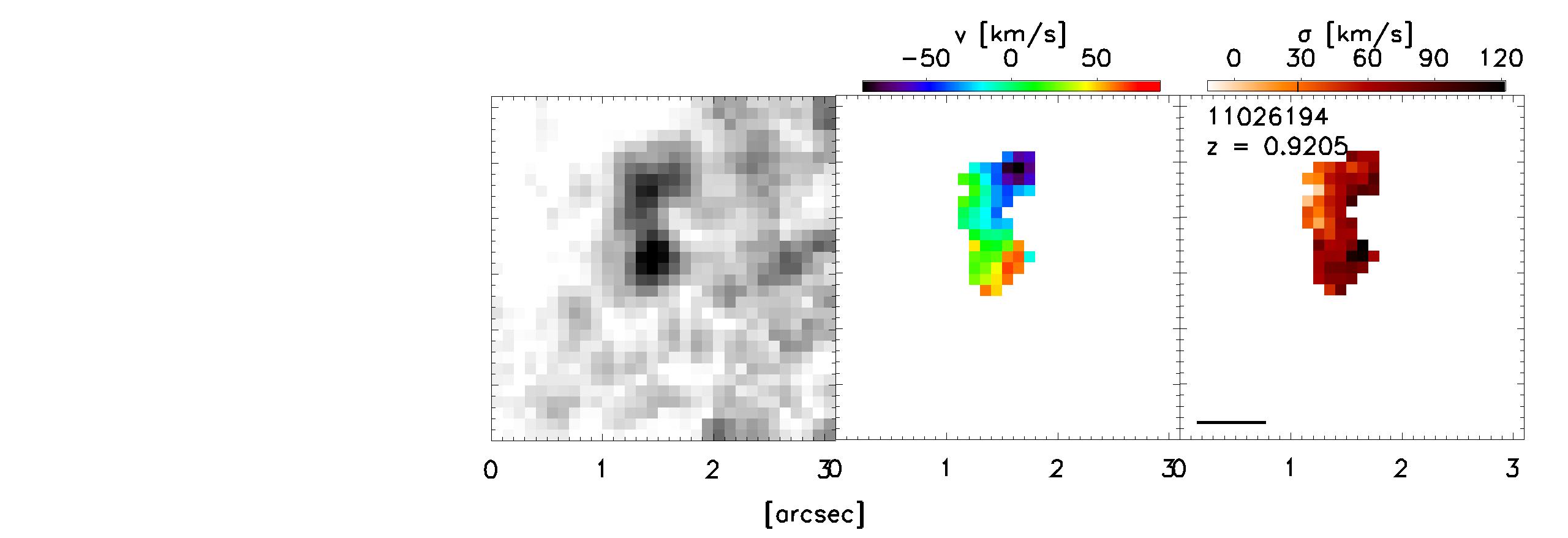}
\includegraphics[width=0.82\textwidth]{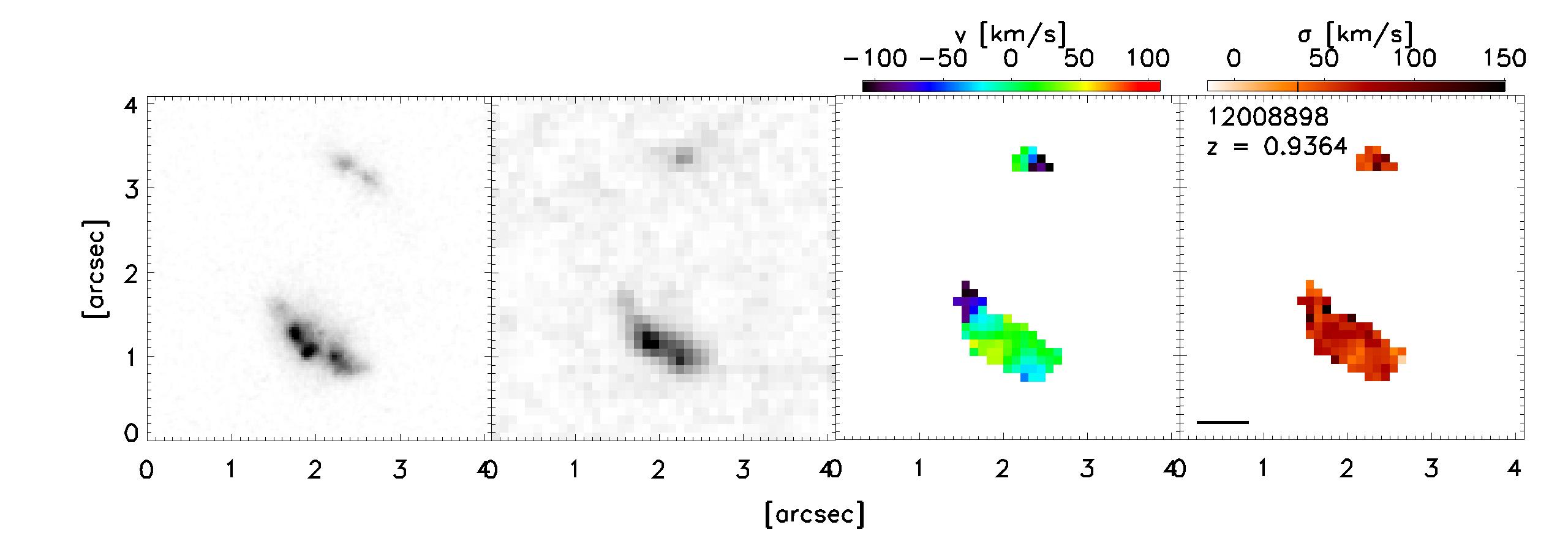}
\includegraphics[width=0.82\textwidth]{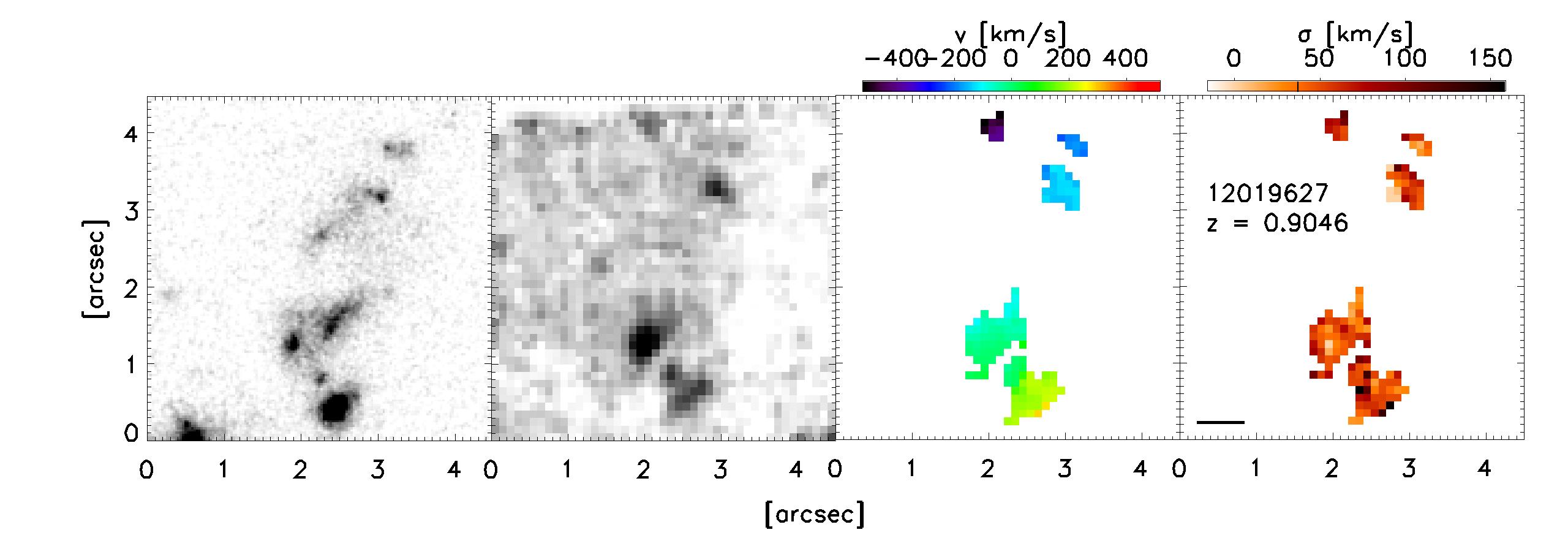}
\includegraphics[width=0.82\textwidth]{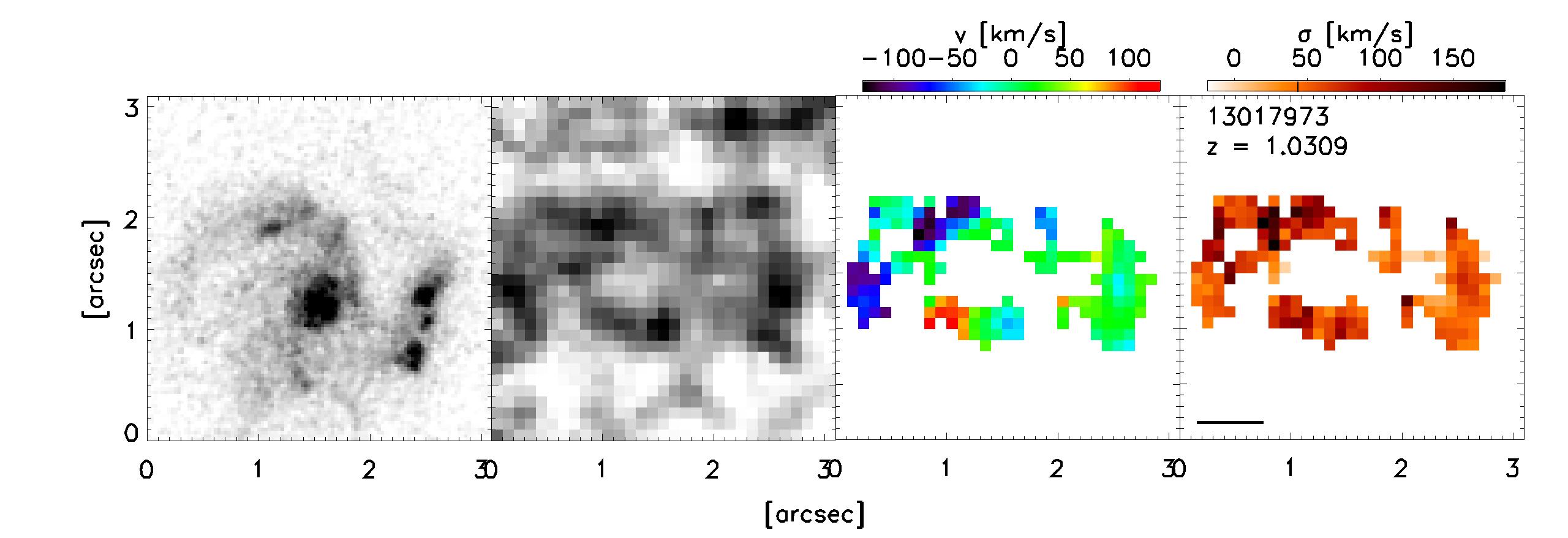}
\caption{}
\end{Contfigure}
\begin{Contfigure}[p]
\centering
\includegraphics[width=0.82\textwidth]{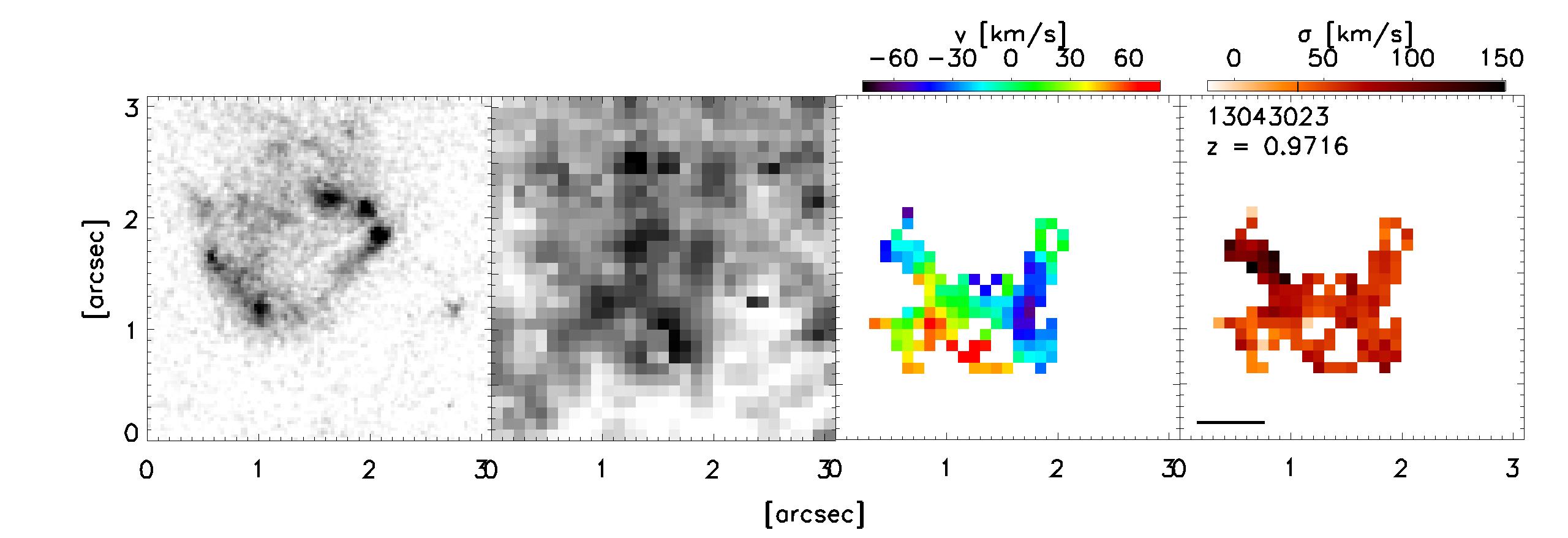}
\includegraphics[width=0.82\textwidth]{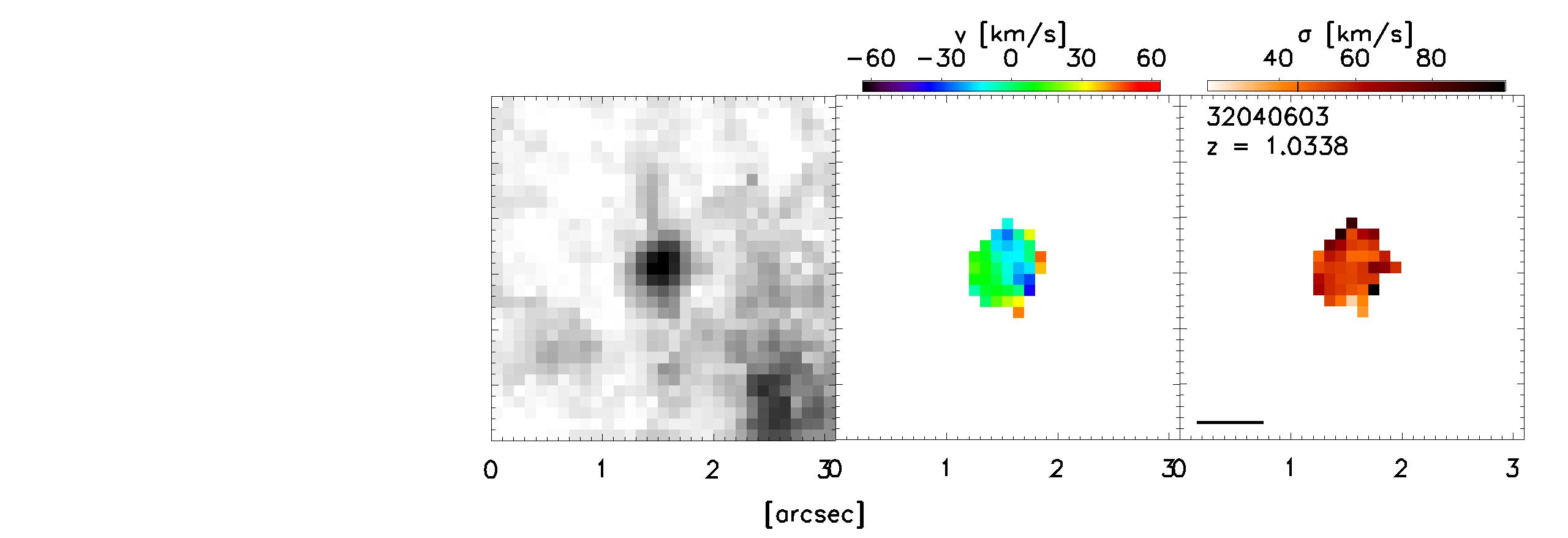}
\includegraphics[width=0.82\textwidth]{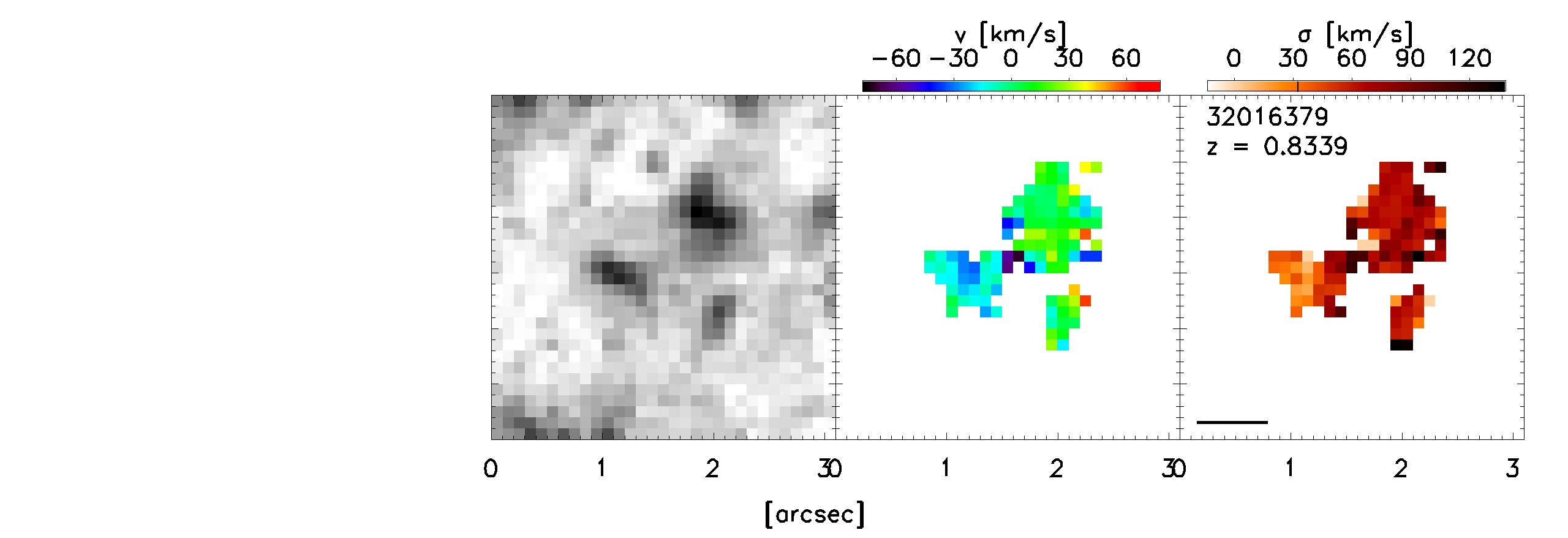}
\includegraphics[width=0.82\textwidth]{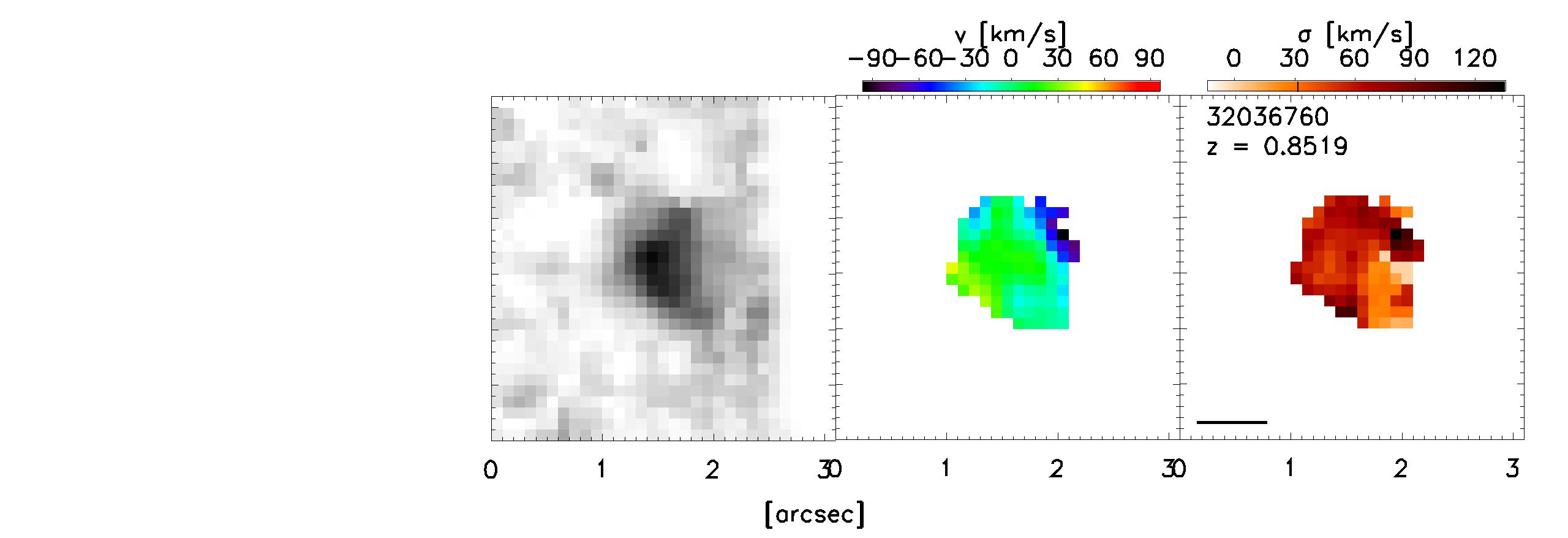}
\caption{}
\end{Contfigure}
\begin{Contfigure}[t]
\centering
\includegraphics[width=0.82\textwidth]{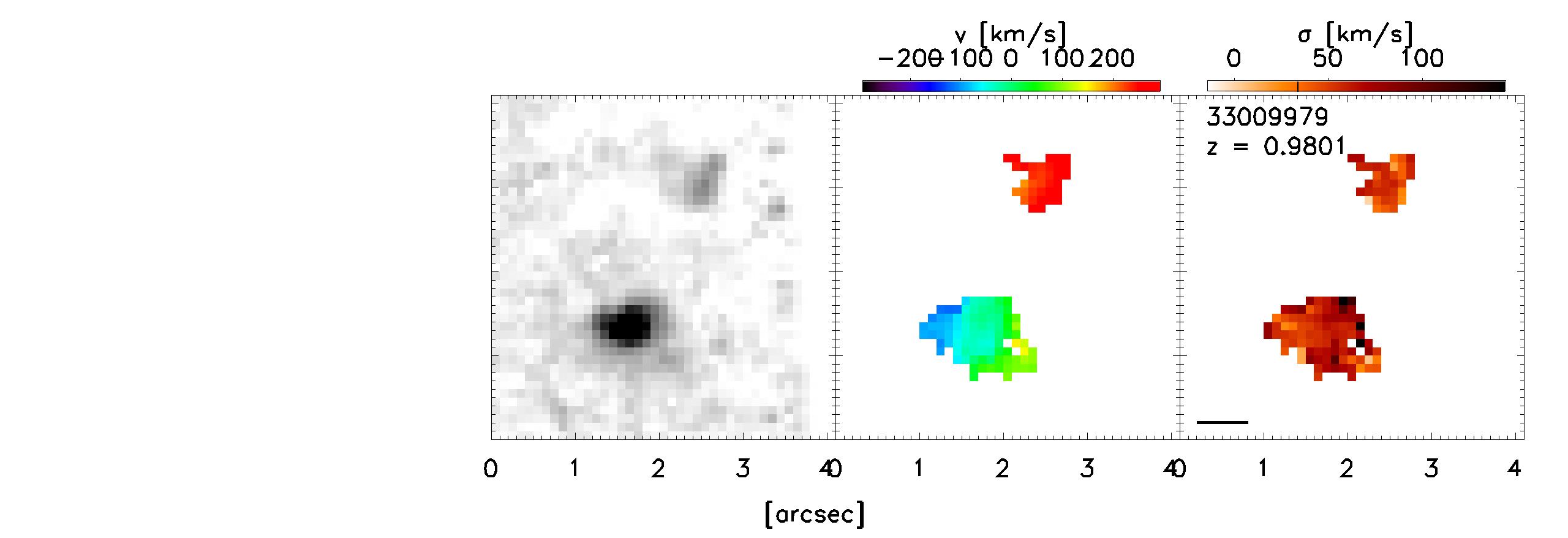}
\caption{}
\end{Contfigure}

\section{1D spectrum}
In this section, spatially integrated 1D spectra of IROCKS samples are shown. When the target is a multiple system, we spatially separate them and make each 1D spectrum.
\begin{figure*}[b]
\centering
\includegraphics[width=0.49\textwidth]{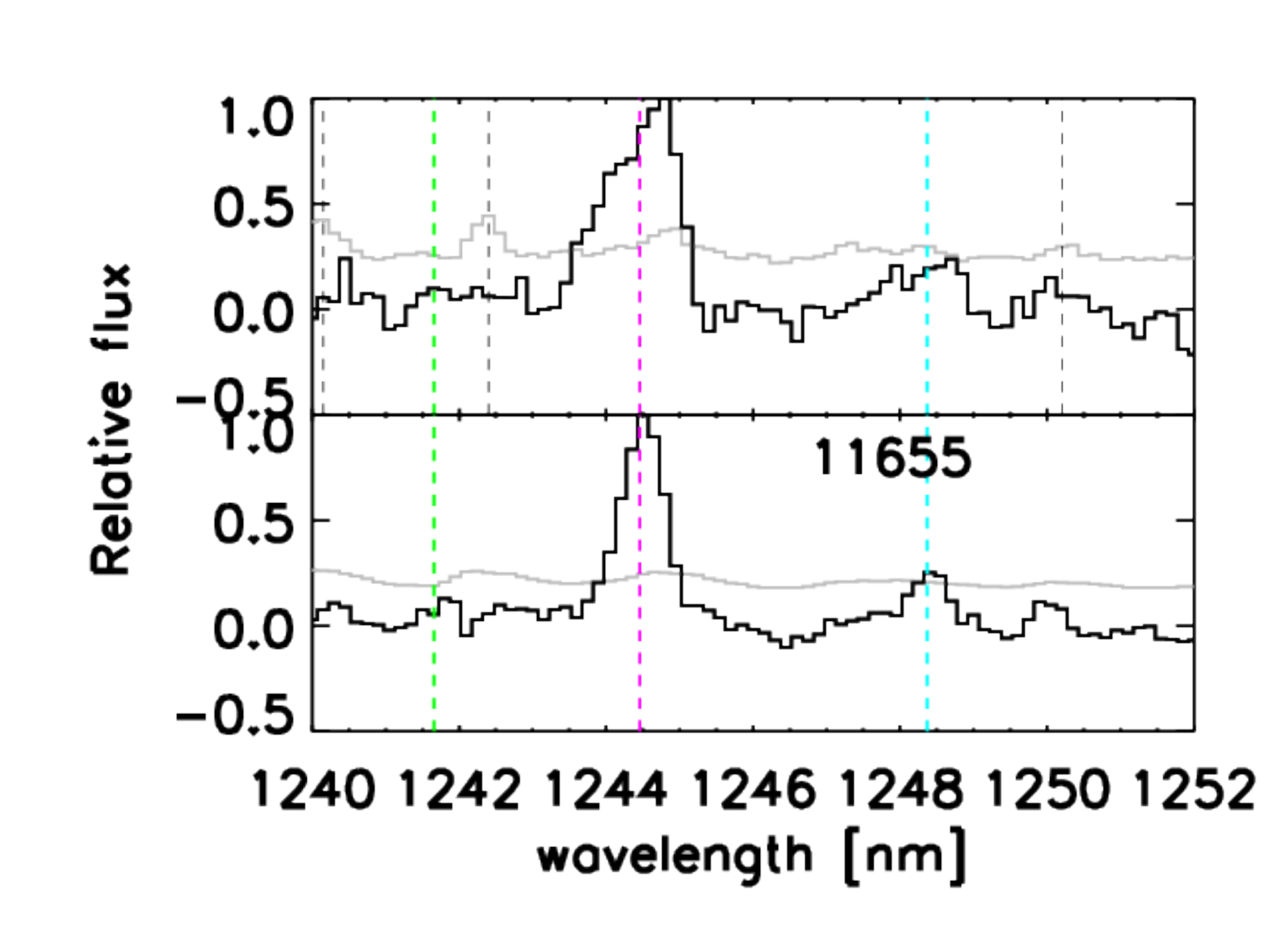}
\includegraphics[width=0.49\textwidth]{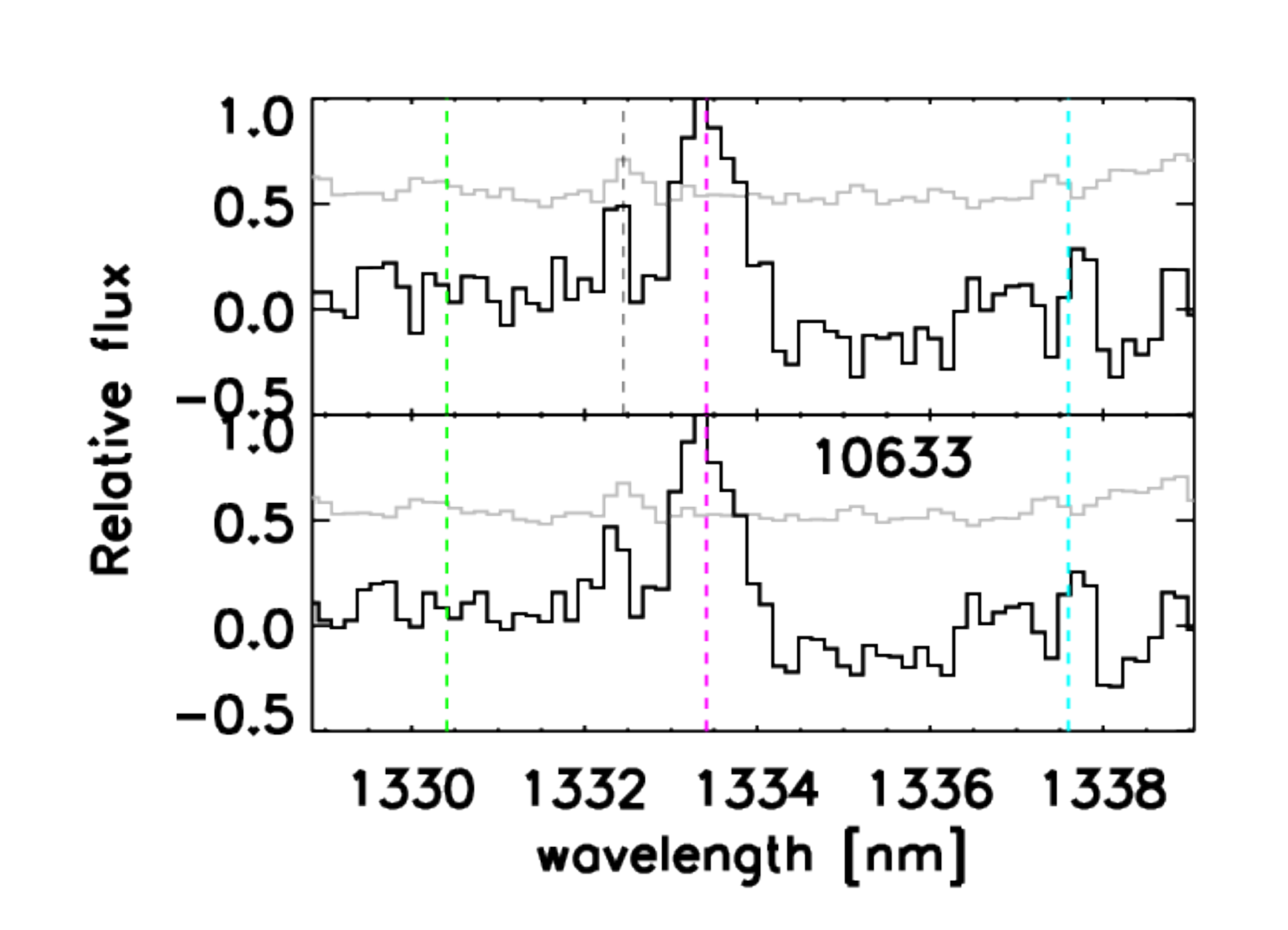}
\includegraphics[width=0.49\textwidth]{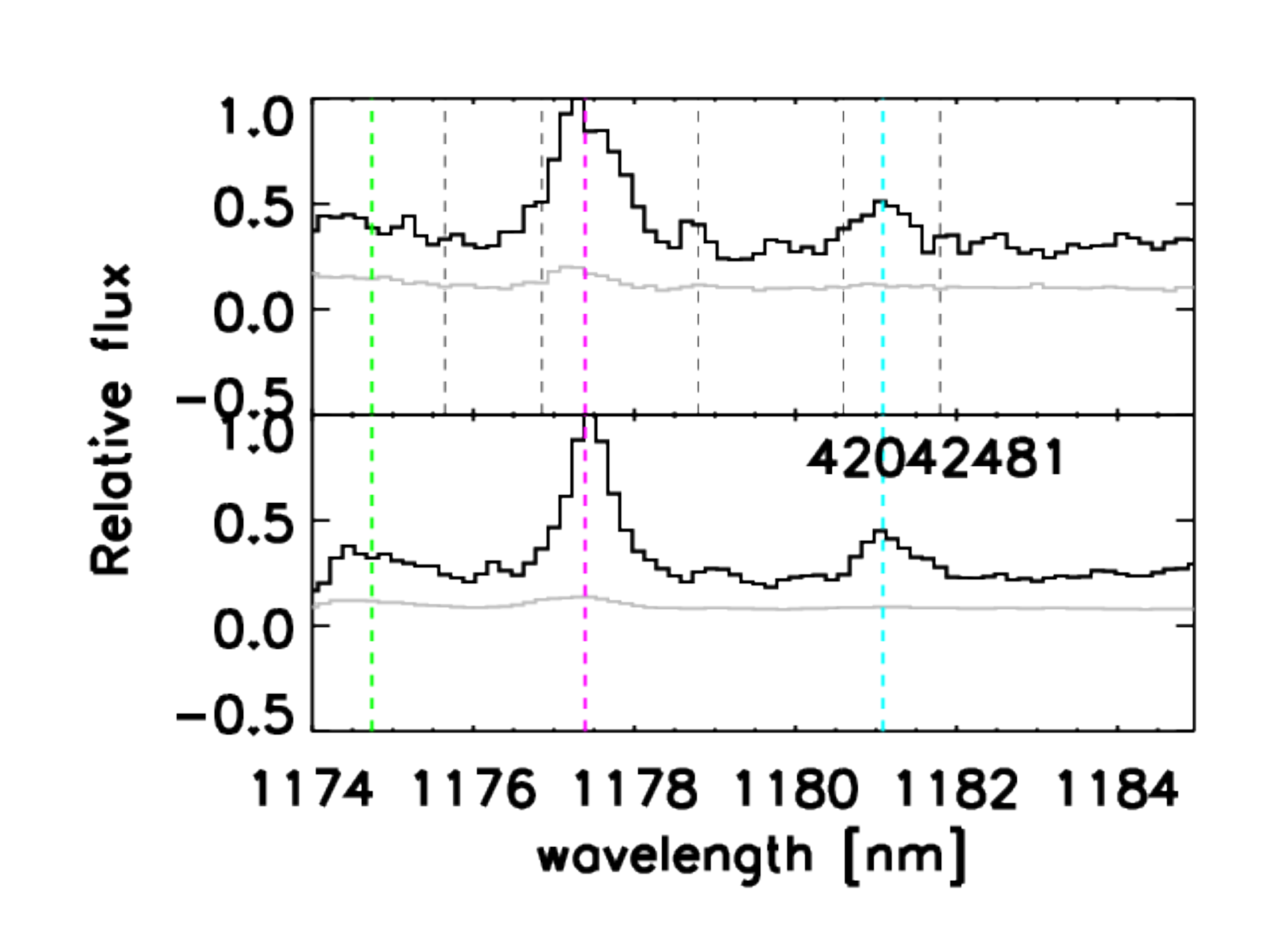}
\includegraphics[width=0.49\textwidth]{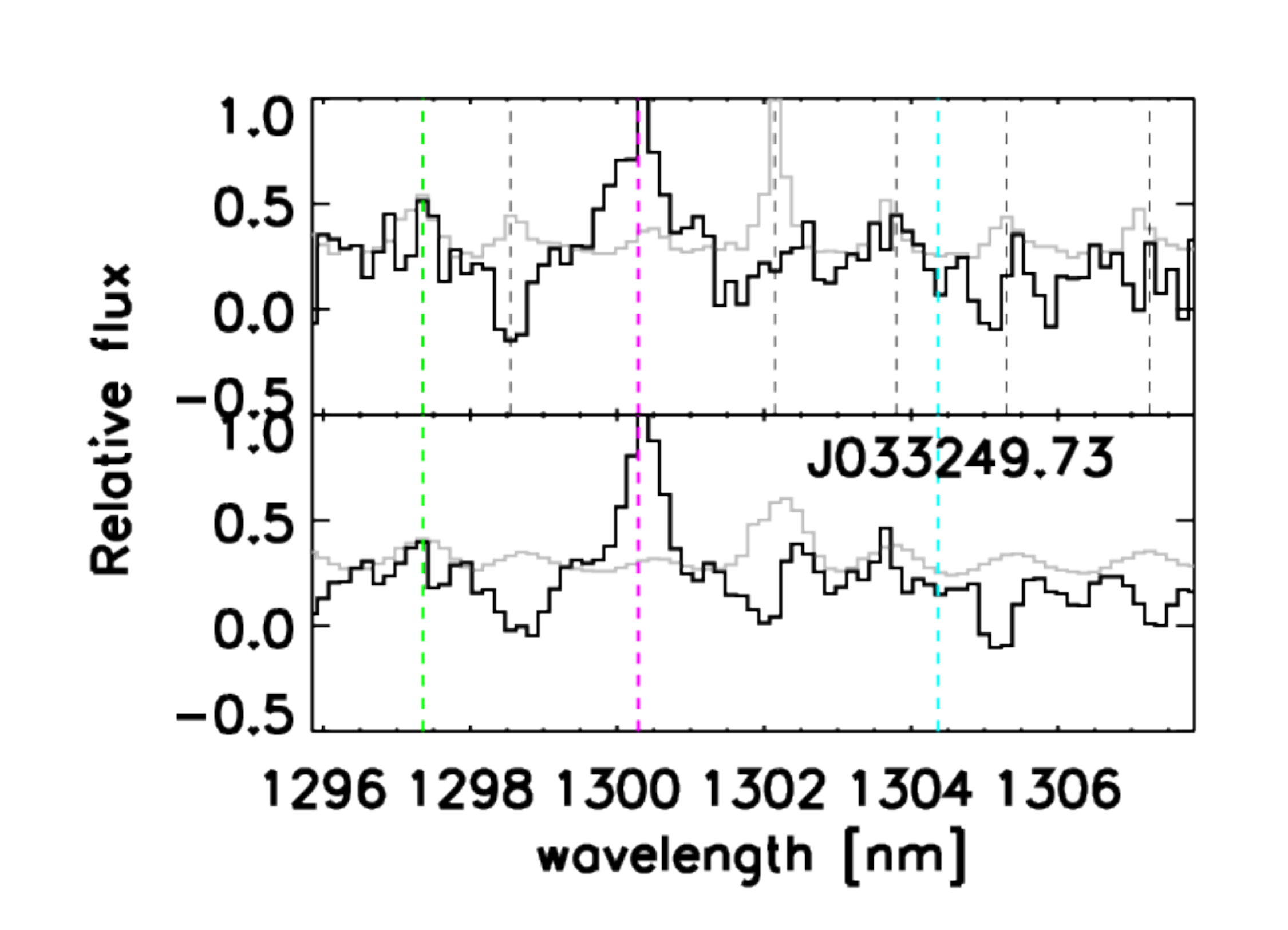}
\caption{Spatially integrated 1D spectra (sum of all spectra in segmentation maps) of each component in IROCKS, covering the spectral region around the redshifted \ha{} emission line. When the integrated spectrum has only one \ha{} peak, the source has only one component and is classified as a single source. When the integrated spectrum has more than one \ha{} peak, the source is classified as multiple, and components are spatially separated. The west component of 7187 still has more than one spectral peak, but different components are difficult to spatially separate, thus it is treated as one component. One $\sigma$ noise is plotted in gray. The magenta dashed line is the location of \ha{} peak, and green and cyan lines are location of [NII]6548 and [NII]6583 based on the centroid of the \ha{} line. Top: spectra in the segmentation map are simply summed up. Dashed black vertical lines are location of strong sky OH lines measured using non-sky-subtracted data. Bottom: spatially integrated spectra in the segmentation map, but individual spectra are shifted so that all Gaussian fitted \ha{} line peaks match at a single redshift, which is the Gaussian peak of the integrated 1D spectrum ($z_{\text{sys}}$ in Table \ref{flux_table}), to increase the line signal.}
\label{1Dspec}
\end{figure*}
\begin{Contfigure}[p]
\centering
\includegraphics[width=0.49\textwidth]{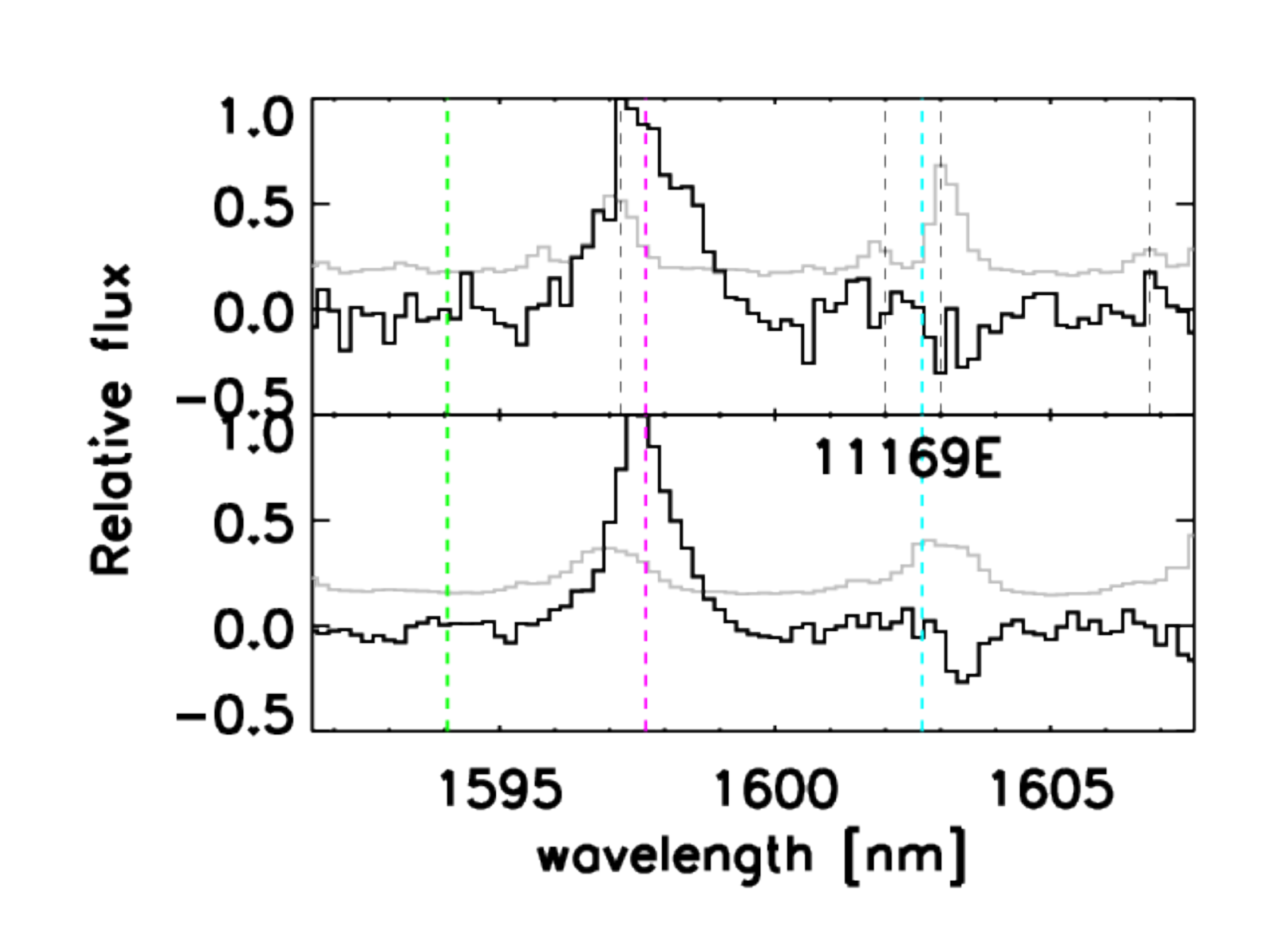}
\includegraphics[width=0.49\textwidth]{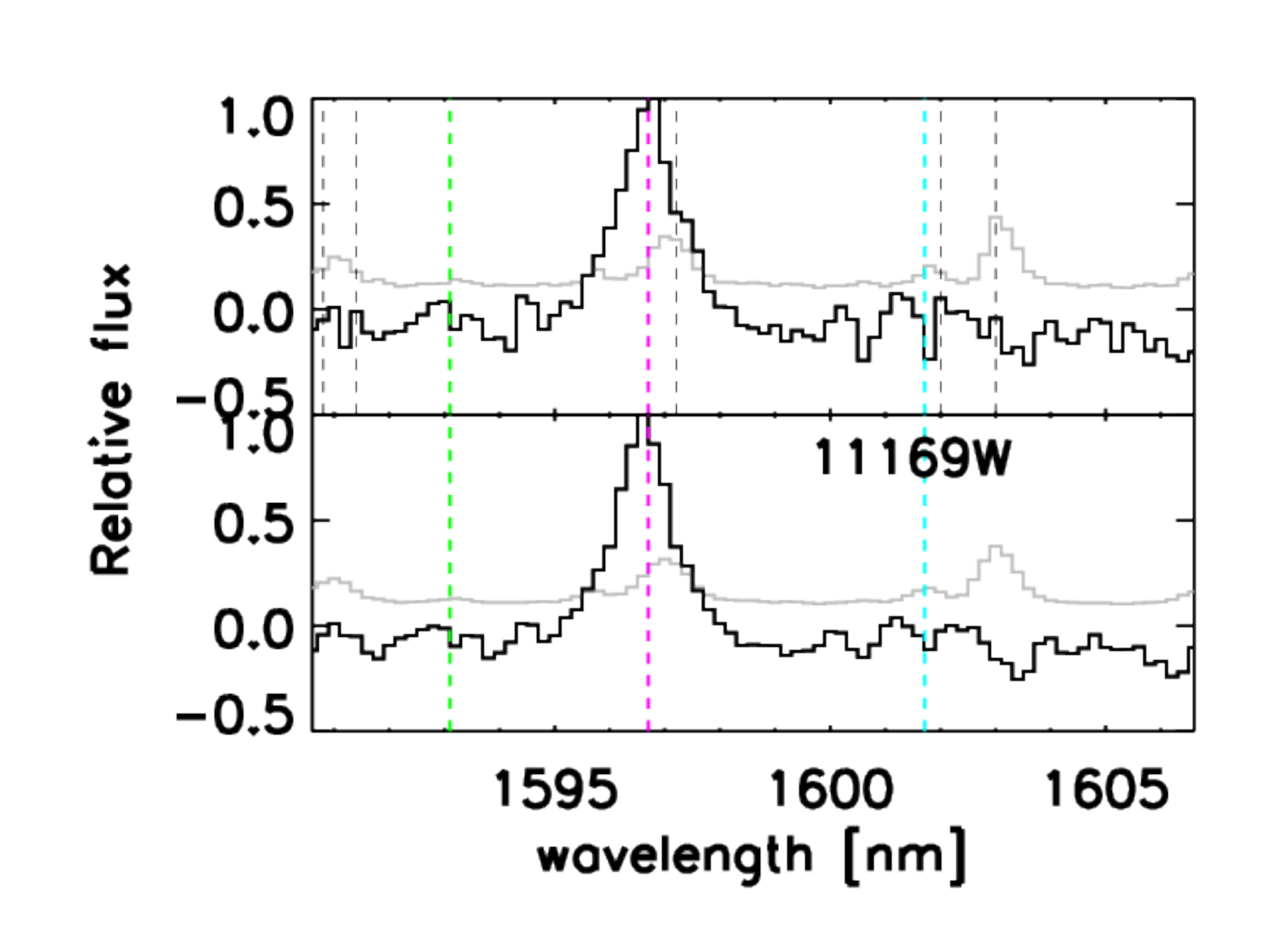}
\includegraphics[width=0.49\textwidth]{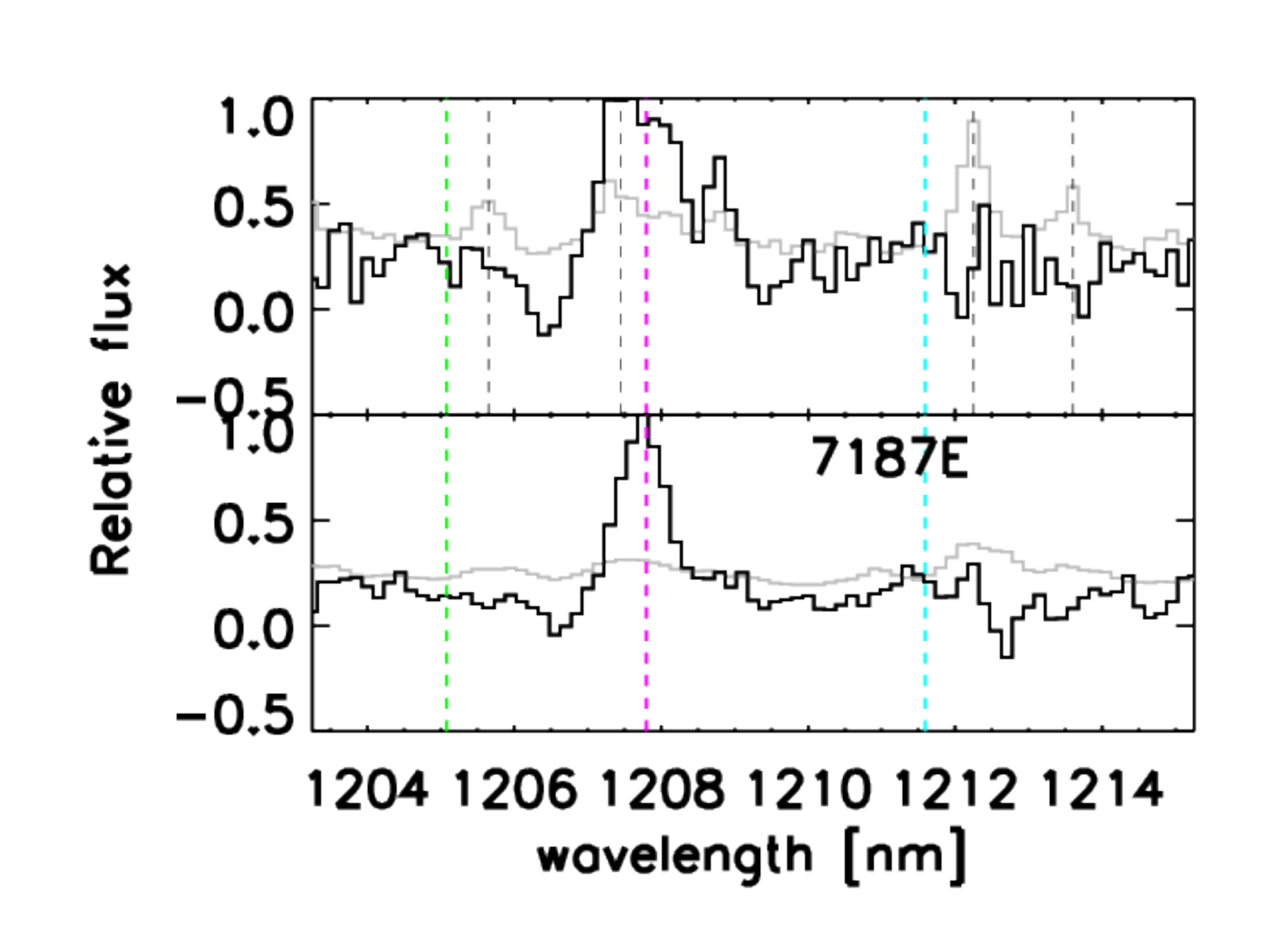}
\includegraphics[width=0.49\textwidth]{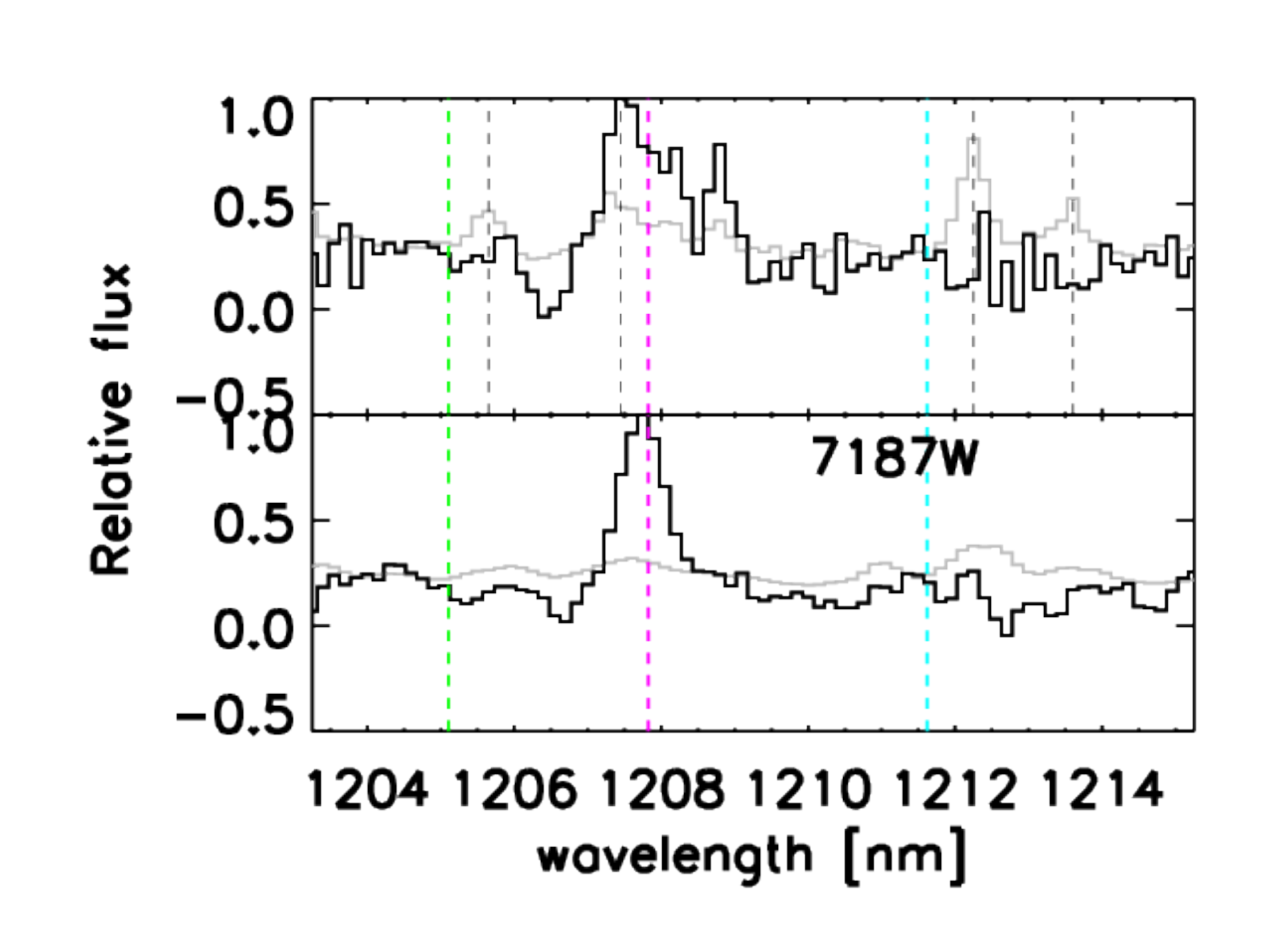}
\includegraphics[width=0.49\textwidth]{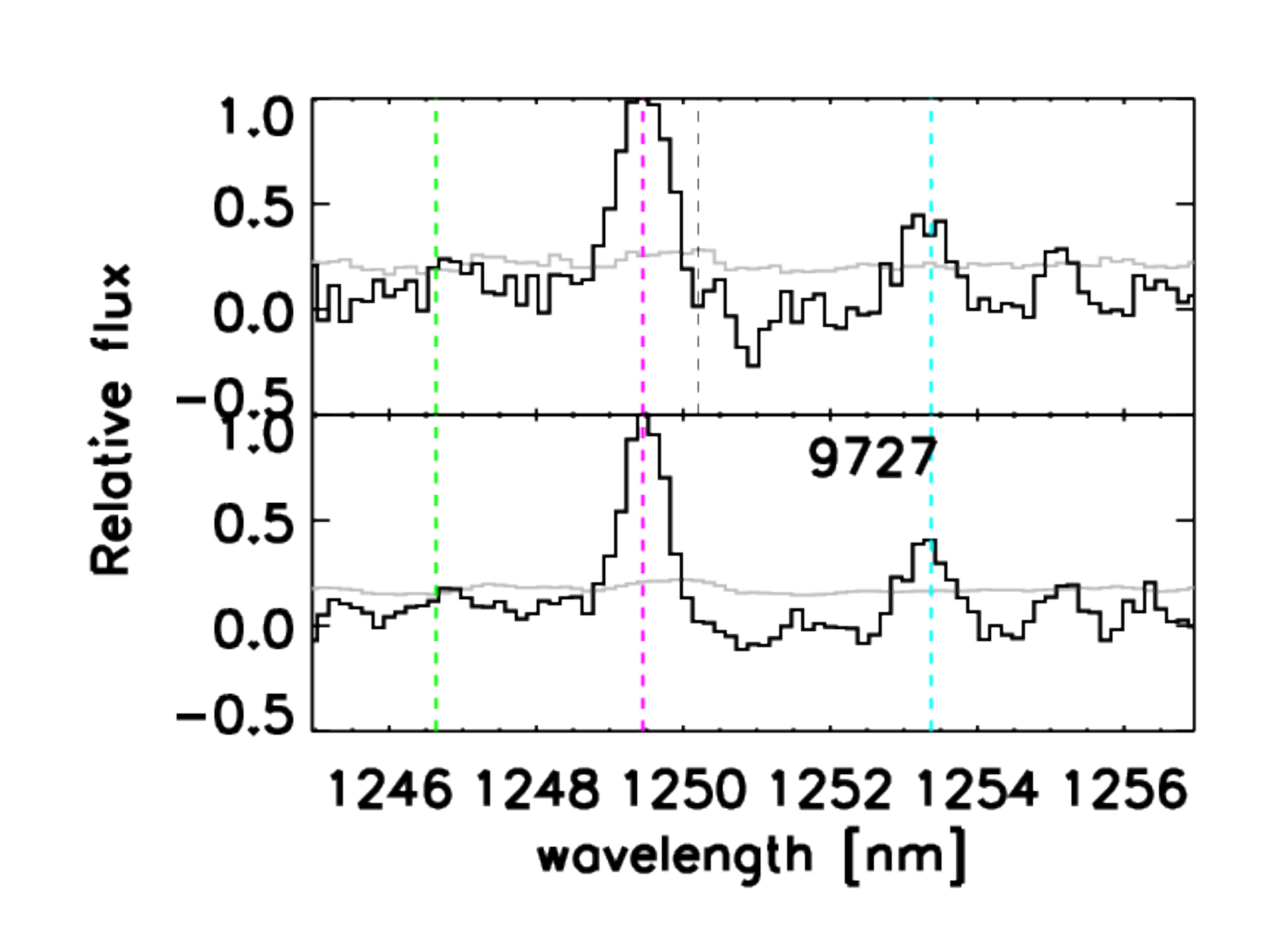}
\includegraphics[width=0.49\textwidth]{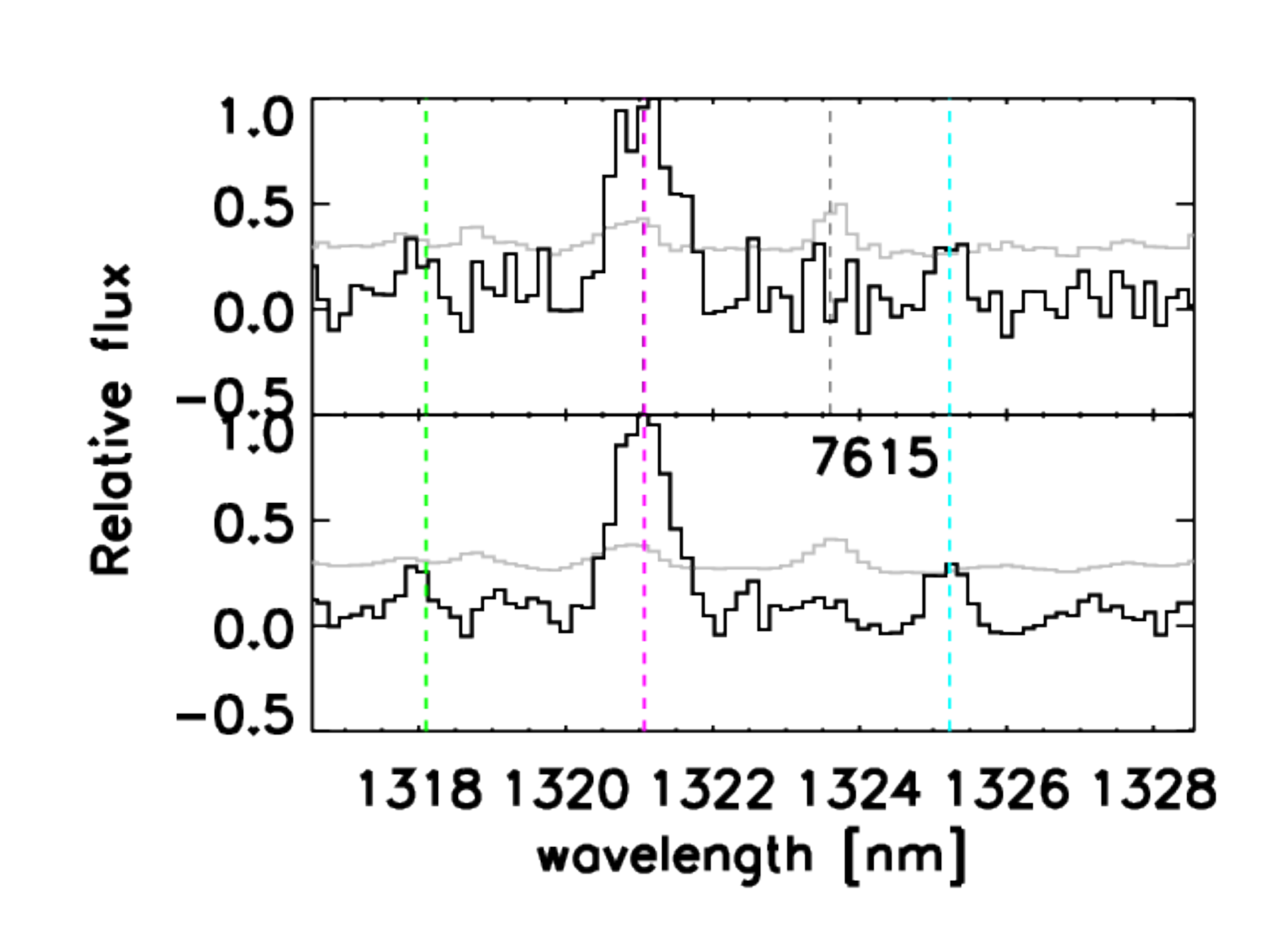}
\caption{}
\end{Contfigure}
\begin{Contfigure}[p]
\centering
\includegraphics[width=0.49\textwidth]{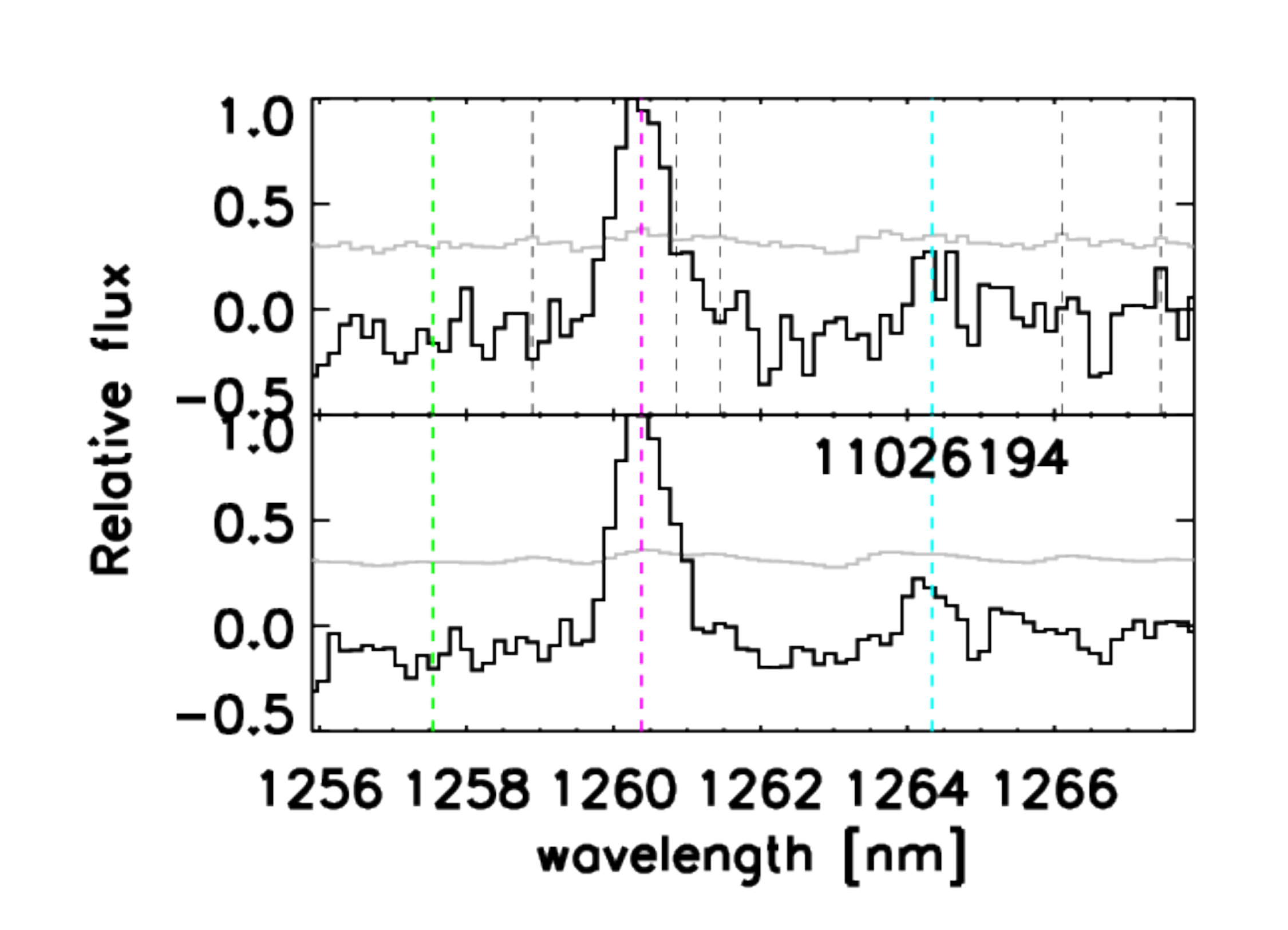}
\includegraphics[width=0.49\textwidth]{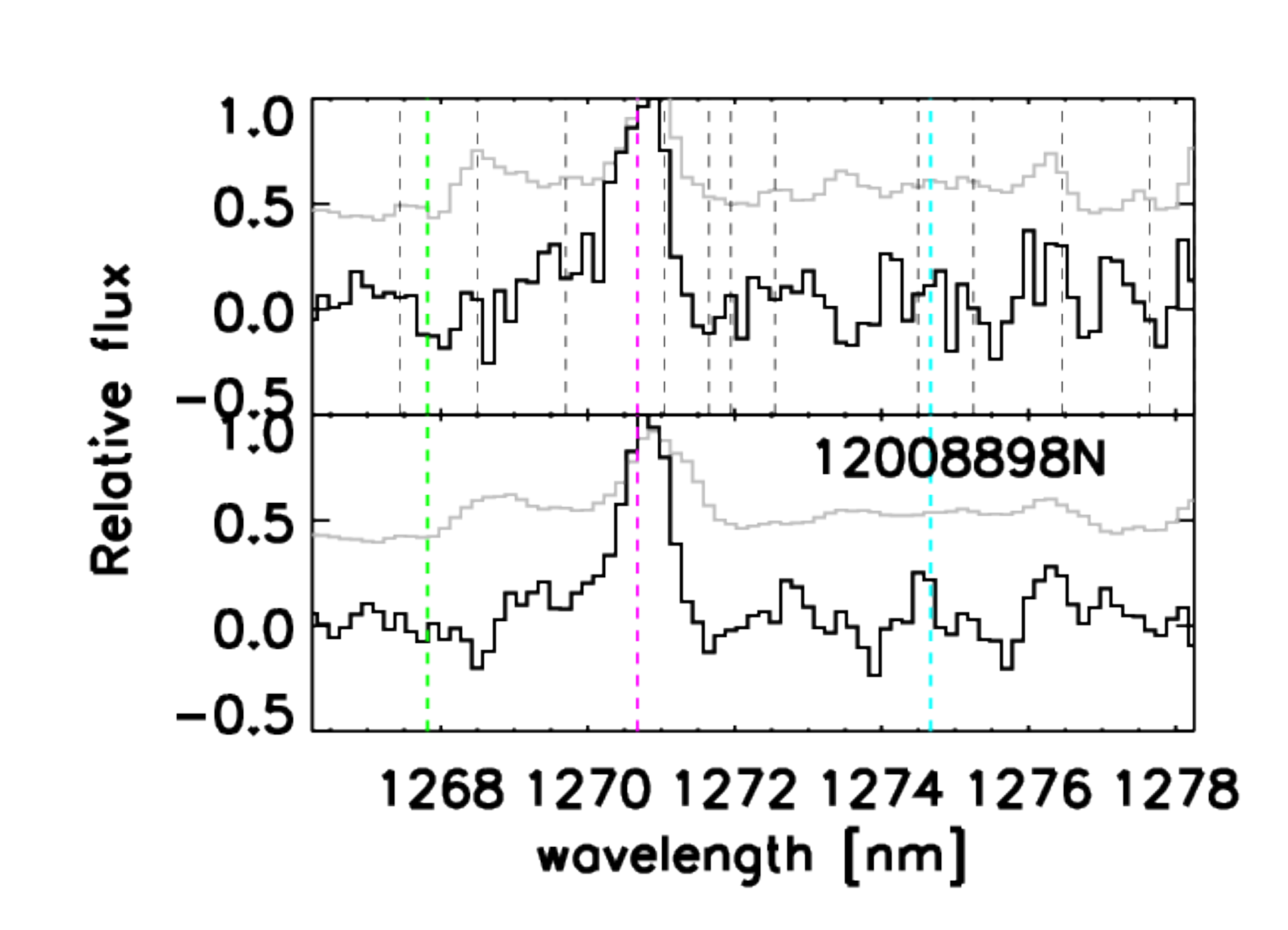}
\includegraphics[width=0.49\textwidth]{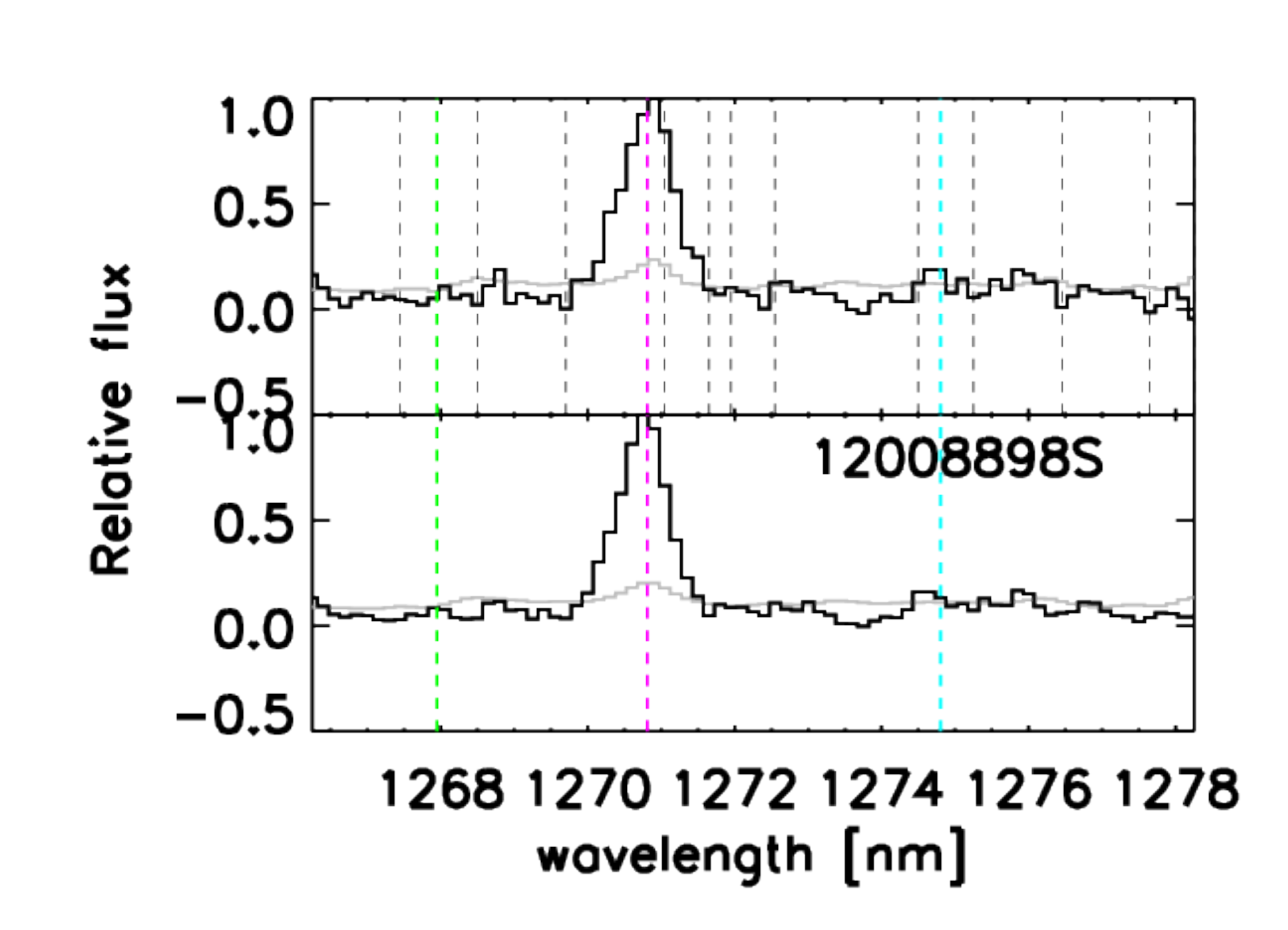}
\includegraphics[width=0.49\textwidth]{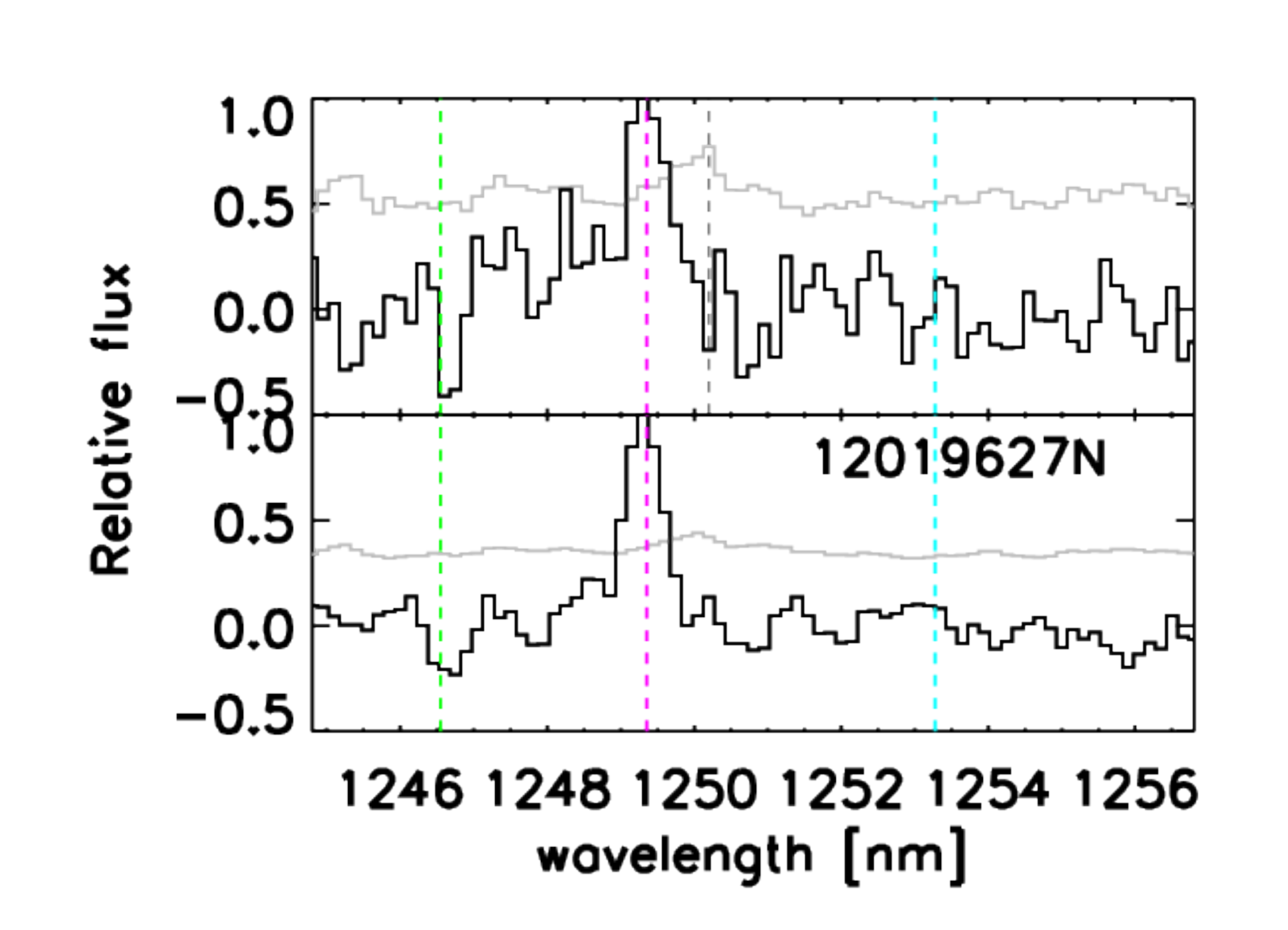}
\includegraphics[width=0.49\textwidth]{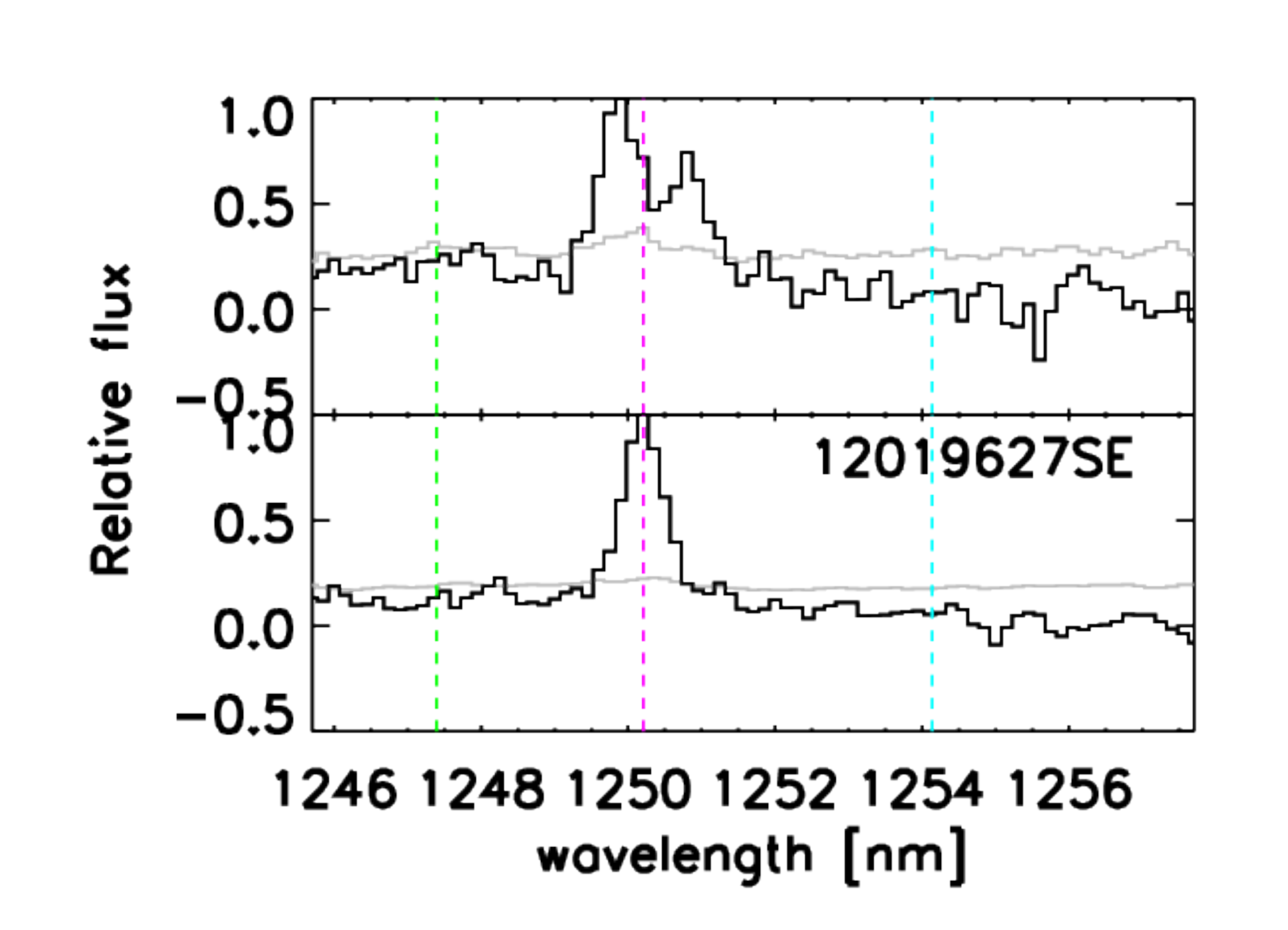}
\includegraphics[width=0.49\textwidth]{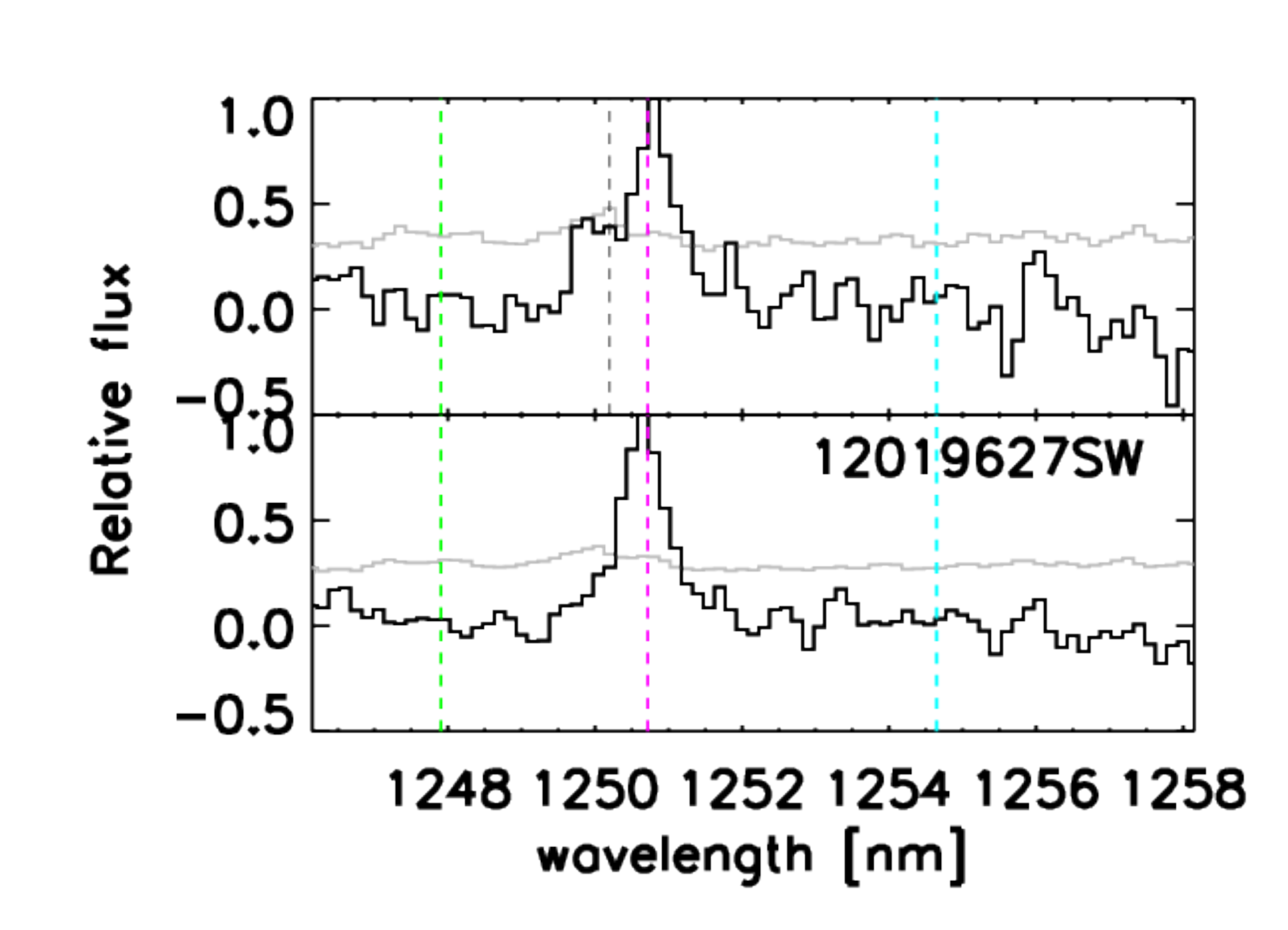}
\caption{}
\end{Contfigure}
\begin{Contfigure}[p]
\centering
\includegraphics[width=0.49\textwidth]{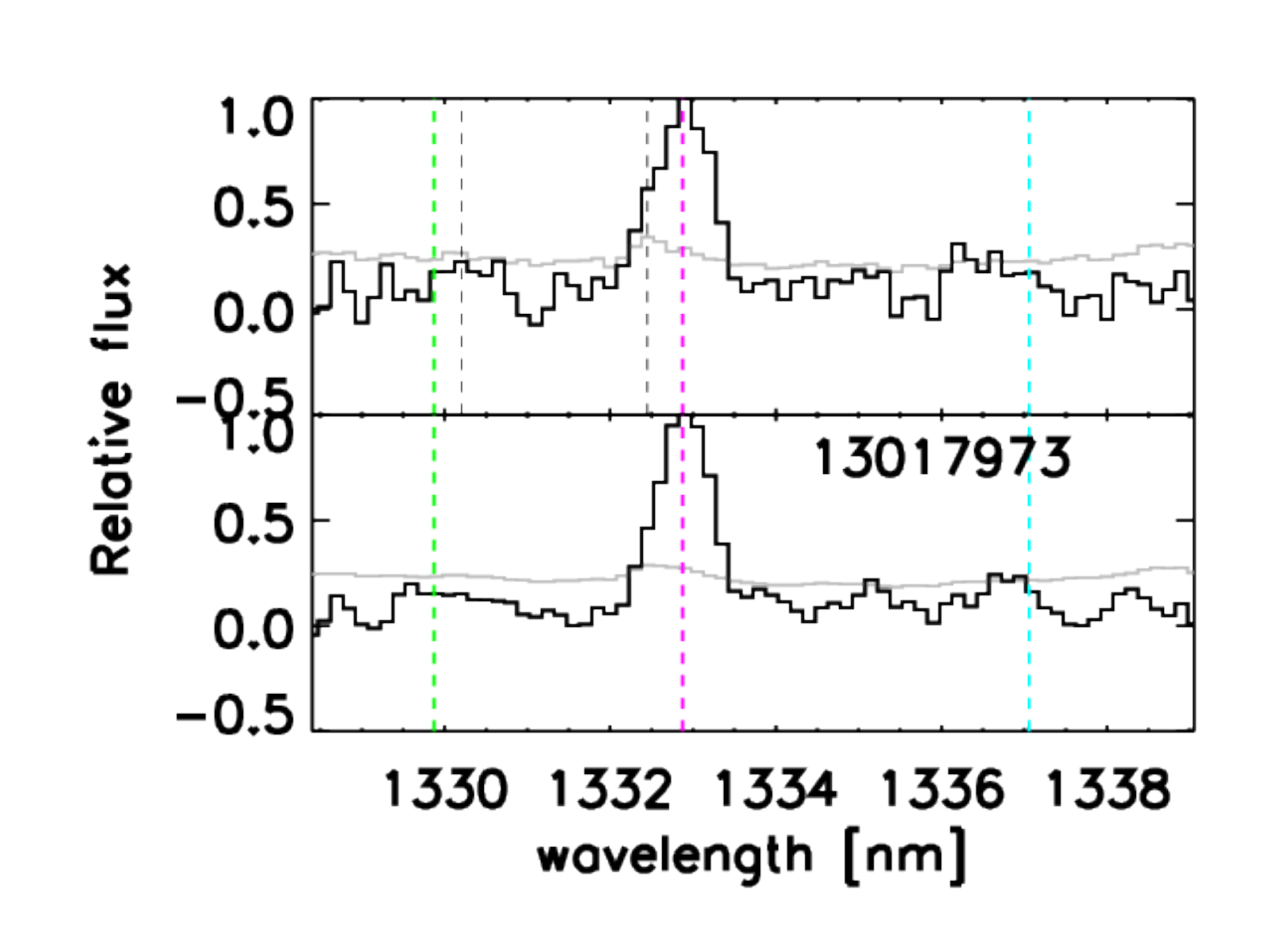}
\includegraphics[width=0.49\textwidth]{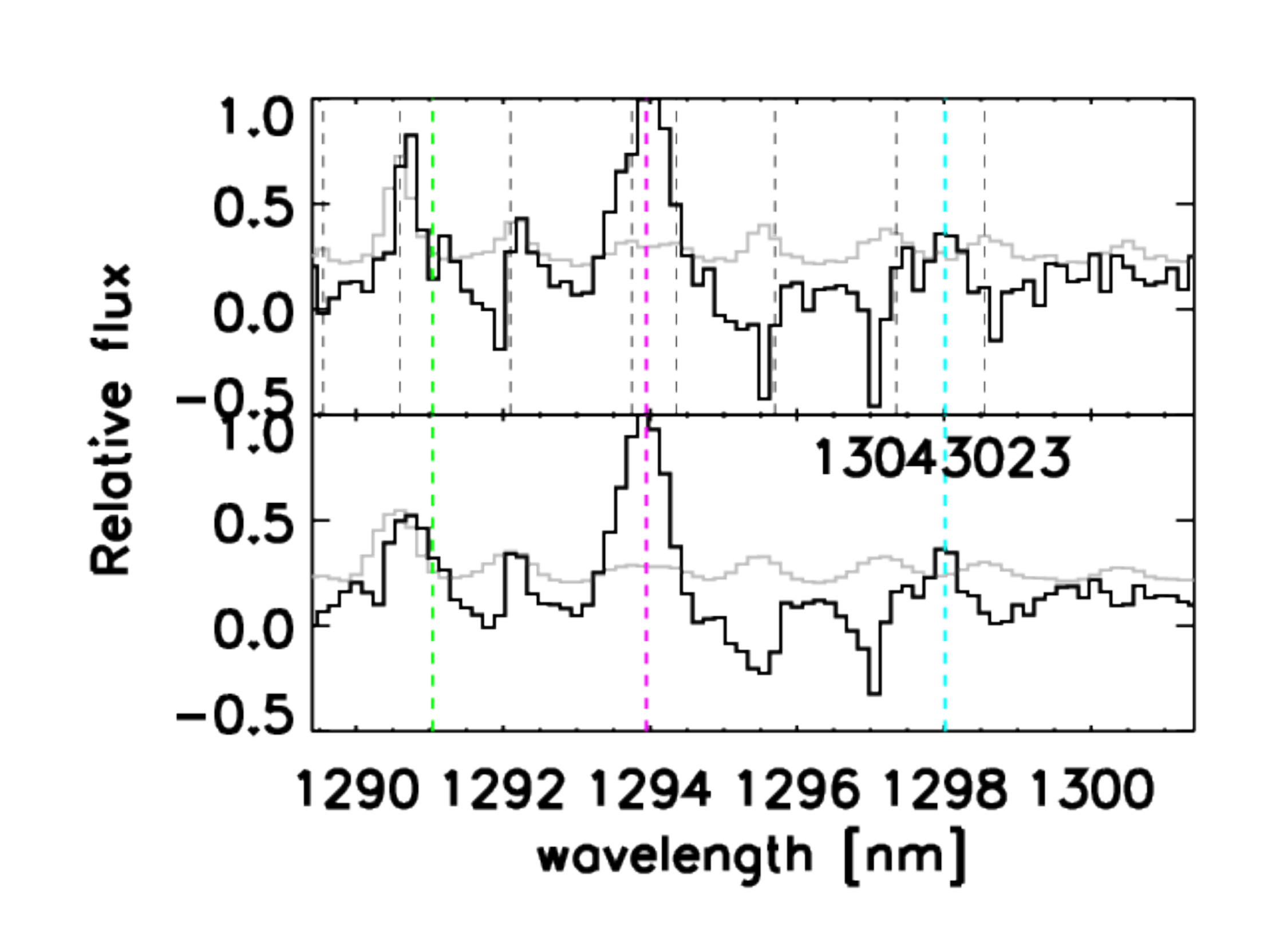}
\includegraphics[width=0.49\textwidth]{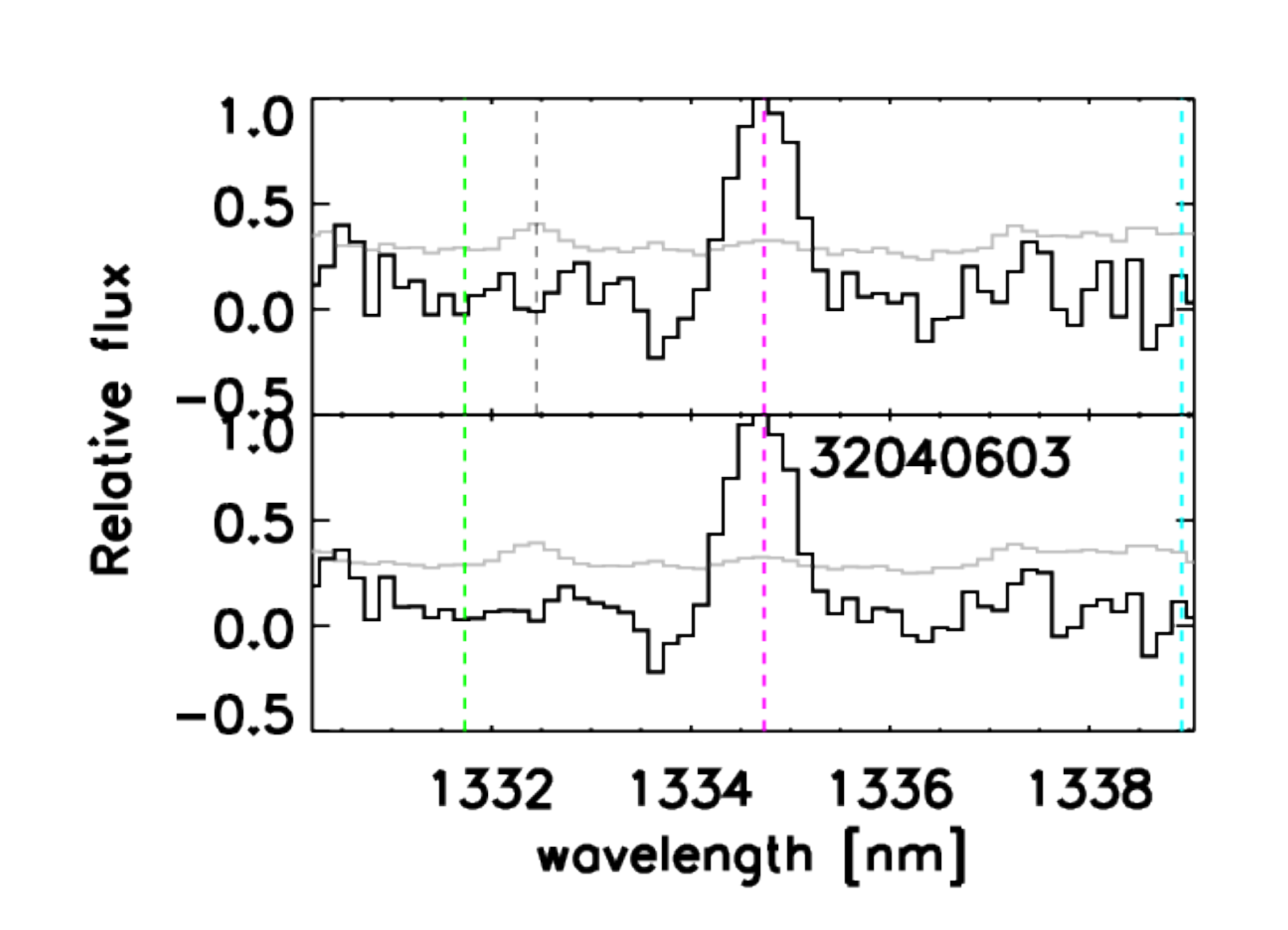}
\includegraphics[width=0.49\textwidth]{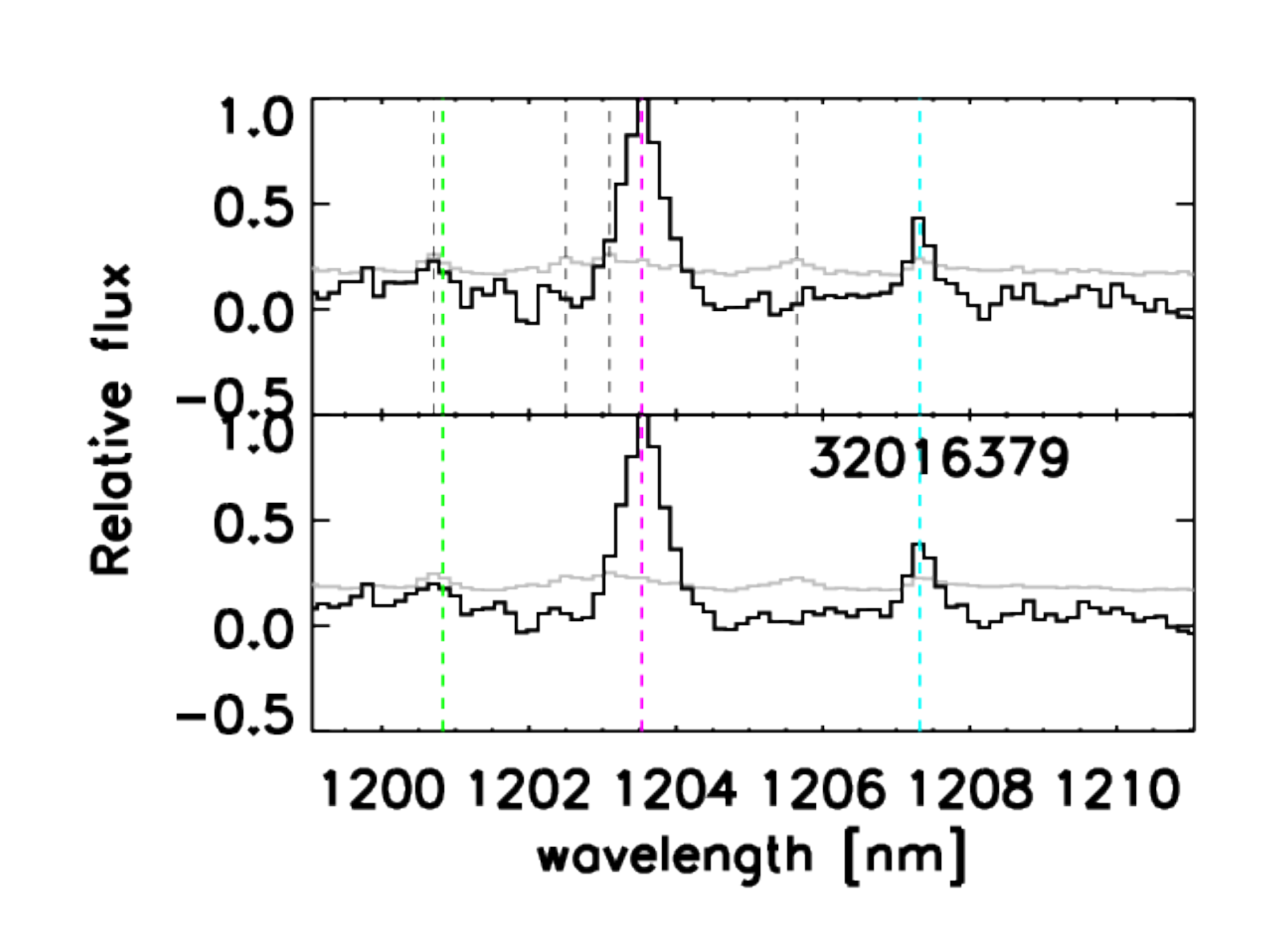}
\includegraphics[width=0.49\textwidth]{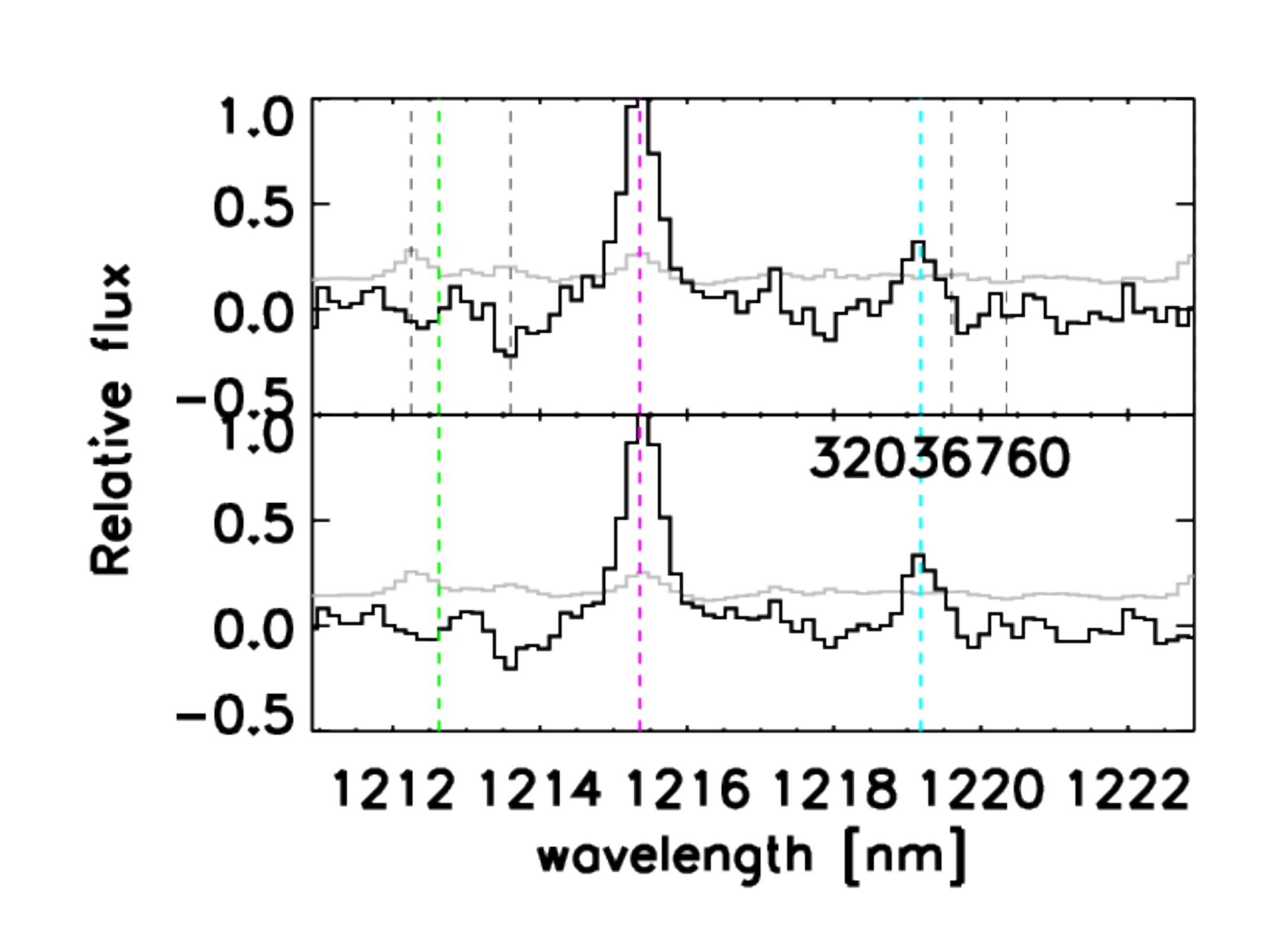}
\includegraphics[width=0.49\textwidth]{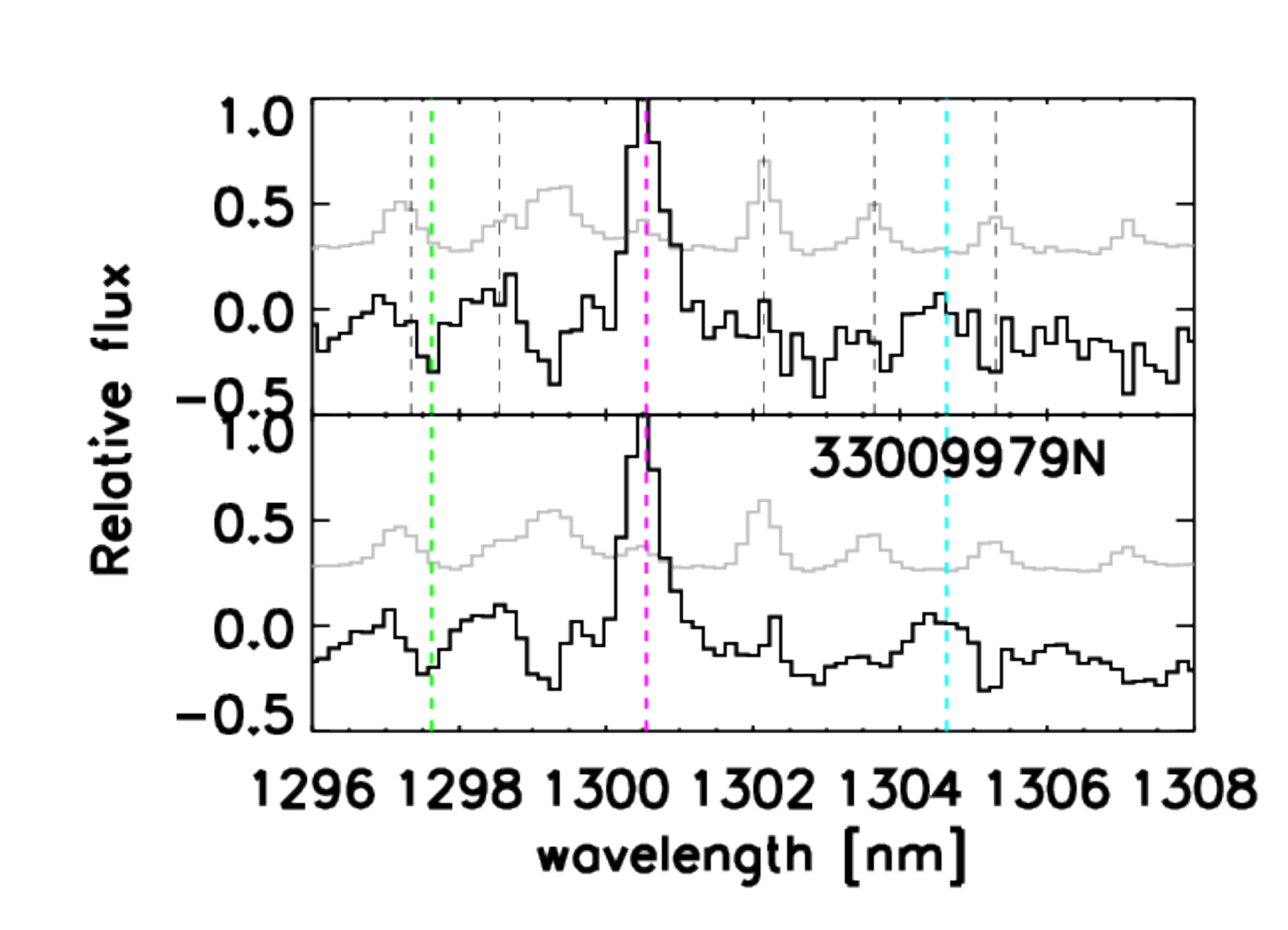}
\caption{}
\end{Contfigure}
\begin{Contfigure}[t]
\centering
\includegraphics[width=0.49\textwidth]{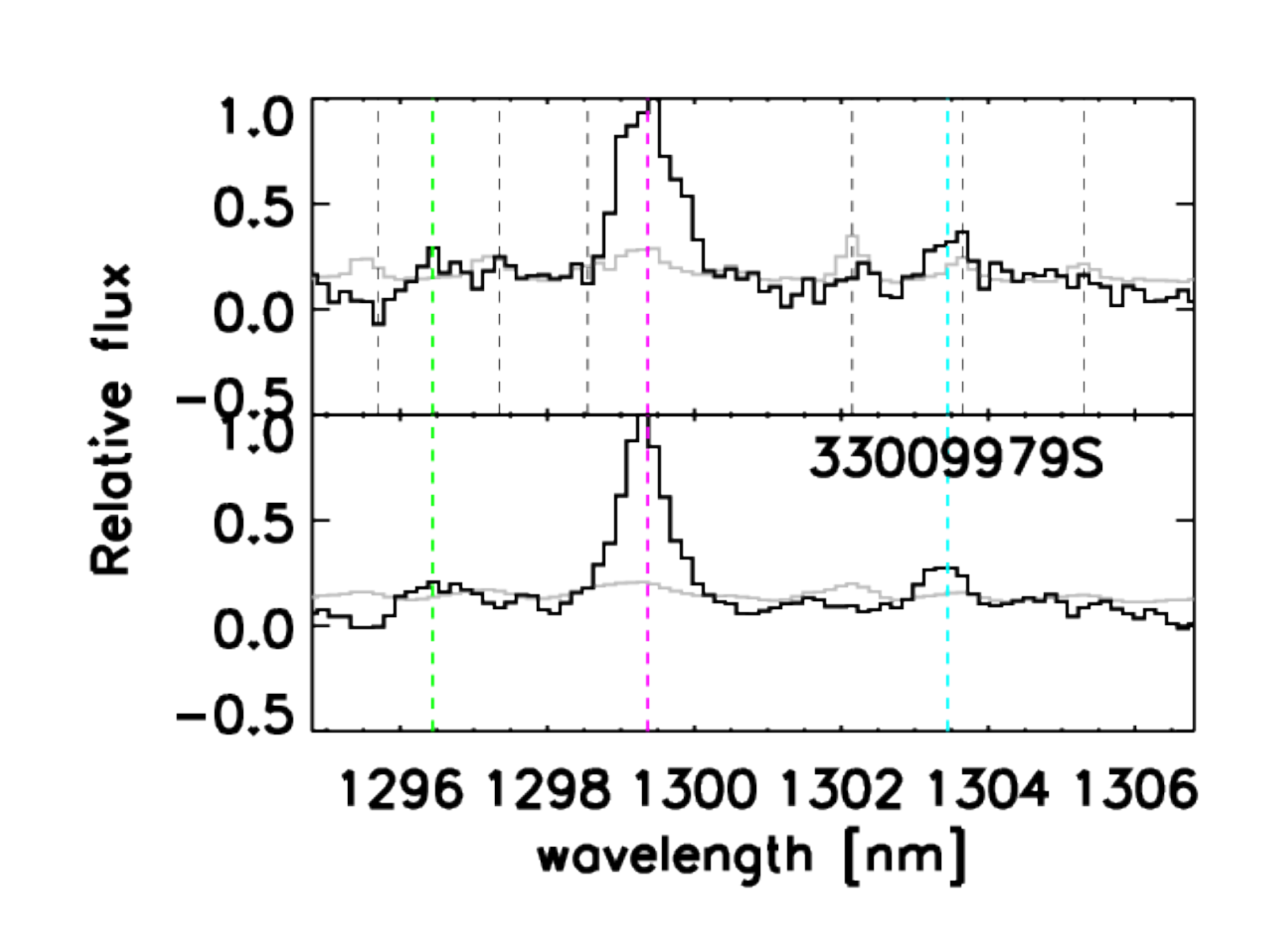}
\caption{}
\end{Contfigure}

\section{Notes on individual galaxies}\label{sec_indiv}
\begin{figure*}[h]
\centering
\includegraphics[width=0.82\textwidth]{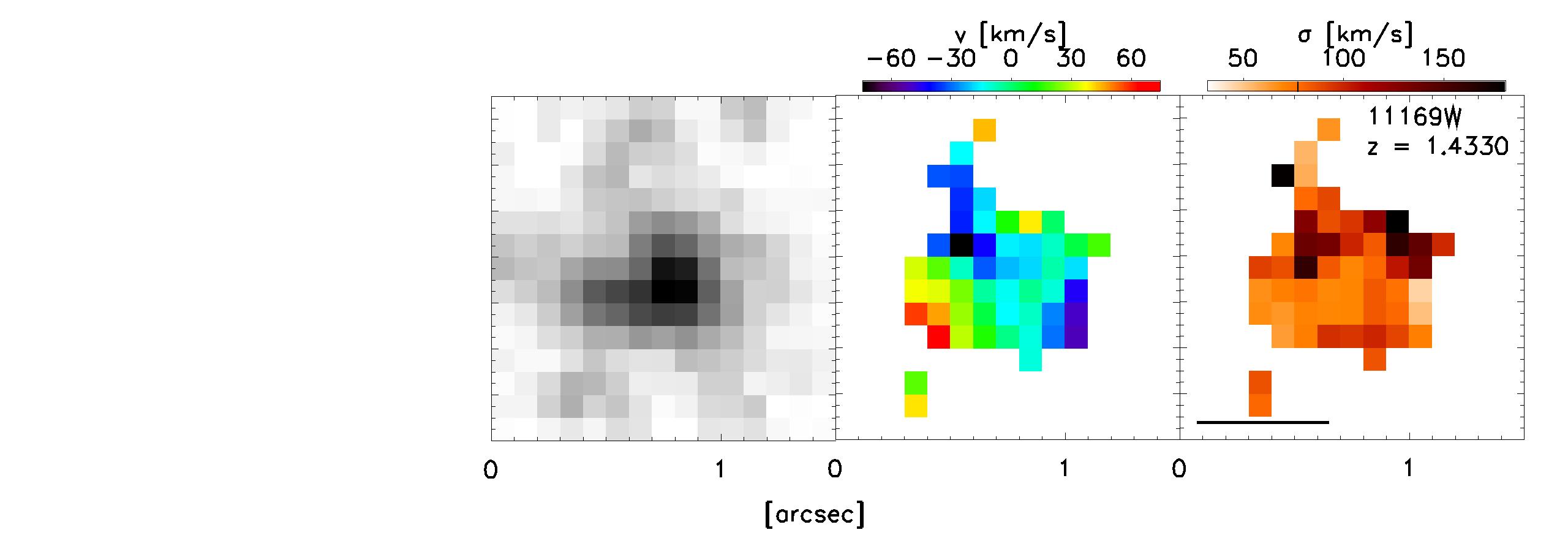}
\includegraphics[width=0.82\textwidth]{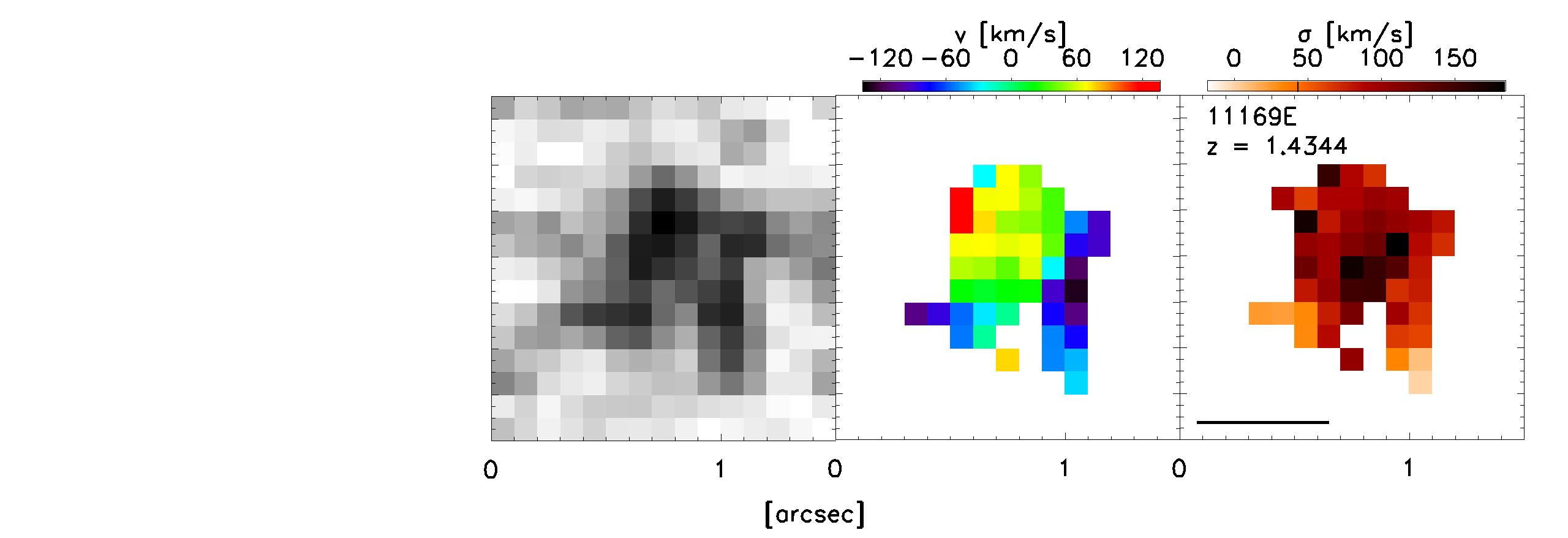}
\caption{\ha{} flux, rotation velocity, and velocity dispersion maps of individually separated components for the galaxies that are classified as "multiple", 11169 (East and West), 7187 (East and West), 12008898 (North and South), 12019627 (North, South-East, and South-West), and 33009979 (North and South). ''Multiple'' galaxies are classified using peaks in their integrated 1D spectra or their spatially well separated components (\S \ref{section_component}).}
\label{11169_comp}
\end{figure*}

In this section we briefly describe the OSIRIS results of individual galaxies. For galaxies that are classified as ''multiple'' in \S \ref{section_component}, their individual \ha{} flux, radial velocity, and velocity dispersion maps are shown in Figure \ref{11169_comp}. 

\subsection{UDS11655}\label{sec_uds111655}
This is one of four IROCKS disk candidates whose HST rest frame UV image shows disk-like morphology with a spiral arm pattern. OSIRIS kinematic map is well fitted by a disk model (P.A. = 125$^\circ$, $V_p$ = 140 km s$^{-1}$), with a small residual $<\Delta>$ = 13.4 km s$^{-1}$. It has \sigmaave{} of 55 km s$^{-1}$, with higher dispersion ($\sim$ 80 km s$^{-1}$) along the rotation axis and lower ($\sim$ 20 km s$^{-1}$) off axis. This is the typical velocity dispersion structure for a disk galaxy. The stellar mass is $\log{M_*/\text{\msun}}$ = 10.22 with estimated gas fraction of $\sim$ 56 \%. The virial mass $\log{M_{\text{vir}}/\text{\msun}}$ = 10.35, the enclosed mass $\log{M_{\text{enc}}/\text{\msun}}$ = 10.49, and the halo mass $\log{M_{\text{halo}}/\text{\msun}}$ = 11.80. The \ha{} detected size is almost the same as HST image, and we align the two by matching the overall structures.

\subsection{UDS10633}\label{sec_uds10633}
This source has the largest stellar mass estimate ($\log{M_*/\text{\msun}}$ = 11.24) among our sample. The rest frame UV image from HST shows a compact source in the north, a bar-like structure in the south, and another compact source in the south-west. The \ha{} does not show all of these components. To align the two image, we match the south tip of the bar in HST to that of the \ha{} image. Due to its unresolved size, \vshear{} value is negligible, but \sigmaave{} is still high (\sigmaave{} = 54.5 km s$^{-1}$). Since some components are not detected, dynamical mass estimates from OSIRIS should be considered lower limits.

\subsection{DEEP2-42042481}\label{sec_42042481}
This is the largest single component galaxy in the IROCKS sample. It is one of four IROCKS disk candidates and has one of the higher \vshear{} (\vshear = 180 km s$^{-1}$) and \vshear/\sigmaave{} (\vshear/\sigmaave{} = 2.70) values in our sample. The OSIRIS kinematic map is well fitted by a disk model (P.A. = 153$^\circ$, $V_p$ = 152 km s$^{-1}$) with a low residual, $<\Delta>$ = 23.6 km s$^{-1}$. The velocity dispersion has a slope that is perpendicular to the rotation axis, which is similar to disk galaxy velocity profiles. It has the stellar mass $\log{M_{*}/\text{\msun}}$ = 10.62 and the lowest gas fraction (22\%) among the four IROCKS disk candidates.

\subsection{J033249.73}\label{sec_j033249}
HST imaging in rest-frame UV shows a compact component in the east connected to a stretched arch component in the west. We match the bright compact source in HST with the bright \ha{} detection to the south-east. The \ha{} kinematic map does not show a velocity gradient, and the velocity dispersion varies across the whole galaxy. The east component has a lower dispersion of \textless 30 km s$^{-1}$ while the west arch component has higher ($\sim$ 100 km s$^{-1}$) dispersion.

\subsection{TKRS11169}\label{sec_tkrs11169}
This is the only source at z $\sim$ 1.4 in our sample. It is classified as AGN by X-ray observation. HST image shows two distinct components in the east and west, and both are resolved by \ha{} and aligned to the HST images. The west component is brighter than the east by a factor of 1.5. Both components have significantly higher \sigmaave{} of $\sim$ 90 km s$^{-1}$ than the rest of the z $\sim$ 1 sources. 

\subsection{TKRS7187}\label{sec_tkrs7187}
HST image shows three brighter spots (east, center, west) and $m=2$ like spiral arm, but \ha{} does not show velocity gradient to support the disk model. The central bright spot in HST is matched with the central nod in \ha{} detection. After separating components using 1D spectrum, the west component still has more than one peak in 1D spectrum that could not be spatially separated. This component has the highest \vshear{} of 240 km s$^{-1}$ in our sample and hence is most likely an interacting system.

\subsection{TKRS9727}\label{sec_tkrs9727}
Among resolved IROCKS sample, this has the largest stellar mass ($\log{M_*/\text{\msun}}$ = 11.0) in our sample. It also has the highest SFR$_{\text{SED}}$ (159 \myr) and HII+ISM corrected SFR (SFR$_{\text{\ha}}^{00}$ = 108 \myr) with the highest $\tau_V$ (3.66). This is one of four IROCKS disk candidates whose HST image at rest frame UV shows $m=2$ face-on spiral galaxy morphology, and OSIRIS kinematic map is well fitted by a disk model (P.A. = 224$^\circ$, $V_p$ = 110 km s$^{-1}$) with the smallest residual $<\Delta>$ = 13.2 km s$^{-1}$. The galaxy has $\sim$ 55\% gas fraction. We align the central bar like feature in HST with the central thick part of \ha{}, and match the HST north-west arm with the two \ha{} nods in north-west and the HST south-east arm with the south-east extended curved feature in \ha. 

\subsection{TKRS7615}\label{sec_tkrs7615}
HST image at rest frame UV show face-on grand design spiral galaxy morphology, and OSIRIS rotation map shows subtle variation/gradient across the galaxy. Velocity dispersion is roughly uniform across the whole galaxy at around 70 km s$^{-1}$. This can be a face-on disk, but the rotation variation detected by OSIRIS is too small to fit a disk model. The bright \ha{} north part is matched with the central nod in HST, and the arch like south-east extended component to the south-west nod in \ha{} are matched to the spiral arm in HST.

\subsection{DEEP2-11026194}\label{sec_11026194}
The \ha{} kinematic map shows velocity gradient (blueshifted at the north and redshifted at the south); however, due to its small detected region, the disk fitting is insufficient to determine if it is disk candidate. The velocity dispersion varies along rotation axis, $\sim$ 20 km s$^{-1}$ in the east to $\sim$ 100 km s$^{-1}$ in the east.

\subsection{DEEP2-12008898}\label{sec_12008898}
This is one of the best detected sources in our observations and used for local rotation and smoothing correction analysis in \S \ref{sec_kine_map}. The system has two distinct components, small one in the north and the large one in the south. Both components are seen both in HST and \ha{}. The north component: while HST shows two nods (northeast and southwest) extended $\sim$ 1 arcsec, \ha{} shows one nod of $\sim$ 0.5 arcsec in northeast. It has some velocity gradient, but it is too under-sampled to fit to a disk model. The velocity dispersion is almost uniform around 60 km s$^{-1}$. The south component: \ha{} and HST extend almost the same size, and three nods are seen in HST while two are seen in \ha{}. The rotation shows redshifts at the center and blueshift at the outside. The velocity dispersion is almost uniform ($\sim$ 70 km s$^{-1}$) over whole galaxy, but slightly lower ($\sim$ 40 km s$^{-1}$) at the center. From the velocity structure and multiple nods, the south component is probably interacting system. We match the northeast nod in HST with the \ha{} north component, and the overall shape of south component. 

\subsection{DEEP2-12019627}\label{sec_12019627}
Both HST and \ha{} show patchy morphology. We separate three \ha{} north nods as the north component and the south ribbon shape part into the south-east and south-west components. The north component: three $\sim$ 0.5'' \ha{} nods are spatially separated but are individually too small to form separate peak in the 1D spectrum, and hence three are together to form a north component. They are also individually too small to see individual rotation. The south-east and -west components: both velocity fields show gradient but are undersampled to fit with disk models. Due to their complicated morphologies, the system is probably an interacting system. Because this source has many components, we align HST and \ha{} detection so that all \ha{} components are on bright part of HST, except the north most component in \ha.

\subsection{DEEP2-13017973}\label{sec_13017973}
This is the only source in our sample detected with the old OSIRIS grating, and therefore the \ha{} emission is slightly noisier than the majority of the sample. HST shows a few distinct knots in a spiral disk-like morpholog, but is quite distinct from the observed \ha{} morphology. We match a few western knots in the HST image with that of the extended component observed in the west in \ha. The rotation map does not show a velocity gradient. This galaxy has the highest uncorrected SFR (42.7 \msun yr$^{-1}$), and the second highest HII+ISM corrected SFR (65.8 \msun yr$^{-1}$) among $z \sim 1$ IROCKS sample. 

\subsection{DEEP2-13043023}\label{sec_13043023}
The HST image shows an irregular morphology with three knots in the north-west region and two in the south-east region. The \ha{} image was matched to the HST arch connecting the south knot to the north-east knots in HST. The rotation map does not show a velocity gradient, and the dispersion is uniform around 50 km s$^{-1}$ except for the north east arm at $\sigma$ = 120 km s$^{-1}$.

\subsection{DEEP2-32040603}\label{sec_32040603}
This is the highest redshift galaxy ($z =$ 1.0338) among IROCKS $z \sim 1$ sample. It can be rotating (blueshift at the northwest and redshift at the southeast), but the surface area of \ha{} is not sufficient to fit a disk model. The velocity dispersion is constant over the galaxy at $\sigma \sim$ 55 km s$^{-1}$.

\subsection{DEEP2-32016379}\label{sec_32016379}
The \ha{} map shows a dual cone-like morphology and the rotation map shows almost no rotation except in the eastern region where it is slightly blueshifted by $\sim$ 20 km s$^{-1}$. The dispersion map is constant $\sigma \sim$ 60 km s$^{-1}$, except the eastern part, $\sigma \sim$ 30 km s$^{-1}$.

\subsection{DEEP2-32036760}\label{sec_32036760}
The \ha{} emission is in a compact single source with only a slight velocity gradient of $\sim$ 60 km s$^{-1}$, and a lower velocity dispersion in the southwest ($\sigma \sim$ 20 km s$^{-1}$) region compared to the entire source of ($\sigma \sim$ 60 km s$^{-1}$). 

\subsection{DEEP2-33009979}\label{sec_33009979}
This system has two well separated components in the north and south. The north component shows a slight velocity gradient, but it is not well fit to an inclined disk model. The velocity dispersion in the north component is uniform at 50 km s$^{-1}$. The southern component is one of the four disk candidates and is well fitted by a disk model (P.A. = 250$^\circ$, $V_p$ = 81.7 km s$^{-1}$) with a velocity residual of $<\Delta>$ = 30.8 km s$^{-1}$. The velocity dispersion is uniform over the galaxy at $\sigma \sim$ 60 km s$^{-1}$. The velocity field deviates near the center compared to the whole galaxy (see Figure \ref{kine_model_figure}), and to help with the disk fitting model we have enforced that the dynamical center is at the \ha{} flux peak. 

\begin{Contfigure}
\centering
\includegraphics[width=0.82\textwidth]{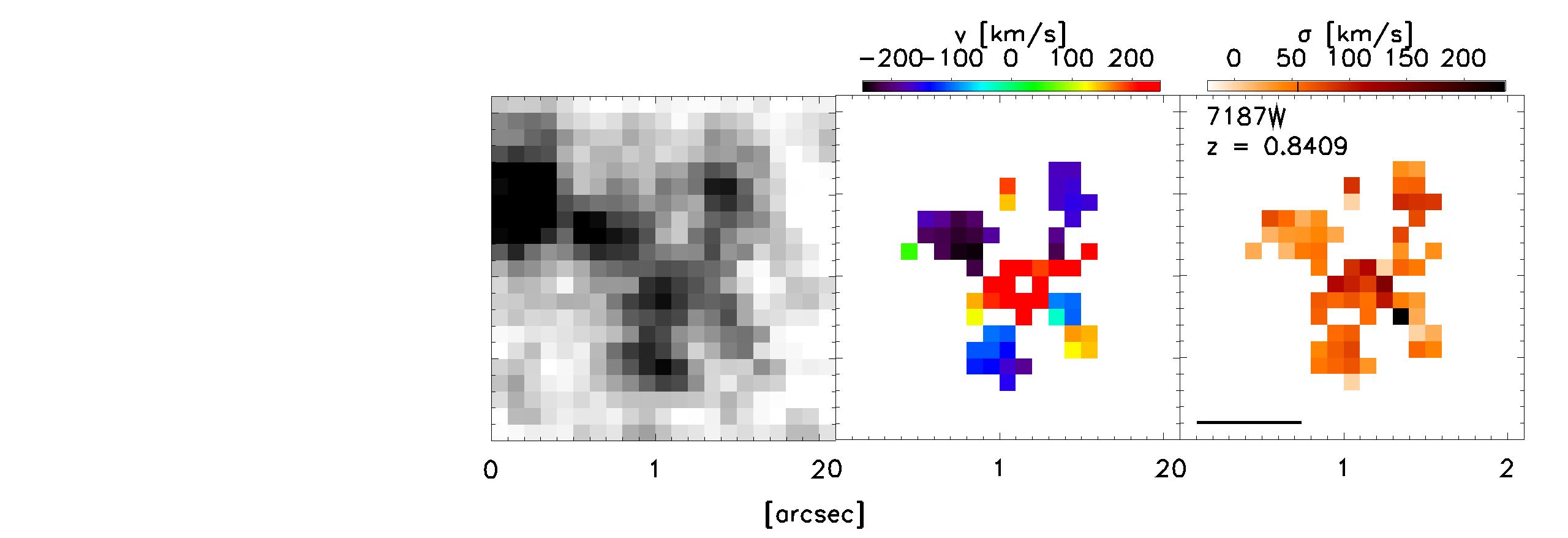}
\includegraphics[width=0.82\textwidth]{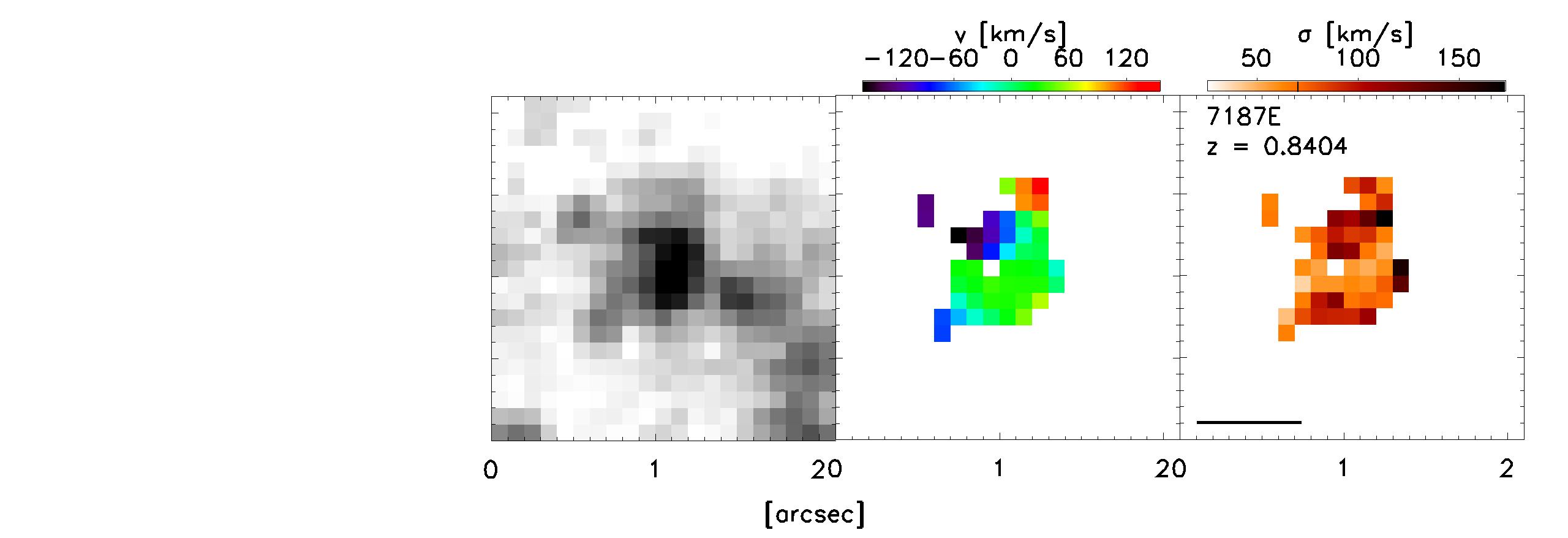}
\includegraphics[width=0.82\textwidth]{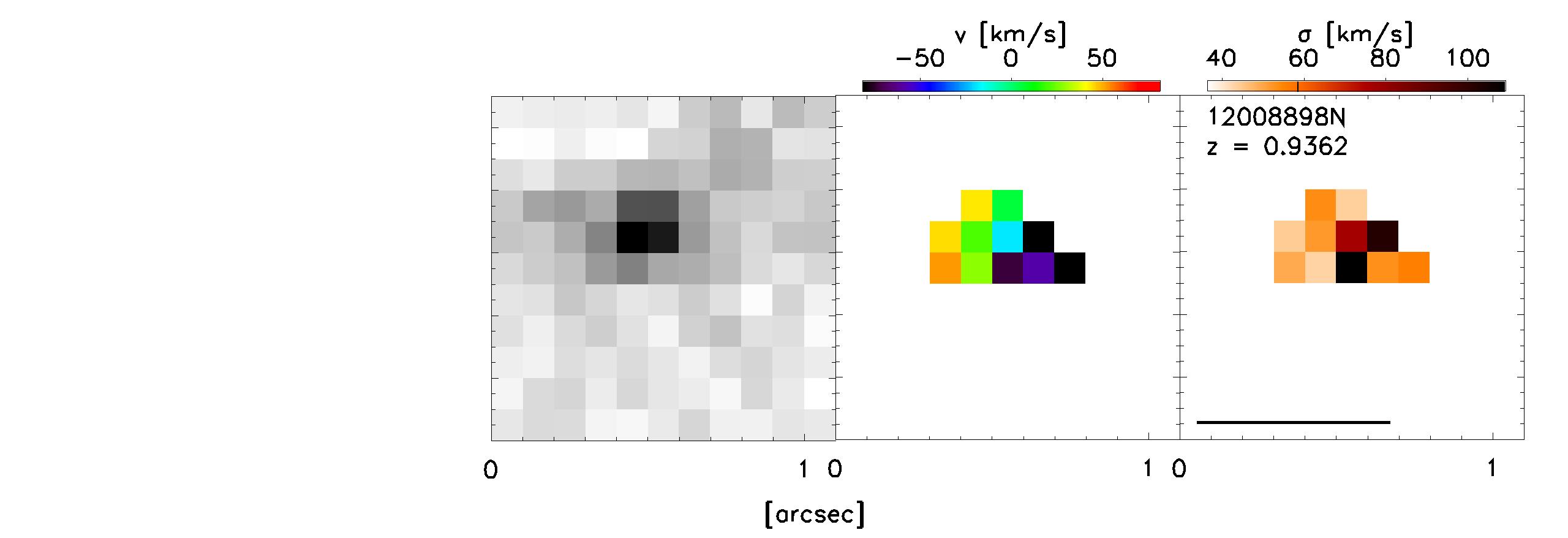}
\includegraphics[width=0.82\textwidth]{convert_image/DEEP12008898_south_combine.jpg}
\caption{}
\end{Contfigure}
\begin{Contfigure}
\centering
\includegraphics[width=0.82\textwidth]{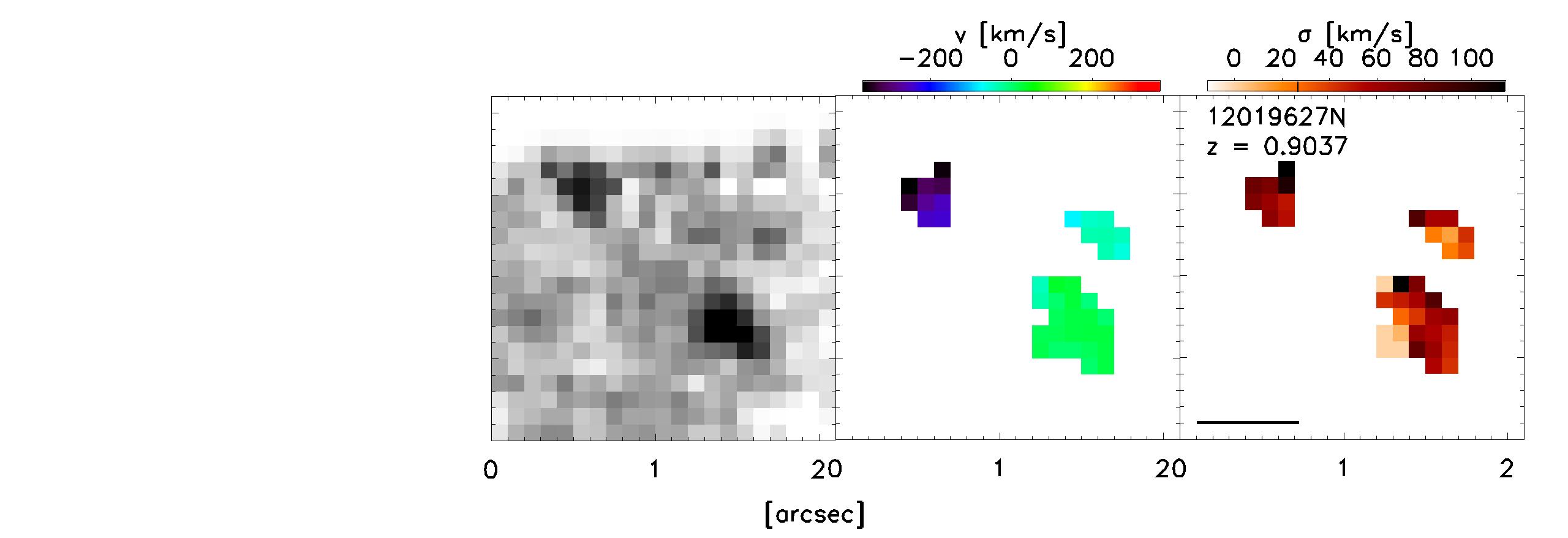}
\includegraphics[width=0.82\textwidth]{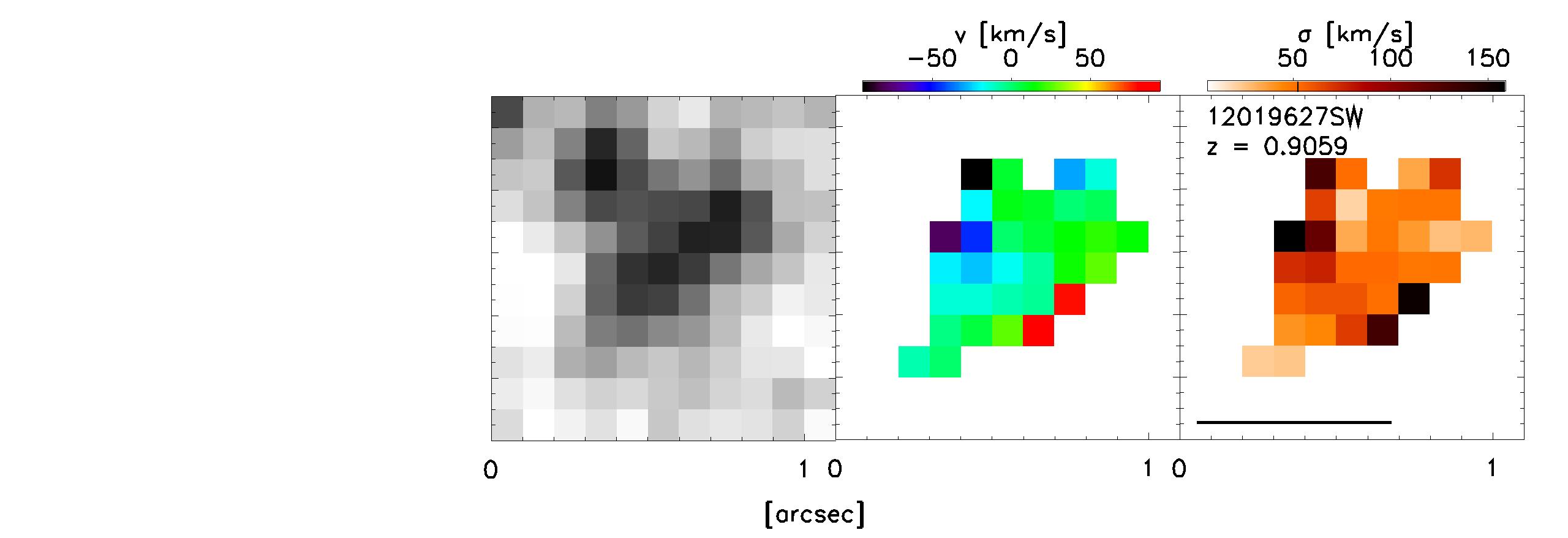}
\includegraphics[width=0.82\textwidth]{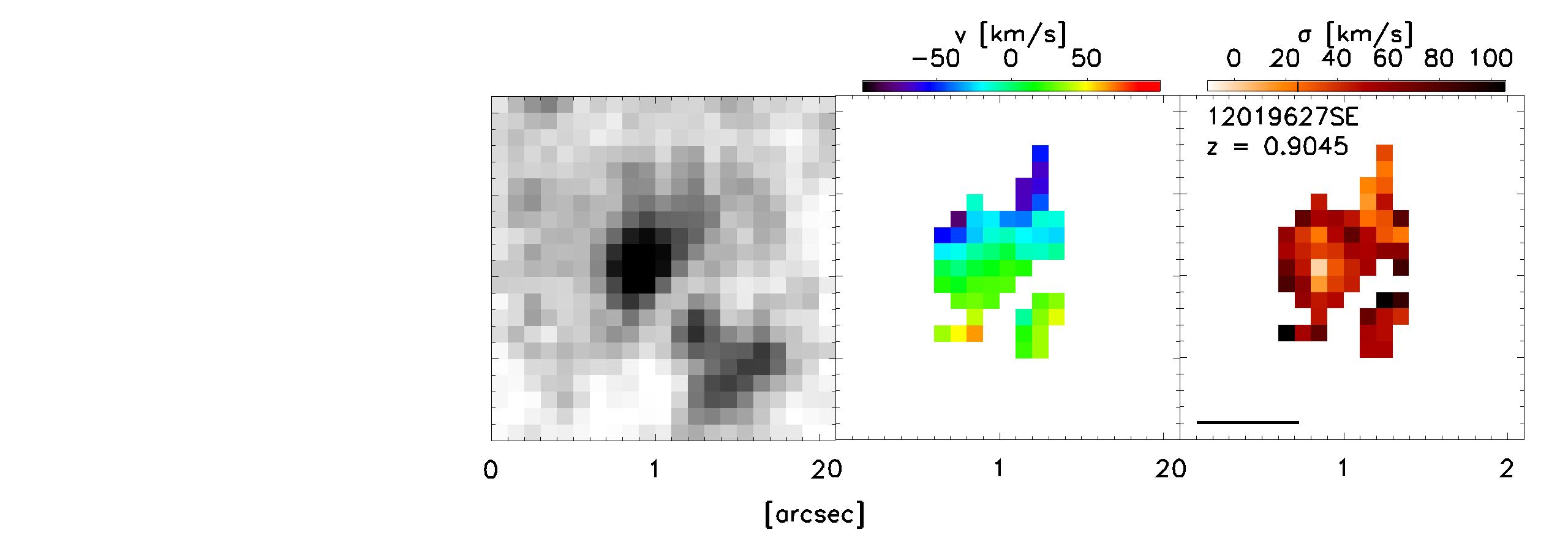}
\includegraphics[width=0.82\textwidth]{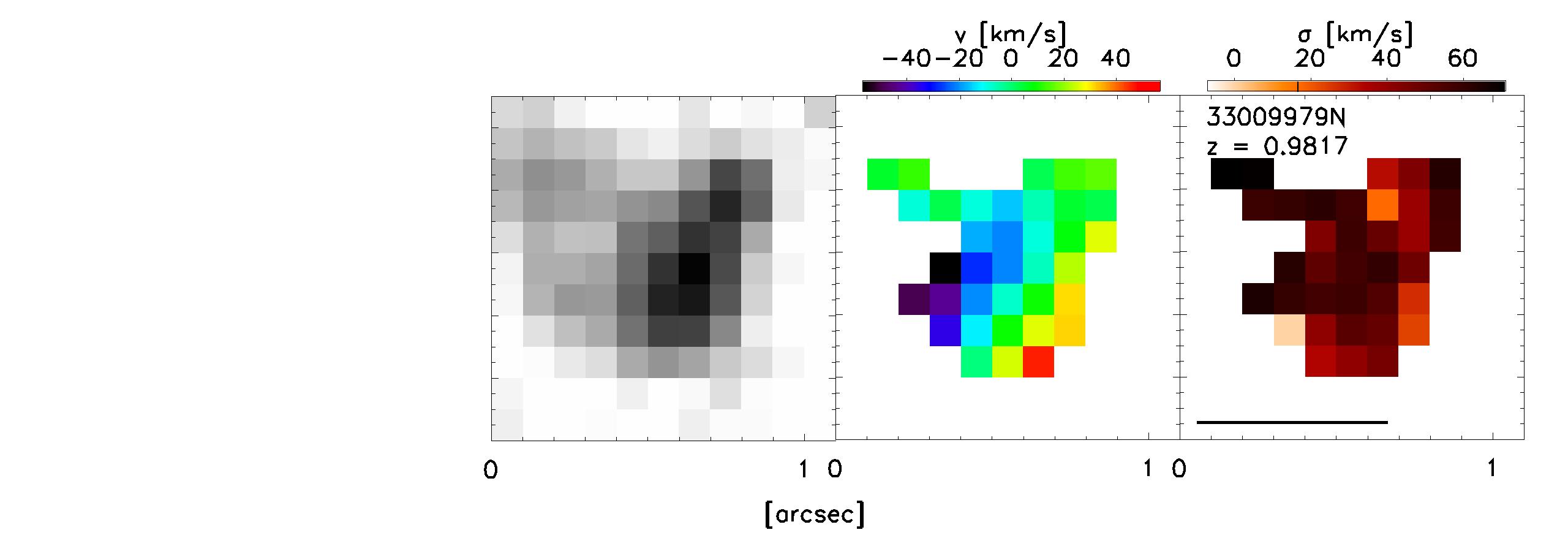}
\includegraphics[width=0.82\textwidth]{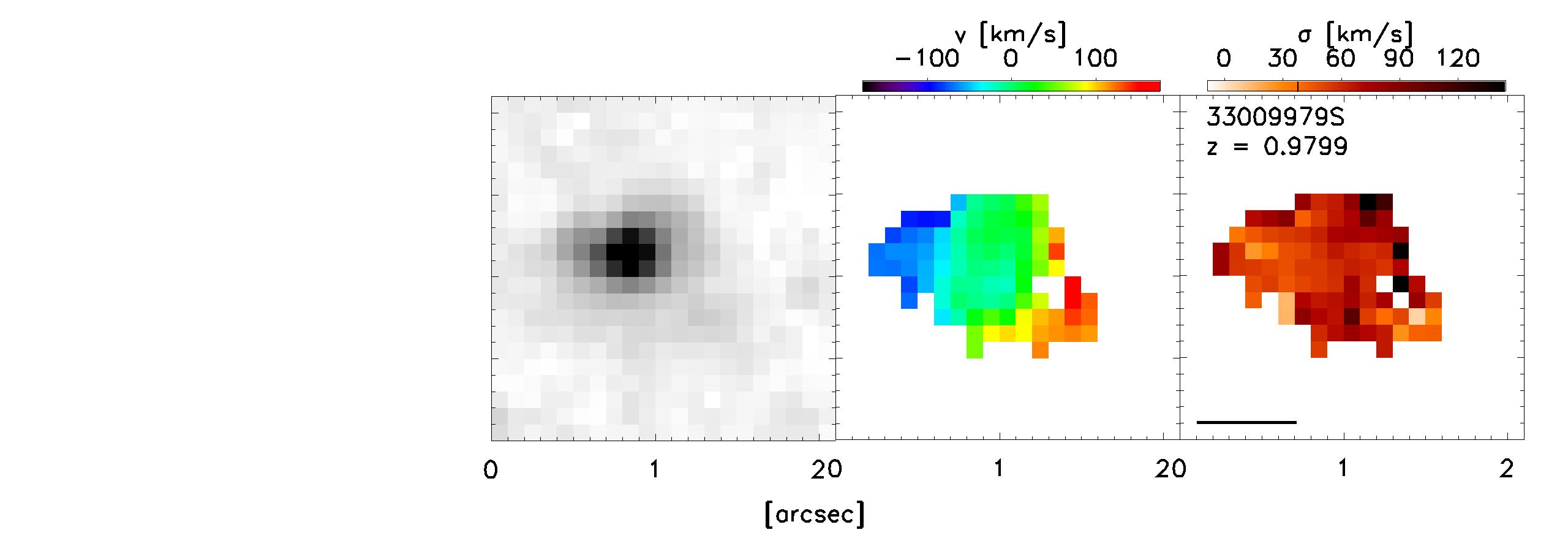}
\caption{}
\end{Contfigure}

\end{document}